\documentclass[11pt]{article}
\usepackage{amssymb,amsmath,amsfonts}
\usepackage{graphicx}
\usepackage{graphics}
\usepackage{eepic,epsfig}

\@addtoreset{equation}{section}

\makeatother

\begin{document}

\title{The generalized Abel-Plana formula with \\
applications to Bessel functions \\
and Casimir effect}
\author{Aram A. Saharian\thanks{%
Email: saharian@ictp.it} \\
\textit{Department of Physics, Yerevan State University} \\
\textit{\ 0025 Yerevan, Armenia,}\\
\textit{and}\\
\textit{The Abdus Salam International Centre for Theoretical Physics} \\
\textit{\ 34014 Trieste, Italy }}
\date{\today}
\maketitle

\begin{abstract}
One of the most efficient methods for the evaluation of the vacuum
expectation values for physical observables in the Casimir effect is based
on using the Abel-Plana summation formula. This enables to derive the
renormalized quantities in a manifestly cutoff independent way and to
present them in the form of strongly convergent integrals. However,
applications of the Abel-Plana formula, in its usual form, are restricted by
simple geometries when the eigenmodes have a simple dependence on quantum
numbers. The author generalized the Abel-Plana formula which essentially
enlarges its application range. Based on this generalization, formulae have
been obtained for various types of series over the zeros of combinations of
Bessel functions and for integrals involving these functions. It has been
shown that these results generalize the special cases existing in
literature. Further, the derived summation formulae have been used to
summarize series arising in the direct mode summation approach to the
Casimir effect for spherically and cylindrically symmetric boundaries, for
boundaries moving with uniform proper acceleration, and in various
braneworld scenarios. This allows to extract from the vacuum expectation
values of local physical observables the parts corresponding to the geometry
without boundaries and to present the boundary-induced parts in terms of
integrals strongly convergent for the points away from the boundaries. As a
result, the renormalization procedure for these observables is reduced to
the corresponding procedure for bulks without boundaries. The present paper
reviews these results. We also aim to collect the results on vacuum
expectation values for local physical observables such as the field square
and the energy-momentum tensor in manifolds with boundaries for various bulk
and boundary geometries.
\end{abstract}

\bigskip

\noindent \textbf{Keywords:} Abel-Plana summation formula; Bessel functions;
Casimir effect

\bigskip

\noindent \textbf{Mathematics Subject Classification 2000:} 81-08, 81T20,
33C10, 40G99

\bigskip

\noindent \textbf{PACS numbers:} 02.30.Gp, 03.70.+k, 04.50.+h,
04.62.+v, 11.10.Kk, 11.27.+d

\bigskip

\tableofcontents

\newpage

\section{Introduction}

In many physical problems we need to consider the model on background of
manifolds with boundaries on which the dynamical variables satisfy some
prescribed boundary conditions. In quantum field theory the imposition of
boundary conditions leads to the modification of the spectrum for the
zero-point fluctuations and results in the shift in the vacuum expectation
values for physical observables. In particular, vacuum forces arise acting
on the constraining boundary. This is the familiar Casimir effect \cite%
{Casi48}. The Casimir effect is among the most interesting consequences of
quantum field theory and is essentially the only macroscopic manifestation
of the nontrivial properties of the physical vacuum. Since the original work
by Casimir many theoretical and experimental works have been done on this
problem, including various bulk and boundary geometries and boundary
conditions (see, for instance, \cite%
{Grib94,Most97,Plun86,Bord99,Bord01,Milt02} and references therein). Many
different approaches have been used: direct mode summation method, Green
function formalism, multiple scattering expansions, heat-kernel series, zeta
function regularization technique, etc. Advanced field-theoretical methods
have been developed for Casimir calculations during the past years \cite%
{Bord95,Grah02,Scha98}. From a general theoretical point of view the main
point here is the unique separation and subsequent removing of the
divergences. Within the framework of the mode summation method in
calculations of the expectation values for physical observables, such as
energy-momentum tensor, one often needs to sum over the values of a certain
function at integer points, and then subtract the corresponding quantity for
unbounded space (usually presented in terms of integrals). Practically, the
sum and integral, taken separately, diverge and some physically motivated
procedure to handle the finite result, is needed (for a discussion of
different methods to evaluate this finite difference see \cite{Bart80}). For
a number of geometries one of the most convenient methods to obtain such
renormalized values of the mode sums is based on the use of the Abel-Plana
summation formula (APF) \cite{Hard91, Erde53, Evgr68,Whit27,Olve74,Henr74}.
The development of this formula dates back to Plana in 1820 \cite{Plan20}
and further has been reconsidered by Abel \cite{Abel23}, Cauchy \cite{Cauc26}
and Kronecker \cite{Kron89}. The history of the APF is discussed in detail
in \cite{Lind05} where interesting applications are given as well. Some
formulae for the gamma and zeta functions are obtained as direct
consequences of the APF (see, for instance, \cite{Whit27,Henr76}). This
summation formula significantly and consistently improves the convergence
and accuracy for the slowly convergent series (see, for instance, \cite%
{Demp90}). The summation method based on the APF is so efficient that was
considered for the development of software for special functions \cite%
{Dahl97}. In \cite{Mama76} the APF has been used for the renormalization of
the scalar field energy-momentum tensor on backgrounds of various Friedmann
cosmological models. Further applications of the APF in physical problems
related to the Casimir effect for flat boundary geometries and topologically
non-trivial spaces with corresponding references can be found in \cite%
{Grib94,Most97}. The application of the APF in these problems allows (i) to
extract in a cutoff independent way the Minkowski vacuum part and (ii) to
obtain for the renormalized part strongly convergent integrals useful, in
particular, for numerical calculations.

The applications of the APF in its usual form is restricted to the problems
where the eigenmodes have simple dependence on quantum numbers and the
normal modes are explicitly known. For the case of curved boundaries and for
mixed boundary conditions the normal modes are given implicitly as zeroes of
the corresponding eigenfunctions or their combinations. The necessity for a
generalization of the Abel-Plana summation procedure arises already in the
case of plane boundaries with Robin or non-local boundary conditions, where
the eigenmodes are given implicitly as solutions of transcendental equation.
In problems with spherically and cylindrically symmetric boundaries the
eigenmodes are the zeroes of the cylinder functions and the expectation
values of physical observables contain series over these zeroes. To include
more general class of problems, in \cite{Sah1} the APF has been generalized
(see also \cite{Sahdis}). The generalized version contains two meromorphic
functions and the APF is obtained by specifying one of them (for other
generalizations of the APF see \cite{Most97,Bart80,Zaya88}). Applying the
generalized formula to Bessel functions, in \cite{Sah1, Sahdis} summation
formulae are obtained for the series over the zeros of various combinations
of these functions (for a review with physical applications see also Ref.
\cite{Saha00Rev,Saha06PoS}). In particular, formulae for the summation of
the Fourier-Bessel and Dini series are derived. It has been shown that
{}from the generalized formula interesting results can be obtained for
infinite integrals involving Bessel functions. The summation formulae
derived from the generalized Abel-Plana formula have been applied for the
evaluation of the vacuum expectation values for local physical observables
in the Casimir effect for plane boundaries with Robin or non-local boundary
conditions \cite{Rome02,Saha06nonloc}, for spherically \cite{Grig1, Grig2,
Grig3, Sah2shert,Saha01} and cylindrically symmetric \cite%
{Sah2,Sah3,Rome01,Saha06cyl} boundaries on the Minkowski bulk for both
scalar and electromagnetic fields. As in the case of the APF, the use of the
generalized formula allows to extract in a manifestly cutoff independent way
the contribution of the unbounded space and to present the boundary-induced
parts in terms of exponentially converging integrals. The case of
cylindrical boundaries in topologically nontrivial background of the cosmic
string is considered in \cite{Beze06cyl,Beze06cylEl}. By making use of the
generalized Abel-Plana formula, the vacuum expectation values of the field
square and the energy-momentum tensor in closely related but more
complicated geometry of a wedge with cylindrical boundary are investigated
in \cite{Reza02,Saha05wedSc,Saha07wedEl} for both scalar and electromagnetic
fields. In \cite{SahaRind1,Avag02,SahaRind2} summation formulae for the
series over the zeroes of the modified Bessel functions with an imaginary
order are derived by using the generalized Abel-Plana formula. This type of
series arise in the evaluation of the vacuum expectation values induced by
plane boundaries uniformly accelerated through the Fulling-Rindler vacuum.
In all considered examples the background spacetime is flat.

For curved backgrounds with boundaries the application of the generalized
Abel-Plana formula extracts from the expectation values the parts which
correspond to the polarization of the vacuum in the situation without
boundaries and the boundary induced parts are presented in terms of
integrals which are convergent for points away from the boundaries. As a
result, the renormalization is necessary for the boundary-free parts only
and this procedure is the same as that in quantum field theory without
boundaries. Examples with both bulk and boundary contributions to the vacuum
polarizations are considered in \cite{Saha03,Saha04Mon,Saha04a,Saha04,Beze06}%
. In these papers the background spacetime is generated by a global monopole
and the one-loop quantum effects for both scalar and fermionic fields are
investigated induced by spherical boundaries concentric with the global
monopole. Another class of problems where the application of the generalized
Abel-Plana formula provides an efficient way for the evaluation of the
vacuum expectation values is considered in \cite{Saha05b,Saha06a}. In these
papers braneworld models with two parallel branes on anti-de Sitter bulk are
discussed. The corresponding mode-sums for physical observables bilinear in
the field contain series over the zeroes of cylinder functions which are
summarized by using the generalized Abel-Plana formula. The geometry of
spherical branes in Rindler-like spacetimes is considered in \cite%
{Saha07RindBr}. In \cite{Saha07Helic} from the generalized Abel-Plana
formula a summation formula is derived over the eigenmodes of a dielectric
cylinder and this formula is applied for the evaluation of the radiation
intensity from a point charge orbiting along a helical trajectory inside the
cylinder.

The present paper reviews these results and is organized as follows. In
Section \ref{sec:GAPF} the generalized Abel-Plana formula is derived and it
is shown that the APF is obtained by a special choice of one of the
meromorphic functions entering in the generalized formula. It is indicated
how to generalize the APF for functions having poles. The applications of
the generalized formula to the Bessel functions are considered in Section %
\ref{sec:ApplBes}. We derive two formulae for the sums over zeros of the
function $AJ_{\nu }(z)+BzJ_{\nu }^{\prime }(z)$, where $J_{\nu }(z)$ is the
Bessel function. Specific examples of applications of the general formulae
are considered. In Section \ref{sec:SumFormBess}, by special choice of the
function $g(z)$ summation formulae are derived for the series over zeros of
the function $J_{\nu }(z)Y_{\nu }(\lambda z)-J_{\nu }(\lambda z)Y_{\nu }(z)$
and similar combinations with Bessel functions derivatives. Special examples
are considered. In Section \ref{sec:SumKi} we consider the application of
the generalized Abel-Plana formula to the series over the zeros of the
modified Bessel functions with an imaginary order. The applications to the
integrals involving Bessel functions and their combinations are discussed in
Section \ref{sec:BessInt}. A number of interesting results for these
integrals are presented. Specific examples of applying these general
formulae are described in Section \ref{sec:BessInt2}. In Section \ref%
{sec:BessInt4}, by using the generalized Abel-Plana formula two theorems are
proved for the integrals involving the function $J_{\nu }(z)Y_{\mu }(\lambda
z)-J_{\mu }(\lambda z)Y_{\nu }(z)$ and their applications are considered.

In the second part of the paper we consider the applications of the
generalized Abel-Plana formula to physical problems, mainly for the
investigation of vacuum expectation values of physical observables on
manifolds with boundaries of different geometries and boundary conditions.
First, in Section \ref{sec:Gen} we describe the general procedure for the
evaluation of the vacuum expectation values of physical observables bilinear
in the field operator. Applications of the APF for the evaluation of the the
vacuum expectation values of the field square and the energy-momentum tensor
for a scalar field in the spacetime with topology $R^{D}\times S^{1}$ and
for the geometry of two parallel Dirichlet and Neumann plates are given in
Section \ref{sec:Topol}. In Section \ref{sec:GlobMonSc}, the summation
formulae over the zeroes of combinations of the cylinder functions are
applied for the investigation of the scalar vacuum densities induced by
spherical boundaries on background of the global monopole spacetime. The
corresponding problem for the fermionic field with bag boundary conditions
is considered in Section \ref{sec:GlobMonFerm}. In the latter case the
eigenmodes are the zeroes of the more complicated combination of the Bessel
function and its derivative. Summation formulae for series over these zeroes
are derived from the generalized Abel-Plana formula. The application to the
electromagnetic Casimir effect for spherical boundaries in background of the
Minkowski spacetime is given in Section \ref{sec:ElSpheric}. Further we
consider problems with cylindrical boundaries. In Section \ref{sec:CosStCyl}
we study the vacuum polarization effects for a scalar field induced by a
cylindrical shell in the cosmic string spacetime assuming Robin boundary
condition on the shell. The corresponding problem for the electromagnetic
field and perfectly conducting cylindrical boundary is considered in Section %
\ref{sec:CylElCos}. Section \ref{sec:TwoCyl} is devoted to the investigation
of the vacuum densities in the region between two coaxial cylindrical
surfaces in the Minkowski spacetime for both scalar and electromagnetic
fields. In Sections \ref{sec:FuRi} and \ref{sec:FulRin2pl} we consider the
Wightman function, the vacuum expectation values of the field square and the
energy-momentum tensor for scalar and electromagnetic fields induced by
plane boundaries uniformly accelerated through the Fulling-Rindler vacuum.
The application of the generalized Abel-Plana formula for the evaluation of
the vacuum densities in braneworld scenarios is described in Section \ref%
{sec:Branes} for parallel branes on background of anti-de Sitter spacetime
and in Section \ref{sec:Rindbrane} for spherical branes in Rindler-like
spacetimes. In Section \ref{sec:RadDiel}, a summation formula for the series
over the eigenmodes of a dielectric cylinder is derived and this formula is
applied for the investigation of the radiation intensity from a charged
particle moving along a helical orbit inside the cylinder. Section \ref%
{sec:Conc} concludes the main results of the paper.

\section{Generalized Abel-Plana formula}

\label{sec:GAPF}

Let $f(z)$ and $g(z)$ be meromorphic functions for $a\leqslant x\leqslant b$
in the complex plane $z=x+iy$. Let us denote by $z_{f,k}$ and $z_{g,k}$ the
poles of $f(z)$ and $g(z)$ in the strip $a<x<b$. We assume that ${\mathrm{%
Im\,}}z_{f,k}\neq 0$ (see, however, the Remark to Lemma).

\noindent \textbf{Lemma.} \textit{If functions $f(z)$ and $g(z)$ satisfy
condition
\begin{equation}
\lim_{h\rightarrow \infty }\int_{a\pm ih}^{b\pm ih}{\left[ g(z)\pm f(z)%
\right] dz}=0,  \label{cor11}
\end{equation}%
then the following formula takes place
\begin{equation}
\int_{a}^{b}{f(x)dx}=R[f(z),g(z)]-\frac{1}{2}\int_{-i\infty }^{+i\infty }{%
\left[ g(u)+{\sigma (z)}f(u)\right] _{u=a+z}^{u=b+z}dz,\;\sigma (z)\equiv
\mathrm{sgn}}({{\mathrm{Im\,}}}z),  \label{cor12}
\end{equation}%
where
\begin{equation}
R[f(z),g(z)]=\pi i\left[ \sum_{k}\underset{z=z_{g,k}}{\mathrm{Res}}%
g(z)+\sum_{k}\sigma (z_{f,k})\underset{z={\mathrm{\,}}z_{f,k}}{\mathrm{Res}}%
f(z)\right] .  \label{cor13}
\end{equation}%
}

\bigskip

\noindent \textbf{Proof.} Consider a rectangle $C_{h}$ with vertices $a\pm
ih $, $b\pm ih$, described in the positive sense. In accordance to the
residue theorem
\begin{equation}
\int_{C_{h}}dz\,g(z)=2\pi i\sum_{k}\underset{z=z_{g,k}}{\mathrm{Res}}g(z),
\label{corp1}
\end{equation}%
where the rhs contains the sum over poles within $C_{h}$. Let $C_{h}^{+}$
and $C_{h}^{-}$ denote the upper and lower halves of this contour. Then one
has
\begin{equation}
\int_{C_{h}}dz\,g(z)=\sum_{\alpha =+,-}\left\{ \int_{C_{h}^{\alpha
}}dz\,[g(z)+\alpha f(z)]-\alpha \int_{C_{h}^{\alpha }}dz\,f(z)\right\} .
\label{corp2}
\end{equation}%
By the same residue theorem
\begin{equation}
\int_{C_{h}^{-}}dz\,f(z)-\int_{C_{h}^{+}}dz\,f(z)=2\int_{a}^{b}f(x)dx-2\pi
i\sum_{k}\sigma (z_{f,k})\underset{z=z_{f,k}}{\mathrm{Res}}f(z).
\label{corp3}
\end{equation}%
Then
\begin{equation}
\int_{C_{h}^{\pm }}dz\,[g(z)\pm f(z)]=\pm \int_{0}^{\pm ih}dz\,[g(u)\pm
f(u)]_{u=a+z}^{u=b+z}\mp \int_{a\pm ih}^{b\pm ih}dz\,[g(z)\pm f(z)].
\label{corp4}
\end{equation}%
Combining these results and allowing $h\rightarrow \infty $ in (\ref{corp1})
one obtains formula (\ref{cor12}). \rule{1.5ex}{1.5ex}

\bigskip

If the functions $f(z)$ and $g(z)$ have poles with ${\mathrm{Re\,}}%
z_{j,k}=a,b$ ($j=f,g$) the contour has to pass round these points on the
right or left, correspondingly.

\bigskip

\noindent \textbf{Remark.} Formula (\ref{cor12}) is also valid when the
function $f(z)$ has real poles $z_{f,n}^{(0)}$, ${\mathrm{Im\,}}%
z_{f,n}^{(0)}=0$, in the region $a<{\mathrm{Re\,}}z<b$ if the main part of
its Laurent expansion near these poles does not contain even powers of $%
z-z_{f,n}^{(0)}$. In this case on the left of formula (\ref{cor12}) the
integral is meant in the sense of the principal value, which exists as a
consequence of the above mentioned condition. For brevity let us consider
the case of a single pole $z=z_{0}$. One has
\begin{eqnarray}
&&\int_{C_{h}^{-}}dz\,f(z)-\int_{C_{h}^{+}}dz\,f(z)=2\int_{a}^{z_{0}-\rho
}dz\,f(z)+2\int_{z_{0}+\rho }^{b}dz\,f(z)  \notag \\
&&-2\pi i\sum_{k}\sigma (z_{f,k})\underset{z=z_{f,k}}{\mathrm{Res}}%
f(z)+\int_{\Gamma _{\rho }^{+}}dz\,f(z)+\int_{\Gamma _{\rho }^{-}}dz\,f(z),
\label{corp5}
\end{eqnarray}%
with contours $\Gamma _{\rho }^{+}$ and $\Gamma _{\rho }^{-}$ being the
upper and lower circular arcs (with center at $z=z_{0}$) joining the points $%
z_{0}-\rho $ and $z_{0}+\rho $. By taking into account that for odd negative
$l$
\begin{equation}
\int_{\Gamma _{\rho }^{+}}dz\,(z-z_{0})^{l}+\int_{\Gamma _{\rho
}^{-}}dz\,(z-z_{0})^{l}=0,  \label{corp6}
\end{equation}%
in the limit $\rho \rightarrow 0$ we obtain the required result. \rule%
{1.5ex}{1.5ex}

\bigskip

In the following, on the left of (\ref{cor12}) we will write ${\mathrm{p.v.}}%
\! \int_{a}^{b}dx\,f(x)$, assuming that this integral converges in the sense
of the principal value. As a direct consequence of Lemma one obtains \cite%
{Sah1}:

\bigskip

\noindent \textbf{Theorem 1.} \textit{\ If in addition to the conditions of
Lemma one has
\begin{equation}
\lim_{b\rightarrow \infty }\int_{b}^{b\pm i\infty }{dz\,\left[ g(z)\pm f(z)%
\right] }=0,  \label{th11}
\end{equation}%
then
\begin{equation}
\lim_{b\rightarrow \infty }\left\{ {\mathrm{p.v.}}\! \int_{a}^{b}{dx\,f(x)}%
-R[f(z),g(z)]\right\} =\frac{1}{2}\int_{a-i\infty }^{a+i\infty }{dz\,\left[
g(z)+{\sigma (z)}f(z)\right] },  \label{th12}
\end{equation}%
where on the left $R[f(z),g(z)]$ is defined as (\ref{cor13}), $a<{\mathrm{%
Re\,}}z_{f,k},{\mathrm{Re\,}}z_{g,k}<b$, and the summation goes over poles $%
z_{f,k}$ and $z_{g,k}$ arranged in order ${\mathrm{Re\,}}z_{j,k}\leqslant {%
\mathrm{Re\,}}z_{j,k+1}$, $j=f,g$. }

\bigskip

\noindent \textbf{Proof.} It is sufficient to insert in general formula (\ref%
{cor12}) $b\rightarrow \infty $ and to use the condition (\ref{th11}). The
order of summation in $R[f(z),g(z)]$ is determined by the choice of the
integration contour $C_{h}$ and by limiting transition $b\rightarrow \infty $%
. \rule{1.5ex}{1.5ex}

\bigskip

We will refer formula (\ref{th12}) as generalized Abel-Plana formula (GAPF)
as for $b=n+a$, $0<a<1$,%
\begin{equation}
g(z)=-if(z)\cot \pi z,  \label{gforAPF}
\end{equation}%
with an analytic function $f(z)$, from (\ref{th12}) the Abel-Plana formula
(APF) \cite{Hard91, Erde53, Evgr68} \ is obtained in the form
\begin{eqnarray}
\lim_{n\rightarrow \infty }\left[ \sum_{k=1}^{n}f(k)-\int_{a}^{n+a}dx\,f(x)%
\right] &=&\frac{1}{2i}\int_{a}^{a-i\infty }dz\,f(z)(\cot \pi z-i)  \notag \\
&&-\frac{1}{2i}\int_{a}^{a+i\infty }dz\,f(z)(\cot \pi z+i).  \label{apsf1}
\end{eqnarray}%
Another form of the APF, given in \cite{Whit27,Olve74}, is obtained directly
from formula (\ref{cor12}) taking the function $g(z)$ in the form (\ref%
{gforAPF}) with an analytic function $f(z)$ and $a=m$, $b=n$ with integers $%
m $ and $n$. In this case the points $z=m$ and $z=n$ are poles of the
function $g(z)$ and in the integral on the right of formula (\ref{cor12})
they should be avoided by small semicircles from the right and from the left
respectively. The integrals over these semicircles give the contribution $%
-f(m)-f(n)$ and one obtains the formula%
\begin{eqnarray}
\sum_{k=m}^{n}f(k) &=&\int_{m}^{n}dx\,f(x)+\frac{1}{2}f(m)+\frac{1}{2}f(n)
\notag \\
&&+i\int_{0}^{\infty }dz\,\frac{f(m+iz)-f(n+iz)-f(m-iz)+f(n-iz)}{e^{2\pi z}-1%
}.  \label{apsf1n}
\end{eqnarray}%
Expanding the functions in the numerator of the last integral over $z$ and
interchanging the order of the summation and integration, the integral is
expressed in terms of the Bernoulli numbers $B_{2j}$. As a result, from (\ref%
{apsf1n}) we obtain the formula%
\begin{equation}
\sum_{k=m}^{n}f(k)=\int_{m}^{n}dx\,f(x)+\frac{1}{2}f(m)+\frac{1}{2}%
f(n)-\sum_{j=1}^{\infty }\frac{B_{2j}}{(2j)!}\left[
f^{(2j-1)}(m)-f^{(2j-1)}(n)\right] ,  \label{EulMac}
\end{equation}%
which is a special case of the Euler-Maclaurin formula \cite{Olve74} (for a
relationship between the APF and Euler-Maclaurin summation formula see also
\cite{Dowl89}).

A useful form of (\ref{apsf1}) may be obtained performing the limit $%
a\rightarrow 0$. By taking into account that the point $z=0$ is a pole for
the integrands and therefore has to be circled by arcs of the small circle $%
C_{\rho }$ on the right and performing $\rho \rightarrow 0$, one obtains
\begin{equation}
\sum_{k=0}^{\infty }f(k)=\int_{0}^{\infty }dx\,f(x)+\frac{1}{2}%
f(0)+i\int_{0}^{\infty }dx\,\frac{f(ix)-f(-ix)}{e^{2\pi x}-1}.  \label{apsf2}
\end{equation}%
Note that, now condition (\ref{cor11}) is satisfied if
\begin{equation}
\lim_{y\rightarrow \infty }e^{-2\pi |y|}|f(x+iy)|=0,  \label{cond1apsf}
\end{equation}%
uniformly in any finite interval of $x$. Formula (\ref{apsf2}) is the most
frequently used form of the APF in physical applications. Another useful
form (in particular for fermionic field calculations) to sum over the values
of an analytic function at half of an odd integer points is obtained from (%
\ref{th12}) taking%
\begin{equation}
g(z)=if(z)\tan \pi z,  \label{gforAPFferm}
\end{equation}%
with an analytic function $f(z)$. This leads to the summation formula (see
also \cite{Grib94,Most97})
\begin{equation}
\sum_{k=0}^{\infty }f(k+1/2)=\int_{0}^{\infty }dx\,f(x)-i\int_{0}^{\infty
}dx\,\frac{f(ix)-f(-ix)}{e^{2\pi x}+1}.  \label{apsf2half}
\end{equation}%
By adding to the rhs of (\ref{apsf2}) the term
\begin{equation}
\pi \sum_{k}\underset{z={\mathrm{\,}}z_{f,k}}{\mathrm{Res}}\frac{e^{i\pi
z\sigma (z_{f,k})}}{\sin \pi z}f(z),  \label{polecase}
\end{equation}%
and to the rhs of (\ref{apsf2half}) the term
\begin{equation}
-i\pi \sum_{k}\sigma (z_{f,k})\underset{z={\mathrm{\,}}z_{f,k}}{\mathrm{Res}}%
\frac{e^{i\pi z\sigma (z_{f,k})}}{\cos \pi z}f(z),  \label{polecaseferm}
\end{equation}%
these formulae can be generalized to functions $f(z)$ that have non-real
poles $z_{f,k}$, ${\mathrm{Re\,}}z_{f,k}>0$. The second generalization with
an application to the problem of diffraction scattering of charged particles
is given in \cite{Ipon70}. Another generalization of the APF, given in \cite%
{Inui03}, is obtained from (\ref{th12}) taking in this formula $%
g(z)=-if(z)\cot \pi (z-\beta )$, $0<\beta <1$, with an analytic function $%
f(z)$.

As a next consequence of (\ref{th12}), a summation formula can be obtained
over the points $z_{k},\,{\mathrm{Re\,}}z_{k}>0$ at which the analytic
function $s(z)$ takes integer values, $s(z_{k})$ is an integer, and $%
s^{\prime }(z_{k})\neq 0$. Taking in (\ref{th12}) $g(z)=-if(z)\cot \pi s(z)$%
, one obtains the following formula \cite{Sah}
\begin{equation}
\sum_{k}\frac{f(z_{k})}{s^{\prime }(z_{k})}=w+\int_{0}^{\infty
}f(x)dx+\int_{0}^{\infty }dx\,\left[ \frac{f(ix)}{e^{-2\pi is(ix)}-1}-\frac{%
f(-ix)}{e^{2\pi is(-ix)}-1}\right] ,  \label{apsf3}
\end{equation}%
where
\begin{equation}
w=\left\{
\begin{array}{ll}
0, & {\text{if}}\quad s(0)\neq 0,\pm 1,\pm 2,\ldots \\
f(0)/[2s^{\prime }(0)], & {\text{if}}\quad s(0)=0,\pm 1,\pm 2,\ldots%
\end{array}%
\right. .  \label{wintval}
\end{equation}%
For $s(z)=z$ we return to the APF in the usual form. An example for the
application of this formula to the Casimir effect is given in \cite{Sah}.

\section{Applications to Bessel functions}

\label{sec:ApplBes}

Formula (\ref{th12}) contains two meromorphic functions and is too general.
In order to obtain more special consequences we have to specify one of them.
As we have seen in the previous section, one of possible ways leads to the
APF. Here we will consider another choices of the function $g(z)$ and will
obtain useful formulae for the series over zeros of Bessel functions and
their combinations, as well as formulae for integrals involving these
functions.

First of all, to simplify the formulae let us introduce the notation
\begin{equation}
\bar{F}(z)\equiv AF(z)+BzF^{\prime }(z)  \label{efnot1}
\end{equation}%
for a given function $F(z)$, where the prime denotes the derivative with
respect to the argument of the function, $A$ and $B$ are real constants. As
a function $g(z)$ in the GAPF let us choose
\begin{equation}
g(z)=i\frac{\bar{Y}_{\nu }(z)}{\bar{J}_{\nu }(z)}f(z),  \label{gebessel}
\end{equation}%
where $J_{\nu }(z)$ and $Y_{\nu }(z)$ are Bessel functions of the first and
second (Neumann function) kind. For the sum and difference on the right of (%
\ref{th12}) one obtains
\begin{equation}
f(z)-(-1)^{k}g(z)=\frac{\bar{H}_{\nu }^{(k)}(z)}{\bar{J}_{\nu }(z)}%
f(z),\quad k=1,2,  \label{gefsum}
\end{equation}%
with $H_{\nu }^{(1)}(z)$ and $H_{\nu }^{(2)}(z)$ being Bessel functions of
the third kind or Hankel functions. For such a choice the integrals (\ref%
{cor11}) and (\ref{th11}) can be estimated by using the asymptotic formulae
for Bessel functions for fixed $\nu $ and $|z|\rightarrow \infty $ (see, for
example, \cite{Watson, abramowiz}). It is easy to see that conditions (\ref%
{cor11}) and (\ref{th11}) are satisfied if the function $f(z)$ is restricted
to one of the following constraints
\begin{equation}
|f(z)|<\varepsilon (x)e^{c|y|}\quad \text{ or }\quad |f(z)|<\frac{Me^{2|y|}}{%
|z|^{\alpha }},\quad z=x+iy,\quad |z|\rightarrow \infty ,  \label{condf}
\end{equation}%
where $c<2$, $\alpha >1$ and $\varepsilon (x)\rightarrow 0$ for $%
x\rightarrow \infty $. Indeed, from asymptotic expressions for Bessel
functions it follows that
\begin{eqnarray}
\left\vert \int_{a\pm ih}^{b\pm ih}dz\,\left[ g(z)\pm f(z)\right]
\right\vert &=&\left\vert \int_{a}^{b}dx\,\frac{\bar{H}_{\nu }^{(j)}(x\pm ih)%
}{\bar{J}_{\nu }(x\pm ih)}f(x\pm ih)\right\vert <\left\{
\begin{array}{l}
M_{1}e^{(c-2)h} \\
M_{1}^{\prime }/h^{\alpha }%
\end{array}%
\right. \\
\left\vert \int_{b}^{b\pm i\infty }dz\,\left[ g(z)\pm f(z)\right]
\right\vert &=&\left\vert \int_{0}^{\infty }dx\,\frac{\bar{H}_{\nu
}^{(j)}(b\pm ix)}{\bar{J}_{\nu }(b\pm ix)}f(b\pm ix)\right\vert <\left\{
\begin{array}{l}
N_{1}\varepsilon (b) \\
N_{1}^{\prime }/b^{\alpha -1}%
\end{array}%
\right.  \label{condgf}
\end{eqnarray}%
with constants $M_{1},\,M_{1}^{\prime },\,N_{1},\,N_{1}^{\prime }$, and $j=1$
($j=2$) corresponds to the upper (lower) sign.

Let us denote by $\lambda _{\nu ,k}\neq 0$, $k=1,2,3\ldots $, the zeros of $%
\bar{J}_{\nu }(z)$ in the right half-plane, arranged in ascending order of
the real part, ${\mathrm{Re\,}}\lambda _{\nu ,k}\leqslant {\mathrm{Re\,}}%
\lambda _{\nu ,k+1}$, (if some of these zeros lie on the imaginary axis we
will take only zeros with positive imaginary part). All these zeros are
simple. Note that for real $\nu >-1$ the function $\bar{J}_{\nu }(z)$ has
only real zeros, except for the case $A/B+\nu <0$ when there are two purely
imaginary zeros \cite{Watson}. By using the Wronskian $W[J_{\nu }(z),Y_{\nu
}(z)]=2/\pi z$, for (\ref{cor13}) one finds
\begin{equation}
R[f(z),g(z)]=2\sum_{k}T_{\nu }(\lambda _{\nu ,k})f(\lambda _{\nu
,k})+r_{1\nu }[f(z)],  \label{rbessel}
\end{equation}%
where we have introduced the notations
\begin{eqnarray}
T_{\nu }(z) &=&\frac{z}{\left( z^{2}-\nu ^{2}\right) J_{\nu
}^{2}(z)+z^{2}J^{\prime }{}_{\nu }^{2}(z)}\,,  \label{teka} \\
r_{1\nu }[f(z)] &=&\pi i\sum_{k}\underset{{\mathrm{Im}}z_{k}>0}{\mathrm{Res}}%
f(z)\frac{\bar{H}_{\nu }^{(1)}(z)}{\bar{J}_{\nu }(z)}-\pi i\sum_{k}\underset{%
{\mathrm{Im}}z_{k}<0}{\mathrm{Res}}f(z)\frac{\bar{H}_{\nu }^{(2)}(z)}{\bar{J}%
_{\nu }(z)}-  \notag \\
&&-\pi \sum_{k}\underset{{\mathrm{Im}}z_{k}=0}{\mathrm{Res}}f(z)\frac{\bar{Y}%
_{\nu }(z)}{\bar{J}_{\nu }(z)}\,.  \label{r1}
\end{eqnarray}%
Here $z_{k}$ ($\neq \lambda _{\nu ,i}$) are the poles of the function $f(z)$
in the region $\mathrm{Re\,}z>a>0$. Substituting (\ref{rbessel}) into (\ref%
{th12}), we obtain that for the function $f(z)$ meromorphic in the
half-plane ${\mathrm{Re\,}}z\geqslant a$ and satisfying condition (\ref%
{condf}), the following formula takes place
\begin{eqnarray}
&&\lim_{b\rightarrow +\infty }\left\{ 2\sum_{k=m}^{n}T_{\nu }(\lambda _{\nu
,k})f(\lambda _{\nu ,k})+r_{1\nu }[f(z)]-{\mathrm{p.v.}}\!
\int_{a}^{b}dx\,f(x)\right\} =  \notag \\
&&\quad =-\frac{1}{2}\int_{a}^{a+i\infty }dz\,f(z)\frac{\bar{H}_{\nu
}^{(1)}(z)}{\bar{J}_{\nu }(z)}-\frac{1}{2}\int_{a}^{a-i\infty }dz\,f(z)\frac{%
\bar{H}_{\nu }^{(2)}(z)}{\bar{J}_{\nu }(z)},  \label{formgen}
\end{eqnarray}%
where ${\mathrm{Re\,}}\lambda _{\nu ,m-1}<a<{\mathrm{Re\,}}\lambda _{\nu ,m}$%
, ${\mathrm{Re\,}}\lambda _{\nu ,n}<b<{\mathrm{Re\,}}\lambda _{\nu ,n+1}$, $%
a<{\mathrm{Re\,}}z_{k}<b$. We will apply this formula to the function $f(z)$
meromorphic in the half-plane ${\mathrm{Re\,}}z\geqslant 0$ taking $%
a\rightarrow 0$. Let us consider two cases separately.

\subsection{Case (a)}

Let $f(z)$ has no poles on the imaginary axis, except possibly at $z=0$, and
\begin{equation}
f(ze^{\pi i})=-e^{2\nu \pi i}f(z)+o(z^{\beta _{\nu }}),\quad z\rightarrow 0
\label{case21}
\end{equation}%
(this condition is trivially satisfied for the function $f(z)=o(z^{\beta
_{\nu }})$), with
\begin{equation}
\beta _{\nu }=\left\{
\begin{array}{ll}
2|{\mathrm{Re\,}}\nu |-1, & \text{ for integer $\nu $,} \\
{\mathrm{Re\,}}\nu +|{\mathrm{Re\,}}\nu |-1, & \text{ for non-integer $\nu $.%
}%
\end{array}%
\right.  \label{betanju}
\end{equation}%
Under this condition, for values $\nu $, for which $\bar{J}_{\nu }(z)$ has
no purely imaginary zeros, the rhs of Eq. (\ref{formgen}) in the limit $%
a\rightarrow 0$ can be presented in the form
\begin{eqnarray}
&&-\frac{1}{\pi }\int_{\rho }^{\infty }{\ }dx\,\frac{\bar{K}_{\nu }(x)}{\bar{%
I}_{\nu }(x)}\left[ e^{-\nu \pi i}f(xe^{\pi i/2})+e^{\nu \pi i}f(xe^{-\pi
i/2})\right]  \notag \\
&&+\int_{\gamma _{\rho }^{+}}dz\,f(z)\frac{\bar{H}_{\nu }^{(1)}(z)}{\bar{J}%
_{\nu }(z)}-\int_{\gamma _{\rho }^{-}}dz\,f(z)\frac{\bar{H}_{\nu }^{(2)}(z)}{%
\bar{J}_{\nu }(z)},  \label{rel1}
\end{eqnarray}%
with $\gamma _{\rho }^{+}$ and $\gamma _{\rho }^{-}$ being upper and lower
halves of the semicircle in the right half-plane with radius $\rho $ and
with center at point $z=0$, described in the positive sense with respect to
this point. In (\ref{rel1}) we have introduced modified Bessel functions $%
I_{\nu }(z)$ and $K_{\nu }(z)$ \cite{abramowiz}. It follows from (\ref%
{case21}) that for $z\rightarrow 0$
\begin{equation}
\frac{\bar{H}_{\nu }^{(1)}(z)}{\bar{J}_{\nu }(z)}f(z)=\frac{\bar{H}_{\nu
}^{(2)}(ze^{-\pi i})}{\bar{J}_{\nu }(ze^{-\pi i})}f(ze^{-\pi i})+o(z^{-1}).
\label{rel2}
\end{equation}%
{}From here for $\rho \rightarrow 0$ one finds
\begin{equation}
D_{\nu }\equiv \int_{\gamma _{\rho }^{+}}dz\,f(z)\frac{\bar{H}_{\nu
}^{(1)}(z)}{\bar{J}_{\nu }(z)}-\int_{\gamma _{\rho }^{-}}dz\,f(z)\frac{\bar{H%
}_{\nu }^{(2)}(z)}{\bar{J}_{\nu }(z)}=-\pi \underset{z=0}{\mathrm{Res}}f(z)%
\frac{\bar{Y}_{\nu }(z)}{\bar{J}_{\nu }(z)}.  \label{rel3}
\end{equation}%
Indeed,
\begin{eqnarray}
D_{\nu } &=&\int_{\gamma _{\rho }^{+}}dz\,f(z)\frac{\bar{H}_{\nu }^{(1)}(z)}{%
\bar{J}_{\nu }(z)}+\int_{\gamma _{1\rho }^{+}}dz\,f(ze^{-\pi i})\frac{\bar{H}%
_{\nu }^{(2)}(ze^{-\pi i})}{\bar{J}_{\nu }(ze^{-\pi i})}  \notag \\
&=&\int_{\gamma _{\rho }^{+}+\gamma _{1\rho }^{+}}dz\,f(z)\frac{\bar{H}_{\nu
}^{(1)}(z)}{\bar{J}_{\nu }(z)}{}+\int_{\gamma _{1\rho }^{+}}dz\,o(z^{-1})
\notag \\
&=&i\int_{\gamma _{\rho }^{+}+\gamma _{1\rho }^{+}}dz\,f(z)\frac{\bar{Y}%
_{\nu }(z)}{\bar{J}_{\nu }(z)}+\int_{\gamma _{1\rho }^{+}}dz\,o(z^{-1}),
\label{rel4}
\end{eqnarray}%
where $\gamma _{1\rho }^{+}$ ($\gamma _{1\rho }^{-}$, see below) is the
upper (lower) half of the semicircle with radius $\rho $ in the left
half-plane with the center at $z=0$ (described in the positive sense). In
the last equality we have used the condition that integral ${\mathrm{p.v.}}%
\! \int_{0}^{b}dx\,f(x)$ converges at lower limit. In a similar way it can
be seen that
\begin{equation}
D_{\nu }=i\int_{\gamma _{\rho }^{-}+\gamma _{1\rho }^{-}}dz\,f(z)\frac{\bar{Y%
}_{\nu }(z)}{\bar{J}_{\nu }(z)}+\int_{\gamma _{1\rho }^{-}}dz\,o(z^{-1}).
\label{rel5}
\end{equation}%
Combining the last two results we obtain (\ref{rel3}) in the limit $\rho
\rightarrow 0$. By using (\ref{formgen}), (\ref{rel1}) and (\ref{rel3}) we
have \cite{Sah1}:

\bigskip

\noindent \textbf{Theorem 2.} \textit{If f(z) is a single valued analytic
function in the half-plane ${\mathrm{Re\,}}z\geqslant 0$ (with possible
branch point at $z=0$) except the poles $z_{k}$ ($\neq \lambda _{\nu ,i}$), $%
{\mathrm{Re\,}}z_{k}>0$ (for the case of function $f(z)$ having purely
imaginary poles see Remark after Theorem 3), and satisfy conditions (\ref%
{condf}) and (\ref{case21}), then in the case of $\nu $ for which the
function $\bar{J}_{\nu }(z)$ has no purely imaginary zeros, the following
formula is valid
\begin{eqnarray}
&&\lim_{b\rightarrow +\infty }\left\{ 2\sum_{k=1}^{n}T_{\nu }(\lambda _{\nu
,k})f(\lambda _{\nu ,k})+r_{1\nu }[f(z)]-{\mathrm{p.v.}}\! \int_{0}^{b}{%
f(x)dx}\right\} ={}  \notag \\
&=&\frac{\pi }{2}\underset{z=0}{\mathrm{Res}}f(z)\frac{\bar{Y}_{\nu }(z)}{%
\bar{J}_{\nu }(z)}-\frac{1}{\pi }\int_{0}^{\infty }dx\frac{\bar{K}_{\nu }(x)%
}{\bar{I}_{\nu }(x)}\left[ e^{-\nu \pi i}f(xe^{\frac{\pi i}{2}})+e^{\nu \pi
i}f(xe^{-\frac{\pi i}{2}})\right] ,  \label{sumJ1}
\end{eqnarray}%
where on the left ${\mathrm{Re\,}}\lambda _{\nu ,n}<b<{\mathrm{Re\,}}\lambda
_{\nu ,n+1}$, $0<{\mathrm{Re\,}}z_{k}<b$, and $\,T_{\nu }(\lambda _{\nu ,k})$
and $\,r_{1\nu }[f(z)]$ are determined by relations (\ref{teka}) and (\ref%
{r1}).}

\bigskip

\noindent Under the condition (\ref{case21}) the integral on the right
converges at lower limit. Recall that we assume the existence of the
integral on the left as well (see section 2). Formula (\ref{sumJ1}) and
analogous ones given below are especially useful for numerical calculations
of the sums over $\lambda _{\nu ,k}$, as under the first conditions in (\ref%
{condf}) the integral on the right converges exponentially fast at the upper
limit. Below, in Section \ref{sec:GlobMonFerm} we will see that summation
formula (\ref{sumJ1}) may be generalized for the case when the coefficients
in (\ref{efnot1}) are functions of $z$.

\bigskip

\noindent \textbf{Remark.} Deriving formula (\ref{sumJ1}) we have assumed
that the function $f(z)$ is meromorphic in the half-plane ${\mathrm{Re\,}}%
z\geqslant 0$ (except possibly at $z=0$). However this formula is valid also
for some functions having branch points on the imaginary axis, for example,
\begin{equation}
f(z)=f_{1}(z)\prod_{l=1}^{k}\left( z^{2}+c_{l}^{2}\right) ^{\pm 1/2},
\label{fbranch1}
\end{equation}%
with meromorphic function $f_{1}(z)$. The proof for (\ref{sumJ1}) in this
case is similar to the one given above with the difference that branch
points $\pm ic_{l}$ have to be circled on the right along contours with
small radii. In view of further applications to the Casimir effect let us
consider the case $k=1$. By taking into account that
\begin{equation}
\left( x^{2}e^{\pm \pi i}+c^{2}\right) ^{1/2}=\left\{
\begin{array}{lll}
\sqrt{c^{2}-x^{2}}, & \text{if} & 0\leqslant x\leqslant c, \\
e^{\pm i\pi /2}\sqrt{x^{2}-c^{2}}, & \text{if} & x\geqslant c,%
\end{array}%
\right.  \label{rel6}
\end{equation}%
from (\ref{sumJ1}) one obtains
\begin{eqnarray}
&&\lim_{b\rightarrow +\infty }\left\{ 2\sum_{k=1}^{n}T_{\nu }(\lambda _{\nu
,k})f(\lambda _{\nu ,k})+r_{1\nu }[f(z)]-{\mathrm{p.v.}}\!
\int_{0}^{b}dx\,f(x)\right\} =\frac{\pi }{2}\underset{z=0}{\mathrm{Res}}f(z)%
\frac{\bar{Y}_{\nu }(z)}{\bar{J}_{\nu }(z)}{}  \notag \\
&&\quad -\frac{1}{\pi }\int_{0}^{c}{\ }dx{}\,\frac{\bar{K}_{\nu }(x)}{\bar{I}%
_{\nu }(x)}\left[ e^{-\nu \pi i}f_{1}(xe^{\pi i/2})+e^{\nu \pi
i}f_{1}(xe^{-\pi i/2})\right] \left( c^{2}-x^{2}\right) ^{\pm 1/2}  \notag \\
&&\quad \mp \frac{i}{\pi }\int_{c}^{\infty }{\ }dx\,\frac{\bar{K}_{\nu }(x)}{%
\bar{I}_{\nu }(x)}\left[ e^{-\nu \pi i}f_{1}(xe^{\pi i/2})-e^{\nu \pi
i}f_{1}(xe^{-\pi i/2})\right] \left( x^{2}-c^{2}\right) ^{\pm 1/2},
\label{sumJbranch}
\end{eqnarray}%
where $f(z)=f_{1}(z)\left( z^{2}+c^{2}\right) ^{\pm 1/2},\,c>0$. In Section %
\ref{sec:GlobMonSc} we apply this formula with an analytic function $%
f_{1}(z) $ to derive the Wightman function and the vacuum expectation values
of the energy-momentum tensor in geometries with spherical boundaries. \rule%
{1.5ex}{1.5ex}

\bigskip

For an analytic function $f(z)$ formula (\ref{sumJ1}) yields
\begin{eqnarray}
&&\sum_{k=1}^{\infty }\frac{2\lambda _{\nu ,k}f(\lambda _{\nu ,k})}{\left(
\lambda _{\nu ,k}^{2}-\nu ^{2}\right) J_{\nu }^{2}(\lambda _{\nu
,k})+\lambda _{\nu ,k}^{2}J_{\nu }^{\prime 2}(\lambda _{\nu ,k})}%
=\int_{0}^{\infty }dx\,f(x)+\frac{\pi }{2}\underset{z=0}{\mathrm{Res}}f(z)%
\frac{\bar{Y}_{\nu }(z)}{\bar{J}_{\nu }(z)}{}  \notag \\
&&{}-\frac{1}{\pi }\int_{0}^{\infty }{\ }dx\,\frac{\bar{K}_{\nu }(x)}{\bar{I}%
_{\nu }(x)}\left[ e^{-\nu \pi i}f(xe^{\pi i/2})+e^{\nu \pi i}f(xe^{-\pi i/2})%
\right] .  \label{sumJ1anal}
\end{eqnarray}%
By taking in this formula $\nu =1/2$, $A=1,\,B=0$ (see notation (\ref{efnot1}%
)), as a particular case we immediately receive the APF in the form (\ref%
{apsf2}). In like manner, substituting $\nu =1/2$, $A=1$, $B=2$, we obtain
APF in the form (\ref{apsf2half}). Consequently, the formula (\ref{sumJ1})
is a generalization of the APF for general $\nu $ (with restrictions given
above) and for functions $f(z)$ having poles in the right half-plane.

For further applications to the Casimir effect in Sections \ref%
{sec:GlobMonSc} and \ref{sec:ElSpheric}, let us choose in (\ref{sumJ1anal})
\begin{equation}
f(z)=F(z)J_{\nu +m}^{2}(zt),\quad t>0,\quad {\mathrm{Re\,}}\nu \geqslant 0,
\label{casapf1}
\end{equation}%
with $m$ being an integer. Now the conditions (\ref{condf}) formulated in
terms of $F(z)$ are of the form
\begin{equation}
|F(z)|<|z|\varepsilon e^{(c-2t)|y|}\quad \text{ or }\quad |F(z)|<\frac{%
Me^{2(1-t)|y|}}{|z|^{\alpha -1}},\;z=x+iy,\;|z|\rightarrow \infty ,
\label{condFcas}
\end{equation}%
with the same notations as in (\ref{condf}). In like manner, from condition (%
\ref{case21}) for $F(z)$ one has
\begin{equation}
F(ze^{\pi i})=-F(z)+o(z^{-2m-1}),\quad z\rightarrow 0.  \label{case21F}
\end{equation}%
Now, as a consequence of (\ref{sumJ1anal}) we obtain that if the conditions (%
\ref{condFcas}) and (\ref{case21F}) are satisfied, then for a function $F(z)$
analytic in the right half-plane, the following formula takes place:
\begin{eqnarray}
&&2\sum_{k=1}^{\infty }T_{\nu }(\lambda _{\nu ,k})F(\lambda _{\nu ,k})J_{\nu
+m}^{2}(\lambda _{\nu ,k}t)=\int_{0}^{\infty }dx\,F(x)J_{\nu +m}^{2}(xt){}
\notag \\
&&{}-\frac{1}{\pi }\int_{0}^{\infty }dx{\,}\frac{\bar{K}_{\nu }(x)}{\bar{I}%
_{\nu }(x)}I_{\nu +m}^{2}(xt)\left[ F(xe^{\pi i/2})+F(xe^{-\pi i/2})\right] ,
\label{sumJ1analcas}
\end{eqnarray}%
for ${\mathrm{Re\,}}\nu \geqslant 0$ and ${\mathrm{Re\,}}\nu +m\geqslant 0$.

\subsection{Case (b)}

Let $f(z)$ be a function satisfying the condition
\begin{equation}
f(xe^{\pi i/2})=-e^{2\nu \pi i}f(xe^{-\pi i/2}),  \label{caseb}
\end{equation}%
for real $x$. It is clear that if $f(z)$ has purely imaginary poles, then
they are complex conjugate: $\pm iy_{k}$, $y_{k}>0$. By (\ref{caseb}) the
rhs of (\ref{formgen}) for $a\rightarrow 0$ and ${\mathrm{arg\,}}\lambda
_{\nu ,k}=\pi /2$ may be written as
\begin{equation}
\sum_{\alpha =+,-}\alpha \left( \int_{\gamma _{\rho }^{\alpha
}}+\sum_{\sigma _{k}=\alpha iy_{k},\alpha \lambda _{\nu ,k}}\int_{C_{\rho
}(\sigma _{k})}\right) \frac{\bar{H}_{\nu }^{(p_{\alpha })}(z)}{\bar{J}_{\nu
}(z)}f(z)dz,  \label{rel7}
\end{equation}%
where $p_{+}=1$, $p_{-}=2$, $C_{\rho }(\sigma _{k})$ denotes the right half
of the circle with radius $\rho $ and with the center at the point $\sigma
_{k}$, described in the positive sense, and the contours $\gamma _{\rho
}^{\pm }$ are the same as in (\ref{rel1}). We have used the fact the purely
imaginary zeros of $\bar{J}_{\nu }(z)$ are complex conjugate numbers, as $%
\bar{J}_{\nu }(ze^{\pi i})=e^{\nu \pi i}\bar{J}_{\nu }(z)$. We have also
used the fact that on the right of (\ref{formgen}) the integrals (with $a=0$%
) along straight segments of the upper and lower imaginary semiaxes are
cancelled, as in accordance of (\ref{caseb}) for ${\mathrm{arg\,\,}}z=\pi /2$
\begin{equation}
\frac{\bar{H}_{\nu }^{(1)}(z)}{\bar{J}_{\nu }(z)}f(z)=\frac{\bar{H}_{\nu
}^{(2)}(ze^{-\pi i})}{\bar{J}_{\nu }(ze^{-\pi i})}f(ze^{-\pi i}).
\label{rel8}
\end{equation}%
Let us show that from (\ref{rel8}) for $z_{0}=x_{0}e^{\pi i/2}$ it follows
that this relation is valid for any $z$ in a small enough region including
this point. Namely, as the function $f(z)\bar{H}_{\nu }^{(p)}(z)/\bar{J}%
_{\nu }(z)$, $p=1,2$ is meromorphic near the point $(-1)^{p+1}x_{0}e^{\pi
i/2}$, there exists a neighborhood of this point where this function is
presented as a Laurent expansion
\begin{equation}
\frac{\bar{H}_{\nu }^{(p)}(z)}{\bar{J}_{\nu }(z)}f(z)=\sum_{n=-n_{0}}^{%
\infty }\frac{a_{n}^{(p)}}{\left[ z-(-1)^{p+1}x_{0}e^{\pi i/2}\right] ^{n}}%
\,.  \label{rel9laur}
\end{equation}%
From (\ref{rel8}) for $z=xe^{\pi i/2}$ one concludes
\begin{equation}
\sum_{n=-n_{0}}^{\infty }\frac{a_{n}^{(1)}e^{-n\pi i/2}}{\left(
x-x_{0}\right) ^{n}}=\sum_{n=-n_{0}}^{\infty }\frac{(-1)^{n}a_{n}^{(2)}e^{-n%
\pi i/2}}{\left( x-x_{0}\right) ^{n}},  \label{rel10}
\end{equation}%
and hence $a_{n}^{(1)}=(-1)^{n}a_{n}^{(2)}$. Our statement follows directly
from here. By this it can be seen that
\begin{equation}
\sum_{\alpha =+,-}\alpha \int_{C_{\rho }(\alpha \sigma _{k})}dz\,\frac{\bar{H%
}_{\nu }^{(p_{\alpha })}(z)}{\bar{J}_{\nu }(z)}f(z)=2\pi i\underset{z=\sigma
_{k}}{\mathrm{Res}}\frac{\bar{H}_{\nu }^{(1)}(z)}{\bar{J}_{\nu }(z)}f(z),
\label{rel11}
\end{equation}%
where $\sigma _{k}=iy_{k},\,\lambda _{\nu ,k}$, ${\mathrm{arg\,}}\lambda
_{\nu ,k}=\pi /2$. Now, by taking into account (\ref{rel3}) and letting $%
\rho \rightarrow 0$ we get \cite{Sah1, Sahdis}:

\bigskip

\noindent \textbf{Theorem 3.} \textit{Let $f(z)$ be a meromorphic function
in the half-plane ${\mathrm{Re\,}}z\geqslant 0$ (except possibly at $z=0$)
with poles $z_{k},\,{\mathrm{Re\,}}z_{k}>0$ and $\pm iy_{k},\,y_{k}>0$, $%
k=1,2,...$ ($\neq \lambda _{\nu ,p}$). If this function satisfies conditions
(\ref{condf}) and (\ref{caseb}) then
\begin{eqnarray}
&&\lim_{b\rightarrow +\infty }\left\{ 2\sum_{k=1}^{n}T_{\nu }(\lambda _{\nu
,k})f(\lambda _{\nu ,k})+r_{1\nu }[f(z)]-{\mathrm{p.v.}}\! \int_{0}^{b}{%
f(x)dx}\right\} =  \notag \\
&&\quad =-\frac{\pi i}{2}\sum_{\eta _{k}=0,iy_{k}}\left( 2-\delta _{0\eta
_{k}}\right) \underset{z=\eta _{k}}{\mathrm{Res}}f(z)\frac{\bar{H}_{\nu
}^{(1)}(z)}{\bar{J}_{\nu }(z)},  \label{sumJ2}
\end{eqnarray}%
where on the left $0<{\mathrm{Re\,}}z_{k}<b$, ${\mathrm{Re\,}}\lambda _{\nu
,n}<b<{\mathrm{Re\,}}\lambda _{\nu ,n+1}$ and }$r_{1\nu }[f(z)]$\textit{\ is
defined by (\ref{r1}).}

\bigskip

\noindent Note that the residue terms in (\ref{rel11}) with $\sigma
_{k}=\lambda _{\nu ,k}$, ${\mathrm{arg\,}}\lambda _{\nu ,k}=\pi /2$, are
equal to $4T_{\nu }(\lambda _{\nu ,k})f(\lambda _{\nu ,k})$ and are included
in the first sum on the left of (\ref{sumJ2}).

\bigskip

\noindent \textbf{Remark.} Let $\pm iy_{k},\,y_{k}>0$, and $\pm \lambda
_{\nu ,k}$, ${\mathrm{arg\,}}\lambda _{\nu ,k}=\pi /2$, be purely imaginary
poles of function $f(z)$ and purely imaginary zeros of $\bar{J}_{\nu }(z)$,
correspondingly. Let function $f(z)$ satisfy condition
\begin{equation}
f(z)=-e^{2\nu \pi i}f(ze^{-\pi i})+o\left( (z-\sigma _{k})^{-1}\right)
,\quad z\rightarrow \sigma _{k},\quad \sigma _{k}=iy_{k},\lambda _{\nu ,k}.
\label{caseaplusb}
\end{equation}%
Now in the limit $a\rightarrow 0$ the rhs of (\ref{formgen}) can be
presented in the form (\ref{rel7}) plus integrals along the straight
segments of the imaginary axis between the poles. Using the arguments
similar those given above, we obtain relation (\ref{rel11}) with an
additional contribution from the last term on the right of (\ref{caseaplusb}%
) in the form $\int_{C_{\rho }(-\sigma _{k})}{o\left( (z-\sigma
_{k})^{-1}\right) }dz$. In the limit $\rho \rightarrow 0$ the latter
vanishes and the sum of the integrals along the straight segments of the
imaginary axis gives the principal value of the integral on the right of (%
\ref{sumJ1}). As a result formula (\ref{sumJ1}) can be generalized for
functions having purely imaginary poles and satisfying condition (\ref%
{caseaplusb}). For this on the right of (\ref{sumJ1}) instead of residue
term we have to write the sum of residues from the rhs of (\ref{sumJ2}) and
take the principal value of the integral on the right. The latter exists due
to condition (\ref{caseaplusb}). \rule{1.5ex}{1.5ex}

\bigskip

An interesting result can be obtained from (\ref{sumJ2}). Let $\lambda _{\mu
,k}^{(1)}$ be zeros of the function $A_{1}J_{\mu }(z)+B_{1}zJ_{\mu }^{\prime
}(z)$ with some real constants $A_{1}$ and $B_{1}$. Let $f(z)$ be an
analytic function in the right half-plane satisfying condition (\ref{caseb})
and $f(z)=o(z^{\beta })$ for $z\rightarrow 0$, where $\beta ={\mathrm{max}}%
(\beta _{\mu },\beta _{\nu })$ (the definition $\beta _{\nu }$ see (\ref%
{betanju})). For this function we get from (\ref{sumJ2}):
\begin{equation}
\sum_{k=1}^{\infty }T_{\mu }(\lambda _{\mu ,k}^{(1)})f(\lambda _{\mu
,k}^{(1)})=\sum_{k=1}^{\infty }T_{\nu }(\lambda _{\nu ,k})f(\lambda _{\nu
,k}),\quad \mu =\nu +m\,.  \label{equalsumJ}
\end{equation}%
For the case of Fourier-Bessel and Dini series this result is given in \cite%
{Watson}.

Let us consider some applications of formula (\ref{sumJ2}) to the special
types of series. First we choose in this formula
\begin{equation}
f(z)=F_{1}(z)J_{\mu }(zt),\quad t>0,  \label{examp1}
\end{equation}%
where the function $F_{1}(z)$ is meromorphic on the right half-plane and
satisfies the conditions
\begin{equation}
|F_{1}(z)|<\varepsilon _{1}(x)e^{(c-t)|y|}\quad \text{or}\quad
|F_{1}(z)|<M|z|^{-\alpha _{1}}e^{(2-t)|y|},\;|z|\rightarrow \infty ,
\label{condex1}
\end{equation}%
with $c<2,\,\alpha _{1}>1/2$, $\varepsilon _{1}(x)=o(\sqrt{x})$ for $%
x\rightarrow +\infty $, and the condition
\begin{equation}
F_{1}(xe^{\pi i/2})=-e^{(2\nu -\mu )\pi i}F_{1}(xe^{-\pi i/2}).
\label{condex12}
\end{equation}%
{}From (\ref{condex1}) it follows that the integral ${\mathrm{p.v.}}\!
\int_{0}^{\infty }dx\,F_{1}(x)J_{\mu }(xt)$ converges at the upper limit
and, hence, in this case formula (\ref{sumJ2}) may be written in the form
\begin{eqnarray}
&&\sum_{k=1}^{\infty }T_{\nu }(\lambda _{\nu ,k})F_{1}(\lambda _{\nu
,k})J_{\mu }(\lambda _{\nu ,k}t)=\frac{1}{2}{\mathrm{p.v.}}\!
\int_{0}^{\infty }dx\,F_{1}(x)J_{\mu }(xt)-\frac{1}{2}r_{1\nu }\left[
F_{1}(z)J_{\mu }(zt)\right] {}  \notag \\
&&\quad -\frac{\pi i}{4}\sum_{\eta _{k}=0,iy_{k}}\left( 2-\delta _{0\eta
_{k}}\right) \underset{z=\eta _{k}}{\mathrm{Res}}F_{1}(z)J_{\mu }(zt)\frac{%
\bar{H}_{\nu }^{(1)}(z)}{\bar{J}_{\nu }(z)}.  \label{sumJ3}
\end{eqnarray}%
Note that the lhs of this formula is known as Dini series of Bessel
functions \cite{Watson}. For example, it follows from here that for $t<1$, ${%
\mathrm{Re\,}}\sigma ,\,{\mathrm{Re\,}}\nu >-1$ one has
\begin{eqnarray}
\sum_{k=1}^{\infty }\frac{T_{\nu }(\lambda _{\nu ,k})}{\lambda _{\nu
,k}^{\sigma }}J_{\sigma +\nu +1}(\lambda _{\nu ,k})J_{\nu }(\lambda _{\nu
,k}t) &=&\frac{1}{2}\int_{0}^{\infty }\frac{dz}{z^{\sigma }}J_{\sigma +\nu
+1}(z)J_{\nu }(zt)  \notag \\
&=&\frac{\left( 1-t^{2}\right) ^{\sigma }t^{\nu }}{2^{\sigma +1}\Gamma
(\sigma +1)},  \label{examp1ap}
\end{eqnarray}%
(for the value of the integral see, e.g., \cite{Watson}). For $B=0$ this
result is given in \cite{Prud86}. In a similar way, taking $\mu =\nu +m$,
\begin{equation}
F_{1}(z)=z^{\nu +m+1}\frac{J_{\sigma }(a\sqrt{z^{2}+z_{1}^{2}})}{\left(
z^{2}+z_{1}^{2}\right) ^{\sigma /2}},\quad a>0,  \label{examp2apF2}
\end{equation}%
with ${\mathrm{Re\,}}\nu \geqslant 0$ and ${\mathrm{Re\,}}\nu +m\geqslant 0$%
, {}from (\ref{sumJ3}) for $a<2-t$, ${\mathrm{Re\,}}\sigma >{\mathrm{Re\,}}%
\nu +m$ one finds
\begin{eqnarray}
&&\sum_{k=1}^{\infty }T_{\nu }(\lambda _{\nu ,k})\lambda _{\nu ,k}^{\nu
+m+1}J_{\nu +m}(\lambda _{\nu ,k}t)\frac{J_{\sigma }(a\sqrt{\lambda _{\nu
,k}^{2}+z_{1}^{2}})}{\left( \lambda _{\nu ,k}^{2}+z_{1}^{2}\right) ^{\sigma
/2}}=  \notag \\
&&\quad =\frac{1}{2}\int_{0}^{\infty }dx\,x^{\nu +m+1}J_{\nu +m}(xt)\frac{%
J_{\sigma }(a\sqrt{x^{2}+z_{1}^{2}})}{\left( x^{2}+z_{1}^{2}\right) ^{\sigma
/2}}  \notag \\
&&\quad =\frac{t^{\nu +1}}{a^{\sigma }}\left( -z_{1}\right) ^{m+1}\frac{%
J_{m+1}(z_{1}\sqrt{a^{2}-t^{2}})}{\left( a^{2}-t^{2}\right) ^{(m+1)/2}}%
,\;a>t,  \label{examp2ap}
\end{eqnarray}%
and the sum is zero when $a<t$. Here we have used the known value for the
Sonine integral \cite{Watson}.

If an addition to (\ref{condex1}), (\ref{condex12}) the function $F_{1}(z)$
satisfies the conditions
\begin{equation}
F_{1}(xe^{\pi i/2})=-e^{\mu \pi i}F_{1}(xe^{-\pi i/2})  \label{condex13}
\end{equation}%
and
\begin{equation}
|F_{1}(z)|<\varepsilon _{1}(x)e^{c_{1}t|y|}\quad \text{or}\quad
|F_{1}(z)|<M|z|^{-\alpha _{1}}e^{t|y|},\quad |z|\rightarrow \infty ,
\label{condex1nor}
\end{equation}%
then formula (\ref{intJform42}) (see below) with $B=0$ may be applied to the
integral on the right of (\ref{sumJ3}). This gives \cite{Sah1, Sahdis}:

\bigskip

\noindent \textbf{Corollary 1.} \textit{Let $F(z)$ be a meromorphic function
in the half-plane ${\mathrm{Re\,}}z\geqslant 0$ (except possibly at $z=0$)
with poles $z_{k},\,{\mathrm{Re\,}}z_{k}>0$ and $\pm iy_{k},\,y_{k}>0$ ($%
\neq \lambda _{\nu ,i}$). If $F(z)$ satisfy condition
\begin{equation}
F(xe^{\pi i/2})=(-1)^{m+1}e^{\nu \pi i}F(xe^{-\pi i/2}),  \label{cor1cond1}
\end{equation}%
with an integer $m$, and one of the inequalities
\begin{equation}
|F(z)|<\varepsilon _{1}(x)e^{a|y|}\quad \text{or}\quad |F(z)|<M|z|^{-\alpha
_{1}}e^{a_{0}|y|},\quad |z|\rightarrow \infty ,  \label{cor1cond2}
\end{equation}%
with $a<{\mathrm{min}}(t,2-t)\equiv a_{0}$, $\varepsilon
_{1}(x)=o(x^{1/2}),\,x\rightarrow +\infty $, $\alpha _{1}>1/2$, the
following formula is valid
\begin{eqnarray}
&&\sum_{k=1}^{\infty }T_{\nu }(\lambda _{\nu ,k})F(\lambda _{\nu ,k})J_{\nu
+m}(\lambda _{\nu ,k}t)=\frac{\pi i}{4}\sum_{\eta _{k}=0,iy_{k},z_{k}}\left(
2-\delta _{0\eta _{k}}\right)  \notag \\
&&\quad \times \underset{z=\eta _{k}}{\mathrm{Res}}\left\{ \left[ J_{\nu
+m}(zt)\bar{Y}_{\nu }(z)-Y_{\nu +m}(zt)\bar{J}_{\nu }(z)\right] \frac{F(z)}{%
\bar{J}_{\nu }(z)}\right\} .  \label{cor1form}
\end{eqnarray}%
}

\bigskip

\noindent Recall that for the imaginary zeros $\lambda _{\nu ,k}$, on the
lhs of (\ref{cor1form}) the zeros with positive imaginary parts enter only.
By using formula (\ref{cor1form}) a number of Fourier-Bessel and Dini series
can be summarized (see below).

\bigskip

\noindent \textbf{Remark.} Formula (\ref{cor1form}) may also be obtained by
considering the integral
\begin{equation}
\frac{1}{\pi }\int_{C_{h}}{\ }dz\,\left[ H_{\nu +m}^{(2)}(zt)\bar{H}_{\nu
}^{(1)}(z)-H_{\nu +m}^{(1)}(zt)\bar{H}_{\nu }^{(2)}(z)\right] \frac{F(z)}{%
\bar{J}_{\nu }(z)},  \label{contint}
\end{equation}%
where $C_{h}$ is a rectangle with vertices $\pm ih,\,b\pm ih$, described in
the positive sense (purely imaginary poles of $F(z)/\bar{J}_{\nu }(z)$ and
the origin are circled by semicircles in the right half-plane with small
radii). This integral is equal to the sum of residues over the poles within $%
C_{h}$ (points $z_{k}$, $\lambda _{\nu ,k}$, ($\mathrm{Re\,}z_{k},\,{\mathrm{%
Re\,}}\lambda _{\nu ,k}>0$)). On the other hand, it follows from (\ref%
{cor1cond1}) that the integrals along the segments of the imaginary axis
cancel each other. The sum of integrals along the conjugate semicircles give
the sum of residues over purely imaginary poles in the upper half-plane. The
integrals along the remaining three segments of $C_{h}$ in accordance with (%
\ref{cor1cond2}) approach to zero in the limit $b,\,h\rightarrow \infty $.
Equating these expressions for (\ref{contint}), one immediately obtains the
result (\ref{cor1form}). \rule{1.5ex}{1.5ex}

\bigskip

{}From (\ref{cor1form}), for $t=1,\,F(z)=J_{\nu }(zx)$, $m=1$, one obtains
\cite{Erde53, Watson}
\begin{equation}
\sum_{k=1}^{\infty }T_{\nu }(\lambda _{\nu ,k})J_{\nu }(\lambda _{\nu
,k}x)J_{\nu +1}(\lambda _{\nu ,k})=\frac{x^{\nu }}{2},\quad 0\leqslant x<1.
\label{examp2}
\end{equation}%
In a similar way, choosing $m=0$, $F(z)=zJ_{\nu }(zx)/\left(
z^{2}-a^{2}\right) $, $B=0$, we obtain the Kneser-Sommerfeld expansion \cite%
{Watson}:
\begin{equation}
\sum_{k=1}^{\infty }\frac{J_{\nu }(\lambda _{\nu ,k}t)J_{\nu }(\lambda _{\nu
,k}x)}{\left( \lambda _{\nu ,k}^{2}-a^{2}\right) J_{\nu +1}^{2}(\lambda
_{\nu ,k})}=\frac{\pi }{4}\frac{J_{\nu }(ax)}{J_{\nu }(a)}\left[ J_{\nu
}(at)Y_{\nu }(a)-Y_{\nu }(at)J_{\nu }(a)\right] ,  \label{examp3}
\end{equation}%
for $0\leqslant x\leqslant t\leqslant 1$. In (\ref{cor1form}) as a function $%
F(z)$ one may choose, for example, the following functions
\begin{eqnarray}
&&\!\!\!z^{\rho -1}\prod_{l=1}^{n}\left( z^{2}+z_{l}^{2}\right) ^{-\mu
_{l}/2}J_{\mu _{l}}(b_{l}\sqrt{z^{2}+z_{l}^{2}}),{}\quad b\leqslant
a_{0},\,b=\sum_{l=1}^{n}b_{l},  \notag \\
&&\qquad {\mathrm{Re\,}}\nu <\sum_{l=1}^{n}{\mathrm{Re\,}}\mu
_{l}+n/2+2p+3/2-m-\delta _{ba_{0}};{}  \label{func1} \\
&&\!\!\!z^{\rho -2n-1}\prod_{l=1}^{n}\left[ 1-J_{0}(b_{l}z)\right] ,{}\quad {%
\mathrm{Re\,}}\nu <2n+2p+3/2-m-\delta _{ba_{0}};  \label{func2n} \\
&&\!\!\!z^{\rho -1}\prod_{l=1}^{n}\left( z^{2}+z_{l}^{2}\right) ^{\mu
_{l}/2}Y_{\mu _{l}}\left( b_{l}\sqrt{z^{2}+z_{l}^{2}}\right) ,\,\mu _{l}>0\,%
\text{-half of an odd integer,}{}  \notag \\
&&\qquad {\mathrm{Re\,}}\nu <-\sum_{l=1}^{n}\mu _{l}+n/2+2p+3/2-m-\delta
_{ba_{0}};{}  \label{func3} \\
&&\!\!\!z^{\rho -1}\prod_{l=1}^{n}z^{|k_{l}|}\left[ J_{\mu
_{l}+k_{l}}(a_{l}z)Y_{\mu _{l}}(b_{l}z)-Y_{\mu _{l}+k_{l}}(a_{l}z)J_{\mu
_{l}}(b_{l}z)\right] ,\quad k_{l}\,\text{- integer,}{}  \notag \\
&&\qquad {\mathrm{Re\,}}\nu <n+2p+3/2-m-\sum |k_{l}|-\delta _{\tilde{a}%
,a_{0}},\;\tilde{a}\equiv \sum_{l=1}^{n}|a_{l}-b_{l}|\leqslant a_{0};
\label{func4}
\end{eqnarray}%
with $\rho =\nu +m-2p$ ($p$ is an integer), as well as any products between
these functions and with $\prod_{l}\left( z^{2}-c_{l}^{2}\right) ^{-p_{l}}$,
provided condition (\ref{cor1cond2}) is satisfied. For example, the
following formulae take place
\begin{eqnarray}
&&\!\!\sum_{k=1}^{\infty }j_{\nu ,k}^{\nu -2}\frac{J_{\nu }(j_{\nu ,k}t)}{%
J_{\nu +1}^{2}(j_{\nu ,k})}\prod_{l=1}^{n}\left[ J_{\mu _{l}}(a_{l}j_{\nu
,k})Y_{\mu _{l}}(b_{l}j_{\nu ,k})-Y_{\mu _{l}}(a_{l}j_{\nu ,k})J_{\mu
_{l}}(b_{l}j_{\nu ,k})\right] {}  \notag \\
&&\qquad =\frac{2^{\nu -2}}{\pi ^{n}t^{\nu }}\left( 1-t^{2\nu }\right)
\prod_{l=1}^{n}\frac{b_{l}^{\mu _{l}}}{\mu _{l}a_{l}^{\mu _{l}}}\left[
1-\left( \frac{a_{l}}{b_{l}}\right) ^{2\mu _{l}}\right] ,\quad 0<t\leqslant
1,  \notag \\
&&\qquad c\equiv \sum_{l=1}^{n}|a_{l}-b_{l}|\leqslant t,\quad
a_{l},b_{l}>0,\quad {\mathrm{Re\,}}\mu _{l}\geqslant 0,\quad {\mathrm{Re\,}}%
\nu <n+3/2-\delta _{ct};  \label{examp4} \\
&&\!\!\sum_{k=1}^{\infty }\frac{J_{\nu }(j_{\nu ,k}t)J_{\nu +1}(\lambda
j_{\nu ,k})}{j_{\nu ,k}^{2n+3}J_{\nu +1}^{2}(j_{\nu ,k})}\prod_{l=1}^{n}%
\left[ 1-J_{0}(b_{l}j_{\nu ,k})\right] =\frac{\lambda ^{\nu +1}\left(
1-t^{2\nu }\right) }{4^{n+1}\nu (\nu +1)t^{\nu }}\prod_{l=1}^{n}b_{l}^{2},
\notag \\
&&\qquad \lambda +\sum_{l=1}^{n}b_{l}\leqslant t\leqslant 1,\quad \lambda
,b_{l}>0;  \label{examp5} \\
&&\!\!\sum_{k=1}^{\infty }\frac{J_{\mu }(j_{\nu ,k}b)J_{\nu +1}(\lambda
j_{\nu ,k})J_{\nu }(j_{\nu ,k}t)}{\left( j_{\nu ,k}^{2}-a^{2}\right) j_{\nu
,k}^{\mu +1}J_{\nu +1}^{2}(j_{\nu ,k})}=\frac{\pi J_{\nu +1}(a\lambda )}{%
4a^{\mu +1}}\left[ Y_{\nu }(a)J_{\nu }(at)-J_{\nu }(a)Y_{\nu }(at)\right] ,
\notag \\
&&\qquad \times \frac{J_{\mu }(ba)}{J_{\nu }(a)},\;\lambda +b\leqslant
t\leqslant 1,\,\lambda ,b>0,\,{\mathrm{Re}}\mu >-7/2+\delta _{\lambda +b,t},
\label{examp6}
\end{eqnarray}%
where $j_{\nu ,k}$ are zeros of the function $J_{\nu }(z)$. The examples of
the series over zeros of Bessel functions we found in literature (see, e.g.,
\cite{Erde53, Watson, Prud86, Magnus}), when the corresponding sum was
evaluated in finite terms, are particular cases of the formulae given in
this section.

\section{Summation formulae over zeros of combinations of cylinder functions}

\label{sec:SumFormBess}

Here we will consider series over the zeros of the function
\begin{equation}
C_{\nu }^{ab}(\eta ,z)\equiv \bar{J}_{\nu }^{(a)}(z)\bar{Y}_{\nu
}^{(b)}(\eta z)-\bar{Y}_{\nu }^{(a)}(z)\bar{J}_{\nu }^{(b)}(\eta z),
\label{bescomb1}
\end{equation}%
where and in what follows for a given function $f(z)$ the quantities with
overbars are defined by the formula
\begin{equation}
\bar{f}^{(j)}(z)=A_{j}f(x)+B_{j}zf^{\prime }(z),\;j=a,b,  \label{barjnot}
\end{equation}%
with constant coefficients $A_{j}$ and $B_{j}$. This type of series arises
in calculations of the vacuum expectation values in confined regions with
boundaries of spherical and cylindrical form. To obtain a summation formula
for these series, in the GAPF we substitute
\begin{equation}
g(z)=\frac{1}{2i}\left[ \frac{\bar{H}_{\nu }^{(1b)}(\eta z)}{\bar{H}_{\nu
}^{(1a)}(z)}+\frac{\bar{H}_{\nu }^{(2b)}(\eta z)}{\bar{H}_{\nu }^{(2a)}(z)}%
\right] \frac{h(z)}{C_{\nu }^{ab}(\eta ,z)},\quad f(z)=\frac{h(z)}{\bar{H}%
_{\nu }^{(1a)}(z)\bar{H}_{\nu }^{(2a)}(z)},  \label{gefcomb}
\end{equation}%
where for definiteness we shall assume that $\eta >1$. The sum and
difference of these functions are given by the formula
\begin{equation}
g(z)-(-1)^{k}f(z)=-i\frac{\bar{H}_{\nu }^{(ka)}(\lambda z)}{\bar{H}_{\nu
}^{(ka)}(z)}\frac{h(z)}{C_{\nu }^{ab}(\eta ,z)},\quad k=1,2.
\label{gefsumnew}
\end{equation}%
The condition for the GAPF written in terms of the function $h(z)$ takes the
form
\begin{equation}
|h(z)|<\varepsilon _{1}(x)e^{c_{1}|y|}\quad |z|\rightarrow \infty ,\quad
z=x+iy,  \label{cond31}
\end{equation}%
where $c_{1}<2(\eta -1)$ and $x^{\delta _{B_{a}0}+\delta
_{B_{b}0}-1}\varepsilon _{1}(x)\rightarrow 0$ for $x\rightarrow +\infty $.
Let $\gamma _{\nu ,k}$ be zeros of the function $C_{\nu }^{ab}(\eta ,z)$ in
the right half-plane. In this section we will assume values of $\nu $, $A_{j}
$, and $B_{j}$ for which all these zeros are real and simple, and the
function $\bar{H}_{\nu }^{(1a)}(z)$ ($\bar{H}_{\nu }^{(2a)}(z)$) has no
zeros in the right half of the upper (lower) half-plane. For real $\nu $ and
$A_{j}$, $B_{j}$ the zeros $\gamma _{\nu ,k}$ are simple. To see this note
that the function $J_{\nu }(tz)\bar{Y}_{\nu }^{(a)}(z)-Y_{\nu }(tz)\bar{J}%
_{\nu }^{(a)}(z)$ is a cylinder function with respect to $t$. Using the
standard result for indefinite integrals containing the product of any two
cylinder functions (see \cite{Watson,abramowiz}), it can be seen that
\begin{equation}
\int_{1}^{\eta }dt\,t\left[ J_{\nu }(tz)\bar{Y}_{\nu }^{(a)}(z)-Y_{\nu }(tz)%
\bar{J}_{\nu }^{(a)}(z)\right] ^{2}=\frac{2}{\pi ^{2}zT_{\nu }^{ab}(\eta ,z)}%
\,,\quad z=\gamma _{\nu ,k},  \label{tekapositive}
\end{equation}%
where we have introduced the notation
\begin{equation}
T_{\nu }^{ab}(\eta ,z)=z\left\{ \frac{\bar{J}_{\nu }^{(a)2}(z)}{\bar{J}_{\nu
}^{(b)2}(\eta z)}\left[ A_{b}^{2}+B_{b}^{2}(\eta ^{2}z^{2}-\nu ^{2})\right]
-A_{a}^{2}-B_{a}^{2}(z^{2}-\nu ^{2})\right\} ^{-1}.  \label{tekaAB}
\end{equation}%
On the other hand
\begin{equation}
\frac{\partial }{\partial z}C_{\nu }^{ab}(\eta ,z)=\frac{2}{\pi T_{\nu
}^{ab}(\eta ,z)}\frac{\bar{J}_{\nu }^{(b)}(\eta z)}{\bar{J}_{\nu }^{(a)}(z)}%
\,,\quad z=\gamma _{\nu ,k}.  \label{CABderivative}
\end{equation}%
Combining the last two results we deduce that for real $\nu $, $A_{j}$, $%
B_{j}$ the derivative (\ref{CABderivative}) is nonzero and, hence, the zeros
$z=\gamma _{\nu ,k}$ are simple. By using this we can see that
\begin{equation}
\underset{z=\gamma _{\nu ,k}}{\mathrm{Res}}g(z)=\frac{\pi }{2i}T_{\nu
}^{ab}(\eta ,\gamma _{\nu ,k})h(\gamma _{\nu ,k}).  \label{rel31}
\end{equation}%
Hence, if the function $h(z)$ is analytic in the half-plane ${\mathrm{Re\,}}%
z\geqslant a>0$ except at the poles $z_{k}$ ($\neq \gamma _{\nu ,i}$) and
satisfies condition (\ref{cond31}), the following formula takes place
\begin{eqnarray}
&&\lim_{b\rightarrow +\infty }\left\{ \frac{\pi ^{2}}{2}\sum_{k=n}^{m}T_{\nu
}^{ab}(\lambda ,\gamma _{\nu ,k})h(\gamma _{\nu ,k})+r_{2\nu }[h(z)]-{%
\mathrm{p.v.\!}}\int_{a}^{b}\frac{h(x)dx}{\bar{J}_{\nu }^{(a)2}(x)+\bar{Y}%
_{\nu }^{(a)2}(x)}\right\}   \notag \\
&&\quad =\frac{i}{2}\int_{a}^{a+i\infty }\frac{\bar{H}_{\nu }^{(1b)}(\lambda
z)}{\bar{H}_{\nu }^{(1a)}(z)}\frac{h(z)}{C_{\nu }^{ab}(\lambda ,z)}dz-\frac{i%
}{2}\int_{a}^{a-i\infty }\frac{\bar{H}_{\nu }^{(2b)}(\lambda z)}{\bar{H}%
_{\nu }^{(2a)}(z)}\frac{h(z)}{C_{\nu }^{ab}(\lambda ,z)}dz,
\label{gapsfcomb}
\end{eqnarray}%
where we have assumed that the integral on the left exists. In formula (\ref%
{gapsfcomb}), $\gamma _{\nu ,n-1}<a<\gamma _{\nu ,n}$, $\gamma _{\nu
,m}<b<\gamma _{\nu ,m+1}$, $a<{\mathrm{Re\,}}z_{k}<b$, ${\mathrm{Re\,}}%
z_{k}\leqslant {\mathrm{Re\,}}z_{k+1}$, and the following notation is
introduced
\begin{eqnarray}
r_{2\nu }[h(z)] &=&\pi \sum_{k}\underset{{\mathrm{Im}}z_{k}=0}{\mathrm{Res}}%
\left[ \frac{\bar{J}_{\nu }^{(a)}(z)\bar{J}_{\nu }^{(b)}(\lambda z)+\bar{Y}%
_{\nu }^{(a)}(z)\bar{Y}_{\nu }^{(b)}(\lambda z)}{\bar{J}_{\nu }^{(a)2}(z)+%
\bar{Y}_{\nu }^{(a)2}(z)}\frac{h(z)}{C_{\nu }^{ab}(\lambda ,z)}\right] {}
\notag \\
&&+\pi \sum_{k,l=1,2}\underset{(-1)^{l}{\mathrm{Im}}z_{k}<0}{\mathrm{Res}}%
\left[ \frac{\bar{H}_{\nu }^{(lb)}(\lambda z)}{\bar{H}_{\nu }^{(la)}(z)}%
\frac{h(z)}{C_{\nu }^{ab}(\lambda ,z)}\right] .  \label{r3}
\end{eqnarray}%
General formula (\ref{gapsfcomb}) is a direct consequence of the GAPF and
will be used as a starting point for further applications in this section.
In the limit $a\rightarrow 0$ one has \cite{Sah1, Sahdis,Saha01}:

\bigskip

\noindent \textbf{Corollary 2.} \textit{Let $h(z)$ be an analytic function
for ${\mathrm{Re\,}}z\geqslant 0$ except the poles $z_{k}$ ($\neq \gamma
_{\nu i}$), ${\mathrm{Re\,}}z_{k}>0$ (with possible branch point $z=0$). If
it satisfies condition (\ref{cond31}) and
\begin{equation}
h(ze^{\pi i})=-h(z)+o(z^{-1}),\quad z\rightarrow 0,  \label{cor3cond1}
\end{equation}%
and the integral
\begin{equation}
{\mathrm{p.v.}}\! \int_{0}^{b}{\frac{h(x)dx}{\bar{J}_{\nu }^{(a)2}(x)+\bar{Y}%
_{\nu }^{(a)2}(x)}}  \label{cor2cond2}
\end{equation}%
exists, then
\begin{eqnarray}
&&\lim_{b\to +\infty }\left\{ \frac{\pi ^{2}}{2}\sum_{k=1}^{m}h(\gamma _{\nu
,k})T_{\nu }^{ab}(\lambda ,\gamma _{\nu ,k})+r_{2\nu }[h(z)]-{\mathrm{p.v.}}%
\! \int_{0}^{b}{\frac{h(x)dx}{\bar{J}_{\nu }^{(a)2}(x)+\bar{Y}_{\nu
}^{(a)2}(x)}}\right\}  \notag \\
&&=-\frac{\pi }{2}\underset{z=0}{\mathrm{Res}}\left[ \frac{h(z)\bar{H}_{\nu
}^{(1b)}(\lambda z)}{C_{\nu }^{ab}(\lambda ,z)\bar{H}_{\nu }^{(1a)}(z)}%
\right] -\frac{\pi }{4}\int_{0}^{\infty }dx\,\Omega _{a\nu }(x,\lambda x){%
\left[ h(xe^{\frac{\pi i}{2}})+h(xe^{-\frac{\pi i}{2}})\right] .}
\label{cor3form}
\end{eqnarray}%
}

\bigskip

\noindent In formula (\ref{cor3form}) we have introduced the notation%
\begin{equation}
\Omega _{a\nu }(x,\lambda x)=\frac{\bar{K}_{\nu }^{(b)}(\lambda x)/\bar{K}%
_{\nu }^{(a)}(x)}{\bar{K}_{\nu }^{(a)}(x)\bar{I}_{\nu }^{(b)}(\lambda x)-%
\bar{K}_{\nu }^{(b)}(\lambda x)\bar{I}_{\nu }^{(a)}(x)}.  \label{Omegaanu}
\end{equation}%
In the following we shall use this formula to derive the renormalized vacuum
energy-momentum tensor for the region between two spherical and cylindrical
surfaces. Note that (\ref{cor3form}) may be generalized for functions $h(z)$
with purely imaginary poles $\pm iy_{k}$, $y_{k}>0$, satisfying the
condition
\begin{equation}
h(ze^{\pi i})=-h(z)+o\left( (z\mp iy_{k})^{-1}\right) ,\quad z\rightarrow
\pm iy_{k}.  \label{cor3cond1plus2}
\end{equation}%
The corresponding formula is obtained from (\ref{cor3form}) by adding
residue terms for $z=iy_{k}$ in the form of (\ref{cor2form}) (see below) and
taking the principal value of the integral on the right. The arguments here
are similar to those for Remark after Theorem 3.

In a way similar to (\ref{sumJ1}), one has another result \cite{Sah1, Sahdis}%
:

\bigskip

\noindent \textbf{Corollary 3.} \textit{Let $h(z)$ be a meromorphic function
in the half-plane ${\mathrm{Re\,}}z\geqslant 0$ (with exception the possible
branch point $z=0$) with poles $z_{k},\,\pm iy_{k}$ ($\neq \gamma _{\nu ,i}$%
), ${\mathrm{Re\,}}z_{k},y_{k}>0$. If this function satisfies the condition
\begin{equation}
h(xe^{\pi i/2})=-h(xe^{-\pi i/2}),  \label{cor2cond1}
\end{equation}%
and the integral (\ref{cor2cond2}) exists, then
\begin{eqnarray}
&&\lim_{b\rightarrow +\infty }\left\{ \frac{\pi ^{2}}{2}\sum_{k=1}^{m}h(%
\gamma _{\nu ,k})T_{\nu }^{ab}(\lambda ,\gamma _{\nu ,k})+r_{2\nu }[h(z)]-{%
\mathrm{p.v.}}\int_{0}^{b}{\frac{h(x)dx}{\bar{J}_{\nu }^{(a)2}(x)+\bar{Y}%
_{\nu }^{(a)2}(x)}}\right\} {}  \notag \\
&&\quad =-\frac{\pi }{2}\sum_{\eta _{k}=0,iy_{k}}\left( 2-\delta _{0\eta
_{k}}\right) \underset{z=\eta _{k}}{\mathrm{Res}}\left[ \frac{\bar{H}_{\nu
}^{(1b)}(\lambda z)}{\bar{H}_{\nu }^{(1a)}(z)}\frac{h(z)}{C_{\nu
}^{ab}(\lambda ,z)}\right] ,  \label{cor2form}
\end{eqnarray}%
where on the left $\gamma _{\nu ,m}<b<\gamma _{\nu ,m+1}$. }

\bigskip

Let us consider a special application of formula (\ref{cor2form}) for $%
A_{j}=1,\,B_{j}=0$. Under the conditions given above, the generalizations of
these results for general $A_{j},B_{j}$ are straightforward.

\bigskip

\noindent \textbf{Theorem 4.} \textit{Let the function $F(z)$ be meromorphic
in the right half-plane ${\mathrm{Re\,}}z\geqslant 0$ (with the possible
exception $z=0$) with poles $z_{k},\pm iy_{k}$ ($\neq \gamma _{\nu ,i}$), $%
y_{k},{\mathrm{Re\,}}z_{k}>0$. If it satisfies condition
\begin{equation}
F(xe^{\pi i/2})=(-1)^{m+1}F(xe^{-\pi i/2}),  \label{th4cond1}
\end{equation}%
with an integer $m$, and one of two inequalities
\begin{equation}
|F(z)|<\varepsilon (x)e^{a_{1}|y|}\quad \text{or}\quad |F(z)|<M|z|^{-\alpha
}e^{a_{2}|y|},\quad |z|\rightarrow \infty ,  \label{th4cond2}
\end{equation}%
with $a_{1}<{\mathrm{min}}(2\lambda -\sigma -1,\sigma -1)\equiv a_{2}$, $%
\sigma >0$, $\varepsilon (x)\rightarrow 0$ for $x\rightarrow +\infty $, $%
\alpha >1$, then
\begin{eqnarray}
&&\sum_{k=1}^{\infty }\frac{\gamma _{\nu ,k}F(\gamma _{\nu ,k})}{J_{\nu
}^{2}(\gamma _{\nu ,k})/J_{\nu }^{2}(\lambda \gamma _{\nu ,k})-1}\left[
J_{\nu }(\gamma _{\nu ,k})Y_{\nu +m}(\sigma \gamma _{\nu ,k})-Y_{\nu
}(\gamma _{\nu ,k})J_{\nu +m}(\sigma \gamma _{\nu ,k})\right] {}  \notag \\
&&\quad =\sum_{\eta _{k}=0,iy_{k},z_{k}}\frac{2-\delta _{0\eta _{k}}}{\pi }%
\underset{z=\eta _{k}}{\mathrm{Res}}\frac{Y_{\nu }(\lambda z)J_{\nu
+m}(\sigma z)-J_{\nu }(\lambda z)Y_{\nu +m}(\sigma z)}{J_{\nu }(z)Y_{\nu
}(\lambda z)-J_{\nu }(\lambda z)Y_{\nu }(z)}F(z).  \label{th4form}
\end{eqnarray}%
}

\bigskip

\noindent \textbf{Proof.} In (\ref{cor2form}) let us choose
\begin{equation}
h(z)=F(z)\left[ J_{\nu }(z)Y_{\nu +m}(\sigma z)-Y_{\nu }(z)J_{\nu +m}(\sigma
z)\right] ,  \label{rel32}
\end{equation}%
which in virtue of (\ref{th4cond2}) satisfies condition (\ref{cond31}).
Condition (\ref{cor2cond1}) is satisfied as well. Hence, $h(z)$ satisfies
the conditions for Corollary 3. The corresponding integral in (\ref{cor2form}%
) with $h(z)$ from (\ref{rel32}) can be evaluated by using formula (\ref%
{intJYth65}) (see below). Putting the value of this integral into (\ref%
{cor2form}) after some manipulations we obtain formula (\ref{th4form}). \rule%
{1.5ex}{1.5ex}

\bigskip

\noindent \textbf{Remark.} Formula (\ref{th4form}) may also be derived by
applying to the contour integral
\begin{equation}
\int_{C_{h}}\frac{Y_{\nu }(\lambda z)J_{\nu +m}(\sigma z)-J_{\nu }(\lambda
z)Y_{\nu +m}(\sigma z)}{J_{\nu }(z)Y_{\nu }(\lambda z)-J_{\nu }(\lambda
z)Y_{\nu }(z)}F(z)dz  \label{altth4}
\end{equation}%
the residue theorem, where $C_{h}$ is a rectangle with vertices $\pm
ih,\,b\pm ih$. Here the proof is similar to that for Remark to Corollary 1.
\rule{1.5ex}{1.5ex}

\bigskip

By using (\ref{cor2form}) a formula similar to (\ref{th4form}) can also be
obtained for series of the type $\sum_{k=1}^{\infty }G(\gamma _{\nu
,k})J_{\mu }(\gamma _{\nu ,k}t)$.

As a function $F(z)$ in (\ref{th4form}) one can choose, for example,

\begin{itemize}
\item function (\ref{func1}) for $\rho =m-2p$, $\sum_{l}b_{l}<a_{2}$, $%
m<2p+\sum_{l}{\mathrm{Re\,}}\mu _{l}+n/2+1$, $p$ - integer;

\item function (\ref{func2n}) for $\rho =m-2p$, $\sum_{l}b_{l}<a_{2}$, $%
m<2p+2n+1$;

\item function (\ref{func4}) for $\rho =m-2p$, $a_{l}>0$, ${\mathrm{Re\,}}%
\mu _{l}\geqslant 0$ (for ${\mathrm{Re\,}}\mu _{l}<0$, $k_{l}>|{\mathrm{Re\,}%
}\mu _{l}|$),\newline
$\sum_{l=1}^{n}|a_{l}-b_{l}|<a_{2}$, $m<2p+n-\sum_{l}|k_{l}|+1$.
\end{itemize}

\noindent Taking $F(z)=1/z,\,m=0$, one obtains
\begin{equation}
\sum_{k=1}^{\infty }\frac{J_{\nu }(\gamma _{\nu ,k})Y_{\nu }(\sigma \gamma
_{\nu ,k})-Y_{\nu }(\gamma _{\nu ,k})J_{\nu }(\sigma \gamma _{\nu ,k})}{%
J_{\nu }^{2}(\gamma _{\nu ,k})/J_{\nu }^{2}(\lambda \gamma _{\nu ,k})-1}=%
\frac{\sigma ^{\nu }}{\pi }\frac{(\lambda /\sigma )^{2\nu }-1}{\lambda
^{2\nu }-1},\quad \lambda \geqslant \sigma >1.  \label{examp31}
\end{equation}%
In a similar way it can be seen that
\begin{eqnarray}
&&\sum_{k=1}^{\infty }\frac{\gamma _{\nu ,k}^{2}\left[ J_{\nu }(\gamma _{\nu
,k})Y_{\nu }(\sigma \gamma _{\nu ,k})-Y_{\nu }(\gamma _{\nu ,k})J_{\nu
}(\sigma \gamma _{\nu ,k})\right] }{\left( \gamma _{\nu ,k}^{2}-c^{2}\right) %
\left[ J_{\nu }^{2}(\gamma _{\nu ,k})/J_{\nu }^{2}(\lambda \gamma _{\nu
,k})-1\right] }=  \notag \\
&&\quad \;=\frac{1}{\pi }\frac{Y_{\nu }(\lambda c)J_{\nu }(\sigma c)-J_{\nu
}(\lambda c)Y_{\nu }(\sigma c)}{J_{\nu }(c)Y_{\nu }(\lambda c)-J_{\nu
}(\lambda c)Y_{\nu }(c)},  \label{examp32} \\
&&\sum_{k=1}^{\infty }\frac{J_{\nu }(\gamma _{\nu ,k})Y_{\nu }(\sigma \gamma
_{\nu ,k})-Y_{\nu }(\gamma _{\nu ,k})J_{\nu }(\sigma \gamma _{\nu ,k})}{%
J_{\nu }^{2}(\gamma _{\nu ,k})/J_{\nu }^{2}(\lambda \gamma _{\nu ,k})-1}%
\prod_{l=1}^{p}\frac{J_{\mu _{l}}(b_{l}\gamma _{\nu ,k})}{\gamma _{\nu
,k}^{\mu _{l}}}=\frac{\sigma ^{\nu }}{\pi }\frac{(\lambda /\sigma )^{2\nu }-1%
}{\lambda ^{2\nu }-1}{}  \notag \\
&&\;\;\times \prod_{l=1}^{p}\frac{2^{-\mu _{l}}b_{l}^{\mu _{l}}}{\Gamma (\mu
_{l}+1)},\quad b\equiv \sum_{1}^{p}b_{l}<\sigma -1,\,{\mathrm{Re\,}}\mu _{l}+%
\frac{p}{2}+1>\delta _{b,\sigma -1},  \label{examp33}
\end{eqnarray}%
where ${\mathrm{Re\,}}c\geqslant 0,\,b_{l}>0$, $\lambda \geqslant \sigma >1$%
, $\mu _{l}\neq -1,-2,\ldots $. Physical applications of the formulae
derived in this section will be considered below.

\section{Summation formulae over the zeros of modified Bessel functions with
an imaginary order}

\label{sec:SumKi}

\subsection{Summation formula over the zeros of the function $\bar{K}_{iz}(%
\protect\eta )$}

We denote by $z=\omega _{n}=\omega _{n}(\eta )$ the zeros of the function%
\begin{equation}
\bar{K}_{iz}(\eta )\equiv AK_{iz}(\eta )+B\eta \partial _{\eta }K_{iz}(\eta )
\label{Kizbar}
\end{equation}%
in the half-plane $\mathrm{Re}\,z>0$ for a given $\eta $:
\begin{equation}
\bar{K}_{i\omega _{n}}(\eta )=0,\quad n=1,2,\ldots  \label{Kzeros}
\end{equation}%
A summation formula for the series over these zeros can be obtained by
making use of the GAPF. For this, as functions $f(z)$ and $g(z)$ in formula (%
\ref{th12}) we choose
\begin{equation}
f(z)=\frac{2i}{\pi }F(z)\sinh \pi z,\quad g(z)=\frac{\bar{I}_{iz}(\eta )+%
\bar{I}_{-iz}(\eta )}{\bar{K}_{iz}(\eta )}F(z),  \label{fgK}
\end{equation}%
with a function $F(z)$ meromorphic in the right half-plane $\mathrm{Re\,}z>0$%
. For the sum and difference of these functions one finds
\begin{equation}
g(z)\pm f(z)=2F(z)\frac{\bar{I}_{\mp iz}(\eta )}{\bar{K}_{iz}(\eta )}.
\label{gpmfK}
\end{equation}%
By using the asymptotic formulae for the modified Bessel functions for large
values of the index, the conditions for the GAPF can be written in terms of
the function $F(z)$ as follows:
\begin{equation}
|F(z)|<\epsilon (|z|)e^{-\pi x}\left( |z|/\eta \right) ^{2|y|},\quad
z=x+iy,\quad x>0,\quad |z|\rightarrow \infty ,  \label{condFKi}
\end{equation}%
where $|z|\epsilon (|z|)\rightarrow 0$ when $|z|\rightarrow \infty $. Let us
denote by $z_{F,k}$, $k=1,2,\ldots ,$ the poles of the function $F(z)$. By
making use of the fact that the zeros $\omega _{k}$ are simple, one finds
\begin{equation}
R[f(z),g(z)]=2\pi i\left\{ \sum_{k}\frac{\bar{I}_{i\omega _{k}}(\eta
)F(\omega _{k})}{\partial _{z}\bar{K}_{iz}(\eta )|_{z=\omega _{k}}}%
+r_{K}[f(z)]\right\} ,  \label{RfgKiom}
\end{equation}%
with the notation
\begin{eqnarray}
r_{K}[f(z)] &=&\sum_{k,\mathrm{Im}\,z_{F,k}\neq 0}\underset{z=z_{F,k}}{%
\mathrm{Res}}\frac{\bar{I}_{-iz\sigma (z_{F,k})}(\eta )}{\bar{K}_{iz}(\eta )}%
F(z)  \notag \\
&&+\sum_{k,\mathrm{Im}\,z_{F,k}=0}\underset{z=z_{F,k}}{\mathrm{Res}}\frac{%
\bar{I}_{iz}(\eta )+\bar{I}_{-iz}(\eta )}{2\bar{K}_{iz}(\eta )}F(z),
\label{rKi}
\end{eqnarray}%
and the function $\sigma (z)$ is defined in (\ref{cor12}). Substituting (\ref%
{RfgKiom}) into (\ref{th12}), for the function $F(z)$ meromorphic in the
half-plane $\mathrm{Re}\,z\geqslant a$ and satisfying the condition (\ref%
{condFKi}) one obtains the following formula
\begin{eqnarray}
&&\lim_{b\rightarrow +\infty }\left\{ \sum_{k=m}^{n}\frac{\bar{I}_{i\omega
_{k}}(\eta )F(\omega _{k})}{\partial _{z}\bar{K}_{iz}(\eta )|_{z=\omega _{k}}%
}-\frac{1}{\pi ^{2}}\mathrm{p.v.}\int_{a}^{b}dxF(x)\sinh \pi
x+r_{K}[f(z)]\right\}  \notag \\
&&\quad =-\frac{1}{2\pi i}\left[ \int_{a}^{a+i\infty }dzF(z)\frac{\bar{I}%
_{-iz}(\eta )}{\bar{K}_{iz}(\eta )}-\int_{a}^{a-i\infty }dzF(z)\frac{\bar{I}%
_{iz}(\eta )}{\bar{K}_{iz}(\eta )}\right] ,  \label{sumformKi}
\end{eqnarray}%
where $\mathrm{Re}\,\omega _{m-1}<a<\mathrm{Re}\,\omega _{m}$, $\mathrm{Re}%
\,\omega _{n}<b<\mathrm{Re}\,\omega _{n+1}$, and in the expression for $%
r_{K}[f(z)]$ the summation goes over the poles with $a<\mathrm{Re}%
\,z_{F,k}<b $. Below we consider this formula in the limit $a\rightarrow 0$,
assuming that the function $F(z)$ is meromorphic in the right half-plane $%
\mathrm{Re}\,z>0$.

From the relation $\bar{K}_{-iz}(\eta )=\bar{K}_{iz}(\eta )$ it follows that
the purely imaginary zeros of the function $\bar{K}_{iz}(\eta )$ are complex
conjugate. We will denote them by $\pm i\omega _{k}^{\prime }$, $\omega
_{k}^{\prime }>0$, and will assume that possible purely imaginary poles of
the function $F(z)$ are also complex conjugate. Let us denote the latter by $%
\pm iy_{k}$, $y_{k}>0$ and assume that near these poles $F(z)$ satisfies the
condition
\begin{equation}
F(ze^{i\pi })=-F(z)+o((z-iy_{k})^{-1}),\quad z\rightarrow iy_{k}.
\label{condFKi1}
\end{equation}%
In the limit $a\rightarrow 0$ the rhs of formula (\ref{sumformKi}) is
presented as the sum of integrals along the straight segments of the
imaginary axis between the purely imaginary poles plus the terms
\begin{equation}
\sum_{\alpha =+,-}\alpha \left( \int_{\gamma _{\rho }^{\alpha
}}+\sum_{\sigma _{k}=\alpha iy_{k},\alpha \omega _{k}^{\prime
}}\int_{C_{\rho }(\sigma _{k})}\right) F(z)\frac{\bar{I}_{-\alpha iz}(\eta )%
}{\bar{K}_{iz}(\eta )}dz,  \label{purelyimpolesKi1}
\end{equation}%
where the contours $\gamma _{\rho }^{\pm }$ and $C_{\rho }(\sigma _{k})$ are
the same as in (\ref{rel7}). Assuming the relation
\begin{equation}
F(ze^{i\pi })=-F(z)+o(1),\quad z\rightarrow i\omega _{k}^{\prime },
\label{condFKi2}
\end{equation}%
in the limit $\rho \rightarrow 0$ \ the integrals in (\ref{purelyimpolesKi1}%
) are expressed in terms of the residues at the corresponding poles. Noting
also that in this limit the sum of integrals over the straight segments of
the imaginary axis gives the principal value of the integral, one obtains
\cite{SahaRind1}
\begin{eqnarray}
&&\lim_{b\rightarrow +\infty }\left\{ \sum_{k=1}^{n}\frac{\bar{I}_{iz}(\eta
)F(z)}{\partial _{z}\bar{K}_{iz}(\eta )}|_{z=\omega _{k},i\omega
_{k}^{\prime }}-\frac{1}{\pi ^{2}}\mathrm{p.v.}\int_{0}^{b}dxF(x)\sinh \pi
x+r_{K}[f(z)]\right\}  \notag \\
&&\quad =-\frac{1}{2\pi }\mathrm{p.v.}\int_{0}^{\infty }dx\,\frac{\bar{I}%
_{x}(\eta )}{\bar{K}_{x}(\eta )}\left[ F(xe^{\frac{\pi i}{2}})+F(xe^{-\frac{%
\pi i}{2}})\right]  \notag \\
&&\quad \quad -\frac{F_{0}\bar{I}_{0}(\eta )}{2\bar{K}_{0}(\eta )}-\sum_{k}%
\underset{z=iy_{k}}{\mathrm{Res}}F(z)\frac{\bar{I}_{-iz}(\eta )}{\bar{K}%
_{iz}(\eta )},  \label{sumformKi2}
\end{eqnarray}%
where $F_{0}=\lim_{z\rightarrow 0}zF(z)$. The application of this formula to
physical problems will be given below in Section \ref{sec:FuRi}. Taking in
formula (\ref{sumformKi2}) $F(z)=\bar{K}_{iz}(\eta )G(z)$, we obtain a
formula relating two types of integrals with the integration over the index
of the modified Bessel functions. In particular, from the latter formula it
follows that for an integer $m$ one has%
\begin{eqnarray}
\int_{0}^{\infty }dx\,x^{2m}\sinh \pi x\bar{K}_{ix}(\eta )\frac{\cosh ax}{%
\sinh bx} &=&\frac{\pi ^{2}\delta _{m0}}{2b}\bar{I}_{0}(\eta )+\frac{\pi ^{2}%
}{b}\sum_{k=1}^{\infty }(-1)^{k+m}\left( \frac{\pi k}{b}\right) ^{2m}  \notag
\\
&&\times \cos \left( \frac{\pi ka}{b}\right) \bar{I}_{\pi k/b}(\eta ),
\label{IntKi}
\end{eqnarray}%
under the condition $|\mathrm{Re}\,b|-|\mathrm{Re}\,a|>\pi /2$. The special
case of this formula with $B=0$ and $m=0$ is given in \cite{Prud86}.

\subsection{Summation formula over the zeros of the function $Z_{iz}(u,v)$}

In this subsection we will derive a summation formula over the zeros $%
z=\Omega _{n}$ of the function
\begin{equation}
Z_{iz}(u,v)=\bar{K}_{iz}^{(a)}(u)\bar{I}_{iz}^{(b)}(v)-\bar{I}_{iz}^{(a)}(u)%
\bar{K}_{iz}^{(b)}(v),  \label{Zi}
\end{equation}%
where we use the notation defined by formula (\ref{barjnot}). As we will see
in the applications (see Section \ref{sec:FulRin2pl}), the vacuum
expectation values of physical observables in the region between two plates
\ uniformly accelerated through the Fulling-Rindler vacuum are expressed in
terms of series over these zeros. To derive a summation formula for this
type of series, we choose in the GAPF the functions
\begin{equation}
\begin{split}
f(z)=& \frac{2i}{\pi }\sinh \pi z\,F(z), \\
g(z)=& \frac{\bar{I}_{iz}^{(b)}(v)\bar{I}_{-iz}^{(a)}(u)+\bar{I}%
_{iz}^{(a)}(u)\bar{I}_{-iz}^{(b)}(v)}{Z_{iz}(u,v)}F(z),
\end{split}
\label{DfgtoAP}
\end{equation}%
with a meromorphic function $F(z)$ having poles $z=z_{k}$ ($\neq \Omega _{n}$%
), $\mathrm{Im}\,z_{k}\neq 0$, in the right half-plane $\mathrm{Re}\,z>0$.
The zeros $\Omega _{n}$ are simple poles of the function $g(z)$. By taking
into account the relation
\begin{equation}
g(z)\pm f(z)=\frac{2\bar{I}_{\mp iz}^{(a)}(u)\bar{I}_{\pm iz}^{(b)}(v)}{%
Z_{iz}(u,v)}F(z),  \label{Dgpmf}
\end{equation}%
for the function $R[f(z),g(z)]$ in the GAPF one obtains
\begin{eqnarray}
R[f(z),g(z)] &=&2\pi i\left[ \sum_{n=1}^{\infty }\frac{\bar{I}_{-i\Omega
_{n}}^{(b)}(v)\bar{I}_{i\Omega _{n}}^{(a)}(u)}{\partial
_{z}Z_{iz}(u,v)|_{z=\Omega _{n}}}F(\Omega _{n})\right.  \notag \\
&&+\left. \sum_{k}\underset{{z=z_{k}}}{\mathrm{Res}}\frac{F(z)}{Z_{iz}(u,v)}%
\bar{I}_{i\sigma (z_{k})z}^{(b)}(v)\bar{I}_{-i\sigma (z_{k})z}^{(a)}(u)%
\right] ,  \label{DRfg}
\end{eqnarray}%
where the zeros $\Omega _{n}$ are arranged in ascending order, and the
function $\sigma (z)$ is defined in formula (\ref{cor12}). Substituting
relations (\ref{Dgpmf}) and (\ref{DRfg}) into the GAPF, as a special case
the following summation formula is obtained \cite{Avag02,SahaRind2}:
\begin{eqnarray}
\sum_{n=1}^{\infty }\frac{\bar{I}_{i\Omega _{n}}^{(a)}(u)\bar{I}_{-i\Omega
_{n}}^{(b)}(v)}{\partial _{z}Z_{iz}(u,v)|_{z=\Omega _{n}}}F(\Omega _{n}) &=&%
\frac{1}{\pi ^{2}}\int_{0}^{\infty }dx\sinh \pi x\,F(x)  \notag \\
&&-\int_{0}^{\infty }dx\frac{F(xe^{\pi i/2})+F(xe^{-\pi i/2})}{2\pi
Z_{x}(u,v)}\bar{I}_{x}^{(a)}(u)\bar{I}_{-x}^{(b)}(v)  \notag \\
&&-\sum_{k}\underset{z=z_{k}}{\mathrm{Res}}\frac{F(z)}{Z_{iz}(u,v)}\bar{I}%
_{i\sigma (z_{k})z}^{(b)}(v)\bar{I}_{-i\sigma (z_{k})z}^{(a)}(u).
\label{Dsumformula}
\end{eqnarray}%
Here the condition for the function $F(z)$ is obtained from the
corresponding condition in the GAPF by using the asymptotic formulae for the
modified Bessel function and has the form
\begin{equation}
|F(z)|<\epsilon (|z|)e^{-\pi x}\left( v/u\right) ^{2|y|},\quad x>0,\;z=x+iy,
\label{condforAPF2pl}
\end{equation}%
for $|z|\rightarrow \infty $, where $|z|\epsilon (|z|)\rightarrow 0$ when $%
|z|\rightarrow \infty $. Formula (\ref{Dsumformula}) can be generalized for
the case when the function $F(z)$ has real poles, under the assumption that
the first integral on the right of this formula converges in the sense of
the principal value. In this case the first integral on the right of Eq. (%
\ref{Dsumformula}) is understood in the sense of the principal value and
residue terms from real poles in the form $\sum_{k}\underset{{z=z_{k}}}{%
\mathrm{Res}}g(z)$, ${\mathrm{Im}\,z}_{k}=0$, have to be added to the
right-hand side of this formula, with $g(z)$ from (\ref{DfgtoAP}).

\section{Applications to integrals involving Bessel functions}

\label{sec:BessInt}

The applications of the GAPF to infinite integrals involving combinations of
Bessel functions lead to interesting results \cite{Sah1, Sahdis}. First of
all one can express integrals over Bessel functions through the integrals
involving modified Bessel functions. Let us substitute in formula (\ref%
{sumJ1})
\begin{equation}
f(z)=F(z)\bar{J}_{\nu }(z).  \label{fint41}
\end{equation}%
For the function $F(z)$ having no poles at $z=\lambda _{\nu ,k}$ the sum
over zeros of $\bar{J}_{\nu }(z)$ is zero. The conditions (\ref{condf}) and (%
\ref{case21}) may be written in terms of $F(z)$ as
\begin{equation}
|F(z)|<\varepsilon _{1}(x)e^{c_{1}|y|}\quad \text{or}\quad
|F(z)|<M|z|^{-\alpha _{1}}e^{|y|},\quad |z|\rightarrow \infty ,
\label{cond42}
\end{equation}%
with $c_{1}<1$, $x^{1/2-\delta _{B0}}\varepsilon _{1}(x)\rightarrow 0$ for $%
x\rightarrow \infty $, $\alpha _{1}>\alpha _{0}=3/2-\delta _{B0}$, and
\begin{equation}
F(ze^{\pi i})=-e^{\nu \pi i}F(z)+o(z^{|{\mathrm{Re}}\nu |-1}).
\label{cond43}
\end{equation}%
Hence, for the function $F(z)$ satisfying conditions (\ref{cond42}) and (\ref%
{cond43}), from (\ref{sumJ1}) it follows that
\begin{eqnarray}
&&{\mathrm{p.v.}}\! \int_{0}^{\infty }dx\,F(x)\bar{J}_{\nu }(x)=r_{1\nu }%
\left[ F(z)\bar{J}_{\nu }(z)\right] +\frac{\pi }{2}\underset{z=0}{\mathrm{Res%
}}F(z)\bar{Y}_{\nu }(z){}  \notag \\
&&\quad +\frac{1}{\pi }\int_{0}^{\infty }dx\,\bar{K}_{\nu }(x)\left[ e^{-\nu
\pi i/2}F(xe^{\pi i/2})+e^{\nu \pi i/2}F(xe^{-\pi i/2})\right] .
\label{intJform41}
\end{eqnarray}%
In expression (\ref{r1}) for $r_{1\nu }[f(z)]$, the points $z_{k}$ are poles
of the meromorphic function $F(z),\,{\mathrm{Re\,}}z_{k}>0$. On the basis of
Remark after Theorem 3, formula (\ref{intJform41}) may be generalized for
the functions $F(z)$ with purely imaginary poles $\pm iy_{k}$, $y_{k}>0$ and
satisfying the condition
\begin{equation}
F(ze^{\pi i})=-e^{\nu \pi i}F(z)+o\left( (z\mp iy_{k})^{-1}\right) ,\quad
z\rightarrow \pm iy_{k}.  \label{cor3cond1plus2n}
\end{equation}%
The corresponding formula is obtained from (\ref{intJform41}) by adding
residue terms for $z=iy_{k}$ in the form of (\ref{intJform42}) (see below)
and taking the principal value of the integral on the right.

The same substitution of (\ref{fint41}) into formula (\ref{sumJ2}), with the
function $F(z)$ satisfying conditions (\ref{cond42}) and
\begin{equation}
F(xe^{\pi i/2})=-e^{\nu \pi i}F(xe^{-\pi i/2}),  \label{cond44}
\end{equation}%
for real $x$, yields the following result
\begin{equation}
{\mathrm{p.v.}}\! \int_{0}^{\infty }dx\,F(x)\bar{J}_{\nu }(x)=r_{1\nu }\left[
F(z)\bar{J}_{\nu }(z)\right] +\frac{\pi i}{2}\sum_{\eta _{k}=0,iy_{k}}\left(
2-\delta _{0\eta _{k}}\right) \underset{z=\eta _{k}}{\mathrm{Res}}F(z)\bar{H}%
_{\nu }^{(1)}(z).  \label{intJform42}
\end{equation}%
In (\ref{r1}), the summation goes over the poles $z_{k}$, ${\mathrm{Re}}%
z_{k}>0$ of the meromorphic function $F(z)$, and $\pm iy_{k},\,y_{k}>0$ are
purely imaginary poles of this function. Recall that the possible real poles
of $F(z)$ are such, that integral on the left of (\ref{intJform42}) exists.

For the functions $F(z)=z^{\nu +1}\tilde{F}(z)$, with $\tilde{F}(z)$ being
analytic in the right half-plane and even along the imaginary axis, $\tilde{F%
}(ix)=\tilde{F}(-ix)$, one obtains
\begin{equation}
\int_{0}^{\infty }dx\,x^{\nu +1}\tilde{F}(x)\bar{J}_{\nu }(x)=0.
\label{examp41}
\end{equation}%
This result for $B=0$ (see (\ref{efnot1})) has been given previously in \cite%
{Schwartz}. Another result of \cite{Schwartz} is obtained from (\ref%
{intJform42}) choosing $F(z)=z^{\nu +1}\tilde{F}(z)/(z^{2}-a^{2})$.

Formulae similar to (\ref{intJform41}) and (\ref{intJform42}) can be derived
for the Neumann function $Y_{\nu }(z)$. Let for the function $F(z)$ the
integral ${\mathrm{p.v.}}\! \int_{0}^{\infty }dx\,F(x)\bar{Y}_{\nu }(x)$
exists. We substitute in formula (\ref{th12})
\begin{equation}
f(z)=Y_{\nu }(z)F(z),\quad g(z)=-iJ_{\nu }(z)F(z),  \label{fgneum41}
\end{equation}%
and consider the limit $a\rightarrow +0$. The terms containing residues may
be presented in the form
\begin{eqnarray}
R[f(z),g(z)] &=&\pi \sum_{k}\underset{{\mathrm{Im}}z_{k}>0}{\mathrm{Res}}%
H_{\nu }^{(1)}(z)F(z)+\pi \sum_{k}\underset{{\mathrm{Im}}z_{k}<0}{\mathrm{Res%
}}H_{\nu }^{(2)}(z)F(z){}  \notag \\
&&+\pi \sum_{k}\underset{{\mathrm{Im}}z_{k}=0}{\mathrm{Res}}J_{\nu
}(z)F(z)\equiv r_{3\nu }[F(z)],  \label{r2}
\end{eqnarray}%
where $z_{k}$ (${\mathrm{Re}}z_{k}>0$) are the poles of $F(z)$ in the right
half-plane. Now the following results can be proved by using (\ref{th12}):

\bigskip

\textit{1) If the meromorphic function $F(z)$ has no poles on the imaginary
axis and satisfies condition (\ref{cond42}), then
\begin{eqnarray}
{\mathrm{p.v.}}\! \int_{0}^{\infty }{dx\,F(x)Y_{\nu }(x)} &=&r_{3\nu }[F(z)]-%
\frac{i}{\pi }\int_{0}^{\infty }{dx\,K_{\nu }(x)}  \notag \\
&&\times {\left[ e^{-\nu \pi i/2}F(xe^{\pi i/2})-e^{\nu \pi i/2}F(xe^{-\pi
i/2})\right] ,}  \label{intYform41}
\end{eqnarray}%
}

\noindent and

\bigskip

\textit{2) If the meromorphic function $F(z)$ satisfies the conditions
\begin{equation}
F(xe^{\pi i/2})=e^{\nu \pi i}F(xe^{-\pi i/2})  \label{cond45}
\end{equation}%
and (\ref{cond42}), then one has
\begin{equation}
{\mathrm{p.v.}}\! \int_{0}^{\infty }{dx\,F(x)Y_{\nu }(x)}=r_{3\nu
}[F(z)]+\pi \sum_{k}\underset{z=iy_{k}}{\mathrm{Res}}H_{\nu }^{(1)}(z)F(z),
\label{intYform42}
\end{equation}%
where $\pm iy_{k},\,y_{k}>0$ are purely imaginary poles of $F(z)$.}

\bigskip

{}From (\ref{intYform42}) it directly follows that for $F(z)=z^{\nu }\tilde{F%
}(z)$, with $\tilde{F}(z)$ being even along the imaginary axis, $\tilde{F}%
(ix)=\tilde{F}(-ix)$, and analytic in the right half-plane one has \cite%
{Schwartz}:
\begin{equation}
\int_{0}^{\infty }dx\,x^{\nu }\tilde{F}(x)Y_{\nu }(x)=0,  \label{examp42}
\end{equation}%
if condition (\ref{cond42}) takes place.

Let us consider more general case. Let the function $F(z)$ satisfy the
condition
\begin{equation}
F(ze^{-\pi i})=-e^{-\lambda \pi i}F(z),  \label{case42}
\end{equation}%
for ${\mathrm{arg\,}}z=\pi /2$. In the GAPF as functions $f(z)$ and $g(z)$
we choose
\begin{eqnarray}
f(z) &=&F(z)\left[ J_{\nu }(z)\cos \delta +Y_{\nu }(z)\sin \delta \right] ,
\notag \\
g(z) &=&-iF(z)[J_{\nu }(z)\sin \delta -Y_{\nu }(z)\cos \delta ],\quad \delta
=(\lambda -\nu )\pi /2,  \label{fgcomb42}
\end{eqnarray}%
with $g(z)-(-1)^{k}f(z)=H_{\nu }^{(k)}(z)F(z)\exp [(-1)^{k}i\delta ]$, $%
k=1,2 $. It can be seen that for such a choice the integral on the rhs of (%
\ref{th12}) for $a\rightarrow 0$ is equal to
\begin{equation}
\pi i\sum_{\eta _{k}=iy_{k}}\underset{z=\eta _{k}}{\mathrm{Res}}H_{\nu
}^{(1)}(z)F(z)e^{-i\delta },  \label{rel42}
\end{equation}%
where $\pm iy_{k},\,y_{k}>0$, as above, are purely imaginary poles of $F(z)$%
. Substituting (\ref{fgcomb42}) into (\ref{th12}) and using (\ref{cor13}) we
obtain \cite{Sah1, Sahdis}:

\bigskip

\noindent \textbf{Corollary 4.} \textit{Let $F(z)$ be a meromorphic function
for ${\mathrm{Re\,}}z\geqslant 0$ (except possibly at $z=0$) with poles $%
z_{k},\,\pm iy_{k}$; $y_{k},{\mathrm{Re\,}}z_{k}>0$. If this function
satisfies conditions (\ref{cond42}) (for $B=0$) and (\ref{case42}) then
\begin{eqnarray}
&&{\mathrm{p.v.}}\! \int_{0}^{\infty }{dx\,F(x)\left[ J_{\nu }(x)\cos \delta
+Y_{\nu }(x)\sin \delta \right] }=\pi i\Bigg\{\sum_{z_{k}}\underset{{\mathrm{%
Im}}z_{k}>0}{\mathrm{Res}}H_{\nu }^{(1)}(z)F(z)e^{-i\delta }  \notag \\
&&\quad -\sum_{z_{k}}\underset{{\mathrm{Im}}z_{k}<0}{\mathrm{Res}}H_{\nu
}^{(2)}(z)F(z)e^{i\delta }-i\sum_{z_{k}}\underset{{\mathrm{Im}}z_{k}=0}{%
\mathrm{Res}}\left[ J_{\nu }(z)\sin \delta -Y_{\nu }(z)\cos \delta \right]
F(z)  \notag \\
&&\quad +\sum_{\eta _{k}=iy_{k}}\underset{z=\eta _{k}}{\mathrm{Res}}H_{\nu
}^{(1)}(z)F(z)e^{-i\delta }\Bigg\},  \label{intJYform43}
\end{eqnarray}%
where it is assumed that the integral on the left exists.}

\bigskip

\noindent In particular for $\delta =\pi n,\,n=0,1,2...$, formula (\ref%
{intJform42}) follows from here in the case $B=0$. One will find many
particular cases of formulae (\ref{intJform42}) and (\ref{intJYform43})
looking at the standard books and tables of known integrals with Bessel
functions (see, e.g., \cite{Erde53, Watson, Magnus,
Prud86,Luke,Gradshteyn,Wheelon,Oberhettinger}). Some special examples are
given in the next section.

\section{Integrals involving Bessel functions: Illustrations of general
formulae}

\label{sec:BessInt2}

In order to illustrate the applications of general formulae from the
previous section, first we consider integrals involving the function $\bar{J}%
_{\nu }(z)$. Let us introduce the functional
\begin{equation}
A_{\nu m}[G(z)]\equiv {\mathrm{p.v.}}\! \int_{0}^{\infty }dz\,z^{\nu
-2m-1}G(z)\bar{J}_{\nu }(z),  \label{Anjuem}
\end{equation}%
where $m$ is an integer. Let $F_{1}(z)$ be an analytic function in the right
half-plane satisfying the condition
\begin{equation}
F_{1}(xe^{\pi i/2})=F_{1}(xe^{-\pi i/2}),\quad F_{1}(0)\neq 0,
\label{cond5F1}
\end{equation}%
(the case when $F_{1}(z)\sim z^{q},\,z\rightarrow 0$, with an integer $q$,
can be reduced to this one by redefinitions of $F_{1}(z)$ and $m$). {}From (%
\ref{intJform42}) the following results can be obtained \cite{Sah1, Sahdis}:
\begin{eqnarray}
A_{\nu m}[F_{1}(z)] &=&A_{\nu m}^{(0)}[F_{1}(z)]\equiv -\frac{\pi (1+{%
\mathrm{sgn\,}}m)}{4(2m)!}\partial _{z}^{2m}\left[ z^{\nu }\bar{Y}_{\nu
}(z)F_{1}(z)\right] _{z=0},  \label{examp51} \\
A_{\nu m}\left[ \frac{F_{1}(z)}{z^{2}-a^{2}}\right] &=&-\frac{\pi }{2}a^{\nu
-2m-2}\bar{Y}_{\nu }(a)F_{1}(a)+A_{\nu m}^{(0)}\left[ \frac{F_{1}(z)}{%
z^{2}-a^{2}}\right] ,\quad  \label{examp52} \\
A_{\nu m}\left[ \frac{F_{1}(z)}{z^{4}-a^{4}}\right] &=&-\frac{a^{\nu -2m-4}}{%
2}\left[ \frac{\pi }{2}\bar{Y}_{\nu }(a)F_{1}(a)-(-1)^{m}\bar{K}_{\nu
}(a)F_{1}(ia)\right] +  \notag \\
&&+A_{\nu m}^{(0)}\left[ \frac{F_{1}(z)}{z^{4}-a^{4}}\right] ,
\label{examp53} \\
A_{\nu m}\left[ \frac{F_{1}(z)}{\left( z^{2}-c^{2}\right) ^{p+1}}\right] &=&%
\frac{\pi i}{2^{p+1}p!}\left( \frac{d}{cdc}\right) ^{p}\left[ c^{\nu
-2m-2}F_{1}(c)H_{\nu }^{(1)}(c)\right]  \notag \\
&&+A_{\nu m}^{(0)}\left[ \frac{F_{1}(z)}{\left( z^{2}-c^{2}\right) ^{p+1}}%
\right]  \label{examp54} \\
A_{\nu m}\left[ \frac{F_{1}(z)}{\left( z^{2}+a^{2}\right) ^{p+1}}\right] &=&%
\frac{(-1)^{m+p+1}}{2^{p}\cdot p!}\left( \frac{d}{ada}\right) ^{p}\left[
a^{\nu -2m-2}K_{\nu }(a)F_{1}(ae^{\pi i/2})\right]  \notag \\
&&+A_{\nu m}^{(0)}\left[ \frac{F_{1}(z)}{\left( z^{2}+a^{2}\right) ^{p+1}}%
\right] ,  \label{examp54ad}
\end{eqnarray}%
and etc. Note that $A_{\nu m}^{(0)}=0$ for $m<0$. In these formulae $a>0$, $%
0<{\mathrm{arg\,}}c<\pi /2$, and we have assumed that ${\mathrm{Re\,}}\nu >0$%
. To secure convergence at the origin the condition ${\mathrm{Re\,}}\nu >m$
should be satisfied. In the last two formulae we have used the identity
\begin{equation}
\left( \frac{d}{dz}\right) ^{p}\left[ \frac{zF(z)}{(z+b)^{p+1}}\right]
_{z=b}=\frac{1}{2^{p+1}}\left( \frac{d}{bdb}\right) ^{p}F(b).  \label{ident1}
\end{equation}%
Note that (\ref{examp54ad}) can also be obtained from (\ref{examp54}) in the
limit ${\mathrm{Re\,}}c\rightarrow 0$. For the case $F_{1}=1,\,m=-1$ in (\ref%
{examp54ad}) see, for example, \cite{Watson}. In (\ref{examp51})-(\ref%
{examp54ad}) as a function $F_{1}(z)$ we can choose:

\begin{itemize}
\item function (\ref{func1}) for $\rho =1$, ${\mathrm{Re\,}}\nu <\sum {%
\mathrm{Re\,}}\mu _{l}+2m+(n+1)/2-\delta _{b1}+\delta _{B0}$, $b=\sum
b_{l}\leqslant 1,\,b_{l}>0$;

\item function (\ref{func2n}) with $\rho =1$, ${\mathrm{Re\,}}\nu
<2(m+n)+1/2-\delta _{b1}+\delta _{B0}$, $b=\sum b_{l}\leqslant 1$;

\item function (\ref{func3}) for $\rho =1$, ${\mathrm{Re\,}}\nu <2m-\sum {%
\mathrm{Re\,}}\mu _{l}+(n+1)/2-\delta _{b1}+\delta _{B0}$, $\mu _{l}>0$ is
half of an odd integer, $b=\sum b_{l}\leqslant 1$;

\item function (\ref{func4}) for ${\mathrm{Re\,}}\nu <2m+n-\sum
|k_{l}|+1/2+\delta _{B0}-\delta _{\tilde{a}1}$, $\tilde{a}=\sum
|a_{l}-b_{l}|\leqslant 1$, $a_{l}\geqslant 0$, $k_{l}$ - integer.
\end{itemize}

\noindent Here we have written the conditions for (\ref{examp51}). The
corresponding ones for (\ref{examp52}), (\ref{examp53}), (\ref{examp54}), (%
\ref{examp54ad}) are obtained by adding on the rhs of inequalities for ${%
\mathrm{Re\,}}\nu $, respectively 2, 4, $2(p+1)$, $2(p+1)$. In (\ref{examp51}%
)-(\ref{examp54ad}) we can also choose any combinations of functions (\ref%
{func1})-(\ref{func4}) with appropriate conditions.

For the evaluation of $A_{\nu m}^{(0)}$ in special cases the following
formula is useful
\begin{equation}
\lim_{z\rightarrow 0}\left( \frac{d}{dz}\right)
^{2m}f_{1}(z)=(2m-1)!!\lim_{z\rightarrow 0}\left( \frac{d}{zdz}\right)
^{m}f_{1}(z),  \label{rel51}
\end{equation}%
valid for the function $f_{1}(z)$ satisfying condition $%
f_{1}(-z)=f_{1}(z)+o(z^{2m})$, $z\rightarrow 0$. From here, for instance, it
follows that for $z\rightarrow 0$
\begin{eqnarray}
\left( \frac{d}{dz}\right) ^{2m}\left[ z^{\nu }Y_{\nu }(bz)F_{1}(z)\right]
&=&-(2m-1)!!\frac{2^{\nu -m}}{\pi b^{\nu -m}}\sum_{k=0}^{m}2^{k}{\binom{m}{k}%
}  \notag \\
&&\times \frac{\Gamma (\nu -m+k)}{b^{2k}}\left( \frac{d}{zdz}\right)
^{k}F_{1}(z),  \label{rel52}
\end{eqnarray}%
where we have used the standard formula for the derivative $(d/zdz)^{n}$ of
cylinder functions (see \cite{abramowiz}). From (\ref{examp51}) one obtains (%
$B=0$)
\begin{eqnarray}
&&\int_{0}^{\infty }dz\,z^{\nu -2m-1}J_{\nu }(z)\prod_{l=1}^{n}\left(
z^{2}+z_{l}^{2}\right) ^{-\mu _{l}/2}J_{\mu _{l}}(b_{l}\sqrt{z^{2}+z_{l}^{2}}%
)=\frac{-\pi }{2^{m+1}m!}  \notag \\
&&\times \left( \frac{d}{zdz}\right) ^{m}\left[ z^{\nu }Y_{\nu
}(z)\prod_{l=1}^{n}\left( z^{2}+z_{l}^{2}\right) ^{-\mu _{l}/2}J_{\mu
_{l}}(b_{l}\sqrt{z^{2}+z_{l}^{2}})\right] _{z=0},  \label{examp55}
\end{eqnarray}%
for $m\geqslant 0$ and the integral is zero for $m<0$. Here ${\mathrm{Re\,}}%
\nu >0$, $b\equiv \sum_{1}^{n}b_{l}\leqslant 1$, $b_{l}>0$, $m<{\mathrm{Re\,}%
}\nu <\sum_{l=1}^{n}{\mathrm{Re\,}}\mu _{l}+2m+(n+3)/2-\delta _{b1}$. In the
particular case $m=0$ we obtain the Gegenbauer integral \cite{Erde53,Watson}%
. In the limit $z_{l}\rightarrow 0$, from (\ref{examp55}) the value of
integral $\int_{0}^{\infty }dz\,z^{\nu -2m-1}J_{\nu
}(z)\prod_{l=1}^{n}z^{-\mu _{l}}J_{\mu _{l}}(z)$ is obtained.

By using (\ref{intJYform43}), formulae similar to (\ref{examp51})-(\ref%
{examp54ad}) may be derived for the integrals of type
\begin{equation}
B_{\nu }[G(z)]\equiv {\mathrm{p.v.}}\! \int_{0}^{\infty }dz\,G(z)\left[
J_{\nu }(z)\cos \delta +Y_{\nu }(z)\sin \delta \right] ,\quad \delta
=(\lambda -\nu )\pi /2.  \label{Bnju}
\end{equation}%
It directly follows from Corollary 4 that for function $F(z)$ analytic for ${%
\mathrm{Re\,}}z\geqslant 0$ and satisfying conditions (\ref{cond42}) and (%
\ref{case42}) the following formulae take place
\begin{eqnarray}
B_{\nu }[F(z)] &=&0  \label{examp56} \\
B_{\nu }\left[ \frac{F(z)}{z^{2}-a^{2}}\right] &=&\pi F(a)\left[ J_{\nu
}(a)\sin \delta -Y_{\nu }(a)\cos \delta \right] /2,\quad  \label{examp57} \\
B_{\nu }\left[ \frac{F(z)}{z^{4}-a^{4}}\right] &=&\frac{\pi }{4a^{3}}F(a)%
\left[ J_{\nu }(a)\sin \delta -Y_{\nu }(a)\cos \delta \right]  \notag \\
&&+\frac{i}{2a^{3}}K_{\nu }(a)F(ia)e^{-i\lambda \pi /2},  \label{examp58} \\
B_{\nu }\left[ \frac{F_{1}(z)}{\left( z^{2}-c^{2}\right) ^{p+1}}\right] &=&%
\frac{\pi i}{2^{p+1}\cdot p!}\left( \frac{d}{cdc}\right) ^{p}\left[
c^{-1}F(c)H_{\nu }^{(1)}(c)\right] e^{-i\delta }  \label{examp59} \\
B_{\nu }\left[ \frac{F_{1}(z)}{\left( z^{2}+a^{2}\right) ^{p+1}}\right] &=&%
\frac{(-1)^{p+1}}{2^{p}\cdot p!}\left( \frac{d}{ada}\right) ^{p}\left[
a^{-1}F(ae^{\pi i/2})K_{\nu }(a)\right] e^{-i\pi \lambda /2},
\label{examp59ad}
\end{eqnarray}%
where $a>0$, $0<{\mathrm{arg\,}}c\leqslant \pi /2$. To obtain the last two
formulae we have used identity (\ref{ident1}). Formula (\ref{examp56})
generalizes the result of \cite{Schwartz} (the cases $\lambda =\nu $ and $%
\lambda =\nu +1$). Taking $F(z)=z^{\lambda -1}$ {}from the last formula we
obtain the result given in \cite{Watson}. In (\ref{examp56})-(\ref{examp59ad}%
), as a function $F(z)$ one can choose (the constraints on the parameters
are written for formula (\ref{examp56}); the corresponding constraints for (%
\ref{examp57}), (\ref{examp58}), (\ref{examp59}), (\ref{examp59ad}) are
obtained from the ones given by adding 2, 4, $2(p+1)$, $2(p+1)$ to the rhs
of inequalities for ${\mathrm{Re\,}}\nu $, correspondingly):

\begin{itemize}
\item function (\ref{func1}) for $\rho =\lambda $, $|{\mathrm{Re\,}}\nu |<{%
\mathrm{Re\,}}\rho <\sum {\mathrm{Re\,}}\mu _{l}+(n+3)/2-\delta _{b1}$, $%
b=\sum_{l}b_{l}\leqslant 1$;

\item function (\ref{func2n}) for $\rho =\lambda $, $|{\mathrm{Re\,}}\nu |<{%
\mathrm{Re\,}}\rho <3/2-\delta _{b1}$, $b=\sum_{l}b_{l}\leqslant 1$;

\item function
\begin{eqnarray}
&&z^{\rho -1}\prod_{l=1}^{n}\left[ J_{\mu _{l}+k_{l}}(a_{l}z)Y_{\mu
_{l}}(b_{l}z)-Y_{\mu _{l}+k_{l}}(a_{l}z)J_{\mu _{l}}(b_{l}z)\right] ,\quad
\lambda =\rho +\sum_{l=1}^{n}k_{l},\,  \notag \\
&&\quad a_{l}>0,\;|{\mathrm{Re\,}}\nu |+\sum |k_{l}|<{\mathrm{Re\,}}\rho
<n+3/2-\delta _{c1},\,  \notag \\
&&\quad c=\sum |a_{l}-b_{l}|\leqslant 1,\,{\mathrm{Re\,}}\mu _{l}\geqslant 0,
\label{func5new}
\end{eqnarray}%
(for ${\mathrm{Re\,}}\mu _{l}<0$ one has $k_{l}>|{\mathrm{Re\,}}\mu _{l}|$).
\end{itemize}

\noindent Any combination of these functions with appropriate conditions on
parameters can be chosen as well.

Now consider integrals which can be expressed via series by using formulae (%
\ref{intJform42}) and (\ref{intJYform43}). In (\ref{intJform42}) let us
choose the function
\begin{equation}
F(z)=\frac{z^{\nu -2m}F_{1}(z)}{\sinh \pi z},  \label{examp510}
\end{equation}%
where $F_{1}(z)$ is the same as in formulae (\ref{examp51})-(\ref{examp54}).
As the points $\pm i,\pm 2i,\ldots $ are simple poles for $F(z)$, from (\ref%
{intJform42}) one obtains
\begin{equation}
\int_{0}^{\infty }dz\,\frac{z^{\nu -2m}\bar{J}_{\nu }(z)}{\sinh {(\pi z)}}%
F_{1}(z)=A_{\nu m}^{(0)}\left[ \frac{zF_{1}(z)}{\sinh (\pi z)}\right] +\frac{%
2}{\pi }\sum_{k=1}^{\infty }(-1)^{m+k}k^{\nu -2m}\bar{K}_{\nu }(k)F_{1}(ik),
\label{intJ5sum}
\end{equation}%
where $A_{\nu m}^{(0)}[f(z)]$ is defined by (\ref{examp51}) and ${\mathrm{%
Re\,}}\nu >m$. The corresponding constraints on $F_{1}(z)$ follow
directly
from (\ref{cond42}). The particular case of this formula when $%
F_{1}(z)=\sinh (az)/z$ and $m=-1$ is given in \cite{Watson}. As a function $%
F_{1}(z)$ here one can choose any of functions (\ref{func1})-(\ref{func4})
with $\rho =1$ and $\tilde{a},\,\sum_{l}b_{l}<1$. From (\ref{intJ5sum}) it
follows that
\begin{eqnarray}
&&\int_{0}^{\infty }dz\,\frac{z^{\nu -2m}}{\sinh {(\pi z)}}J_{\nu
}(z)\prod_{l=1}^{n}z^{-\mu _{l}}I_{\mu _{l}}(b_{l}z)=A_{\nu m}^{(0)}\left[
\frac{z}{\sinh (\pi z)}\prod_{l=1}^{n}z^{-\mu _{l}}I_{\mu _{l}}(b_{l}z)%
\right]  \notag \\
&&+\frac{2}{\pi }\sum_{k=1}^{\infty }(-1)^{m+k}K_{\nu
}(k)\prod_{l=1}^{n}k^{-\mu _{l}}J_{\mu
_{l}}(b_{l}k),\;b_{l}>0,\,\sum_{l=1}^{n}b_{l}<\pi ,\,{\mathrm{Re}}\nu >m.
\label{examp511}
\end{eqnarray}%
In a similar way the following formula can be derived from (\ref{intJYform43}%
):
\begin{equation}
\int_{0}^{\infty }dz\,\frac{zF(z)}{\sinh (\pi z)}\left[ J_{\nu }(z)\cos
\delta +Y_{\nu }(z)\sin \delta \right] =\frac{2i}{\pi }e^{-i\lambda \pi
/2}\sum_{k=1}^{\infty }(-1)^{k}kK_{\nu }(k)F(ik),  \label{intJY5sum}
\end{equation}%
where $\lambda $ is defined by relation (\ref{case42}). The constraints on
the function $F(z)$ immediately follow from Corollary 4. Instead of this
function we can choose functions (\ref{func1}), (\ref{func2n}), (\ref{func4}%
).

As it has been mentioned above, adding the residue terms $\pi i\underset{%
z=iy_{k}}{\mathrm{Res}}F(z)\bar{H}_{\nu }^{(1)}(z)$ to the rhs of (\ref%
{intJform41}) this formula may be generalized for functions having purely
imaginary poles $\pm iy_{k}$, $y_{k}>0$, provided condition (\ref%
{cor3cond1plus2n}) is satisfied. As an application let us choose
\begin{equation}
F(z)=\frac{z^{\nu }F_{1}(z)}{e^{2\pi z/b}-1},\quad F_{1}(-z)=F_{1}(z),\quad
b>0,  \label{exampforpole}
\end{equation}%
with an analytic function $F_{1}(z)$. Function (\ref{exampforpole})
satisfies condition (\ref{cor3cond1plus2n}) and has poles $\pm ikb$, $%
k=0,1,2\ldots $. The additional constraint directly follows from (\ref%
{cond42}). Then one obtains
\begin{equation}
\int_{0}^{\infty }dx\,\frac{x^{\nu }J_{\nu }(x)}{e^{2\pi x/b}-1}F_{1}(x)=%
\frac{2}{\pi }\sideset{}{'}{\sum}_{k=0}^{\infty }(bk)^{\nu }K_{\nu
}(bk)F_{1}(ibk)-\frac{1}{\pi }\int_{0}^{\infty }dx\,x^{\nu }K_{\nu
}(x)F_{1}(ix),  \label{exampforpoleform}
\end{equation}%
where the prime indicates that the $m=0$ term is to be halved. For the
particular case $F_{1}(z)=1$, using the relation
\begin{equation}
\sideset{}{'}{\sum}_{k=0}^{\infty }(bk)^{\nu }K_{\nu }(bk)=\frac{\sqrt{\pi }%
}{b}2^{\nu }\Gamma (\nu +1/2)\sideset{}{'}{\sum}_{n=0}^{\infty }\left[ (2\pi
n/b)^{2}+1\right] ^{-\nu -1/2},  \label{relipole}
\end{equation}%
and the known value for the integral on the right, we immediately obtain the
result given in \cite{Watson}. Relation (\ref{relipole}) can be proved by
using the formulae (for the integral representation of McDonald function see
\cite{Watson})
\begin{equation}
K_{\nu }(z)=\frac{2^{\nu }\Gamma (\nu +1/2)}{\sqrt{\pi }z^{\nu }}%
\int_{0}^{\infty }\frac{\cos zt\,dt}{(t^{2}+1)^{\nu +1/2}},\quad
\sum_{k=-\infty }^{+\infty }e^{ikz}=2\pi \sum_{n=-\infty }^{+\infty }\delta
(z-2\pi n),  \label{relipole1}
\end{equation}%
where $\delta (z)$ is the Dirac delta function.

\section{Formulae for integrals involving $J_\protect\protect\nu %
(z)Y_\protect\protect\mu (\protect\lambda z)-Y_\protect\protect\nu %
(z)J_\protect\protect\mu (\protect\lambda z)$}

\label{sec:BessInt4}

In this section we will consider applications of the GAPF to integrals
involving the function $J_{\nu }(z)Y_{\mu }(\lambda z)-Y_{\nu }(z)J_{\mu
}(\lambda z)$. In formula (\ref{th12}) we substitute
\begin{equation}
f(z)=-\frac{1}{2i}F(z)\sum_{l=1}^{2}(-1)^{l}\frac{H_{\mu }^{(l)}(\lambda z)}{%
H_{\nu }^{(l)}(z)},\quad g(z)=\frac{1}{2i}F(z)\sum_{l=1}^{2}\frac{H_{\mu
}^{(l)}(\lambda z)}{H_{\nu }^{(l)}(z)}\,.  \label{fg61}
\end{equation}%
For definiteness we consider the case $\lambda >1$ (for $\lambda <1$ the
expression for $g(z)$ has to be chosen with opposite sign). Conditions (\ref%
{cor11}) and (\ref{th11}) are satisfied if the function $F(z)$ is
constrained by one of the following two inequalities
\begin{equation}
|F(z)|<\varepsilon (x)e^{c|y|},\quad c<\lambda -1,\,\varepsilon
(x)\rightarrow 0,\,x\rightarrow +\infty ,  \label{cond61}
\end{equation}%
or
\begin{equation}
|F(z)|<M|z|^{-\alpha }e^{(\lambda -1)|y|},\quad \alpha >1,\,|z|\rightarrow
\infty ,\,z=x+iy.  \label{cond62}
\end{equation}%
Then, from (\ref{th12}) it follows that for the function $F(z)$ meromorphic
in ${\mathrm{Re\,}}z\geqslant a>0$ one has
\begin{eqnarray}
&&{\mathrm{p.v.}}\! \int_{0}^{\infty }dx\,\frac{J_{\nu }(x)Y_{\mu }(\lambda
x)-J_{\mu }(\lambda x)Y_{\nu }(x)}{J_{\nu }^{2}(x)+Y_{\nu }^{2}(x)}%
F(x)=r_{1\mu \nu }[F(z)]  \notag \\
&&\quad +\frac{1}{2i}\left[ \int_{a}^{a+i\infty }dz\,F(z)\frac{H_{\mu
}^{(1)}(\lambda z)}{H_{\nu }^{(1)}(z)}-\int_{a}^{a-i\infty }dz\,F(z)\frac{%
H_{\mu }^{(2)}(\lambda z)}{H_{\nu }^{(2)}(z)}\right] ,  \label{intJY61}
\end{eqnarray}%
where we have introduced the notation
\begin{eqnarray}
r_{1\mu \nu }[F(z)] &=&\frac{\pi }{2}\sum_{k}\underset{{\mathrm{Im}}z_{k}=0}{%
\mathrm{Res}}\left[ F(z)\sum_{l=1}^{2}\frac{H_{\mu }^{(l)}(\lambda z)}{%
H_{\nu }^{(l)}(z)}\right]  \notag \\
&&+\pi \sum_{k}\sum_{l=1}^{2}\underset{(-1)^{l}{\mathrm{Im}}z_{k}<0}{\mathrm{%
Res}}\left[ F(z)\frac{H_{\mu }^{(l)}(\lambda z)}{H_{\nu }^{(l)}(z)}\right] .
\label{r1mn}
\end{eqnarray}%
The most important case for the applications is the limit $a\rightarrow 0$.
The following statements take place \cite{Sah1, Sahdis}:

\bigskip

\noindent \textbf{Theorem 5.} \textit{Let the function $F(z)$ be meromorphic
for ${\mathrm{Re\,}}z\geqslant 0$ (except the possible branch point $z=0$)
with poles $z_{k},\,\pm iy_{k}$ ($y_{k},{\mathrm{Re\,}}z_{k}>0$). If this
function satisfies conditions (\ref{cond61}) or (\ref{cond62}) and
\begin{equation}
F(xe^{\pi i/2})=-e^{(\mu -\nu )\pi i}F(xe^{-\pi i/2}),  \label{condth65}
\end{equation}%
then, for values of $\nu $ for which the function $H_{\nu }^{(1)}(z)$ ($%
H_{\nu }^{(2)}(z)$) has no zeros for $0\leqslant \mathrm{arg\,}z\leqslant
\pi /2$ ($-\pi /2\leqslant \mathrm{arg\,\,}z\leqslant 0$), the following
formula is valid
\begin{eqnarray}
&&{\mathrm{p.v.}}\! \int_{0}^{\infty }{dx\,\frac{J_{\nu }(x)Y_{\mu }(\lambda
x)-Y_{\nu }(x)J_{\mu }(\lambda x)}{J_{\nu }^{2}(x)+Y_{\nu }^{2}(x)}F(x)}%
=r_{1\mu \nu }[F(z)]+  \notag \\
&&\quad +\frac{\pi }{2}\sum_{\eta _{k}=0,iy_{k}}\left( 2-\delta _{0\eta
_{k}}\right) \underset{z=\eta _{k}}{\mathrm{Res}}F(z)\frac{H_{\mu
}^{(1)}(\lambda z)}{H_{\nu }^{(1)}(z)},  \label{intJYth65}
\end{eqnarray}%
where it is assumed that the integral on the left exists.}

\bigskip

\noindent \textbf{Proof.} From condition (\ref{condth65}) it follows that
for ${\mathrm{arg\,}}z=\pi /2$ one has
\begin{equation}
\frac{H_{\mu }^{(1)}(\lambda z)}{H_{\nu }^{(1)}(z)}F(z)=\frac{H_{\mu
}^{(2)}(\lambda ze^{-\pi i})}{H_{\nu }^{(2)}(ze^{-\pi i})}F(ze^{-\pi i}),
\label{rel61}
\end{equation}%
and that the possible purely imaginary poles of $F(z)$ are conjugate: $\pm
iy_{k},\,y_{k}>0$. Hence, on the rhs of (\ref{intJY61}), in the limit $%
a\rightarrow 0$ the term in the square brackets may be presented in the form
(it can be seen in a way similar to that used for (\ref{rel11}))
\begin{equation}
\sum_{\alpha =+,-}\left( \int_{\gamma _{\rho }^{\alpha
}}+\sum_{k}\int_{C_{\rho }(\alpha iy_{k})}\right) dz\,\frac{H_{\mu
}^{(p_{\alpha })}(\lambda z)}{H_{\nu }^{(p_{\alpha })}(z)}F(z),
\label{rel62}
\end{equation}%
with the same notations as in (\ref{rel11}). By using (\ref{rel61}) and the
condition that the integral converges at the origin, we obtain
\begin{equation}
\sum_{\alpha =+,-}\int_{\Omega _{\rho }^{\alpha }(\eta _{k})}dz\,\frac{%
H_{\mu }^{(p_{\alpha })}(\lambda z)}{H_{\nu }^{(p_{\alpha })}(z)}F(z)=\left(
2-\delta _{0\eta _{k}}\right) \pi i\,\underset{z=\eta _{k}}{\mathrm{Res}}%
\frac{H_{\mu }^{(1)}(\lambda z)}{H_{\nu }^{(1)}(z)}F(z),  \label{rel63}
\end{equation}%
where $\Omega _{\rho }^{\pm }(0)=\gamma _{\rho }^{\pm }$, $\Omega _{\rho
}^{\pm }(iy_{k})=C_{\rho }(\pm iy_{k})$. By using this relation, {}from (\ref%
{intJY61}) we receive formula (\ref{intJYth65}). \rule{1.5ex}{1.5ex}

\bigskip

Note that one can also write the residue at $z=0$ in the form
\begin{equation}
\underset{z=0}{\mathrm{Res}}\frac{H_{\mu }^{(1)}(\lambda z)}{H_{\nu
}^{(1)}(z)}F(z)=\underset{z=0}{\mathrm{Res}}\frac{J_{\nu }(z)J_{\mu
}(\lambda z)+Y_{\nu }(z)Y_{\mu }(\lambda z)}{J_{\nu }^{2}(z)+Y_{\nu }^{2}(z)}%
F(z).  \label{rel64}
\end{equation}%
Integrals of type (\ref{intJYth65}) which we have found in literature (see,
e.g., \cite{Erde53,Prud86,Gradshteyn}) are special cases of this formula.
For example, taking $F(z)=J_{\nu }(z)Y_{\nu +1}(\lambda ^{\prime }z)-Y_{\nu
}(z)J_{\nu +1}(\lambda ^{\prime }z)$ for the integral on the left in (\ref%
{intJYth65}) we obtain $-\lambda ^{-\nu }\lambda ^{\prime }{}^{-\nu -1}$ for
$\lambda ^{\prime }<\lambda $ and $\lambda ^{\nu }\lambda ^{\prime }{}^{-\nu
-1}-\lambda ^{-\nu }\lambda ^{\prime }{}^{-\nu -1}$ for $\lambda ^{\prime
}>\lambda $ (see \cite{Erde53}). By taking $z^{2m+1}/(z^{2}+a^{2})$, $%
z^{2m+1}/(z^{2}-c^{2})$ as $F(z)$, for $\mu =\nu $ and integer $m\geqslant 0$
one obtains
\begin{eqnarray}
\int_{0}^{\infty }dx\,\frac{J_{\nu }(x)Y_{\nu }(\lambda x)-Y_{\nu }(x)J_{\nu
}(\lambda x)}{J_{\nu }^{2}(x)+Y_{\nu }^{2}(x)}\frac{x^{2m+1}}{x^{2}+a^{2}}
&=&(-1)^{m}a^{2m}\frac{\pi }{2}\frac{K_{\nu }(\lambda a)}{K_{\nu }(a)},
\label{examp61} \\
{\mathrm{p.v.}}\! \int_{0}^{\infty }dx\,\frac{J_{\nu }(x)Y_{\nu }(\lambda
x)-Y_{\nu }(x)J_{\nu }(\lambda x)}{J_{\nu }^{2}(x)+Y_{\nu }^{2}(x)}\frac{%
x^{2m+1}}{x^{2}-c^{2}} &=&\frac{\pi }{2}c^{2m}  \notag \\
\quad \times \frac{J_{\nu }(c)J_{\nu }(\lambda c)+Y_{\nu }(c)Y_{\nu
}(\lambda c)}{J_{\nu }^{2}(c)+Y_{\nu }^{2}(c)}, &&  \label{examp62}
\end{eqnarray}%
where ${\mathrm{Re\,}}a>0$, $c>0$,$\,\lambda >1$. Special cases of this
formula for $\nu =m=0$ are given in \cite{Erde53}. In (\ref{examp61}) taking
the limit $a\rightarrow 0$ and choosing $m=0$, we obtain the integral of
this type given in \cite{Prud86}. In (\ref{intJYth65}) as a function $F(z)$
we can choose (\ref{func1}), (\ref{func2n}), (\ref{func4}) (the
corresponding conditions for parameters directly follow from (\ref{cond61})
or (\ref{cond62})) with $\rho =\mu -\nu -2m$ ($m$ is an integer), as well as
any products between them and with $%
\prod_{l=1}^{n}(z^{2}-c_{l}^{2})^{-k_{l}} $. For instance,
\begin{eqnarray}
&&\int_{0}^{\infty }\frac{dx}{x}\frac{J_{\nu }(x)Y_{\nu }(\lambda x)-Y_{\nu
}(x)J_{\nu }(\lambda x)}{J_{\nu }^{2}(x)+Y_{\nu }^{2}(x)}\prod_{l=1}^{n}%
\frac{J_{\mu _{l}}(b_{l}\sqrt{x^{2}+z_{l}^{2}})}{\left(
x^{2}+z_{l}^{2}\right) ^{\mu _{l}/2}}=\frac{\pi }{2\lambda ^{\nu }}  \notag
\\
&&\quad \times \prod_{l=1}^{n}z^{-\mu _{l}}J_{\mu
_{l}}(b_{l}z_{l}),\;b_{l},\,{\mathrm{Re\,}}\nu >0,\,{\mathrm{Re\,}}%
z_{l}\geqslant 0,\,\lambda >1,\,  \notag \\
&&\quad \sum_{l=1}^{n}{\mathrm{Re\,}}\mu _{l}+n/2+1>\delta _{b,\lambda
-1},\,b\equiv \sum_{l=1}^{n}b_{l}\leqslant \lambda -1.  \label{examp63}
\end{eqnarray}

As another consequence of formula (\ref{intJY61}) one has:

\bigskip

\noindent \textbf{Theorem 6.} \textit{Let $F(z)$ be meromorphic in the right
half-plane (with possible exception $z=0$) with poles $z_{k},\,{\mathrm{Re\,}%
}z_{k}>0$, and satisfy conditions (\ref{cond61}) or (\ref{cond62}) and
\begin{equation}
F(ze^{\pi i})=-e^{(\mu -\nu )\pi i}F(z)+o(z^{|{\mathrm{Re}}\mu |-|{\mathrm{Re%
}}\nu |-1}),\quad z\rightarrow 0,  \label{condth61}
\end{equation}%
then for values of $\nu $ for which the function $H_{\nu }^{(1)}(z)$ ($%
H_{\nu }^{(2)}(z)$) has no zeros for $0\leqslant \mathrm{arg\,}z\leqslant
\pi /2$ ($-\pi /2\leqslant \mathrm{arg\,\,}z\leqslant 0$) the following
formula takes place
\begin{eqnarray}
&&{\mathrm{p.v.}}\! \int_{0}^{\infty }{dx\,\frac{J_{\nu }(x)Y_{\mu }(\lambda
x)-Y_{\nu }(x)J_{\mu }(\lambda x)}{J_{\nu }^{2}(x)+Y_{\nu }^{2}(x)}F(x)}%
=r_{1\mu \nu }[F(z)]+\frac{\pi }{2}\underset{z=0}{\mathrm{Res}}\frac{H_{\mu
}^{(1)}(\lambda z)}{H_{\nu }^{(1)}(z)}F(z)  \notag \\
&&+\frac{1}{2}\int_{0}^{\infty }{dx\,\frac{K_{\mu }(\lambda x)}{K_{\nu }(x)}%
\left[ e^{(\nu -\mu )\pi i/2}F(xe^{\pi i/2})+e^{(\mu -\nu )\pi
i/2}F(xe^{-\pi i/2})\right] },\;\lambda >1,  \label{th6form}
\end{eqnarray}%
provided the integral on the left exists.}

\bigskip

\noindent \textbf{Proof.} This result immediately follows {}from (\ref%
{intJY61}) in the limit $a\to 0$ and from (\ref{rel63}) with $\eta _k=0$.
\rule{1.5ex}{1.5ex}

\bigskip

For example, by using (\ref{th6form}) one obtains
\begin{eqnarray}
&&\int_{0}^{\infty }dx\frac{J_{\nu }(x)Y_{\mu }(\lambda x)-Y_{\nu }(x)J_{\mu
}(\lambda x)}{J_{\nu }^{2}(x)+Y_{\nu }^{2}(x)}\prod_{l=1}^{n}J_{\mu
_{l}}(b_{l}x)=\cos \mu _{s}\int_{0}^{\infty }dx\,\frac{K_{\mu }(\lambda x)}{%
K_{\nu }(x)}  \notag \\
&&\times \prod_{l=1}^{n}I_{\mu _{l}}(b_{l}x),\;\sum_{l=1}^{n}{\mathrm{Re\,}}%
\mu _{l}+|{\mathrm{Re\,}}\nu |>|{\mathrm{Re\,}}\mu
|-1,\,b=\sum_{l=1}^{n}b_{l}\leqslant \lambda -1,\,b_{l}>0,\,  \notag \\
&&n>\delta _{b,\lambda -1},\,\mu _{s}\equiv \nu -\mu +\sum_{l=1}^{n}\mu _{l}.
\label{examp64}
\end{eqnarray}%
Such relations are useful in numerical calculations of the integrals on the
left as the integrand on the right at infinity goes to zero exponentially
fast.

We have considered formulae for integrals containing $J_{\nu }(z)Y_{\mu
}(\lambda z)-Y_{\nu }(z)J_{\mu }(\lambda z)$. Similar results can be
obtained for integrals with the functions $J_{\nu }^{\prime }(z)Y_{\mu
}(\lambda z)-Y_{\nu }^{\prime }(z)J_{\mu }(\lambda z)$ and $J_{\nu }^{\prime
}(z)Y_{\mu }^{\prime }(\lambda z)-Y_{\nu }^{\prime }(z)J_{\mu }^{\prime
}(\lambda z)$.

\section{Wightman function, the vacuum expectation values of the field
square and the energy-momentum tensor on manifolds with boundaries}

\label{sec:Gen}

In this and following sections we will consider applications of the GAPF to
the problems in quantum field theory with boundaries, mainly to the Casimir
effect. The Casimir effect is one of the most interesting macroscopic
manifestations of the non-trivial structure of the vacuum state in quantum
field theory \cite{Grib94,Most97,Plun86,Bord99,Bord01,Milt02}. The effect is
a phenomenon common to all systems characterized by fluctuating quantities
and results from changes in the vacuum fluctuations of a quantum field that
occur because of the imposition of boundary conditions or the choice of
topology. It may have important implications on all scales, from
cosmological to subnuclear, and has become in recent decades an increasingly
popular topic in quantum field theory. In particular, the recent
measurements of the Casimir forces between macroscopic bodies (see, for
instance, \cite{Bord01,Lamo05}) provide a sensitive test for constraining
the parameters of long-range interactions predicted by modern unification
theories of fundamental interactions \cite{Kuzm82}. In addition to its
fundamental interest the Casimir effect also plays an important role in the
fabrication and operation of nano- and micro-scale mechanical systems \cite%
{Bord01}.

Before to consider special problems we first give general introduction for
the procedure of the evaluation of the vacuum expectation values (VEVs) of
the functions bilinear in the field on manifolds with boundaries. Consider a
real scalar field $\varphi $ with curvature coupling parameter $\zeta $ on a
$(D+1)$-dimensional background spacetime $\mathcal{M}$ with boundary $%
\partial \mathcal{M}$ described by the metric tensor $g_{ik}$. The
corresponding field equation in the bulk has the form
\begin{equation}
\left( \nabla _{i}\nabla ^{i}+m^{2}+\zeta R\right) \varphi =0,\quad
\label{mfieldeq}
\end{equation}%
where $R$ is the scalar curvature for the background spacetime, $m$ is the
mass for the field quanta, $\nabla _{i}$ the covariant derivative operator
associated with the metric $g_{ik}$. On the boundary of the manifold the
field satisfies some prescribed boundary condition $F[\varphi (x)]|_{x\in
\partial \mathcal{M}}=0$. Here and below we adopt the conventions of Birrell
and Davies \cite{Birr82} for the metric signature and the curvature tensor.
The values of the curvature coupling parameter $\zeta =0$ and $\zeta =\zeta
_{D}$ with $\zeta _{D}\equiv (D-1)/4D$ correspond to the most important
special cases of the minimal and conformal couplings. It is convenient for
later use to write the corresponding metric energy-momentum tensor (EMT) in
the form
\begin{equation}
T_{ik}=\nabla _{i}\varphi \nabla _{k}\varphi +\left[ \left( \zeta -\frac{1}{4%
}\right) g_{ik}\nabla _{l}\nabla ^{l}-\zeta \nabla _{i}\nabla _{k}-\zeta
R_{ik}\right] \varphi ^{2},  \label{mTik}
\end{equation}%
with $R_{ik}$ being the Ricci tensor. This form of the EMT differs from the
standard one by the term $\varphi g_{ik}\left( \nabla _{l}\nabla
^{l}+m^{2}+\zeta R\right) \varphi /2$ which vanishes on solutions of the
field equation (see, for instance, \cite{Saha04Surf}) . It can be easily
checked that the following relation for the trace of the EMT\ takes place:%
\begin{equation}
T_{i}^{i}=D(\zeta -\zeta _{D})\nabla _{i}\nabla ^{i}\varphi
^{2}+m^{2}\varphi ^{2}.  \label{TraceSc}
\end{equation}%
Note that on manifolds with boundaries the EMT in addition to bulk part (\ref%
{mTik}) contains a contribution located on the boundary. For arbitrary bulk
and boundary geometries the expression for the surface EMT is derived in
\cite{Saha04Surf} by using the standard variational procedure for the action
with boundary term. For solutions of Eq. (\ref{mfieldeq}), $\varphi _{1}$
and $\varphi _{2}$, the scalar product is defined by the relation%
\begin{equation}
(\varphi _{1},\varphi _{2})=-i\int_{\Sigma }d\Sigma \,\sqrt{-g_{\Sigma }}%
n^{i}\left[ \varphi _{1}\partial _{i}\varphi _{2}^{\ast }-\left( \partial
_{i}\varphi _{1}\right) \varphi _{2}^{\ast }\right] ,  \label{mscproduct}
\end{equation}%
where $d\Sigma $ is the volume element on a given spacelike hypersurface $%
\Sigma $, $n^{i}$ is the timelike unit vector normal to this hypersurface,
and $g_{\Sigma }$ is the determinant of the induced metric. The quantization
of the field can be done by using the standard canonical quantization
procedure (see, for instance, \cite{Grib94,Birr82,Full89,Wald94}).

Let $\{\varphi _{\sigma }(x),\varphi _{\sigma }^{\ast }(x)\}$ is a complete
orthonormal set of classical solutions to the field equation satisfying the
boundary condition,%
\begin{equation}
\left( \varphi _{\sigma }(x),\varphi _{\sigma ^{\prime }}(x)\right) =\delta
_{\sigma \sigma ^{\prime }},\;F[\varphi _{\sigma }(x)]|_{x\in \partial
\mathcal{M}}=0,  \label{morthnorm}
\end{equation}%
where the collective index $\sigma $ can contain both discrete and
continuous components. On the right of the orthonormalization condition, $%
\delta _{\sigma \sigma ^{\prime }}$ is understood as the Kronecker symbol
for discrete components and the Dirac delta function for continuous ones. In
quantum field theory we expand the field operator in terms of the complete
set:%
\begin{equation}
\varphi (x)=\sum_{\sigma }[a_{\sigma }\varphi _{\sigma }(x)+a_{\sigma
}^{+}\varphi _{\sigma }^{\ast }(x)],  \label{mphiexpansion}
\end{equation}%
with annihilation and creation operators $a_{\sigma }$, $a_{\sigma }^{+}$,
satisfying the commutation relation $\left[ a_{\sigma },a_{\sigma ^{\prime
}}^{+}\right] =\delta _{\sigma \sigma ^{\prime }}$. The vacuum state $%
|0\rangle $ is defined by the relation $a_{\sigma }|0\rangle =0$ for any $%
\sigma $. Our main interest below will be the VEVs of the field square and
the EMT. These functions give comprehensive insight into vacuum fluctuations
and are among the most important quantities characterizing the properties of
the quantum vacuum. Though the corresponding operators are local, due to the
global nature of the vacuum, the VEVs describe both local and global
properties of the bulk and carry an important information about the
background geometry. In addition to describing the physical structure of the
quantum field at a given point, the VEV of the EMT acts as the source of
gravity in the Einstein equations, and therefore plays an important role in
modelling a self-consistent dynamics involving the gravitational field \cite%
{Birr82}.

The VEVs of the field square and the EMT can be obtained from a two-point
function taking the coincidence limit in combination with the
renormalization procedure. As a two-point function we will take the positive
frequency Wightman function defined as the VEV%
\begin{equation}
W(x,x^{\prime })=\langle 0|\varphi (x)\varphi (x^{\prime })|0\rangle .
\label{mWF}
\end{equation}%
Our choice of this function is related to that it also determines the
response of the Unruh-DeWitt type particle detector at a given state of
motion \cite{Birr82}. Substituting into formula (\ref{mWF}) the expansion of
the field operator (\ref{mphiexpansion}) and using the commutation relation
for the annihilation and creation operators, the following mode-sum formula
is obtained%
\begin{equation}
W(x,x^{\prime })=\sum_{\sigma }\varphi _{\sigma }(x)\varphi _{\sigma }^{\ast
}(x^{\prime }).  \label{mmodesumWF}
\end{equation}%
We can evaluate the VEVs of the field square and the energy-momentum tensor
as the following coincidence limits:%
\begin{eqnarray}
\langle 0|\varphi ^{2}(x)|0\rangle &=&\lim_{x^{\prime }\rightarrow
x}W(x,x^{\prime }),  \label{mphi2VEV} \\
\langle 0|T_{ik}(x)|0\rangle &=&\lim_{x^{\prime }\rightarrow x}\partial
_{i}\partial _{k}^{\prime }W(x,x^{\prime })  \notag \\
&&+\left[ \left( \zeta -\frac{1}{4}\right) g_{ik}\nabla _{l}\nabla
^{l}-\zeta \nabla _{i}\nabla _{k}-\zeta R_{ik}\right] \langle 0|\varphi
^{2}(x)|0\rangle ,  \label{mTikVEV}
\end{eqnarray}%
where $\partial _{i}$ is the derivative with respect to $x^{i}$ and $%
\partial _{k}^{\prime }$ is the derivative with respect to $x^{\prime k}$.
Similar formulae can be written for fermionic and electromagnetic fields.
The expressions on the right of formulae (\ref{mphi2VEV}), (\ref{mTikVEV})
are formal as they diverge and to obtain a finite physical result some
renormalization procedure is needed. The consideration of the factors in the
field bilinear products at different spacetime points corresponds to the
point-splitting regularization procedure. An alternative way is to introduce
in the mode-sums of the field square and the energy-momentum tensor a cutoff
function, which makes them convergent, and to remove this function after the
renormalization. Other regularization methods for the VEVs of the field
square and the EMT are based on the use of the Green function (for the Green
function method in the Casimir effect calculations see \cite{Milt02} and
references therein) and the local zeta function \cite{More97} (see also \cite%
{Byts03}). The discussion for the relations between various regularization
techniques can be found in \cite{Grib94,Most97,Milt02,Birr82,Byts03}.

The important thing is that, for points away from boundaries the divergences
in the VEVs of the field square and the EMT are the same as those on
background of manifolds without boundaries and, hence, for these points the
corresponding renormalization procedure is also the same. However, the local
VEVs diverge for the points on the boundary. These surface divergences are
well known in quantum field theory with boundaries and are investigated for
various types of bulk and boundary geometries (see, for example, \cite%
{Birr82,Full89,Bali78,Deutsch,Kennedy,Syma81}). A powerful tool for studying
one-loop divergences in the VEVs is the heat-kernel expansion \cite%
{Byts03,Kirs01,Vass03}. In general, the physical quantities in problems with
boundary conditions can be classified into two main groups. The first group
includes quantities which do not contain surface divergences. For these
quantities the renormalization procedure is the same as in quantum field
theory without boundaries and they can be evaluated by boundary condition
calculations. The contribution of the higher modes into the boundary induced
effects in these quantities is suppressed by parameters already present in
the idealized model. Examples of such quantities are the energy density and
the vacuum stresses at the points away from the boundary and the interaction
forces between disjoint bodies. For quantities from the second group, such
as the energy density on the boundary and the total vacuum energy, the
contribution of the arbitrary higher modes is dominant and they contain
divergences which cannot be eliminated by the standard renormalization
procedure of quantum field theory without boundaries. Of course, the model
where the physical interaction is replaced by the imposition of boundary
conditions on the field for all modes is an idealization. The appearance of
divergences in the process of the evaluation of physical quantities of the
second type indicate that more realistic physical model should be employed
for their evaluation. In literature on the Casimir effect different
field-theoretical approaches have been discussed to extract the finite parts
from the diverging quantities. However, in the physical interpretation of
these results it should be taken into account that these terms are only a
part of the full expression of the physical quantity and the terms which are
divergent in the idealized model can be physically essential and their
evaluation needs a more realistic model. It seems plausible that such
effects as surface roughness, or the microstructure of the boundary on small
scales can introduce a physical cutoff needed to produce finite values for
surface quantities.

Below we are interested in the VEVs of the field square and the EMT at
points away from boundaries. These VEVs belong to the first group of
quantities and, hence, they are well-defined within the framework of
standard renormalization procedure of quantum field theory without
boundaries. We expect that the same results will be obtained in the model
where instead of externally imposed boundary condition the fluctuating field
is coupled to a smooth background potential that implements the boundary
condition in a certain limit \cite{Grah02}. For a scalar field we will
assume that on the boundary the field satisfies Robin boundary condition
\begin{equation}
\left( \tilde{A}+\tilde{B}n^{i}\nabla _{i}\right) \varphi (x)=0,\;x\in
\partial \mathcal{M},  \label{mrobcond}
\end{equation}%
where $\tilde{A}$ and $\tilde{B}$ are constants, and $n^{i}$ is the unit
inward-pointing normal to the boundary. Of course, all results will depend
only on the ratio of the coefficients in (\ref{mrobcond}). However, to keep
the transition to Dirichlet and Neumann cases transparent we will write the
boundary condition in the form (\ref{mrobcond}). Robin type conditions are
an extension of Dirichlet and Neumann boundary conditions and appear in a
variety of situations, including the considerations of vacuum effects for a
confined charged scalar field in external fields, spinor and gauge field
theories, quantum gravity, supergravity and braneworld scenarios \cite%
{Saha05b,Ambj83b,Espo97,Gher00}. In some geometries, Robin boundary
conditions may be useful for depicting the finite penetration of the field
into the boundary with the 'skin-depth' parameter related to the Robin
coefficient \cite{Most85,Lebe01}. In examples with fermionic fields we will
consider the bag boundary condition and in the case of the electromagnetic
field we assume perfect conductor boundary conditions.

\section{Examples for the application of the Abel-Plana formula}

\label{sec:Topol}

\subsection{Casimir effect in $R^{D}\times S^{1}$}

First we consider an example of the application of the standard Abel-Plana
summation formula for the evaluation of the vacuum characteristics in the
flat spacetime assuming that one of the spatial coordinates, the coordinate $%
x^{D}=y$, is compactified to a circle with the length $a$ and the scalar
field satisfies periodic boundary condition $\varphi (y+a)=\varphi (y)$ (for
the Casimir effect in topologically non-trivial spaces and its role in
cosmological models see \cite{Most97,Bord01,Milt02,Eliz06} and references
therein). The corresponding normalized eigenfunctions have the form%
\begin{equation}
\varphi _{\sigma }=\frac{e^{ik_{n}y+i\mathbf{k}\cdot \mathbf{x}-i\omega
_{n}t}}{[2\omega _{n}a(2\pi )^{D-1}]^{1/2}},\;k_{n}=\frac{2\pi }{a}%
n,\;n=0,\pm 1,\pm 2,\ldots ,  \label{eigtopol}
\end{equation}%
and $\omega _{n}^{2}=k_{n}^{2}+k^{2}+m^{2}$. The vector $\mathbf{x}%
=(x^{1},x^{2},\ldots ,x^{D-1})$ specifies the uncompactified spatial
dimensions and $k=|\mathbf{k}|$. From mode-sum formula (\ref{mmodesumWF})
for the Wightman function one finds
\begin{equation}
W_{\mathrm{P}}(x,x^{\prime },a)=\frac{1}{a}\int d\mathbf{k\,}\frac{e^{i%
\mathbf{k}\cdot \Delta \mathbf{x}}}{(2\pi )^{D-1}}\sideset{}{'}{\sum}%
_{n=0}^{\infty }\frac{e^{-i\omega _{n}\Delta t}}{\omega _{n}}\cos
(k_{n}\Delta y),  \label{WF1topol}
\end{equation}%
where $\Delta \mathbf{x=x-x}^{\prime }$, $\Delta t=t-t^{\prime }$, $\Delta
y=y-y^{\prime }$. We apply to the series over $n$ the APF in the form (\ref%
{apsf2}). The corresponding conditions are satisfied if $|\Delta y|+|\Delta
t|<2a$. In particular, this is the case in the coincidence limit assuming
that $|y|<a$. It is easily seen that the contribution coming from the first
integral on the right of the APF is the Wightman function for the Minkowski
spacetime, $W_{\mathrm{M}}(x,x^{\prime })$. As a result, the Wightman
function is written in the form%
\begin{equation}
W_{\mathrm{P}}(x,x^{\prime },a)=W_{\mathrm{M}}(x,x^{\prime })+2\int d\mathbf{%
k\,}\frac{e^{i\mathbf{k}\cdot \Delta \mathbf{x}}}{(2\pi )^{D}}%
\int_{k_{m}}^{\infty }dx\,\frac{\cosh (x\Delta y)}{e^{ax}-1}\frac{\cosh
(\Delta t\sqrt{x^{2}-k_{m}^{2}})}{\sqrt{x^{2}-k_{m}^{2}}},  \label{WF2topol}
\end{equation}%
with the notation $k_{m}=\sqrt{k^{2}+m^{2}}$. In the second term on the
right of this formula we first integrate over the angular part of the vector
$\mathbf{k}$. Further, we introduce polar coordinates in the plane $(k,x)$
and integrate over the polar angle. The corresponding integrals can be found
in \cite{Prud86}. In this way we obtain the formula%
\begin{eqnarray}
W_{\mathrm{P}}(x,x^{\prime },a) &=&W_{\mathrm{M}}(x,x^{\prime })+\frac{(2\pi
)^{-\frac{D}{2}}}{(\Delta z)^{\frac{D-2}{2}}}\int_{m}^{\infty }dx\,\frac{%
(x^{2}-m^{2})^{\frac{D-2}{4}}}{e^{ax}-1}  \notag \\
&&\times \cosh (x\Delta y)J_{D/2-1}(\Delta z\sqrt{x^{2}-m^{2}}),
\label{WF3topol}
\end{eqnarray}%
where $\Delta z=[|\Delta \mathbf{x}|^{2}-(\Delta t)^{2}]^{1/2}$. By using
the expansion $(e^{ax}-1)^{-1}=\sum_{n=1}^{\infty }e^{-anx}$, in (\ref%
{WF3topol}) the separate terms in the series are explicitly integrated and
we find an equivalent representation%
\begin{equation}
W_{\mathrm{P}}(x,x^{\prime },a)=\frac{m^{D-1}}{(2\pi )^{\frac{D+1}{2}}}%
\sum_{n=-\infty }^{+\infty }\frac{K_{(D-1)/2}(u_{n})}{u_{n}^{(D-1)/2}},
\label{WF4topol}
\end{equation}%
with $u_{n}=m[|\Delta \mathbf{x}|^{2}+(\Delta y+an)^{2}-(\Delta
t)^{2}]^{1/2} $. The $n=0$ term in this formula is the Minkowskian Wightman
function. Formula (\ref{WF4topol}) presents the Wightman function as an
image sum of the corresponding Minkowskian functions.

The renormalized VEV of the field square is obtained subtracting from the
Wightman function the Minkowskian part and taking the coincidence limit. As
we have already separated the latter, this procedure here is trivial and one
obtains%
\begin{eqnarray}
\langle \varphi ^{2}\rangle _{\mathrm{ren}}^{\mathrm{(P)}} &=&\frac{2^{1-D}}{%
\pi ^{D/2}\Gamma (D/2)}\int_{m}^{\infty }dx\,\frac{(x^{2}-m^{2})^{D/2-1}}{%
e^{ax}-1}  \notag \\
&=&\frac{2m^{D-1}}{(2\pi )^{(D+1)/2}}\sum_{n=1}^{\infty }\frac{%
K_{(D-1)/2}(man)}{(man)^{(D-1)/2}}.  \label{phi2topol}
\end{eqnarray}%
In the similar way, for the components of the vacuum EMT we find (no
summation over $i$)%
\begin{eqnarray}
\langle T_{i}^{i}\rangle _{\mathrm{ren}}^{\mathrm{(P)}} &=&-\frac{2^{1-D}}{%
\pi ^{D/2}D\Gamma (D/2)}\int_{m}^{\infty }dx\,\frac{(x^{2}-m^{2})^{D/2}}{%
e^{ax}-1}  \notag \\
&=&-\frac{2m^{D+1}}{(2\pi )^{(D+1)/2}}\sum_{n=1}^{\infty }\frac{%
K_{(D+1)/2}(man)}{(man)^{(D+1)/2}},  \label{Tiitopol}
\end{eqnarray}%
for $i=0,1,\ldots ,D-1$, and the component $\langle T_{D}^{D}\rangle _{%
\mathrm{ren}}^{\mathrm{(P)}}$ is obtained from the trace relation (\ref%
{TraceSc}): $\langle T_{l}^{l}\rangle _{\mathrm{ren}}^{\mathrm{(P)}%
}=m^{2}\langle \varphi ^{2}\rangle _{\mathrm{ren}}^{\mathrm{(P)}}$. We have
considered periodic boundary conditions. For the field with antiperiodic
conditions (twisted scalar field) we have $\varphi (y+a)=-\varphi (y)$. Now,
in the expressions of the eigenfunctions $k_{n}=2\pi (n+1/2)/a$. The
corresponding VEVs are evaluated in the way similar to that for the periodic
case with the application of the APF in the form (\ref{apsf2half}). Already
on these simple examples we have seen two important features of the
application of the APF. First, this enables to extract from the VEVs the
part corresponding to the geometry with decompactified dimensions and
second, to present the parts induced by the non-trivial topology in terms of
rapidly convergent integrals.

\subsection{Casimir effect for parallel plates with Dirichlet and Neumann
boundary conditions}

In considered examples, due to the homogeneity of the background spacetime
the VEVs of the field square and the EMT do not depend on coordinates. This
is not the case in the Minkowski spacetime with boundaries. The presence of
boundaries breaks the homogeneity and leads to coordinate-dependent VEVs.
Simplest examples of this kind are scalar fields with Dirichlet and Neumann
boundary conditions on two parallel plates with distance $a$. Assuming that
the plates are located at $y=0$ and $y=a$, the corresponding eigenfunctions
are obtained from (\ref{eigtopol}) by the replacement $e^{ik_{n}y}%
\rightarrow \sqrt{2}\sin (k_{n}y)$ with $k_{n}=\pi n/a$, $n=1,2,\ldots $,
for Dirichlet case, and by the replacement $e^{ik_{n}y}\rightarrow \sqrt{%
2-\delta _{n0}}\cos (k_{n}y)$ with $k_{n}=\pi n/a$, $n=0,1,2,\ldots $, for
Neumann one. Now, applying to the series over $n$ in the mode-sum for the
Wightman function $W_{\mathrm{J}}(x,x^{\prime })$ (with $\mathrm{J=D,N}$ for
Dirichlet and Neumann scalars) the APF (\ref{apsf2}), we can see that the
term with the first integral on the right of the APF corresponds to the
Wightman function $W_{\mathrm{J}}^{(0)}(x,x^{\prime })$ for the geometry of
a single plate at $y=0$, and the term with the second integral is the part
induced by the presence of the second plate at $y=a$. It can be seen that
the following relations take place%
\begin{eqnarray}
W_{\mathrm{J}}^{(0)}(x,x^{\prime }) &=&W_{\mathrm{M}}(x,x^{\prime })+\delta
_{\mathrm{J}}W_{\mathrm{M}}(x,x_{\mathrm{I}}^{\prime }),  \label{WJ0} \\
W_{\mathrm{J}}(x,x^{\prime }) &=&W_{\mathrm{P}}(x,x^{\prime },2a)+\delta _{%
\mathrm{J}}W_{\mathrm{P}}(x,x_{\mathrm{I}}^{\prime },2a),  \label{WJ}
\end{eqnarray}%
with $\delta _{\mathrm{D}}=-1$, $\delta _{\mathrm{N}}=1$, and $x_{\mathrm{I}%
}^{\prime }$ is the image for the point $x^{\prime }$ with respect to the
plate at $y=0$: $x_{\mathrm{I}}^{\prime }=(t,\mathbf{x}^{\prime },-y^{\prime
})$.

For the geometry of a single plate at $y=0$, by using the Wightman function (%
\ref{WJ0}) one finds%
\begin{equation}
\langle \varphi ^{2}\rangle _{\mathrm{J,ren}}^{(0)}=\frac{\delta _{\mathrm{J}%
}m^{D-1}}{(2\pi )^{(D+1)/2}}\frac{K_{(D-1)/2}(2m|y|)}{(2m|y|)^{(D-1)/2}},
\label{phi20J}
\end{equation}%
for the VEV of the field square, and (no summation over $i$)
\begin{equation}
\langle T_{i}^{i}\rangle _{\mathrm{J,ren}}^{(0)}=\frac{4\delta _{\mathrm{J}%
}m^{D+1}(\zeta -\zeta _{D})}{(2\pi )^{(D+1)/2}}\left[ \frac{%
K_{(D+1)/2}(2m|y|)}{(2m|y|)^{(D+1)/2}}-\frac{K_{(D+3)/2}(2m|y|)}{%
(2m|y|)^{(D-1)/2}}\right] +\frac{m^{2}}{D}\langle \varphi ^{2}\rangle _{%
\mathrm{J,ren}}^{(0)},  \label{Tik20J}
\end{equation}%
$i=0,1,\ldots ,D-1$, $\langle T_{D}^{D}\rangle _{\mathrm{J,ren}}^{(0)}=0$,
for the VEV of the EMT. In particular, for a conformally coupled massless
scalar the latter vanishes. For points on the plate the VEVs diverge with
the leading terms%
\begin{equation}
\langle \varphi ^{2}\rangle _{\mathrm{J,ren}}^{(0)}\approx \frac{\delta _{%
\mathrm{J}}\Gamma ((D-1)/2)}{(4\pi )^{(D+1)/2}|y|^{D-1}},\;\langle
T_{i}^{k}\rangle _{\mathrm{J,ren}}^{(0)}\approx -\frac{\delta
_{i}^{k}D\delta _{\mathrm{J}}(\zeta -\zeta _{D})}{2^{D}\pi
^{(D+1)/2}|y|^{D+1}}\Gamma \left( \frac{D+1}{2}\right) ,  \label{phi2Tiknear}
\end{equation}%
for $i=0,1,\ldots ,D-1$. Note that the expressions on the right of these
formulae are the VEVs of the field square and the EMT for a massless scalar.

For the geometry of two plates at $y=0$ and $y=a$ the VEVs are presented in
the form%
\begin{equation}
\langle \varphi ^{2}\rangle _{\mathrm{J,ren}}=\langle \varphi ^{2}\rangle _{%
\mathrm{J,ren}}^{(0)}+\frac{2^{1-D}}{\pi ^{D/2}\Gamma (D/2)}\int_{m}^{\infty
}dx\,\frac{(x^{2}-m^{2})^{D/2-1}}{e^{2ax}-1}\left[ 1+\delta _{\mathrm{J}%
}\cosh (2xy)\right] ,  \label{phi2J}
\end{equation}%
for the field square, and (no summation over $i$)%
\begin{eqnarray}
\langle T_{i}^{i}\rangle _{\mathrm{J,ren}} &=&\langle T_{i}^{i}\rangle _{%
\mathrm{J,ren}}^{(0)}-\frac{2^{1-D}}{\pi ^{D/2}\Gamma (D/2)}\int_{m}^{\infty
}dx\,\frac{(x^{2}-m^{2})^{D/2-1}}{e^{2ax}-1}  \notag \\
&&\times \left\{ \frac{x^{2}-m^{2}}{D}+\delta _{\mathrm{J}}\left[ 4(\zeta
-\zeta _{D})x^{2}-m^{2}/D\right] \cosh (2xy)\right\} ,  \label{TiiJ} \\
\langle T_{D}^{D}\rangle _{\mathrm{J,ren}} &=&\frac{2^{1-D}}{\pi
^{D/2}\Gamma (D/2)}\int_{m}^{\infty }dx\,x^{2}\frac{(x^{2}-m^{2})^{D/2-1}}{%
e^{2ax}-1},  \label{TDDJ}
\end{eqnarray}%
$i=0,1,\ldots ,D-1$, for the EMT. Note that by using the formula
\begin{eqnarray}
\int_{m}^{\infty }dx\,x^{p}\frac{(x^{2}-m^{2})^{D/2-1}}{e^{2ax}-1}\cosh
(2xy) &=&(-1)^{p}\frac{m^{D+p-1}}{\sqrt{\pi }}2^{\frac{D-3}{2}}\Gamma \left(
\frac{D}{2}\right)  \notag \\
&&\times \sideset{}{'}{\sum}_{n=-\infty }^{\infty }\partial _{z}^{p}\frac{%
K_{(D-1)/2}(z)}{z^{(D-1)/2}}|_{z=2m|an+y|},  \label{Int2DN}
\end{eqnarray}%
where the prime means that the $n=0$ term should be omitted, we can present
the corresponding results in terms of series containing the McDonald
function. The vacuum forces acting on the plates are determined by the
component $\langle T_{D}^{D}\rangle _{\mathrm{J,ren}}$ and they are the same
for Dirichlet and Neumann scalars. Formula (\ref{TDDJ}) for the vacuum
forces acting on the plates can also be obtained by differentiating the
corresponding vacuum energy from \cite{Ambj83a} (see also \cite{Milt02}).
The forces are attractive for all values of the interplate distance. In
particular, in the case of a massless field for the modulus of the force
acting per unit surface of the plate one finds%
\begin{equation}
F_{\mathrm{J}}=\frac{D\zeta _{\mathrm{R}}(D+1)}{(4\pi )^{(D+1)/2}a^{D+1}}%
\Gamma \left( \frac{D+1}{2}\right) ,\;\mathrm{J=D,N},  \label{ForceDN}
\end{equation}%
where $\zeta _{\mathrm{R}}(x)$ is the Riemann zeta function. Local analysis
of a quantum scalar field confined within rectangular cavities is presented
in \cite{Acto94}.

In the dimension $D=3$ with perfect conductor boundary condition on two
parallel plates, the electromagnetic field is presented as a set of two
types of modes corresponding to massless Dirichlet and Neumann scalars. The
corresponding VEV\ for the EMT does not depend on coordinates and in the
region between the plates is given by \cite{Brow69}%
\begin{equation}
\langle T_{i}^{k}\rangle _{\mathrm{El,ren}}=-\frac{\pi ^{2}}{720a^{4}}%
\mathrm{diag}(1,1,1,-3).  \label{TikElPlates}
\end{equation}%
Due to the cancellation between Dirichlet and Neumann modes, the
electromagnetic EMT is uniform in the region between the plates and vanishes
in the outside regions.

\section{Casimir effect for parallel plates with Robin boundary conditions}

\label{sec:Robplates}

As we have demonstrated in the previous section the application of the APF
provides an efficient way for the investigation of the boundary-induced
effects. Here we will see that already in the case of two parallel plate
geometry with Robin boundary conditions this procedure needs a
generalization \cite{Rome02}. This can be done by using the GAPF. As before,
we assume that the plates are located at $y=a_{1}=0$ and $y=a_{2}=a$, and
the field obeys boundary conditions (\ref{mrobcond}) with coefficients $%
\tilde{A}_{j}$, $\tilde{B}_{j}$ for the plate at $y=a_{j}$. In the region
between the plates the eigenfunctions are presented in the form
\begin{equation}
\varphi _{\sigma }=\beta (k_{y})e^{i\mathbf{k}\cdot \mathbf{x}-i\omega
(k_{y})t}\cos \left( k_{y}y+\alpha (k_{y})\right) ,  \label{eigfunc2plR}
\end{equation}%
where $\omega (k_{y})\equiv \sqrt{k_{y}^{2}+k^{2}+m^{2}}$, and $\alpha
(k_{y})$ is defined by the relation
\begin{equation}
e^{2i\alpha (k_{y})}\equiv \frac{i\beta _{1}k_{y}-1}{i\beta _{1}k_{y}+1}%
,\;\beta _{j}=(-1)^{j-1}\tilde{B}_{j}/\tilde{A}_{j}.  \label{alfjR}
\end{equation}%
The corresponding eigenvalues for $k_{y}$ are obtained from the boundary
conditions and are solutions of the following transcendental equation:
\begin{equation}
F(z)\equiv \left( 1-b_{1}b_{2}z^{2}\right) \sin z-\left( b_{1}+b_{2}\right)
z\cos z=0,\;z=k_{y}a,\;b_{j}=\beta _{j}/a.  \label{eigvaluesR}
\end{equation}%
The expression for the coefficient $\beta (k_{y})$ in (\ref{eigfunc2plR}) is
obtained from the normalization condition:
\begin{equation}
\beta ^{-2}(k_{y})=(2\pi )^{D-1}a\omega \left[ 1+\frac{\sin (k_{y}a)}{k_{y}a}%
\cos (k_{y}a+2\alpha (k_{y}))\right] .  \label{bet2plR}
\end{equation}%
The eigenvalue equation (\ref{eigvaluesR}) has an infinite set of real zeros
which we will denote by $k_{y}=\lambda _{n}/a$, $n=1,2,\ldots $. In
addition, depending on the values of the coefficients in the boundary
conditions, this equation has two or four complex conjugate purely imaginary
zeros $\pm iy_{l}$, $y_{l}>0$ (see \cite{Rome02}).

Substituting eigenfunctions (\ref{eigfunc2plR}) into mode-sum formula (\ref%
{mmodesumWF}), for the positive-frequency Wightman function in the region
between two plates one finds
\begin{eqnarray}
W(x,x^{\prime }) &=&\int d\mathbf{k}\,e^{i\mathbf{k}\cdot \Delta \mathbf{x}%
}\sum_{k_{y}=\lambda _{n}/a,iy_{l}/a}\beta ^{2}(k_{y})e^{-i\omega
(k_{y})\Delta t}  \notag \\
&&\times \cos (k_{y}y+\alpha (k_{y}))\cos (k_{y}y^{\prime }+\alpha (k_{y})).
\label{WF2plR}
\end{eqnarray}%
To obtain the summation formula over the zeros $\lambda _{n}$, in the GAPF
as a function $g(z)$ we choose%
\begin{equation}
g(z)=-i\left[ \left( 1-b_{1}b_{2}z^{2}\right) \cos z+\left(
b_{1}+b_{2}\right) z\sin z\right] \frac{f(z)}{F(z)}.  \label{gz2plR}
\end{equation}%
Substituting (\ref{gz2plR}) into (\ref{th12}) and taking the limit $%
a\rightarrow 0$, under the assumption that the poles $\pm iy_{l}$ are
excluded by small semicircles on the right half-plane, one obtains the
following summation formula \cite{Rome02}
\begin{eqnarray}
\sum_{z=\lambda _{n},iy_{l}}\frac{\pi f(z)}{1+\cos (z+2\alpha )\sin z/z} &=&%
\frac{\pi f(0)/2}{b_{1}+b_{2}-1}+\int_{0}^{\infty }dzf(z)  \notag \\
&&+i\int_{0}^{\infty }dt\frac{f(te^{\pi i/2})-f(te^{-\pi i/2})}{\frac{%
(b_{1}t-1)(b_{2}t-1)}{(b_{1}t+1)(b_{2}t+1)}e^{2t}-1}  \notag \\
&&-\frac{\theta (b_{1})}{2b_{1}}\left[ h(e^{\frac{\pi i}{2}%
}/b_{1})+h(c_{1}e^{-\frac{\pi i}{2}}/b_{1})\right] ,  \label{sumformula2plR}
\end{eqnarray}%
where $h(z)\equiv \left( b_{1}^{2}z^{2}+1\right) f(z)$. For the case $%
b_{2}=0 $, $b_{1}<0$, with an analytic function $f(z)$ this formula is
derived in \cite{Lebe01}.

For the summation over the eigenmodes $\lambda _{n}$ in (\ref{WF2plR}) as a
function $f(z)$ in (\ref{sumformula2plR}) we take
\begin{equation}
f(z)\equiv \frac{e^{-i\omega (z/a)\Delta t}}{a\omega (z/a)}\cos (zy/a+\alpha
(z/a))\cos (zy^{\prime }/a+\alpha (z/a)),  \label{fzR}
\end{equation}%
with first-order poles at $z=\pm i/b_{j}$. By making use of the definition
for $\alpha (k)$ we see that $e^{2i\alpha (0)}=-1$, and hence $\cos (2\alpha
(0))=-1$, which implies that $f(0)=0$. The resulting Wightman function from (%
\ref{WF2plR}) is found to be
\begin{eqnarray}
W(x,x^{\prime }) &=&W_{0}(x,x^{\prime })+\frac{4}{(2\pi )^{D}}\int d\mathbf{k%
}\,e^{i\mathbf{k}\cdot \Delta \mathbf{x}}\int_{k_{m}}^{\infty }du  \notag \\
&&\times \frac{\cosh (yu+\tilde{\alpha}(u))\cosh (y^{\prime }u+\tilde{\alpha}%
(u))}{\frac{(\beta _{1}u-1)(\beta _{2}u-1)}{(\beta _{1}t+1)(\beta _{2}t+1)}%
e^{2au}-1}\frac{\cosh \left[ \Delta t\sqrt{u^{2}-k_{m}^{2}}\right] }{\sqrt{%
u^{2}-k_{m}^{2}}},  \label{WF2pl1R}
\end{eqnarray}%
where $k_{m}=\sqrt{k^{2}+m^{2}}$ and the function $\tilde{\alpha}_{j}(t)$ is
defined by the relation%
\begin{equation}
e^{2\tilde{\alpha}(u)}\equiv \frac{\beta _{1}u-1}{\beta _{1}u+1}.
\label{alfajtildeR}
\end{equation}%
In formula (\ref{WF2pl1R}),%
\begin{eqnarray}
W_{0}(x,x^{\prime }) &=&W_{M}(x,x^{\prime })+\int \frac{d\mathbf{k}}{(2\pi
)^{D}}e^{i\mathbf{k}\cdot \Delta \mathbf{x}}\int_{0}^{\infty }du\frac{%
e^{-i\omega (u)\Delta t}}{\omega (u)}\cos \left[ u(y+y^{\prime })+2\alpha (u)%
\right]  \notag \\
&&+\frac{\theta (\beta _{1})e^{-(y+y^{\prime })/\beta _{1}}}{(2\pi
)^{D-1}\beta _{1}}\int d\mathbf{k}\frac{\exp (i\mathbf{k}\cdot \Delta
\mathbf{x}-i\Delta t\sqrt{k_{m}^{2}-1/\beta _{1}^{2}})}{\sqrt{%
k_{m}^{2}-1/\beta _{1}^{2}}},  \label{W0R}
\end{eqnarray}%
is the Wightman function for a single plate located at $y=0$ and $%
W_{M}(x,x^{\prime })$ is the Wightman function in the Minkowski spacetime
without boundaries. The last term on the right comes from the bound state
present in the case $\beta _{1}>0$. To escape the instability of the vacuum
state, we will assume that $m\beta _{1}\geqslant 1$. On taking the
coincidence limit, for the VEV of the field square we obtain the formula
\begin{equation}
\langle \varphi ^{2}\rangle _{\mathrm{R,ren}}=\langle \varphi ^{2}\rangle _{%
\mathrm{R,ren}}^{(0)}+4\frac{(4\pi )^{-D/2}}{\Gamma (D/2)}\int_{m}^{\infty
}dt\frac{(t^{2}-m^{2})^{D/2-1}}{\frac{(\beta _{1}t-1)(\beta _{2}t-1)}{(\beta
_{1}t+1)(\beta _{2}t+1)}e^{2at}-1}\cosh ^{2}(ty+\tilde{\alpha}(t)),
\label{phi22plR}
\end{equation}%
where%
\begin{equation}
\langle \varphi ^{2}\rangle _{\mathrm{R,ren}}^{(0)}=\frac{(4\pi )^{-D/2}}{%
\Gamma (D/2)}\int_{m}^{\infty }dt\,(t^{2}-m^{2})^{D/2-1}e^{-2ty}\frac{\beta
_{1}t+1}{\beta _{1}t-1},  \label{phi21plR}
\end{equation}%
is the corresponding VEV in the region $y>0$ for a single plate at $y=0$.
The surface divergences on the plate at $y=0$ are contained in this term.
The second term on the right of formula (\ref{phi22plR}) is finite at $y=0$
and is induced by the second plate located at $y=a$. This term diverges at $%
y=a$. The corresponding divergence is the same as that for the geometry of a
single plate located at $y=a$. Note that in obtaining (\ref{phi21plR}) from (%
\ref{W0R}) we have written the cos function in the second integral
term on the right of (\ref{W0R}) as a sum of exponentials and have
rotated the integration contour by the angle $\pi /2$ and by $-\pi
/2$ for separate exponentials. For $\beta _{1}>0$ the
corresponding integrals have poles $\pm i/\beta _{1}$ on the
imaginary axis and the contribution from these poles cancel the
part coming from the last term on the right of (\ref{W0R}).

The VEV of the EMT is evaluated by formula (\ref{mTikVEV}). By taking into
account formulae (\ref{WF2pl1R}), (\ref{phi22plR}), for the region between
the plates one finds (no summation over $l$, see \cite{Rome02} for the case
of a massless field)
\begin{equation}
\langle T_{i}^{l}\rangle _{\mathrm{R,ren}}=\langle T_{i}^{l}\rangle _{%
\mathrm{R,ren}}^{(0)}+2\delta _{i}^{l}\frac{(4\pi )^{-D/2}}{\Gamma (D/2)}%
\int_{m}^{\infty }dt\,\frac{(t^{2}-m^{2})^{D/2-1}}{\frac{(\beta
_{1}t-1)(\beta _{2}t-1)}{(\beta _{1}t+1)(\beta _{2}t+1)}e^{2at}-1}f_{l}(t,y),
\label{Til2plR}
\end{equation}%
where%
\begin{equation}
f_{l}(t,x)=\left[ 4t^{2}\left( \zeta _{D}-\zeta \right) +m^{2}/D\right]
\cosh \left( 2ty+2\tilde{\alpha}(t)\right) -(t^{2}-m^{2})/D,  \label{fiRob}
\end{equation}%
for $l=0,1,\ldots ,D-1,$ and $f_{D}(t,y)=t^{2}$. In formula (\ref{Til2plR}),
\begin{eqnarray}
\langle T_{i}^{l}\rangle _{\mathrm{R,ren}}^{(0)} &=&\delta _{i}^{l}\frac{%
(4\pi )^{-D/2}}{\Gamma \left( D/2\right) }\int_{m}^{\infty
}dt(t^{2}-m^{2})^{D/2-1}e^{-2yt}  \notag \\
&&\times \frac{\beta _{1}t+1}{\beta _{1}t-1}\left[ 4\left( \zeta _{D}-\zeta
\right) t^{2}+m^{2}/D\right] ,  \label{Til1plR}
\end{eqnarray}%
for $i=0,1,\ldots ,D-1,$ and $\langle T_{D}^{D}\rangle _{\mathrm{R,ren}%
}^{(0)}=0$, are the VEVs in the region $y>0$ induced by a single plate at $%
y=0$, and the second term on the right is the part of the energy-momentum
tensor induced by the presence of the second plate. In \cite{SahaRind2},
formulae (\ref{phi22plR}) and (\ref{Til2plR}) are obtained as limiting cases
of the corresponding results for the geometry of two parallel plates
uniformly accelerated through the Fulling-Rindler vacuum (see below). For a
conformally coupled massless scalar field the vacuum EMT is uniform and
traceless. Note that we have investigated the vacuum densities in the bulk.
For Robin boundary conditions in addition to this part there is a
contribution to the EMT located on the plates.

Vacuum forces acting on the plates are determined by $\langle
T_{D}^{D}\rangle _{\mathrm{R,ren}}$. This component is uniform and, hence,
is finite on the plates. The latter property is a consequence of the high
symmetry of the problem and is not valid for curved boundaries. In
dependence of the values for the coefficients $\beta _{j}$ the vacuum forces
can be both attractive or repulsive. In particular, the vacuum forces are
repulsive in the case of Dirichlet boundary condition on one plate and
Neumann boundary condition on the other. In the case of a massless scalar
field for the modulus of the corresponding force from (\ref{Til2plR}) with $%
i=l=D$ one finds $F_{\mathrm{DN}}=(1-2^{-D})F_{\mathrm{D}}$, where $F_{%
\mathrm{D}}$ is given by formula (\ref{ForceDN}). Taking $D=3$ and summing
over the polarization states, from here we obtain the result of \cite{Boye74}
for the electromagnetic Casimir force between two parallel plates one of
which is a perfect conductor and the other is infinitely permeable.

Series over the zeros of the function $F(z)$ from (\ref{eigvaluesR}) appear
also related to the Casimir effect for two parallel plates with non-local
boundary conditions
\begin{equation}
n_{(j)}^{\nu }\partial _{\nu }\varphi (x^{\mu })+\int d\mathbf{x}_{\parallel
}^{\prime }\,f_{j}(|\mathbf{x}_{\parallel }-\mathbf{x}_{\parallel }^{\prime
}|)\varphi (x^{\prime \mu })=0,\;y=a_{j},  \label{Boundcond2}
\end{equation}%
where $n_{(j)}^{\nu }$ is the inward-pointing unit normal to the boundary at
$y=a_{j}$. In the region between the plates the corresponding eigenfunctions
have the form (\ref{eigfunc2plR}) and the eigenvalues for $k$ are zeros of
the function $F(ka)$ in (\ref{eigvaluesR}), where now the coefficients are
given by the formula $1/b_{j}=(-1)^{j-1}aF_{j}(k_{\parallel })$ with
\begin{equation}
F_{j}(k_{\parallel })\equiv \int d\mathbf{x}_{\parallel }\,f_{j}(|\mathbf{x}%
_{\parallel }|)e^{i\mathbf{k}_{\parallel }\mathbf{x}_{\parallel }}=\frac{%
(2\pi )^{(D-1)/2}}{k_{\parallel }^{(D-3)/2}}\int_{0}^{\infty
}du\,u^{(D-1)/2}f_{j}(u)J_{(D-3)/2}(uk_{\parallel }).  \label{Fj}
\end{equation}%
The corresponding VEVs are investigated in \cite{Saha06nonloc}.

\section{Scalar vacuum densities for spherical boundaries in the global
monopole bulk}

\label{sec:GlobMonSc}

Historically, the investigation of the Casimir effect for a spherical shell
was motivated by the Casimir semiclassical model of an electron. In this
model Casimir suggested that Poincare stress, to stabilize the charged
particle, could arise {}from vacuum quantum fluctuations and the fine
structure constant can be determined by a balance between the Casimir force
(assumed attractive) and the Coulomb repulsion. However, as it has been
shown by Boyer \cite{Boyer}, the electromagnetic Casimir energy for the
perfectly conducting sphere is positive, implying a repulsive force. This
result has later been reconsidered by a number of authors \cite{DaviesSph,
Bali78, MiltonSph}. More recently new methods have been developed for this
problem including direct mode summation techniques \cite{Leseduarte,
Nesterenko,Sphere} (see also \cite{Most97,Bord01,Milt02} and references
therein). However, the main part of the studies was focused on global
quantities such as the total energy and the vacuum forces acting on the
sphere. The investigation of the energy distribution inside a perfectly
reflecting spherical shell was made in \cite{Olaussen1} in the case of QED
and in \cite{Olaussen2} for QCD. The distribution of the other components
for the electromagnetic EMT inside as well as outside the shell can be
obtained from the results of \cite{Brevik1, Brevik2}. In these papers
consideration was carried out in terms of Schwinger's source theory. In \cite%
{Grig1, Grig2, Grig3} the calculations of the the VEVs for the EMT
components inside and outside the perfectly conducting spherical shell are
based on the GAPF. The same method was used in \cite{Saha01} for the
evaluation of the Wightman function and the VEVs for the EMT of a massive
scalar field with general curvature coupling and obeying Robin boundary
condition on spherically symmetric boundaries in D-dimensional Minkowski
space. In \cite{Schv05} the local VEVs outside a spherical boundary and the
corresponding energy conditions are investigated using the calculational
framework developed in \cite{Grah03}. More complicated problems with
spherical boundaries in the background spacetime of a global monopole are
considered in \cite{Saha03,Saha04Mon,Saha04a,Saha04,Beze06} for both scalar
and fermionic fields. As the corresponding results in the Minkowski bulk are
obtained as special cases, here we discuss the application of the GAPF for
the geometry of spherical boundaries in the global monopole bulk.

Global monopoles are spherically symmetric topological defects created due
to phase transition when a global symmetry is spontaneously broken and they
have important role in cosmology and astrophysics. The global monopole was
first introduced by Sokolov and Starobinsky \cite{Soko77}. A few years
later, the gravitational effects of the global monopole were considered in
\cite{Barr89}, where a solution is presented which describes a global
monopole at large radial distances. Neglecting the small size of the
monopole core, in the hyperspherical polar coordinates $(r,\vartheta ,\phi
)\equiv (r,\theta _{1},\theta _{2},\ldots \theta _{n},\phi )$, $n=D-2$, the
corresponding metric tensor generalized to an arbitrary number of spatial
dimensions can be approximately given by the line element
\begin{equation}
ds^{2}=dt^{2}-dr^{2}-\alpha ^{2}r^{2}d\Omega _{D}^{2},  \label{mmetric}
\end{equation}%
where $d\Omega _{D}^{2}$ is the line element on a surface of the unit sphere
in $D$-dimensional Euclidean space, the parameter $\alpha $ is related to
the symmetry breaking energy scale in the theory. The solid angle
corresponding to (\ref{mmetric}) is $\alpha ^{2}S_{D}$ with $S_{D}=2\pi
^{D/2}/\Gamma (D/2)$ being the total area of the surface of the unit sphere
in $D$-dimensional Euclidean space. The nonzero components of the Ricci
tensor and Ricci scalar for the metric corresponding to line element (\ref%
{mmetric}) are given by expressions (no summation over $i$)
\begin{equation}
R_{i}^{i}=n\frac{1-\alpha ^{2}}{\alpha ^{2}r^{2}},\quad R=n(n+1)\frac{%
1-\alpha ^{2}}{\alpha ^{2}r^{2}},  \label{mRictens}
\end{equation}%
where the indices $i=2,3,\ldots ,D$ correspond to the coordinates $\theta
_{1},\theta _{2},\ldots ,\phi $, respectively. Note that for $\alpha \neq 1$
the geometry is singular at the origin (point-like monopole, for vacuum
polarization effects by a global monopole with finite core see \cite%
{Beze06Fin}). In this section we will consider the applications of the GAPF
for the evaluation of the Wightman function, VEVs of the field square and
the EMT for a scalar field with general curvature coupling parameter under
the presence of spherical boundaries concentric with the global monopole.

\subsection{Region inside a single sphere}

For the region inside a sphere with radius $a$ the complete set of solutions
to (\ref{mfieldeq}) is specified by the set of quantum numbers $\sigma
=(\lambda ,m_{k})$, where $m_{k}=(m_{0}\equiv l,m_{1},\ldots ,m_{n})$ and $%
m_{1},m_{2},\ldots ,m_{n}$ are integers such that $0\leqslant
m_{n-1}\leqslant m_{n-2}\leqslant \cdots \leqslant m_{1}\leqslant l$, $%
-m_{n-1}\leqslant m_{n}\leqslant m_{n-1}$. The corresponding eigenfunctions
have the form
\begin{equation}
\varphi _{\sigma }(x)=\beta _{\sigma }r^{-n/2}J_{\nu _{l}}(\lambda
r)Y(m_{k};\vartheta ,\phi )e^{-i\omega t},\,\,\omega =\sqrt{\lambda
^{2}+m^{2}},\;l=0,1,2,\ldots ,  \label{meigfunc}
\end{equation}%
where the order of the Bessel function is given by the formula%
\begin{equation}
\nu _{l}=\frac{1}{\alpha }\left[ \left( l+\frac{n}{2}\right) ^{2}+(1-\alpha
^{2})n\left( (n+1)\zeta -\frac{n}{4}\right) \right] ^{1/2},  \label{nuel}
\end{equation}%
and $Y(m_{k};\vartheta ,\phi )$ is the hyperspherical harmonic of degree $l$%
. The coefficient $\beta _{\sigma }$ is found from the normalization
condition and is equal to
\begin{equation}
\beta _{\sigma }^{2}=\frac{\lambda T_{\nu _{l}}(\lambda a)}{N(m_{k})\omega
a\alpha ^{D-1}},  \label{normcoef}
\end{equation}%
with $T_{\nu }(z)$ defined by (\ref{teka}). From boundary condition (\ref%
{mrobcond}) for eigenfunctions (\ref{meigfunc}), we see that the eigenvalues
for $\lambda a$ are zeros of the function $\bar{J}_{\nu _{l}}(\lambda a)$
with the barred notation from (\ref{efnot1}) and with the coefficients
\begin{equation}
A=\tilde{A}-nB/2,\quad B=n^{1}\tilde{B}/a,  \label{eigenmodes}
\end{equation}%
where $n^{1}=-1$ for the region inside the sphere. In the notations of
Section 3, for the eigenvalues of $\lambda $ and $\omega $ one finds%
\begin{equation}
\lambda =\lambda _{\nu _{l},k}/a,\;\omega =\sqrt{\lambda _{\nu
_{l},k}^{2}/a^{2}+m^{2}}.  \label{momegalam}
\end{equation}%
Substituting the functions (\ref{meigfunc}) into (\ref{mmodesumWF}) and
using the addition formula for the hyperspherical harmonics \cite{Erde53},
for the Wightman function we obtain
\begin{eqnarray}
W(x,x^{\prime }) &=&\frac{(rr^{\prime })^{-n/2}}{naS_{D}\alpha ^{D-1}}%
\sum_{l=0}^{\infty }(2l+n)C_{l}^{n/2}(\cos \theta )  \notag \\
&&\times \sum_{k=1}^{\infty }\frac{\lambda _{\nu _{l},k}T_{\nu _{l}}(\lambda
_{\nu _{l},k})}{\sqrt{\lambda _{\nu _{l},k}^{2}+m^{2}a^{2}}}J_{\nu
_{l}}(\lambda _{\nu _{l},k}r/a)J_{\nu _{l}}(\lambda _{\nu _{l},k}r^{\prime
}/a)e^{-i\omega _{\nu _{l},k}\Delta t},  \label{fieldmodesum1}
\end{eqnarray}%
where $\Delta t=t-t^{\prime }$. In this formula, $C_{p}^{q}(x)$ is the
Gegenbauer polynomial of degree $p$ and order $q$, and $\theta $ is the
angle between the directions $(\vartheta ,\phi )$ and $(\vartheta ^{\prime
},\phi ^{\prime })$. To sum over $k$ we will use summation formula (\ref%
{sumJ1}), taking%
\begin{equation}
f(z)=zJ_{\nu _{l}}(zr/a)J_{\nu _{l}}(zr^{\prime }/a)e^{-i\Delta t\sqrt{%
z^{2}/a^{2}+m^{2}}}.  \label{mf1z}
\end{equation}%
For $|z|/a<m$ this function satisfies the relation $f(ze^{-\pi
i/2})=-e^{-2\nu \pi i}f(ze^{\pi i/2})$. As a result, the integral on the
right of formula (\ref{sumJ1}) over the interval $(0,ma)$ vanishes.
Conditions (\ref{condf}) are satisfied if $r+r^{\prime }+|\Delta t|<2a$.
After the application of formula (\ref{sumJ1}) the Wightman function is
presented in the form \cite{Saha03}
\begin{equation}
W^{(a)}(x,x^{\prime })=W_{\mathrm{m}}(x,x^{\prime })+\langle \varphi
(x)\varphi (x^{\prime })\rangle _{\mathrm{b}},  \label{unregWight1}
\end{equation}%
where
\begin{equation}
W_{\mathrm{m}}(x,x^{\prime })=\frac{\alpha ^{1-D}}{2nS_{D}}%
\sum_{l=0}^{\infty }\frac{2l+n}{(rr^{\prime })^{n/2}}C_{l}^{n/2}(\cos \theta
)\int_{0}^{\infty }dz\,\frac{ze^{-i\Delta t\sqrt{z^{2}+m^{2}}}}{\sqrt{%
z^{2}+m^{2}}}J_{\nu _{l}}(zr)J_{\nu _{l}}(zr^{\prime }),  \label{Mink}
\end{equation}%
and
\begin{eqnarray}
\langle \varphi (x)\varphi (x^{\prime })\rangle _{\mathrm{b}} &=&-\frac{%
\alpha ^{1-D}}{\pi nS_{D}}\sum_{l=0}^{\infty }\frac{2l+n}{(rr^{\prime
})^{n/2}}C_{l}^{n/2}(\cos \theta )\int_{m}^{\infty }dz\,z\frac{\bar{K}_{\nu
_{l}}(az)}{\bar{I}_{\nu _{l}}(az)}  \notag \\
&&\times \frac{I_{\nu _{l}}(zr)I_{\nu _{l}}(zr^{\prime })}{\sqrt{z^{2}-m^{2}}%
}\cosh (\Delta t\sqrt{z^{2}-m^{2}}).  \label{Wightbound}
\end{eqnarray}%
The contribution of the term (\ref{Mink}) to the VEV does not depend on the
sphere radius, whereas the contribution of the term (\ref{Wightbound})
vanishes in the limit $a\rightarrow \infty $. It follows from here that
expression (\ref{Mink}) is the Wightman function in the unbounded global
monopole space (for the corresponding Euclidean Green function see \cite%
{Mazz91,Beze02}). This can also be seen by explicit evaluation of the
mode-sum using the eigenfunctions for the global monopole spacetime without
boundaries. As we see, the application of the GAPF allows to extract from
the bilinear field product the contribution due to the unbounded monopole
spacetime, and the term (\ref{Wightbound}) can be interpreted as the part of
the VEV induced by the spherical boundary.

The VEV of the field square is obtained by using formula (\ref{mphi2VEV}).
For $0<r<a$ the divergences are contained in the first summand on the right
of Eq. (\ref{unregWight1}) only. Hence, the renormalization procedure for
the local characteristics of the vacuum, such as field square and EMT, is
the same as for the global monopole geometry without boundaries. This
procedure is discussed in a number of papers \cite{Mazz91,Beze02,Hisc90} and
is a useful illustration for the general renormalization prescription on
curved backgrounds. Here we are interested in the parts of the VEVs induced
by the presence of a spherical shell. Using the relation $%
C_{l}^{n/2}(1)=\Gamma (l+n)/\Gamma (n)l!$, for the corresponding boundary
part in the VEV of the field square we get
\begin{equation}
\langle \varphi ^{2}\rangle _{\mathrm{b}}=-\frac{\alpha ^{1-D}}{\pi
r^{n}S_{D}}\sum_{l=0}^{\infty }D_{l}\int_{m}^{\infty }dz\ \frac{\bar{K}_{\nu
_{l}}(az)}{\bar{I}_{\nu _{l}}(az)}\frac{zI_{\nu _{l}}^{2}(rz)}{\sqrt{%
z^{2}-m^{2}}},  \label{regsquare}
\end{equation}%
where
\begin{equation}
D_{l}=(2l+D-2)\frac{\Gamma (l+D-2)}{\Gamma (D-1)\,l!}  \label{Dlang}
\end{equation}%
is the degeneracy of each angular mode with given $l$. As it has been noted
before, expression (\ref{regsquare}) is finite for all values $0<r<a$. For a
given $l$ and large $z$ the integrand behaves as $e^{2z(r/a-1)}/z$ and,
hence, the integral converges when $r<a$. For large values $l$, introducing
a new integration variable $y=z/\nu _{l}$ in the integral of Eq. (\ref%
{regsquare}) and using the uniform asymptotic expansions for the modified
Bessel functions \cite{abramowiz}, it can be seen that the both integral and
sum are convergent for $r<a$ and diverge at $r=a$. For points near the
sphere the leading term of the corresponding asymptotic expansion over the
distance from the boundary has the form
\begin{equation}
\langle \varphi ^{2}\rangle _{\mathrm{b}}\approx \frac{(1-2\delta
_{B0})\Gamma \left( \frac{D-1}{2}\right) }{(4\pi )^{(D+1)/2}(a-r)^{D-1}}.
\label{asympleadterm}
\end{equation}%
This term does not depend on the mass and on the parameter $\alpha $ and is
the same as that for a sphere on the Minkowski bulk. As the purely
gravitational part is finite for $r=a$, we see that near the sphere surface
the VEV of the field square is dominated by the boundary-induced part. The
boundary-induced VEV (\ref{regsquare}) is zero at the sphere center for $%
(\alpha ^{-2}-1)\zeta >0$, non-zero constant for $(\alpha ^{-2}-1)\zeta =0$,
and is infinite for $(\alpha ^{-2}-1)\zeta <0$.

Substituting the Wightman function (\ref{unregWight1}) into Eq. (\ref%
{mTikVEV}), for the VEV of the EMT inside the spherical shell one finds \cite%
{Saha03}
\begin{equation}
\langle 0|T_{i}^{k}|0\rangle =\langle T_{i}^{k}\rangle _{\mathrm{m}}+\langle
T_{i}^{k}\rangle _{\mathrm{b}},  \label{diagEMT}
\end{equation}%
where $\langle T_{i}^{k}\rangle _{\mathrm{m}}$ is the corresponding VEV for
the monopole geometry when the boundary is absent, and the part $\langle
T_{i}^{k}\rangle _{\mathrm{b}}$ is induced by the presence of the spherical
shell. From Eqs. (\ref{Mink}), (\ref{Wightbound}), and (\ref{regsquare}),
for the boundary-induced parts of the EMT components one obtains (no
summation over $i$)
\begin{equation}
\langle T_{i}^{k}\rangle _{\mathrm{b}}=-\frac{\alpha ^{1-D}\delta _{i}^{k}}{%
2\pi r^{n}S_{D}}\sum_{l=0}^{\infty }D_{l}\int_{m}^{\infty }dz\,z^{3}\frac{%
\bar{K}_{\nu _{l}}(za)}{\bar{I}_{\nu _{l}}(za)}\frac{F_{\nu _{l}}^{(i)}\left[
I_{\nu _{l}}(zr)\right] }{\sqrt{z^{2}-m^{2}}},\quad r<a,  \label{q1in}
\end{equation}%
where for a given function $f(y)$ we have introduced the notations
\begin{eqnarray}
F_{\nu _{l}}^{(0)}\left[ f(y)\right] &=&(1-4\zeta )\left[ f^{^{\prime }2}(y)-%
\frac{n}{y}f(y)f^{\prime }(y)\right.  \notag \\
&&\left. +\left( \frac{\nu _{l}^{2}}{y^{2}}-\frac{1+4\zeta -2(mr/y)^{2}}{%
1-4\zeta }\right) f^{2}(y)\right] ,  \label{Fineps} \\
F_{\nu _{l}}^{(1)}\left[ f(y)\right] &=&f^{^{\prime }2}(y)+\frac{\tilde{%
\zeta }}{y}f(y)f^{\prime }(y)-\left( 1+\frac{\nu _{l}^{2}+\tilde{\zeta }n/2}{%
y^{2}}\right) f^{2}(y),\;  \label{Finperad} \\
F_{\nu _{l}}^{(i)}\left[ f(y)\right] &=&(4\zeta -1)f^{^{\prime }2}(y)-\frac{%
\tilde{\zeta }}{y}f(y)f^{\prime }(y)  \notag \\
&&+\left[ 4\zeta -1+\frac{\nu _{l}^{2}(1+\tilde{\zeta })+\tilde{\zeta }n/2}{%
(n+1)y^{2}}\right] f^{2}(y),  \label{Finpeaz}
\end{eqnarray}%
with $\tilde{\zeta }=4(n+1)\zeta -n$ and in (\ref{Finpeaz}) $i=2,3,\ldots ,D$%
. It can be seen that components (\ref{q1in}) satisfy the continuity
equation $\nabla _{k}\langle T_{i}^{k}\rangle _{\mathrm{b}}=0$ and are
finite for $0<r<a$. They are zero at the sphere center for $n\zeta (\alpha
^{-2}-1)>1$, are non-zero constants for $n\zeta (\alpha ^{-2}-1)=1$, and are
infinite otherwise. These singularities appear because the geometrical
characteristics of global monopole spacetime are divergent at the origin.
However, note that the corresponding contribution to the total energy of the
vacuum inside a sphere coming from $\langle T_{0}^{0}\rangle _{\mathrm{b}}$
is finite due to the factor $r^{D-1}$ in the volume element.

Expectation values (\ref{q1in}) diverge at the sphere surface, $r\rightarrow
a$. The corresponding asymptotic behavior can be found using the uniform
asymptotic expansions for the modified Bessel functions, and the leading
terms are determined by the relations
\begin{equation}
\langle T_{0}^{0}\rangle _{\mathrm{b}}\sim \langle T_{2}^{2}\rangle _{%
\mathrm{b}}\sim \frac{Da\langle T_{1}^{1}\rangle _{\mathrm{b}}}{(D-1)(a-r)}%
\sim -\frac{D\Gamma ((D+1)/2)(\zeta -\zeta _{D})}{2^{D}\pi
^{(D+1)/2}(a-r)^{D+1}}\left( 1-2\delta _{B0}\right) .  \label{epsinasimp}
\end{equation}%
As for the field square, these terms do not depend on mass and parameter $%
\alpha $, and are the same as for a spherical shell in the Minkowski bulk.
In general, asymptotic series can be developed in powers of the distance
from the boundary. The corresponding subleading coefficients will depend on
the mass, Robin coefficient, and parameter $\alpha $. Due to the surface
divergencies near the sphere surface the total vacuum EMT is dominated by
the boundary-induced part $\langle T_{i}^{k}\rangle _{\mathrm{b}}$.

\subsection{Vacuum densities in the region between two spheres}

\label{msec2:out}

In this subsection we consider the VEVs of the field square and the EMT on
background of the geometry described by (\ref{mmetric}), assuming that the
field satisfies Robin boundary condition (\ref{mrobcond}) on two concentric
spherical boundaries with radii $a$ and $b$, $a<b$. We will consider the
general case when the coefficients in the boundary conditions for the inner
and outer spheres are different and will denote them by $\tilde{A}_{j}$ and $%
\tilde{B}_{j}$ with $j=a,b$. For the region between two spheres the
corresponding eigenfunctions can be obtained from formula (\ref{meigfunc})
with the replacement%
\begin{equation}
J_{\nu _{l}}(\lambda r)\rightarrow g_{\nu _{l}}(\lambda a,\lambda r)\equiv
J_{\nu _{l}}(\lambda r)\bar{Y}_{\nu _{l}}^{(a)}(\lambda a)-\bar{J}_{\nu
_{l}}^{(a)}(\lambda a)Y_{\nu _{l}}(\lambda r),  \label{meigfunc2sph}
\end{equation}%
where the functions with overbars are defined by (\ref{barjnot}) with
\begin{equation}
A_{j}=\tilde{A}_{j}-B_{j}n/2,\quad B_{j}=n_{j}\tilde{B}_{j}/j,\;j=a,b,%
\;n_{a}=1,\;n_{b}=-1.  \label{barnotab}
\end{equation}%
Substituting eigenfunctions (\ref{meigfunc}) into the normalization
condition and using the value for the standard integral involving the square
of a cylinder function \cite{Prud86}, one finds
\begin{equation}
\beta _{\sigma }^{2}=\frac{\pi ^{2}\lambda T_{\nu _{l}}^{ab}(b/a,\lambda a)}{%
4N(m_{k})\omega a\alpha ^{D-1}},  \label{betalf2sph}
\end{equation}%
where $T_{\nu }^{ab}(\eta ,z)$ is defined by (\ref{tekaAB}).

The functions chosen in the form (\ref{meigfunc2sph}) satisfy the boundary
condition on the inner sphere. From the boundary condition at $r=b$ one
obtains that the corresponding eigenmodes are solutions to the equation
\begin{equation}
C_{\nu _{l}}^{ab}(b/a,\lambda a)=0,  \label{eigmodesab}
\end{equation}%
with the function $C_{\nu }^{ab}(\eta ,z)$ defined by formula (\ref{bescomb1}%
). Hence, in the region between two spheres the eigenvalues for $\lambda $
and $\omega $ are
\begin{equation}
\lambda =\gamma _{\nu _{l},k}/a,\;\omega =\sqrt{\gamma _{\nu
_{l},k}^{2}/a^{2}+m^{2}}.  \label{mlambom2sph}
\end{equation}

Substituting the eigenfunctions into mode-sum (\ref{mmodesumWF}) and using
the addition formula for the spherical harmonics, for the Wightman function
one finds
\begin{equation}
W(x,x^{\prime })=\frac{\pi ^{2}(rr^{\prime })^{-n/2}}{4naS_{D}\alpha ^{D-1}}%
\sum_{l=0}^{\infty }(2l+n)C_{l}^{n/2}(\cos \theta )\sum_{k=1}^{\infty
}h(\gamma _{\nu _{l},k})T_{\nu _{l}}^{ab}(b/a,\gamma _{\nu _{l},k}),
\label{fieldmodesum1ab}
\end{equation}%
with the function
\begin{equation}
h(z)=\frac{ze^{-i\Delta t\sqrt{z^{2}/a^{2}+m^{2}}}}{\sqrt{z^{2}+m^{2}a^{2}}}%
g_{\nu _{l}}(z,zr/a)g_{\nu _{l}}(z,zr^{\prime }/a).  \label{hab}
\end{equation}%
To sum over $k$ we will use summation formula (\ref{cor3form}). The
corresponding conditions for function (\ref{hab}) are satisfied if $%
r+r^{\prime }+|\Delta t|<2b$. In particular, this is the case in the
coincidence limit for the region under consideration. For the Wightman
function one obtains
\begin{eqnarray}
W(x,x^{\prime }) &=&\frac{\alpha ^{1-D}}{2nS_{D}}\sum_{l=0}^{\infty }\frac{%
2l+n}{(rr^{\prime })^{n/2}}C_{l}^{n/2}(\cos \theta )  \notag \\
&&\times \Bigg\{\frac{1}{a}\int_{0}^{\infty }\frac{h(z)dz}{\bar{J}_{\nu
_{l}}^{(a)2}(z)+\bar{Y}_{\nu _{l}}^{(a)2}(z)}-\frac{2}{\pi }\int_{m}^{\infty
}dz\,\frac{z\Omega _{a\nu }(az,bz)}{\sqrt{z^{2}-a^{2}}}  \notag \\
&&\times G_{\nu _{l}}^{(a)}(az,zr)G_{\nu _{l}}^{(a)}(az,zr^{\prime })\cosh
(\Delta t\sqrt{z^{2}-m^{2}})\Bigg\},  \label{unregWightab}
\end{eqnarray}%
where we have introduced the notation%
\begin{equation}
G_{\nu }^{(j)}(z,y)=I_{\nu }(y)\bar{K}_{\nu }^{(j)}(z)-\bar{I}_{\nu
}^{(j)}(z)K_{\nu }(y),\;j=a,b,  \label{Geab}
\end{equation}%
and the function $\Omega _{a\nu }(az,bz)$ is defined by relation (\ref%
{Omegaanu}). In the limit $b\rightarrow \infty $ the second integral on the
right of (\ref{unregWightab}) tends to zero, whereas the first one does not
depend on $b$. It follows from here that the term with the first integral in
the figure braces corresponds to the Wightman function for the region
outside a single sphere with radius $a$ on background of the global monopole
geometry. The latter we will denote by $W^{(a)}(x,x^{\prime })$. So, the
Wightman function in the region between two spheres is presented in the form
\cite{Saha04a}
\begin{eqnarray}
W(x,x^{\prime }) &=&W^{(a)}(x,x^{\prime })-\frac{\alpha ^{1-D}}{\pi nS_{D}}%
\sum_{l=0}^{\infty }\frac{2l+n}{(rr^{\prime })^{n/2}}C_{l}^{n/2}(\cos \theta
)\int_{m}^{\infty }dz\frac{z\Omega _{a\nu _{l}}(az,bz)}{\sqrt{z^{2}-m^{2}}}
\notag \\
&&\times G_{\nu _{l}}^{(a)}(az,rz)G_{\nu _{l}}^{(a)}(az,r^{\prime }z)\cosh
(\Delta t\sqrt{z^{2}-m^{2}}).  \label{regWightab1}
\end{eqnarray}%
To extract from the function $W^{(a)}(x,x^{\prime })$\ the part induced by
the presence of the sphere we use the identity
\begin{equation}
\frac{g_{\nu }(z,y)g_{\nu }(z,y^{\prime })}{\bar{J}_{\nu }^{2}(z)+\bar{Y}%
_{\nu }^{2}(z)}=J_{\nu }(y)J_{\nu }(y^{\prime })-\frac{1}{2}\sum_{s=1}^{2}%
\frac{\bar{J}_{\nu }(z)}{\bar{H}_{\nu }^{(s)}(z)}H_{\nu }^{(s)}(y)H_{\nu
}^{(s)}(y^{\prime }).  \label{relab}
\end{equation}%
The contribution from the first term on the right of this relation with $%
y=xr/a$ and $y^{\prime }=xr^{\prime }/a$ gives the Wightman function $W_{%
\mathrm{m}}(x,x^{\prime })$ for the geometry without boundaries. In the part
coming from the second term we rotate the contour for the integration over $z
$ by the angle $\pi /2$ for $s=1$ and by the angle $-\pi /2$ for $s=2$.
Introducing the modified Bessel functions, one finds
\begin{eqnarray}
W^{(a)}(x,x^{\prime }) &=&W_{\mathrm{m}}(x,x^{\prime })-\frac{\alpha ^{1-D}}{%
\pi nS_{D}}\sum_{l=0}^{\infty }\frac{2l+n}{(rr^{\prime })^{n/2}}%
C_{l}^{n/2}(\cos \theta )\int_{m}^{\infty }dz\,z  \notag \\
&&\times \frac{\bar{I}_{\nu _{l}}(az)}{\bar{K}_{\nu _{l}}(az)}\frac{K_{\nu
_{l}}(zr)K_{\nu _{l}}(zr^{\prime })}{\sqrt{z^{2}-m^{2}}}\cosh (\Delta t\sqrt{%
z^{2}-m^{2}}).  \label{regWightout}
\end{eqnarray}%
Note that in the coincidence limit, $x^{\prime }=x$, the second summand on
the right hand-side of (\ref{regWightab1}) will give a finite result for $%
a\leqslant r<b$, and is divergent on the boundary $r=b$. It can be seen that
for the case of two spheres the Wightman function in the intermediate region
can also be presented in the form
\begin{eqnarray}
W(x,x^{\prime }) &=&W^{(b)}(x,x^{\prime })-\frac{\alpha ^{1-D}}{\pi nS_{D}}%
\sum_{l=0}^{\infty }\frac{2l+n}{(rr^{\prime })^{n/2}}C_{l}^{n/2}(\cos \theta
)  \label{regWightab2} \\
&\times &\int_{m}^{\infty }dz\frac{z\Omega _{b\nu _{l}}(az,bz)}{\sqrt{%
z^{2}-m^{2}}}G_{\nu _{l}}^{(b)}(bz,rz)G_{\nu _{l}}^{(b)}(bz,r^{\prime
}z)\cosh (\Delta t\sqrt{z^{2}-m^{2}}),  \notag
\end{eqnarray}%
with $W^{(b)}(x,x^{\prime })$ being the Wightman function for the vacuum
inside a single sphere with radius $b$, and%
\begin{equation}
\Omega _{b\nu }(x,\lambda x)=\frac{\bar{I}_{\nu }^{(a)}(x)/\bar{I}_{\nu
}^{(b)}(\lambda x)}{\bar{K}_{\nu }^{(a)}(x)\bar{I}_{\nu }^{(b)}(\lambda x)-%
\bar{K}_{\nu }^{(b)}(\lambda x)\bar{I}_{\nu }^{(a)}(x)}.  \label{Omegatilde}
\end{equation}%
In the coincidence limit, the second summand on the right of formula (\ref%
{regWightab2}) is finite for $a<r\leqslant b$ and diverges on the boundary $%
r=a$.

The VEV of the field square is obtained computing the Wightman function in
the coincidence limit $x^{\prime }\rightarrow x$. In the region between the
spheres, from (\ref{regWightab1}) and (\ref{regWightab2}) we obtain two
equivalent forms \cite{Saha04a}
\begin{equation}
\langle 0|\varphi ^{2}|0\rangle =\langle \varphi ^{2}\rangle ^{(j)}-\frac{%
\alpha ^{1-D}}{\pi nS_{D}r^{n}}\sum_{l=0}^{\infty }D_{l}\int_{m}^{\infty }dz%
\frac{z\Omega _{j\nu _{l}}(az,bz)}{\sqrt{z^{2}-m^{2}}}G_{\nu
_{l}}^{(j)2}(jz,rz),  \label{fieldsqform1}
\end{equation}%
with $j=a,b$, and $\langle \varphi ^{2}\rangle ^{(j)}$ is the VEV for the
case of a single sphere with radius $j$.

Using the Wightman function from (\ref{regWightab1}) and the VEV for the
field square from (\ref{fieldsqform1}), the components of the vacuum EMT can
be presented in two equivalent forms corresponding to $j=a,b$ (no summation
over $i$) \cite{Saha04a}:%
\begin{equation}
\langle 0|T_{i}^{k}|0\rangle =\langle T_{i}^{k}\rangle ^{(j)}-\frac{\alpha
^{1-D}\delta _{i}^{k}}{2\pi r^{n}S_{D}}\sum_{l=0}^{\infty
}D_{l}\int_{m}^{\infty }dz\frac{z^{3}\Omega _{j\nu _{l}}(az,bz)}{\sqrt{%
z^{2}-m^{2}}}F_{\nu _{l}}^{(i)}[G_{\nu _{l}}^{(j)}(jz,rz)].  \label{mTik2sph}
\end{equation}%
Here $\langle T_{i}^{k}\rangle ^{(j)}$ is the vacuum EMT for the case of a
single sphere with radius $j$ and the functions $F_{\nu }^{(i)}\left[ f(y)%
\right] $ are defined by relations (\ref{Fineps})-(\ref{Finpeaz}). In the
region outside a single sphere the expressions for the boundary-induced
parts are obtained from the corresponding formula for the interior region
given in the previous subsection by the replacements $I\rightleftarrows K$.
The scalar vacuum densities for spherical boundaries in the Minkowski
spacetime (see \cite{Saha01}) are obtained from the formulae in this section
as a special case taking $\alpha =1$. Note that in this case $\nu _{l}=l+n/2$%
.

The vacuum force per unit surface of the sphere at $r=j$ is determined by
the ${}_{1}^{1}$ -- component of the vacuum EMT at this point. For the
region between two spheres the corresponding effective pressures can be
presented as a sum of two terms:%
\begin{equation}
p^{(j)}=p_{1}^{(j)}+p_{\mathrm{(int)}}^{(j)},\quad j=a,b.  \label{pjsphere}
\end{equation}%
The first term on the right is the pressure for a single sphere at $r=j$
when the second sphere is absent. This term is divergent due to the surface
divergences in the VEVs of local physical observables. Here we will be
concerned with the second term on the right of Eq. (\ref{pjsphere}) which is
the pressure induced by the presence of the second sphere, and can be termed
as an interaction force. Unlike to the single shell parts, this term is free
from renormalization ambiguities and is determined by the last term on the
right of formula (\ref{mTik2sph}). From the formula for the EMT, for the
interaction parts of the vacuum forces per unit surface one finds \cite%
{Saha04a}
\begin{eqnarray}
p_{\mathrm{(int)}}^{(j)} &=&\frac{n_{j}\alpha ^{1-D}}{2\pi j^{D-1}S_{D}}%
\sum_{l=0}^{\infty }D_{l}\int_{m}^{\infty }\frac{zdz}{\sqrt{z^{2}-m^{2}}}%
\left[ 1-\frac{\tilde{\zeta }\tilde{A}_{j}B_{j}}{%
A_{j}^{2}-B_{j}^{2}(z^{2}j^{2}+\nu _{l}^{2})}\right]  \notag \\
&&\times \frac{\partial }{\partial j}\ln \left\vert 1-\frac{\bar{I}_{\nu
_{l}}^{(a)}(az)}{\bar{I}_{\nu _{l}}^{(b)}(bz)}\frac{\bar{K}_{\nu
_{l}}^{(b)}(bz)}{\bar{K}_{\nu _{l}}^{(a)}(az)}\right\vert ,\qquad j=a,b,
\label{pjintsphere}
\end{eqnarray}%
where $\tilde{\zeta }$ is defined after formula (\ref{Finpeaz}). The
expression on the right of this formula is finite for all non-zero distances
between the shells. For Dirichlet and Neumann scalars the second term in the
square brackets is zero. In the case of Dirichlet scalar the interaction
forces are always attractive. For the general Robin case the interaction
force can be either attractive or repulsive in dependence on the
coefficients in the boundary conditions.

\section{Casimir effect for a fermionic field in a global monopole
background with spherical boundaries}

\label{sec:GlobMonFerm}

\subsection{Vacuum energy-momentum tensor inside a spherical shell}

\label{subsec:GlobMonFermIn}

Our interest in this section will be the VEV of the EMT for a fermionic
field with mass $M$ induced by spherical boundaries in the global monopole
spacetime. The dynamics of the field on a curved spacetime is governed by
the Dirac equation
\begin{equation}
i\gamma ^{l}(\partial _{l}+\Gamma _{l})\psi -M\psi =0\ ,  \label{Diraceq}
\end{equation}%
where $\gamma ^{l}$ are the Dirac matrices and $\Gamma _{l}=\gamma
_{k}\nabla _{l}\gamma ^{k}/4$ is the spin connection with $\nabla _{l}$
being the standard covariant derivative operator. We will assume that the
field satisfies bag boundary condition
\begin{equation}
\left( 1+i\gamma ^{l}n_{\mathrm{b}l}\right) \psi =0\ ,\quad r=a,
\label{boundcond1}
\end{equation}%
on the sphere with radius $a$, $n_{\mathrm{b}l}$ is the outward-pointing
normal to the boundary. Expanding the field operator in terms of a complete
set of single-particle states $\{\psi _{\beta }^{(+)},\psi _{\beta }^{(-)}\}$
and making use of the standard anticommutation relations, for the VEV of the
EMT one finds the following mode-sum formula
\begin{equation}
\left\langle 0\left\vert T_{ik}\right\vert 0\right\rangle =\sum_{\beta
}T_{ik}\{\bar{\psi}_{\beta }^{(-)}(x),\psi _{\beta }^{(-)}(x)\}.
\label{modesumFerm}
\end{equation}%
For the geometry under consideration the eigenfunctions are specified by the
set of quantum numbers $\beta =(kjm\sigma )$, where $j=1/2,3/2,\ldots $
determines the value of the total angular momentum, $m=-j,\ldots ,j$ is its
projection, and $\sigma =0,1$ corresponds to two types of eigenfunctions
with different parities. These functions have the form
\begin{eqnarray}
&&\psi _{\beta }^{(\pm )}=A_{\sigma }\frac{e^{-i\omega t}}{\sqrt{r}}\left(
\begin{array}{c}
Z_{\nu _{\sigma }}(kr)\Omega _{jl_{\sigma }m}(\theta ,\varphi ) \\
in_{\sigma }Z_{\nu _{\sigma }+n_{\sigma }}(kr)\frac{k(\hat{n}\cdot \vec{%
\sigma})}{\omega +M}\Omega _{jl_{\sigma }m}(\theta ,\varphi )%
\end{array}%
\right) ,\quad n_{\sigma }=(-1)^{\sigma },  \label{eigfuncFerm} \\
&&\omega =\pm E,\quad E=\sqrt{k^{2}+M^{2}}\ ,\quad \nu _{\sigma }=\frac{j+1/2%
}{\alpha }-\frac{n_{\sigma }}{2}
\end{eqnarray}%
where $\hat{n}=\vec{r}/r$, $\vec{\sigma}=(\sigma ^{1},\sigma ^{2},\sigma
^{3})$ with the curved space Pauli matrices $\sigma ^{k}$. In Eq. (\ref%
{eigfuncFerm}), $Z_{\nu }(x)$ represents a cylinder function of the order $%
\nu $ and $\Omega _{jl_{\sigma }m}(\theta ,\varphi )$ are the standard
spinor spherical harmonics with $l_{\sigma }=j-n_{\sigma }/2$. As in
mode-sum formula (\ref{modesumFerm}) the negative frequency modes are
employed, in the discussion below we will consider the eigenfunctions (\ref%
{eigfuncFerm}) with the lower sign.

For the region inside a spherical shell one has $Z_{\nu }(x)=J_{\nu }(x)$.
The imposition of the boundary condition on the eigenfunctions (\ref%
{eigfuncFerm}) leads to the following equations for the eigenvalues
\begin{equation}
\tilde{J}_{\nu _{\sigma }}(ka)=0\ ,  \label{boundinFerm}
\end{equation}%
where for a given function $F(z)$ we use the notation
\begin{equation}
\tilde{F}(z)\equiv zF^{\prime }(z)+(\mu _{a}-\sqrt{z^{2}+\mu _{a}^{2}}%
-n_{\sigma }\nu )F(z)\ ,\quad \sigma =0,\ 1,  \label{tildenotFerm}
\end{equation}%
with $\mu _{a}=Ma$. Let us denote by $\lambda _{\nu _{\sigma },s}^{\mathrm{f}%
}=ka$, $s=1,2,\ldots ,$ the roots to equation (\ref{boundinFerm}) in the
right half-plane, arranged in ascending order. By using the standard
integral for the Bessel functions, for the normalization coefficient one
finds%
\begin{equation}
A_{\sigma }^{2}=\frac{z}{2\alpha ^{2}a^{2}}\frac{\sqrt{z^{2}+\mu _{a}^{2}}%
+\mu _{a}}{\sqrt{z^{2}+\mu _{a}^{2}}}T_{\nu _{\sigma }}^{\mathrm{f}}(z)\ ,\
\ z=\lambda _{\nu _{\sigma },s}^{\mathrm{f}}\ \ ,  \label{normcoeffFerm}
\end{equation}%
with
\begin{equation}
T_{\nu }^{\mathrm{f}}(z)=\frac{z}{J_{\nu }^{2}(z)}\left[ z^{2}+(\mu
-n_{\sigma }\nu )(\mu _{a}-\sqrt{z^{2}+\mu _{a}^{2}})+\frac{z^{2}}{2\sqrt{%
z^{2}+\mu _{a}^{2}}}\right] ^{-1}\ .  \label{r1Ferm}
\end{equation}

Substituting eigenfunctions (\ref{eigfuncFerm}) into Eq. (\ref{modesumFerm}%
), the summation over the quantum number $m$ can be done by using the
standard summation formula for the spherical harmonics. For the EMT
components one finds (no summation over $i$)
\begin{equation}
\langle 0|T_{i}^{i}|0\rangle =-\sum_{j=1/2}^{\infty }\frac{2j+1}{8\pi \alpha
^{2}a^{3}r}\sum_{\sigma =0,1}\sum_{s=1}^{\infty }T_{\nu _{\sigma }}^{\mathrm{%
f}}(z)f_{\sigma \nu _{\sigma }}^{(i)}\left[ z,J_{\nu _{\sigma }}(zr/a)\right]
_{z=\lambda _{\nu _{\sigma },s}^{\mathrm{f}}},  \label{qrinFerm}
\end{equation}%
where we have introduced the notations
\begin{eqnarray}
f_{\sigma \nu }^{(0)}\left[ z,g_{\nu }(y)\right] &=&z\left[ (\sqrt{z^{2}+\mu
_{a}^{2}}-\mu _{a})g_{\nu }^{2}(y)+(\sqrt{z^{2}+\mu _{a}^{2}}+\mu
_{a})g_{\nu +n_{\sigma }}^{2}(y)\right] ,  \label{fnuepsfe2} \\
f_{\sigma \nu }^{(1)}\left[ z,g_{\nu }(y)\right] &=&\frac{z^{3}}{\sqrt{%
z^{2}+\mu _{a}^{2}}}\left[ g_{\nu }^{2}(y)+g_{\nu +n_{\sigma }}^{2}(y)-\frac{%
2\nu +n_{\sigma }}{y}g_{\nu }(y)g_{\nu +n_{\sigma }}(y)\right] ,
\label{fnupfe2} \\
f_{\sigma \nu }^{(i)}\left[ z,g_{\nu }(y)\right] &=&\frac{z^{3}(2\nu
+n_{\sigma })}{2y\sqrt{z^{2}+\mu _{a}^{2}}}g_{\nu }(y)g_{\nu +n_{\sigma
}}(y)\ ,\;i=2,3.  \label{fnupperpfe2}
\end{eqnarray}

In order to obtain a summation formula for series over the zeros $\lambda
_{\nu ,s}^{\mathrm{f}}$, in the GAPF as a function $g(z)$ we choose
\begin{equation}
g(z)=i\frac{\tilde{Y}_{\nu }(z)}{\tilde{J}_{\nu }(z)}f(z)\ ,
\label{gebesselFerm}
\end{equation}%
with a function $f(z)$ analytic in the right half-plane $\mathrm{Re\,}z>0$.
By making use of the asymptotic formulae for the Bessel functions for large
values of the argument, the conditions for the GAPF can be written in terms
of the function $f(z)$ as follows:
\begin{equation}
|f(z)|<\epsilon (x)e^{c|y|}\ ,\quad z=x+iy,\quad |z|\rightarrow \infty \ ,
\label{condfFerm}
\end{equation}%
where $c<2$ and $\epsilon (x)\rightarrow 0$ for $x\rightarrow \infty $. By
using the Wronskian relation for the Bessel functions, one can see that $%
\tilde{Y}(\lambda _{\nu _{\sigma },s}^{\mathrm{f}})=2/[\pi \tilde{J}(\lambda
_{\nu _{\sigma },s}^{\mathrm{f}})]$. This allows to present the residue term
coming from the poles of the function $g(z)$ in the form
\begin{equation}
\pi i\underset{z=\lambda _{\nu ,s}^{\mathrm{f}}}{\mathrm{Res}}g(z)=T_{\nu }^{%
\mathrm{f}}(\lambda _{\nu ,s}^{\mathrm{f}})f(\lambda _{\nu ,s}^{\mathrm{f}}).
\label{rbesselFerm}
\end{equation}%
Substituting (\ref{gebesselFerm}) and (\ref{rbesselFerm}) into the GAPF (\ref%
{th12}) and taking in this formula the limit $a\rightarrow 0$ (the branch
points $z=\pm i\mu _{a}$ are avoided by semicircles of small radius), we
obtain that for a function $f(z)$ analytic in the half-plane $\mathrm{Re\,}%
z>0$ and satisfying condition (\ref{condfFerm}) the following formula takes
place \cite{Saha04}
\begin{eqnarray}
&&{}\lim_{b\rightarrow +\infty }\left[ \sum_{s=1}^{n}T_{\nu }^{\mathrm{f}%
}(\lambda _{\nu ,s}^{\mathrm{f}})f(\lambda _{\nu ,s}^{\mathrm{f}%
})-\int_{0}^{b}dx\,f(x)\right] =\frac{\pi }{2}\underset{z=0}{\mathrm{Res}}%
f(z)\frac{\tilde{Y}_{\nu }(z)}{\tilde{J}_{\nu }(z)}  \notag \\
&&{}-\frac{1}{\pi }\int_{0}^{\infty }dx\left[ e^{-\nu \pi i}f(xe^{\pi i/2})%
\frac{K_{\nu }^{(+)}(x)}{I_{\nu }^{(+)}(x)}+e^{\nu \pi i}f(xe^{-\pi i/2})%
\frac{K_{\nu }^{(-)}(x)}{I_{\nu }^{(-)}(x)}\right] \ ,  \label{sumJ1Ferm}
\end{eqnarray}%
where on the left $\lambda _{\nu ,n}^{\mathrm{f}}<b<\lambda _{\nu ,n+1}^{%
\mathrm{f}}$. In formula (\ref{sumJ1Ferm})\bigskip\ we use the notations
\begin{equation}
F^{(\pm )}(z)=zF^{\prime }(z)+(\mu _{a}-\sqrt{z^{2}e^{\pm \pi i}+\mu _{a}^{2}%
}-n_{\sigma }\nu )F(z)  \label{Fbarpm}
\end{equation}%
for a given function $F(z)$.

We apply formula (\ref{sumJ1Ferm}) to the series over $s$ in formula (\ref%
{qrinFerm}) for the VEVs of the energy density and vacuum stresses. As it
can be seen from expressions (\ref{fnuepsfe2})--(\ref{fnupperpfe2}), the
corresponding functions $f(z)$ satisfy the relation
\begin{equation}
e^{-\nu \pi i}f(xe^{\pi i/2})=-e^{\nu \pi i}f(xe^{-\pi i/2})\ ,\quad \mathrm{%
for}\quad 0\leqslant x<\mu _{a}\ .  \label{relforf}
\end{equation}%
By taking into account that for these values $x$ one has $%
F^{(+)}(x)=F^{(-)}(x)$, we conclude that the part of the integral on the
right of Eq. (\ref{sumJ1Ferm}) over the interval $(0,\mu _{a})$ vanishes.

As a result, after the application of summation formula (\ref{sumJ1Ferm}),
the components of the vacuum EMT can be presented in the form
\begin{equation}
\langle 0|T_{i}^{k}|0\rangle =\langle T_{i}^{k}\rangle _{\mathrm{m}}+\langle
T_{i}^{k}\rangle _{\mathrm{b}},  \label{qm+qb}
\end{equation}%
where $\langle T_{i}^{k}\rangle _{\mathrm{m}}$ does not depend on the radius
of the sphere $a$ and is the contribution due to unbounded global monopole
spacetime. The corresponding quantities for the massless case are
investigated in \cite{Beze99}. The second term on the right of formula (\ref%
{qm+qb}) is induced by the presence of the spherical shell and can be
presented in the form (no summation over $i$) \cite{Saha04}
\begin{equation}
\langle T_{i}^{i}\rangle _{\mathrm{b}}=\frac{1}{\pi ^{2}\alpha ^{2}a^{3}r}%
\sum_{l=1}^{\infty }l\int_{\mu _{a}}^{\infty }\frac{x^{3}dx}{\sqrt{x^{2}-\mu
_{a}^{2}}}\frac{W\left[ I_{\nu }(x),K_{\nu }(x)\right] }{W\left[ I_{\nu
}(x),I_{\nu }(x)\right] }F_{\nu }^{(i)}\left[ x,I_{\nu }(xr/a)\right] \ ,
\label{qb}
\end{equation}%
with
\begin{eqnarray}
F_{\nu }^{(0)}\left[ x,I_{\nu }(y)\right] &=&\left( 1-\frac{\mu _{a}^{2}}{%
x^{2}}\right) \left\{ I_{\nu -1}^{2}(y)-I_{\nu }^{2}(y)-\mu _{a}\frac{I_{\nu
-1}^{2}(y)+I_{\nu }^{2}(y)}{W\left[ I_{\nu }(x),K_{\nu }(x)\right] }\right\}
,  \label{Fnueps} \\
F_{\nu }^{(1)}\left[ x,I_{\nu }(y)\right] &=&I_{\nu -1}^{2}(y)-I_{\nu
}^{2}(y)-\frac{2\nu -1}{y}I_{\nu }(y)I_{\nu -1}(y).  \label{Fnupperp}
\end{eqnarray}%
Here and below $l=j+1/2$, $\nu \equiv \nu _{1}=l/\alpha +1/2$, and for given
functions $f(x)$ and $g(x)$ we use the notation
\begin{eqnarray}
W\left[ f(x),g(x)\right] &=&\left[ xf^{\prime }(x)+(\mu _{a}+\nu )f(x)\right]
\notag \\
&&\times \left[ xg^{\prime }(x)+(\mu _{a}+\nu )g(x)\right] +(x^{2}-\mu
_{a}^{2})f(x)g(x).  \label{Wnot}
\end{eqnarray}%
It can be easily checked that for a massless spinor field the
boundary-induced part of the vacuum EMT is traceless and the trace anomalies
are contained only in the purely global monopole part without boundaries.
These expressions diverge in a non-integrable manner as the boundary is
approached. The energy density and azimuthal pressure vary, to leading
order, as the inverse cube of the distance from the sphere. At the sphere
center the boundary parts vanish for the global monopole spacetime ($\alpha
<1$) and are finite for the Minkowski spacetime ($\alpha =1$). In the limit
of strong gravitational field, corresponding to small values of the
parameter $\alpha $, describing the solid angle deficit, the
boundary-induced part of the vacuum EMT is suppressed by the factor $\exp
[-(2/\alpha )|\ln (r/a)|]$ and the corresponding vacuum stresses are
strongly anisotropic: $\left\langle T_{1}^{1}\right\rangle _{b}\sim \alpha
\left\langle T_{2}^{2}\right\rangle _{b}$. Having the components of the
energy-momentum tensor we can find the corresponding fermionic condensate $%
\langle 0|\bar{\psi}\psi |0\rangle $ making use of formula $T_{\mu }^{\mu }=M%
\bar{\psi}\psi $ for the trace of the EMT. Note that the formula for the VEV
of the EMT in the exterior region ($r>a$) is obtained from (\ref{qb}) by the
replacement $I_{\nu }\rightarrow K_{\nu }$ in the denominator and in the
argument of the function $F_{\nu }^{(i)}$ \cite{Saha04}.

Fermionic vacuum densities induced by a spherical shell in the Minkowski
bulk are obtained from the results of this section as a special case with $%
\alpha =1$. The previous investigations on the spinor Casimir effect for a
spherical boundary (see, for instance, \cite%
{Most97,Plun86,Bord01,Milt02,Bend76} and references therein) consider mainly
global quantities, such as total vacuum energy. For the case of a massless
spinor the density of the fermionic condensate $\langle 0|\bar{\psi}\psi
|0\rangle $ is investigated in \cite{Milt81ferm} (see also \cite{Milt02}).

\subsection{Vacuum expectation values of the energy-momentum tensor between
two spherical shells}

\label{subsec:vevemtfe2}

In this subsection we consider the region between two spherical shells
concentric with the global monopole on which the fermionic field obeys bag
boundary conditions:%
\begin{equation}
\left( 1+i\gamma ^{l}n_{l}^{(w)}\right) \psi \big|_{r=w}=0\ ,\;w=a,b,
\label{boundcondfe2}
\end{equation}%
where $a$ and $b$ are the radii for the spheres, $a<b$, $n_{l}^{(w)}=n^{(w)}%
\delta _{l}^{1}$ is the outward-pointing normal to the boundaries. Here and
below we use the notations $n^{(a)}=-1$, $n^{(b)}=1$. The corresponding
eigenfunctions have the form (\ref{eigfuncFerm}). Note that in terms of the
function $Z_{\nu _{\sigma }}(kr)$ the boundary conditions (\ref{boundcondfe2}%
) take the form%
\begin{equation}
Z_{\nu _{\sigma }}(kw)=-Z_{\nu _{\sigma }+n_{\sigma }}(kw)\frac{%
n^{(w)}n_{\sigma }k}{\sqrt{k^{2}+M^{2}}-M},  \label{boundcondZfe2}
\end{equation}%
with $w=a,b$. In the region between two spherical shells, the function $%
Z_{\nu _{\sigma }}(kr)$ is a linear combination of the Bessel and Neuamnn
functions. The coefficient in this linear combination is determined from the
boundary condition (\ref{boundcondZfe2}) on the sphere $r=a$ and one obtains%
\begin{equation}
Z_{\nu _{\sigma }}(kr)=g_{\nu _{\sigma }}^{(a)}(ka,kr)\equiv J_{\nu _{\sigma
}}(kr)\tilde{Y}_{\nu _{\sigma }}^{(a)}(ka)-Y_{\nu _{\sigma }}(kr)\tilde{J}%
_{\nu _{\sigma }}^{(a)}(ka),  \label{gnukakrfe2}
\end{equation}%
where for a given function $F(z)$ we use the notation
\begin{equation}
\tilde{F}^{(w)}(z)\equiv zF^{\prime }(z)+[n^{(w)}(\mu _{w}-\sqrt{z^{2}+\mu
_{w}^{2}})-n_{\sigma }\nu _{\sigma }]F(z)\ ,  \label{tildenotfe2}
\end{equation}%
with $w=a,b$, and $\mu _{w}=Mw$. Now, from the boundary condition on the
outer sphere one finds that the eigenvalues for $k$ are solutions to the
equation%
\begin{equation}
C_{\nu _{\sigma }}^{\mathrm{f}}(b/a,ka)\equiv \tilde{J}_{\nu _{\sigma
}}^{(a)}(ka)\tilde{Y}_{\nu _{\sigma }}^{(b)}(kb)-\tilde{Y}_{\nu _{\sigma
}}^{(a)}(ka)\tilde{J}_{\nu _{\sigma }}^{(b)}(kb)=0.  \label{Cnufe2}
\end{equation}%
Below we denote by $\gamma _{\nu _{\sigma },s}^{\mathrm{f}}=ka$, $%
s=1,2,\ldots ,$ the positive roots to equation (\ref{Cnufe2}), arranged in
ascending order, $\gamma _{\nu _{\sigma },s}^{\mathrm{f}}<\gamma _{\nu
_{\sigma },s+1}^{\mathrm{f}}$. Substituting the eigenfunctions into the
normalization integral and using the standard integrals for cylinder
functions (see, for instance, \cite{Prud86}), for the normalization
coefficient of the negative frequency modes one finds%
\begin{equation}
A_{\sigma }^{2}=\frac{\pi ^{2}k(\sqrt{k^{2}+M^{2}}-M)}{8\alpha ^{2}a\sqrt{%
k^{2}+M^{2}}}T_{\nu }^{\mathrm{f}ab}(\eta ,ka),\;ka=\gamma _{\nu _{\sigma
},s}^{\mathrm{f}},  \label{Acoeffe2}
\end{equation}%
where we have introduced the notation
\begin{equation}
T_{\nu }^{\mathrm{f}ab}(\eta ,z)=z\left[ \frac{\tilde{J}_{\nu }^{(a)2}(z)}{%
\tilde{J}_{\nu }^{(b)2}(\eta z)}D_{\nu }^{(b)}-D_{\nu }^{(a)}\right] ^{-1},
\label{tekaABfe2}
\end{equation}%
with%
\begin{equation}
D_{\nu }^{(w)}=w^{2}\left[ k^{2}+(M-E)(M-n_{\sigma }\nu n^{(w)}/w)+\frac{%
n^{(w)}k^{2}}{2wE}\right] ,  \label{Dnuj}
\end{equation}%
and $ka=z$. Substituting eigenfunctions (\ref{eigfuncFerm}) with (\ref%
{gnukakrfe2}) into Eq. (\ref{modesumFerm}), the summation over the quantum
number $m$ can be done by using the standard summation formula for the
spherical harmonics. For the EMT components one finds (no summation over $i$%
)
\begin{equation}
\langle 0|T_{i}^{k}|0\rangle =-\pi \delta _{i}^{k}\sum_{j=1/2}^{\infty }%
\frac{2j+1}{32\alpha ^{2}a^{3}r}\sum_{\sigma =0,1}\sum_{s=1}^{\infty }T_{\nu
_{\sigma }}^{\mathrm{f}ab}(\eta ,z)f_{\sigma \nu _{\sigma }}^{(i)}[z,g_{\nu
_{\sigma }}^{(a)}(z,zr/a)]_{z=\gamma _{\nu _{\sigma },s}^{\mathrm{f}}},
\label{qrinfe2}
\end{equation}%
where the functions $f_{\sigma \nu }^{(i)}\left[ z,g_{\nu }(y)\right] $ are
defined by formulae (\ref{fnuepsfe2})-(\ref{fnupperpfe2}) with
\begin{equation}
g_{\nu +n_{\sigma }}(y)=g_{\nu +n_{\sigma }}^{(a)}(z,y)\equiv J_{\nu
+n_{\sigma }}(y)\tilde{Y}_{\nu }^{(a)}(z)-Y_{\nu +n_{\sigma }}(y)\tilde{J}%
_{\nu }^{(a)}(z).  \label{gnupl}
\end{equation}
The VEV given by formula (\ref{qrinfe2}) is divergent and needs some
regularization procedure. To make it finite we can introduce a cutoff
function $\Phi _{\lambda }(z)$, $z=\gamma _{\nu _{\sigma },s}$, with the
cutoff parameter $\lambda $, which decreases sufficiently fast with
increasing $z$ and satisfies the condition $\Phi _{\lambda }\rightarrow 1$, $%
\lambda \rightarrow 0$.

To evaluate the VEV of the EMT we need to sum series over the zeros of the
function $C_{\nu }^{\mathrm{f}}(\eta ,z)$. A summation formula for this type
of series can be obtained by making use of the GAPF. In the GAPF as
functions $g(z)$ and $f(z)$ we choose
\begin{equation}
g(z)=\frac{1}{2i}\left[ \frac{\tilde{H}_{\nu }^{(1b)}(\eta z)}{\tilde{H}%
_{\nu }^{(1a)}(z)}+\frac{\tilde{H}_{\nu }^{(2b)}(\eta z)}{\tilde{H}_{\nu
}^{(2a)}(z)}\right] \frac{h(z)}{C_{\nu }^{\mathrm{f}}(\eta ,z)},\quad f(z)=%
\frac{h(z)}{\tilde{H}_{\nu }^{(1a)}(z)\tilde{H}_{\nu }^{(2a)}(z)},
\label{gefcombfe2}
\end{equation}%
with the sum and difference
\begin{equation}
g(z)-(-1)^{n}f(z)=-i\frac{\tilde{H}_{\nu }^{(na)}(\eta z)}{\tilde{H}_{\nu
}^{(na)}(z)}\frac{h(z)}{C_{\nu }^{\mathrm{f}}(\eta ,z)},\quad n=1,2.
\label{gefsumnewfe2}
\end{equation}%
The conditions for the GAPF written in terms of the function $h(z)$ are as
follows
\begin{equation}
|h(z)|<\varepsilon _{1}(x)e^{c|y|},\quad |z|\rightarrow \infty ,\quad z=x+iy,
\label{cond31fe2}
\end{equation}%
where $c<2(\eta -1)$ and $\varepsilon _{1}(x)/x\rightarrow 0$ for $%
x\rightarrow +\infty $. To find the residues of the function $g(z)$ at the
poles $z=\gamma _{\nu ,s}^{\mathrm{f}}$ we need the derivative
\begin{equation}
\frac{\partial }{\partial z}C_{\nu }^{\mathrm{f}}(\eta ,z)=\frac{4}{\pi
T_{\nu }^{\mathrm{f}ab}(\eta ,z)}\frac{\tilde{J}_{\nu }^{(b)}(\eta z)}{%
\tilde{J}_{\nu }^{(a)}(z)}\,,\quad z=\gamma _{\nu ,s}^{\mathrm{f}}.
\label{CABderivativeFerm}
\end{equation}%
By using this relation it can be seen that
\begin{equation}
\underset{z=\gamma _{\nu ,s}^{\mathrm{f}}}{\mathrm{Res}}g(z)=\frac{\pi }{4i}%
T_{\nu }^{\mathrm{f}ab}(\eta ,\gamma _{\nu ,s}^{\mathrm{f}})h(\gamma _{\nu
,s}^{\mathrm{f}}).  \label{rel31fe2}
\end{equation}%
Let $h(z)$ be an analytic function for $\mathrm{Re\,}z\geqslant 0$ except
possible branch points on the imaginary axis. By using the GAPF, it can be
seen that if this function satisfies condition (\ref{cond31fe2}),
\begin{equation}
h(ze^{\pi i})=-h(z)+o(z^{-1}),\quad z\rightarrow 0,  \label{cor3cond1fe2}
\end{equation}%
and the integral
\begin{equation}
\int_{0}^{L}\frac{h(x)dx}{\tilde{J}_{\nu }^{(a)2}(x)+\tilde{Y}_{\nu
}^{(a)2}(x)}  \label{cor2cond2fe2}
\end{equation}%
exists, then \cite{Beze06}
\begin{eqnarray}
&&\lim_{L\rightarrow +\infty }\left[ \frac{\pi ^{2}}{4}\sum_{s=1}^{m}h(%
\gamma _{\nu ,s}^{\mathrm{f}})T_{\nu }^{\mathrm{f}ab}(\eta ,\gamma _{\nu
,s}^{\mathrm{f}})-\int_{0}^{L}\frac{h(x)dx}{\tilde{J}_{\nu }^{(a)2}(x)+%
\tilde{Y}_{\nu }^{(a)2}(x)}\right] =  \notag \\
&&\quad =-\frac{\pi }{2}\underset{z=0}{\mathrm{Res}}\left[ \frac{h(z)\tilde{H%
}_{\nu }^{(1b)}(\eta z)}{C_{\nu }^{\mathrm{f}}(\eta ,z)\tilde{H}_{\nu
}^{(1a)}(z)}\right] -\frac{\pi }{4}\sum_{\beta =\pm }\int_{0}^{\infty
}dx\,\Omega _{a\nu }^{(\beta )}(x,\eta x)h(xe^{\beta \pi i/2}),
\label{cor3formfe2}
\end{eqnarray}%
where $\gamma _{\nu ,m}^{\mathrm{f}}<L<\gamma _{\nu ,m+1}^{\mathrm{f}}$.
Here the function $\Omega _{a\nu }^{(\beta )}(x,\eta x)$ is defined as%
\begin{equation}
\Omega _{a\nu }^{(\beta )}(x,\eta x)=\frac{K_{\nu }^{(b\beta )}(\eta
x)/K_{\nu }^{(a\beta )}(x)}{K_{\nu }^{(a\beta )}(x)I_{\nu }^{(b\beta )}(\eta
x)-I_{\nu }^{(a\beta )}(x)K_{\nu }^{(b\beta )}(\eta x)},  \label{Omafe2}
\end{equation}%
and for a given function $F(z)$\ we use the notation
\begin{equation}
F^{(w\pm )}(z)=zF^{\prime }(z)+[n^{(w)}(\mu _{w}-\sqrt{z^{2}e^{\pm \pi
i}+\mu _{w}^{2}})-n_{\sigma }\nu ]F(z)\ .  \label{Fbarpmfe2}
\end{equation}

Now we apply to the sum over $s$ in (\ref{qrinfe2}) the summation formula (%
\ref{cor3formfe2}). As a function $h(z)$ in this formula we take $%
h(z)=f_{\sigma \nu _{\sigma }}^{(q)}[z,g_{\nu _{\sigma }}^{(a)}(z,zr/a)]\Phi
_{\lambda }(z)$. The function $f(z)=f_{\sigma \nu _{\sigma }}^{(q)}[z,g_{\nu
_{\sigma }}^{(a)}(z,zr/a)]$ satisfies the relation
\begin{equation}
f(xe^{\pi i/2})=-f(xe^{-\pi i/2})\ ,\quad \mathrm{for}\quad 0\leqslant
x\leqslant \mu _{a}\ .  \label{relforffe2}
\end{equation}%
By taking into account that for these values $x$ one has $%
F^{(w+)}(wx/a)=F^{(w-)}(wx/a)$, we conclude that in this case the part of
the integral on the right of Eq. (\ref{cor3formfe2}) over the interval $%
(0,\mu _{a})$ vanishes after removing the cutoff. As a result the components
of the vacuum EMT can be presented in the form \cite{Beze06}%
\begin{equation}
\langle 0|T_{i}^{k}|0\rangle =\langle T_{i}^{k}\rangle _{1a}+\langle
T_{i}^{k}\rangle _{ab},  \label{qr1fe2}
\end{equation}%
with separate parts
\begin{equation}
\langle T_{i}^{k}\rangle _{1a}=\frac{-\delta _{i}^{k}}{8\pi \alpha ^{2}a^{3}r%
}\sum_{j=1/2}^{\infty }(2j+1)\sum_{\sigma =0,1}\int_{0}^{\infty }dx\frac{%
f_{\sigma \nu _{\sigma }}^{(i)}[x,g_{\nu _{\sigma }}^{(a)}(x,xr/a)]}{\tilde{J%
}_{\nu _{\sigma }}^{(a)2}(x)+\tilde{Y}_{\nu _{\sigma }}^{(a)2}(x)},
\label{q1outfe2}
\end{equation}%
and%
\begin{equation}
\langle T_{i}^{k}\rangle _{ab}=\frac{-\delta _{i}^{k}}{2\pi ^{2}\alpha ^{2}r}%
\sum_{l=1}^{\infty }l\sum_{\beta =\pm }\int_{M}^{\infty }dx\Omega _{a\nu
}^{(\beta )}(ax,bx)F_{1\nu }^{(i\beta )}[ax,G_{\nu }^{(a\beta )}(ax,rx)],
\label{qabrlfe2}
\end{equation}%
where $\nu \equiv \nu _{1}=l/\alpha +1/2$, and the notation (\ref{Fbarpmfe2}%
) is specified to%
\begin{equation}
F^{(w\pm )}(z)=zF^{\prime }(z)+(\nu +n^{(w)}\mu _{w}\mp in^{(w)}\sqrt{%
z^{2}-\mu _{w}^{2}})F(z)\ .  \label{Fwpmfe2}
\end{equation}%
In formula (\ref{qabrlfe2}) we have introduced the notations%
\begin{eqnarray}
F_{1\nu }^{(0\beta )}[x,G_{\nu }^{(a\beta )}(ax,y)] &=&x\left[ (\sqrt{%
x^{2}-M^{2}}+\beta iM)G_{\nu }^{(a\beta )2}(ax,y)\right.  \notag \\
&&\left. -(\sqrt{x^{2}-M^{2}}-\beta iM)G_{\nu -1}^{(a\beta )2}(ax,y)\right]
\label{Fepsfe2} \\
F_{1\nu }^{(1\beta )}[x,G_{\nu }^{(a\beta )}(ax,y)] &=&\frac{x^{3}}{\sqrt{%
x^{2}-M^{2}}}\left[ G_{\nu }^{(a\beta )2}(ax,y)-G_{\nu -1}^{(a\beta
)2}(ax,y)\right.  \notag \\
&&\left. +\frac{2\nu -1}{y}G_{\nu }^{(a\beta )}(ax,y)G_{\nu -1}^{(a\beta
)}(ax,y)\right]  \label{Fpfe2} \\
F_{\sigma \nu }^{(2\beta )}[x,G_{\nu }^{(a\beta )}(ax,y)] &=&-\frac{(2\nu
-1)x^{3}}{2y\sqrt{x^{2}-M^{2}}}G_{\nu }^{(a\beta )}(ax,y)G_{\nu -1}^{(a\beta
)}(ax,y)\ ,  \label{Fpperpfe2}
\end{eqnarray}%
where
\begin{eqnarray}
G_{\nu }^{(w\pm )}(x,y) &=&I_{\nu }(y)K_{\nu }^{(w\pm )}(x)-K_{\nu
}(y)I_{\nu }^{(w\pm )}(x),\;w=a,b,  \label{Gnufe2} \\
G_{\nu -1}^{(w\pm )}(x,y) &=&I_{\nu -1}(y)K_{\nu }^{(w\pm )}(x)+K_{\nu
-1}(y)I_{\nu }^{(w\pm )}(x).  \label{Gnunsigfe2}
\end{eqnarray}%
In these formulae, for a given function $F(z)$ we use the notation $F^{(w\pm
)}(z)$ defined by formula (\ref{Fbarpmfe2}).

As it has been shown in Ref. \cite{Saha04}, the term (\ref{q1outfe2})
presents the vacuum EMT in the case of a single spherical shell with radius $%
a$ in the region $r>a$. After the subtraction of the part coming from the
global monopole geometry without boundaries, this term is presented in the
form%
\begin{equation}
\langle T_{i}^{k}\rangle _{1a}=\langle T_{i}^{k}\rangle _{\mathrm{m}%
}+\langle T_{i}^{k}\rangle _{a},  \label{q1rfe2}
\end{equation}%
where the part%
\begin{equation}
\langle T_{i}^{k}\rangle _{a}=\frac{-\delta _{i}^{k}}{2\pi ^{2}\alpha ^{2}r}%
\sum_{l=1}^{\infty }l\sum_{\beta =\pm }\int_{M}^{\infty }dx\frac{I_{\nu
}^{(a\beta )}(ax)}{K_{\nu }^{(a\beta )}(ax)}F_{1\nu }^{(i\beta )}\left[
x,K_{\nu }(rx)\right] \ ,  \label{qarfe2}
\end{equation}%
is induced by a single sphere with radius $a$ in the region $r>a$. This
quantity diverges on the boundary $r=a$ with the leading divergence $%
(r-a)^{-3}$ for the energy density and the azimuthal stress, and $(r-a)^{-2}$
for the radial stress.

Note that by using the identities%
\begin{eqnarray}
&&\frac{I_{\nu }^{(a\beta )}(ax)}{K_{\nu }^{(a\beta )}(ax)}F_{1\nu
}^{(i\beta )}\left[ x,K_{\nu }(rx)\right] =\frac{K_{\nu }^{(b\beta )}(bx)}{%
I_{\nu }^{(b\beta )}(bx)}F_{1\nu }^{(i\beta )}\left[ x,I_{\nu }(rx)\right]
\notag \\
&&+\sum_{w=a,b}n^{(w)}\Omega _{w\nu }^{(\beta )}(ax,bx)F_{1\nu }^{(q\beta
)}[x,G_{\nu }^{(w\beta )}(wx,rx)],  \label{ident2fe2}
\end{eqnarray}%
with the notation
\begin{equation}
\Omega _{b\nu }^{(\beta )}(ax,bx)=\frac{I_{\nu }^{(a\beta )}(ax)/I_{\nu
}^{(b\beta )}(bx)}{K_{\nu }^{(a\beta )}(ax)I_{\nu }^{(b\beta )}(bx)-I_{\nu
}^{(a\beta )}(ax)K_{\nu }^{(b\beta )}(bx)},  \label{Ombfe2}
\end{equation}%
the vacuum EMT in the region between the spheres can also be presented in
the form%
\begin{equation}
\langle T_{i}^{k}\rangle =\langle T_{i}^{k}\rangle _{m}+\langle
T_{i}^{k}\rangle _{b}+\langle T_{i}^{k}\rangle _{ba},  \label{qr2fe2}
\end{equation}%
where%
\begin{equation}
\langle T_{i}^{k}\rangle _{ba}=\frac{-\delta _{i}^{k}}{2\pi ^{2}\alpha ^{2}r}%
\sum_{l=1}^{\infty }l\sum_{\beta =\pm }\int_{M}^{\infty }dx\,\Omega _{b\nu
}^{(\beta )}(ax,bx)F_{1\nu }^{(i\beta )}[x,G_{\nu }^{(b\beta )}(bx,rx)],
\label{qbarfe2}
\end{equation}%
and the quantities $\left\langle T_{i}^{k}\right\rangle _{b}$ are the vacuum
densities induced by a single shell with radius $b$ in the region $r<b$. The
formula for the latter is obtained from (\ref{qarfe2}) by the replacements $%
a\rightarrow b$, $I\rightleftarrows K$ and coincides with the result derived
in the previous subsection. The surface divergences in vacuum expectation
values of the EMT are the same as those for a single sphere when the second
sphere is absent. In particular, the term $\left\langle
T_{i}^{k}\right\rangle _{ab}$ ($\left\langle T_{i}^{k}\right\rangle _{ba}$)
is finite on the outer (inner) sphere. In the formulae above taking $\alpha
=1$ we obtain the corresponding quantities for a spinor field in the
Minkowski bulk. In this case $\nu =l+1/2$ and the Bessel modified functions
are expressed in terms of elementary functions.

As in the case of a scalar field, the vacuum forces acting on the spheres
can be presented in the form of the sum of self-action and interaction
parts. The latter is determined by formulae (\ref{qabrlfe2}) and (\ref%
{qbarfe2}) with $i=k=1$ substituting $r=a$ and $r=b$ respectively. By making
use of the properties of the modified Bessel functions, the interaction
force on the sphere $r=w$ is presented in the form%
\begin{equation}
p_{\mathrm{(int)}}^{(w)}=\frac{n^{(w)}}{\pi ^{2}\alpha ^{2}w^{2}}\frac{%
\partial }{\partial w}\sum_{l=1}^{\infty }l\int_{M}^{\infty }dx\frac{x}{%
\sqrt{x^{2}-M^{2}}}\ln \left\vert 1-\frac{I_{\nu }^{(a+)}(ax)K_{\nu
}^{(b+)}(bx)}{K_{\nu }^{(a+)}(ax)I_{\nu }^{(b+)}(bx)}\right\vert ,
\label{pwint1}
\end{equation}%
where $w=a,b$, and $n^{(w)}$ is defined after formula (\ref{boundcondfe2}).
As it has been shown in \cite{Beze06}, these forces can also be obtained
from the Casimir energy by standard differentiation with respect to the
sphere radii.

\section{Electromagnetic Casimir densities for conducting spherical
boundaries}

\label{sec:ElSpheric}

\subsection{Energy-momentum tensor inside a spherical shell}

\label{subsec:ElintSphere}

In this section we consider the application of the GAPF to the investigation
of the VEV for the EMT of the electromagnetic field induced by perfectly
conducting spherical boundaries in the Minkowski spacetime. For the region
inside a perfectly conducting sphere with radius $a$, in the Coulomb gauge
the corresponding eigenfunctions for the vector potential are presented in
the form
\begin{equation}
\mathbf{A}_{\sigma }=\beta _{\sigma }e^{-i\omega t}\left\{
\begin{array}{ll}
\omega j_{l}(\omega r)\mathbf{X}_{lm}(\theta ,\varphi ), & \text{$\lambda =0$%
} \\
\nabla \times \left[ j_{l}(\omega r)\mathbf{X}_{lm}(\theta ,\varphi )\right]
, & {\text{$\lambda =1$}}%
\end{array}%
\right. ,\quad \sigma =(\omega lm\lambda ),  \label{eigfuncins}
\end{equation}%
where $\lambda =0$ and 1 correspond to the spherical waves of magnetic and
electric type (TE and TM modes respectively). They describe photon with
definite values of the total angular momentum $l$, its projection $m$,
energy $\omega $ and parity $(-1)^{l+\lambda +1}$. Here the spherical vector
harmonics have the form
\begin{equation}
\mathbf{X}_{lm}(\theta ,\varphi )=-i\frac{\mathbf{r}\times \nabla }{\sqrt{%
l(l+1)}}Y_{lm}(\theta ,\varphi ),\quad l\neq 0,  \label{vecspharm}
\end{equation}%
with $Y_{lm}(\theta ,\varphi )$ being the spherical functions, and $j_{l}(x)=%
\sqrt{\pi /2x}J_{l+1/2}(x)$ is the spherical Bessel function. The
coefficients $\beta _{\sigma }$ are determined by the normalization
condition
\begin{equation}
\int dV\,{\mathbf{A}_{\sigma }\cdot \mathbf{A}_{\sigma ^{\prime }}^{\ast }}=%
\frac{2\pi }{\omega }\delta _{\sigma \sigma ^{\prime }},
\label{normElintSph}
\end{equation}%
where the integration goes over the region inside the sphere. Using the
standard formulae for the spherical vector harmonics and spherical Bessel
functions (see, for example, \cite{Jackson}), for the normalization
coefficient one finds
\begin{subequations}
\begin{equation}
\beta _{\sigma }^{2}=8T_{\nu }(\omega a)/\omega a,\quad \nu =l+1/2,
\label{betElSphint}
\end{equation}%
where $T_{\nu }(z)$ is defined in (\ref{teka}).

Inside the perfectly conducting sphere the photon energy levels are
quantized by the standard boundary conditions:
\end{subequations}
\begin{equation}
\mathbf{n}\times \mathbf{E}=0,\quad \mathbf{n}\cdot \mathbf{B}=0,\quad r=a,
\label{ElBCcond}
\end{equation}%
where $\mathbf{E}$ and $\mathbf{B}$ are the electric and magnetic fields and
$\mathbf{n}$ is the normal to the boundary. They lead to the following
eigenvalue equations with respect to $\omega $:
\begin{equation}
\partial _{r}^{\lambda }\left[ rj_{l}(\omega r)\right] _{r=a}=0,\quad \text{$%
\lambda =0,1.$}  \label{TEmode}
\end{equation}%
It is well known that these equations have an infinite number of real simple
roots.

By substituting the eigenfunctions into the mode-sum formula
\begin{equation}
\langle 0|T_{ik}|0\rangle =\sum_{\sigma }T_{ik}\left\{ {\mathbf{A}_{\sigma }}%
,{\mathbf{A}_{\sigma }^{\ast }}\right\} ,  \label{emtgform}
\end{equation}%
with the standard expression for the electromagnetic EMT and after the
summation over $m$ by using the standard formulae for the vector spherical
harmonics (see, for example, \cite{Jackson}), one obtains
\begin{equation}
\langle 0|T_{i}^{k}|0\rangle =\mathrm{diag}\left( \varepsilon
,\,-p,\,-p_{\bot },\,-p_{\bot }\right)  \label{emt1}
\end{equation}%
(index values 1,2,3 correspond to the spherical coordinates $r,\theta
,\varphi $ with the origin at the sphere center). Here the energy density, $%
\varepsilon $, effective pressures in transverse, $p_{\perp }$, and radial, $%
p$, directions are determined by the relations
\begin{equation}
q(a,r)=\sum_{\omega l\lambda }\frac{2l+1}{4\pi ^{2}a}\omega ^{3}T_{\nu
}(\omega a)D_{l}^{(q)}[j_{l}(\omega r)],\quad q=\varepsilon ,\,p,\,p_{\perp
},  \label{tr0}
\end{equation}%
where the following notations are introduced
\begin{equation}
D_{l}^{(q)}[f(y)]=\left\{
\begin{array}{l}
\left[ yf(y)\right] ^{\prime 2}/y^{2}+[1+l(l+1)/y^{2}]f^{2}(y),\quad
q=\varepsilon , \\
l(l+1)f^{2}(y)/y^{2},\quad q=p_{\perp },%
\end{array}%
\right.  \label{Dl}
\end{equation}%
and $p=\varepsilon -2p_{\perp }$. In the sum (\ref{tr0}), $\omega $ takes a
discrete set of values determined by equations (\ref{TEmode}).

The VEVs (\ref{tr0}) are infinite. The renormalization of $\langle
0|T_{ik}|0\rangle $ in flat spacetime is affected by subtracting from this
quantity its singular part $\langle 0_{\mathrm{M}}|T_{ik}|0_{\mathrm{M}%
}\rangle $, which is precisely the value it would have if the boundary is
absent. Here $|0_{\mathrm{M}}\rangle $ is the amplitude for the Minkowski
vacuum state. To evaluate the finite difference between these two infinities
we will introduce a cutoff function $\psi _{\mu }(\omega )$, which makes the
sums finite and satisfies the condition $\psi _{\mu }(\omega )\rightarrow
1,\,\mu \rightarrow 0$. After the subtraction we will allow $\mu \rightarrow
0$:
\begin{equation}
\langle T_{ik}\rangle _{\mathrm{ren}}=\lim_{\mu \rightarrow 0}\left[ \langle
0|T_{ik}|0\rangle -\langle 0_{\mathrm{M}}|T_{ik}|0_{\mathrm{M}}\rangle %
\right] .  \label{reg}
\end{equation}%
Hence, we consider the following finite quantities
\begin{equation}
q(\mu ,a,r)=\sum_{l=1}^{\infty }\frac{2l+1}{4\pi ^{2}a^{4}}\sum_{\lambda
=0,1}\sum_{k=1}^{\infty }j_{\nu ,k}^{(\lambda )3}T_{\nu }(j_{\nu
,k}^{(\lambda )})\psi _{\mu }(j_{\nu ,k}^{(\lambda
)}/a)D_{l}^{(q)}[j_{l}(j_{\nu ,k}^{(\lambda )}x)],  \label{q2}
\end{equation}%
where $x=r/a$, and $\omega =j_{\nu ,k}^{(\lambda )}/a$ are solutions to the
eigenvalue equations (\ref{TEmode}) for $\lambda =0,1$, respectively. The
summation over $k$ in (\ref{q2}) can be done by using formula (\ref%
{sumJ1anal}) and taking $A=1,B=0$ for TE modes ($\lambda =0$) and $A=1,B=2$
for TM modes ($\lambda =1$) (recall that in (\ref{sumJ1}) $\lambda _{\nu ,k}$
are zeros of $\bar{J}_{\nu }(z)$ with barred quantities defined as (\ref%
{efnot1})). Let us substitute in formula (\ref{sumJ1anal})
\begin{equation}
f(z)=z^{3}\psi _{\mu }(z/a)D_{l}^{(q)}[j_{l}(zx)],  \label{ftoAPF}
\end{equation}%
with $D_{l}^{(q)}[f(y)]$ defined from (\ref{Dl}). We will assume the class
of cutoff functions for which the function (\ref{ftoAPF}) satisfies the
conditions for Theorem 2, uniformly with respect to $\mu $ (the
corresponding restrictions for $\psi _{\mu }$ can be easily found from these
conditions using the asymptotic formulae for the Bessel functions). Below
for simplicity we will consider the functions without poles. In this case,
function (\ref{ftoAPF}) is analytic on the right-half plane of the complex
variable $z$. The discussion on the conditions to cutoff functions, under
which the difference between divergent sum and integral exists and has a
finite value independent any further details of cutoff function, can be
found in \cite{Bart80}. For TE and TM modes by choosing the constants $A$
and $B$ as mentioned above one obtains
\begin{eqnarray}
q &=&\frac{1}{8\pi ^{2}}\sum_{l=1}^{\infty }(2l+1)\left\{ 2\int_{0}^{\infty
}d\omega \,\omega ^{3}\psi _{\mu }(\omega )D_{l}^{(q)}[j_{l}(\omega
r)]\right.  \notag \\
&&\left. -\frac{1}{r^{2}}\int_{0}^{\infty }dz\,z\chi _{\mu }(z)\left[ \frac{%
e_{l}(az)}{s_{l}(az)}+\frac{e_{l}^{\prime }(az)}{s_{l}^{\prime }(az)}\right]
F_{l}^{(q)}[s_{l}(zr)]\right\} ,  \label{q3}
\end{eqnarray}%
where the functions $F_{l}^{(q)}[f(y)]$ are defined as
\begin{eqnarray}
F_{l}^{(\varepsilon )}[f(y)] &=&f^{\prime 2}(y)+[l(l+1)/y^{2}-1]f^{2}(y),
\label{Feps} \\
F_{l}^{(p_{\perp })}[f(y)] &=&l(l+1)\frac{f^{2}(y)}{y^{2}},\quad \chi _{\mu
}(y)=\left[ \psi _{\mu }(iy)+\psi _{\mu }(-iy)\right] /2,  \label{chisph}
\end{eqnarray}%
and $F_{l}^{(p)}[f(y)]=F_{l}^{(\varepsilon )}[f(y)]-2F_{l}^{(p_{\perp
})}[f(y)]$. In (\ref{q3}) we have introduced the Ricatti-Bessel functions of
the imaginary argument,
\begin{equation}
s_{l}(z)=\sqrt{\frac{\pi z}{2}}I_{\nu }(z),\quad e_{l}(z)=\sqrt{\frac{2z}{%
\pi }}K_{\nu }(z),\,\nu =l+1/2.  \label{RicBes}
\end{equation}%
As $\langle 0_{\mathrm{M}}|T_{ik}|0_{\mathrm{M}}\rangle =\lim_{a\rightarrow
\infty }\langle 0|T_{ik}|0\rangle $, the first integral in (\ref{q3})
represents the vacuum EMT for empty Minkowski spacetime:
\begin{equation}
q_{\mathrm{M}}=\frac{1}{4\pi ^{2}}\sum_{l=1}^{\infty }(2l+1)\int_{0}^{\infty
}d\omega \,\omega ^{3}\psi _{\mu }(\omega )D_{l}^{(q)}[j_{l}(\omega r)].
\label{EMTMin}
\end{equation}%
This expression can be further simplified. For example, in the case of the
energy density one has
\begin{eqnarray}
\varepsilon _{\mathrm{M}} &=&\frac{1}{4\pi ^{2}}\sum_{l=1}^{\infty
}\int_{0}^{\infty }d\omega \,\omega ^{3}\psi _{\mu }(\omega )\left[
lj_{l+1}^{2}(\omega r)+(l+1)j_{l-1}^{2}(\omega r)+(2l+1)j_{l}^{2}(\omega r)%
\right] {}  \notag \\
&=&\frac{1}{2\pi ^{2}}\int_{0}^{\infty }d\omega \,\omega ^{3}\psi _{\mu
}(\omega )\sum_{l=0}^{\infty }(2l+1)j_{l}^{2}(\omega r)=\int_{0}^{\infty
}d\omega \,\omega ^{3}\psi _{\mu }(\omega ).  \label{epsMin}
\end{eqnarray}

As we see, the use of the GAPF allows to extract from infinite quantities
the divergent part without specifying the form of the cutoff function. Now
the renormalization of the EMT is equivalent to omitting the first summand
in (\ref{q3}), which corresponds to the contribution of the spacetime
without boundaries. For $r<a$ the second term on the right (\ref{q3}) is
finite in the limit $\mu \rightarrow 0$ and for the renormalized components
one obtains
\begin{equation}
q_{\mathrm{ren}}(a,r)=-\sum_{l=1}^{\infty }\frac{2l+1}{8\pi ^{2}r^{2}}%
\int_{0}^{\infty }dz\,z\left[ \frac{e_{l}(az)}{s_{l}(az)}+\frac{%
e_{l}^{\prime }(az)}{s_{l}^{\prime }(az)}\right] F_{l}^{(q)}[s_{l}(zr)],%
\quad r<a.  \label{regq}
\end{equation}%
From here it is obvious the independence of the renormalized quantities on
the specific form of the cutoff, on the class of functions for which (\ref%
{ftoAPF}) satisfies conditions for (\ref{sumJ1anal}). The derivation of the
vacuum densities (\ref{regq}) given above uses the GAPF to summarize
mode-sums and is based on \cite{Grig1, Grig2}. One can see that these
formulae for the case of exponential cutoff function may also be obtained
from the results of \cite{Brevik1, Brevik2}, where the Green function method
is used.

We obtained the renormalized values (\ref{regq}) by introducing a cutoff
function and subsequently subtracting the contribution due to the unbounded
space. The GAPF in the form (\ref{sumJ1anal}) allows to obtain immediately
this finite difference. It should be noted that by using the GAPF in the
form (\ref{sumJ1}) we can derive the expressions for the renormalized
azimuthal pressure without introducing any special cutoff function. To see
this note that for $x<1$ the function (\ref{ftoAPF}) with $q=p_{\perp }$ and
$\psi _{\mu }=1$ satisfies conditions of Theorem 2. It follows from here
that we can apply formula (\ref{sumJ1}) directly to the corresponding sum
over $\omega $ in (\ref{tr0}) or over $k$ in (\ref{q2}) without introducing
a cutoff function. This immediately yields to formula (\ref{regq}) for $%
q=p_{\perp }$ with $\psi _{\mu }=1$.

{}The VEV given by formula (\ref{regq}){} diverges on the sphere surface, $%
r=a$, and this divergence is due to the contribution of large $l$ (note that
the integral over $z$ converges at $r=a$). The corresponding asymptotic
behavior can be found by using the uniform asymptotic expansions for the
Bessel functions and the leading terms have the form
\begin{equation}
\varepsilon \approx 2p_{\perp }\approx \frac{-1}{30\pi ^{2}a(a-r)^{3}},\quad
p\approx \frac{-1}{60\pi ^{2}a^{2}(a-r)^{2}}\,.  \label{nearsurface}
\end{equation}%
These surface divergences originate in the unphysical nature of perfect
conductor boundary conditions and are well-known in quantum field theory
with boundaries. As we have mentioned before, they are investigated in
detail for various types of fields and general shape of the boundary. Eqs. (%
\ref{nearsurface}) are particular cases of the asymptotic expansions for the
VEV near the smooth boundary given in \cite{Deutsch}. In reality the
expectation values for the EMT components will attain a limiting value on
the conductor surface, which will depend on the molecular details of the
conductor. {}From the asymptotic expansions given above it follows that the
main contributions to $q_{\mathrm{ren}}(r)$ are due to the frequencies $%
\omega <(a-r)^{-1}$. Hence we expect that formulae (\ref{regq}) are valid
for real conductors up to distances $r$ for which $(a-r)^{-1}\ll \omega _{0}$%
, with $\omega _{0}$ being the characteristic frequency, such that for $%
\omega >\omega _{0}$ the conditions for perfect conductivity fail.

At the sphere center, in (\ref{regq}) $l=1$ multipole contributes only and
we obtain \cite{Grig2,Olaussen1}
\begin{equation}
\varepsilon (0)=3p(0)=3p_{\perp }(0)\approx -0.0381a^{-4}.  \label{atcenter}
\end{equation}%
At the center the equation of state for the electromagnetic vacuum is the
same as that for blackbody radiation. Note that the corresponding results
obtained by using the uniform asymptotic expansions for the Bessel functions
\cite{Brevik1, Brevik2} are in good agreement with (\ref{atcenter}).

The components of the renormalized EMT satisfy continuity equation $\nabla
_{k}T_{i}^{k}=0$, which for the spherical geometry takes the form
\begin{equation}
p^{\prime }(r)+2(p-p_{\perp })/r=0.  \label{ElSpIntconteq}
\end{equation}%
{} From here by using the zero trace condition the following integral
relations may be obtained%
\begin{equation}
p(r)=\frac{1}{r^{3}}\int_{0}^{r}dt\,t^{2}\varepsilon (t)=\frac{2}{r^{2}}%
\int_{0}^{r}dt\,tp_{\perp }(t),  \label{intrel7}
\end{equation}%
where the integration constant is determined from relations (\ref{atcenter})
at the sphere center. It follows from the first relation that the total
energy within a sphere with radius $r$ is equal to
\begin{equation}
E(r)=4\pi \int_{0}^{r}dt\,t^{2}\varepsilon (t)=3V(r)p(r),  \label{thermrel}
\end{equation}%
where $V(r)$ is the corresponding volume.

The distribution for the vacuum energy density and pressures inside the
perfectly conducting sphere can be obtained from the results of the
numerical calculations given in \cite{Brevik1, Brevik2}. In their
calculations Brevik and Kolbenstvedt used the uniform asymptotic expansions
of the Ricatti-Bessel functions for large values of the order. In \cite%
{Grig1, Grig2} (see also \cite{Sahdis}) the corresponding quantities are
calculated on the base of the exact relations for these functions and the
accuracy of the numerical results in \cite{Brevik1, Brevik2} is estimated ($%
\approx 5\%$). The simple approximation formulae are presented with the same
accuracy as asymptotic expressions. Note that inside the sphere all
quantities $\varepsilon ,\,p,\,p_{\perp }$ are negative.

By the method similar to that used in this subsection, the VEVs
for gluon fields can be evaluated in the bag model for hadrons
\cite{Chod74}. In this model the vacuum outside the spherical bag
is a perfect color magnetic conductor. The color electric and
magnetic fields are confined inside the
bag and satisfy the boundary conditions $\mathbf{n}\cdot \mathbf{E}=0$, $%
\mathbf{n}\times \mathbf{B}=0$ on its boundary (for the role of the Casimir
effect in the bag model see \cite{Most97,Plun86,Milt02}).

\subsection{Electromagnetic vacuum in the region between two spherical shells%
}

\label{subsec:El2Sph}

Now let us consider the electromagnetic vacuum in the region between two
concentric conducting spherical shells with radii $a$ and $b$, $a<b$. In the
Coulomb gauge the complete set of solutions to the Maxwell equations can be
written in the form similar to (\ref{eigfuncins}):%
\begin{equation}
\mathbf{A}_{\sigma }=\beta _{\sigma }\frac{e^{-i\omega t}}{\sqrt{4\pi }}%
\left\{
\begin{array}{ll}
\omega g_{0l}(\omega a,\omega r)\mathbf{X}_{lm}, & \text{$\lambda =0$} \\
\nabla \times \left[ g_{1l}(\omega a,\omega r)\mathbf{X}_{lm}\right] , &
\text{$\lambda =1$}%
\end{array}%
\right. ,  \label{eigfuncab}
\end{equation}%
where as above the values $\lambda =0$ and $\lambda =1$ correspond to the
waves of magnetic (TE modes) and electric (TM modes) type,
\begin{equation}
g_{\lambda l}(x,y)=\left\{
\begin{array}{ll}
j_{l}(y)n_{l}(x)-j_{l}(x)n_{l}(y), & \text{$\lambda =0$} \\
j_{l}(y)[xn_{l}(x)]^{^{\prime }}-[xj_{l}(x)]^{^{\prime }}n_{l}(y), & \text{$%
\lambda =1$}%
\end{array}%
\right. ,  \label{gl8}
\end{equation}%
with $n_{l}(x)$ being the spherical Neumann function. From the boundary
conditions at surfaces $r=a$ and $r=b$ one finds that the possible energy
levels of the photon are solutions to the following equations
\begin{equation}
\partial _{r}^{\lambda }\left[ rg_{\lambda l}(\omega a,\omega r)\right]
_{r=b}=0,\quad \lambda =0,1.  \label{modesab}
\end{equation}%
All roots of these equations are real and simple \cite{abramowiz}.

The coefficient $\beta _{\sigma }$ in (\ref{eigfuncab}) is determined from
the normalization condition (\ref{normElintSph}), where now the integration
goes over the region between spherical shells, $a\leqslant r\leqslant b$. By
using the standard relations for the spherical Bessel functions it can be
presented in the form
\begin{eqnarray}
\beta _{\sigma }^{2} &=&\omega a\left[ \frac{aj_{l}^{2}(\omega a)}{%
bj_{l}^{2}(\omega b)}-1\right] ^{-1},\quad \lambda =0,  \label{normab} \\
\beta _{\sigma }^{2} &=&\frac{1}{\omega a}\left\{ \frac{b\left[ \omega
aj_{l}(\omega a)\right] ^{^{\prime }2}}{a\left[ \omega bj_{l}(\omega b)%
\right] ^{^{\prime }2}}\left[ 1-\frac{l(l+1)}{\omega ^{2}b^{2}}\right] -1+%
\frac{l(l+1)}{\omega ^{2}a^{2}}\right\} ^{-1},\quad \lambda =1.
\label{normab1}
\end{eqnarray}%
From the mode-sum formula (\ref{emtgform}) with the functions (\ref%
{eigfuncab}) as a complete set of solutions one obtains the VEV in the form (%
\ref{emt1}) with
\begin{equation}
q=\frac{1}{8\pi }\sum_{\omega l\lambda }(2l+1)\omega ^{4}\beta _{\alpha
}^{2}D_{l}^{(q)}[g_{\lambda l}(\omega a,\omega r)],\quad q=\varepsilon
,\,p,\,p_{\perp },  \label{epspeab}
\end{equation}%
where the frequencies $\omega $ are solutions to equations (\ref{modesab})
and the functions $D_{l}^{(q)}[f(y)]$ are defined by formulae (\ref{Dl})
with $f(y)=g_{\lambda l}(\omega a,y)$. It is easy to see that the eigenvalue
equations (\ref{modesab}) can be written in terms of the function $C_{\nu
}^{ab}$, defined by (\ref{bescomb1}), as
\begin{equation}
C_{\nu }^{ab}(\eta ,\omega a)=0,\quad \nu =l+1/2,\,\eta
=b/a,\,A=1/(1+\lambda ),\,B=\lambda ,\,\lambda =0,1.  \label{modesabC}
\end{equation}%
By this choice of constants $A_{j}$ and $B_{j}$ the normalization
coefficients (\ref{normab}) and (\ref{normab1}) are related to the function $%
T_{\nu }^{ab}$ from (\ref{tekaAB}) by the formula
\begin{equation}
\beta _{\sigma }^{2}=T_{\nu }^{ab}(\eta ,\omega a).  \label{normabT}
\end{equation}%
This allows us to use the formulae from Section \ref{sec:SumFormBess} for
the summation over the eigenmodes.

As above, to regularize infinite quantities (\ref{epspeab}) we introduce a
cutoff function $\psi _{\mu }(\omega )$ and consider the finite quantities
\begin{equation}
q=\sum_{l=1}^{\infty }\frac{2l+1}{8\pi a^{4}}\sum_{\lambda
=0}^{1}\sum_{k=1}^{\infty }z^{4}T_{\nu }^{AB}(\eta ,z)\psi _{\mu
}(z/a)D_{l}^{(q)}[g_{\lambda l}(z,zx)]|_{z=\gamma _{\nu ,k}^{(\lambda )}},
\label{qab}
\end{equation}%
where $x=r/a$, and $\omega a=\gamma _{\nu ,k}^{(\lambda )}$ are solutions to
equations (\ref{modesab}) or (\ref{modesabC}). In order to sum over $k$ we
will use formula (\ref{cor3form}) with
\begin{equation}
h(z)=z^{4}\psi _{\mu }(z/a)D_{l}^{(q)}[g_{\lambda l}(z,zx)],  \label{hzab}
\end{equation}%
assuming a class of cutoff functions for which (\ref{hzab}) satisfies
conditions (\ref{cond31}) and (\ref{cor3cond1}) uniformly with respect to $%
\mu $. The corresponding restrictions on $\psi _{\mu }$ can be obtained
using the asymptotic formulae for Bessel functions. From (\ref{qab}), by
applying to the sum over $k$ formula (\ref{cor3form}), for the EMT
components one obtains
\begin{eqnarray}
q &=&\sum_{l=1}^{\infty }\frac{2l+1}{8\pi ^{2}a^{4}}\sum_{\lambda
=0,1}\left\{ \int_{0}^{\infty }dz\,z^{3}\psi _{\mu }(z/a)\frac{%
D_{l}^{(q)}[g_{\lambda l}(z,zx)]}{\Omega _{1\lambda l}(z)}\right.  \notag \\
&&\left. +\frac{1}{x^{2}}\int_{0}^{\infty }\ dz\,\frac{e_{l}^{(\lambda
)}(\eta z)}{e_{l}^{(\lambda )}(z)}\frac{z\chi _{\mu
}(z/a)F_{l}^{(q)}[G_{\lambda l}(z,zx)]}{\left[ \partial _{y}^{\lambda
}G_{\lambda l}(z,y)\right] _{y=z\eta }}\right\} ,  \label{sum1kab}
\end{eqnarray}%
where we use the notations
\begin{eqnarray}
e_{l}^{(\lambda )}(y) &\equiv &\partial _{y}^{\lambda }e_{l}(y),\quad
s_{l}^{(\lambda )}(y)\equiv \partial _{y}^{\lambda }s_{l}(y)
\label{RBdernot} \\
G_{\lambda l}(x,y) &=&e_{l}^{(\lambda )}(x)s_{l}(y)-e_{l}(y)s_{l}^{(\lambda
)}(x),
\end{eqnarray}%
for the Riccati-Bessel functions,
\begin{equation}
\Omega _{1\lambda l}(z)=\left\{
\begin{array}{ll}
j_{l}^{2}(z)+n_{l}^{2}(z), & \lambda =0 \\
\left[ zj_{l}(z)\right] ^{^{\prime }2}+[zn_{l}(z)]^{^{\prime }2}, & \lambda
=1%
\end{array}%
\right. ,  \label{Omega8}
\end{equation}%
and the functions $F_{l}^{(q)}[f(y)]$ with $f(y)=G_{\lambda l}(z,y)$ are
defined by relations (\ref{Feps}), (\ref{chisph}).

In (\ref{sum1kab}), taking the limit $b\rightarrow \infty $ for fixed $a$
and $r$, we obtain the VEV for the EMT components outside a single
conducting spherical shell with radius $a$. In this limit the second
integral on the right of formula (\ref{sum1kab}) tends to zero, whereas the
first one does not depend on $b$. Hence, one obtains
\begin{equation}
q_{b\rightarrow \infty }=\sum_{l=1}^{\infty }\frac{2l+1}{8\pi ^{2}a^{4}}%
\sum_{\lambda =0,1}\int_{0}^{\infty }dz\,z^{3}\psi _{\mu }(z/a)\frac{%
D_{l}^{(q)}[g_{\lambda l}(z,zx)]}{\Omega _{1\lambda l}(z)}.  \label{qa1}
\end{equation}%
For the renormalization of expressions (\ref{qa1}) we subtract the
Minkowskian part, namely expression (\ref{EMTMin}). It can be seen that
\begin{equation}
\frac{D_{l}^{(q)}[g_{\lambda l}(z,zx)]}{\Omega _{1\lambda l}(z)}%
-D_{l}^{(q)}[j_{l}(zx)]=-\frac{1}{2}\sum_{m=1,2}\frac{\partial _{z}^{\lambda
}\left[ zj_{l}(z)\right] }{\partial _{z}^{\lambda }[zh_{l}^{(m)}(z)]}%
D_{l}^{(q)}[h_{l}^{(m)}(zx)],  \label{rel81}
\end{equation}%
with $h_{l}^{(m)}(z)$, $\,m=1,2$ being the spherical Hankel functions. Now,
for the corresponding $z$-integral we rotate the integration contour in the
complex $z$-plane by the angle $\pi /2$ for the term with $m=1$ and by the
angle $-\pi /2$ for the term with $m=2$. By introducing the Ricatti-Bessel
functions (\ref{RicBes}) and for points with $r>a$ removing the cutoff, for
the renormalized components of the vacuum EMT outside the sphere we find
\begin{equation}
q_{\mathrm{ren}}(a,r)=-\sum_{l=1}^{\infty }\frac{2l+1}{8\pi ^{2}r^{2}}%
\int_{0}^{\infty }dz\,z\left[ \frac{s_{l}(az)}{e_{l}(az)}+\frac{%
s_{l}^{\prime }(az)}{e_{l}^{\prime }(az)}\right] F_{l}^{(q)}[e_{l}(zr)],%
\quad r>a,  \label{qreg8}
\end{equation}%
where the functions $F_{l}^{(q)}[f(y)]$ are defined by formulae (\ref{Feps}%
), (\ref{chisph}). The exterior mode-sum consideration given in this section
follows \cite{Grig1,Grig3}. For the case of exponential cutoff function
formula (\ref{qreg8}) can also be obtained from the results \cite%
{Brevik1,Brevik2}, where the Green function formalism was used. Note that
the expressions for the exterior components are obtained from the interior
ones replacing $s_{l}\rightleftarrows e_{l}$.

Expressions (\ref{qreg8}) diverge on the sphere. The leading terms of these
divergences may be found using the uniform asymptotic expansions for the
modified Bessel functions for large values of the order and these terms are
given by the same formulae as those for the interior region (see (\ref%
{nearsurface})). In particular, the cancellation of interior and exterior
leading divergent terms occurs in calculating the total energy and force
acting on the sphere. The same cancellations take place for the next
subleading divergent terms as well. For distances far from the sphere one
finds
\begin{equation}
p_{\perp }\approx \frac{a^{3}}{4\pi ^{2}r^{7}}\int_{0}^{\infty
}dz\,z^{2}e_{1}^{2}(z)=\frac{5a^{3}}{16\pi ^{2}r^{7}},\quad \varepsilon
\approx -4p\approx \frac{a^{3}}{2\pi ^{2}r^{7}},\quad r\gg a.
\label{fardist}
\end{equation}%
The results of the numerical calculations for the vacuum EMT components
outside the sphere are given in \cite{Grig2,Brevik1,Brevik2}. In \cite%
{Brevik1,Brevik2} calculations are carried out by using the uniform
asymptotic expansions for the Riccati-Bessel functions. The accuracy of this
approximation is estimated in \cite{Grig2}, where exact relations are used
in numerical calculations. Simple approximating formulae with the same
accuracy as those for the asymptotic calculations are presented as well. In
the exterior region the energy density and azimuthal pressure are positive,
and radial pressure is negative.

Note that the continuity equation (\ref{conteq}) may now be written in the
following integral form
\begin{equation}
p(r)=\frac{1}{r^{3}}\int_{\infty }^{r}dt\,t^{2}\varepsilon (t)=\frac{2}{r^{2}%
}\int_{\infty }^{r}dt\,tp_{\perp }(t).  \label{intrel8}
\end{equation}%
From (\ref{intrel7}) and (\ref{intrel8}) it follows that
\begin{equation}
E(a)=\int dV\,\varepsilon (r)=4\pi a^{3}[p(a-)-p(a+)],  \label{totenergy8}
\end{equation}%
where $E(a)$ is the total vacuum energy for a spherical shell with radius $a$%
, $p(a\pm )=\lim_{r\rightarrow 0}p(a\pm r)$. By using the expressions for $%
p(r)$ given above, one can obtain the following formula for the total energy
(the same result can also be obtained by integrating the energy density)
\begin{eqnarray}
E(a) &=&-\sum_{l=1}^{\infty }\frac{2l+1}{2\pi a}\int_{0}^{\infty }dz\chi
_{\mu }(z/a)z\frac{[s_{l}(z)e_{l}(z)]^{\prime }}{s_{l}^{\prime
}(z)e_{l}^{\prime }(z)}\left[ \frac{s_{l}^{\prime }(z)e_{l}^{\prime }(z)}{%
s_{l}(z)e_{l}(z)}+\frac{l(l+1)}{z^{2}}+1\right]  \notag \\
&=&-\sum_{l=1}^{\infty }\frac{2l+1}{2\pi a}\int_{0}^{\infty }dz\chi _{\mu
}(z/a)z\frac{d}{dz}\ln \left\{ 1-\left[ s_{l}(z)e_{l}(z)\right] ^{^{\prime
}2}\right\} ,  \label{toten}
\end{eqnarray}%
where we have restored the cutoff function. By taking the cutting function $%
\psi _{\mu }(\omega )=e^{-\mu \omega }$ one obtains the expression for the
Casimir energy of the sphere derived in \cite{MiltonSph} by the Green
function method. Note that in this method the factor $\psi _{\mu
}(iz/a)=e^{-i\omega \mu }$ appears automatically as a result of the point
splitting procedure. The evaluation of (\ref{toten}) leads to the result $%
E=0.092353/2a$ for the Casimir energy of a spherical conducting shell \cite%
{Bali78,Boyer,DaviesSph,MiltonSph,Leseduarte,Nesterenko}. This corresponds
to the repulsive vacuum force on the sphere.

As we have seen, the VEV of the EMT in the region between two concentric
perfectly conducting surfaces is given by (\ref{sum1kab}). Using this
formula the components of the vacuum EMT can be presented in the form
\begin{equation}
q(a,b,r)=q(a,r)+q^{(ab)}(r),\quad q=\varepsilon ,p_{\perp },p,\quad a<r<b,
\label{q9}
\end{equation}%
where $q(a,r)$ are the corresponding quantities outside a single sphere of
radius $a$ given by (\ref{qa1}), and \cite{Sah2shert}
\begin{equation}
q^{(ab)}(r)=\sum_{l=1}^{\infty }\frac{2l+1}{8\pi ^{2}r^{2}}\sum_{\lambda
=0,1}\int_{0}^{\infty }dz\,z\frac{F_{l}^{(q)}[G_{\lambda
l}(az,rz)]e_{l}^{(\lambda )}(bz)/e_{l}^{(\lambda )}(az)}{e_{l}^{(\lambda
)}(az)s_{l}^{(\lambda )}(bz)-e_{l}^{(\lambda )}(bz)s_{l}^{(\lambda )}(az)}.
\label{qab92}
\end{equation}%
In (\ref{q9}) the dependence on $b$ is contained in the summand $q^{(ab)}(r)$
only. This quantity is finite for $a\leqslant r<b$ and the renormalization
of $q(a,b,r)$ is equivalent to the renormalization of the first summand.

It can be seen that the quantities (\ref{q9}) may also be written in the
form
\begin{equation}
q(a,b,r)=q(b,r)+q^{(ba)}(r),\quad q=\varepsilon ,p_{\perp },p,\quad a<r<b,
\label{q9in}
\end{equation}%
where
\begin{equation}
q^{(ba)}(r)=\sum_{l=1}^{\infty }\frac{2l+1}{8\pi ^{2}r^{2}}\sum_{\lambda
=0,1}\int_{0}^{\infty }dz\,z\frac{F_{l}^{(q)}[G_{\lambda
l}(bz,rz)]s_{l}^{(\lambda )}(az)/s_{l}^{(\lambda )}(bz)}{e_{l}^{(\lambda
)}(az)s_{l}^{(\lambda )}(bz)-e_{l}^{(\lambda )}(bz)s_{l}^{(\lambda )}(az)}.
\label{qab93}
\end{equation}%
In (\ref{q9in}), $q^{(ba)}(r)\rightarrow 0$ when $a\rightarrow 0$ and $%
q(b,r) $ coincides with the corresponding quantities inside a single
conducting shell with radius $b$. Note that in (\ref{qab93}) the sum and
integral are convergent for $a<r\leqslant b$.

As we have already mentioned the vacuum energy density and stresses inside
and outside a single shell may be obtained from the expressions $q(a,b,r)$
in limiting cases $a\rightarrow 0$ or $b\rightarrow \infty $. It can be seen
that in the limit $a,b\rightarrow \infty $ with fixed $h=b-a$, we obtain the
standard result for the Casimir parallel plate configuration with $%
\varepsilon =-\pi ^{2}/720h^{4}$.

Let us present the quantities $q=\varepsilon ,p,p_{\perp }$ in the form
\begin{equation}
q=q(a,r)+q(b,r)+\Delta q(a,b,r),\quad a<r<b,  \label{qsum9}
\end{equation}%
where the interference term may be written in two ways
\begin{equation}
\Delta q(a,b,r)=q^{(ab)}(a,b,r)-q(b,r)=q^{(ba)}(a,b,r)-q(a,r).  \label{int92}
\end{equation}%
This term is finite for all $a\leqslant r\leqslant b$. Near the surface $r=a$
it is convenient to use the first presentation in (\ref{int92}), as for $%
r\rightarrow a$ both summands in this formula are finite. For the same
reason the second presentation is convenient for calculations near the
surface $r=b$.

So far we have considered the electromagnetic vacuum in the region between
two perfectly conducting spherical surfaces. Consider now a system
consisting of two concentric thin spherical shells with radii $a$ and $b$, $%
a<b$. In this case the VEV for the EMT components may be written in the form
\begin{equation}
q(a,b,r)=q(a,r)\theta (a-r)+q(b,r)\theta (r-b)+[q(a,r)+q^{(ab)}(r)]\theta
(r-a)\theta (b-r),  \label{bettwosph}
\end{equation}%
where $\theta (x)$ is the unit step function. By using the continuity
equation for the EMT, it is easy to see that the total Casimir energy for
the system under consideration can be presented in the form
\begin{equation}
E(a,b)=E(a)+E(b)+4\pi \left[ b^{3}p^{(ba)}(b)-a^{3}p^{(ab)}(a)\right] ,
\label{Casenergytwosph}
\end{equation}%
where $E(j)$ is the Casimir energy for a single sphere with radius $\,j=a,b$%
. As follows from (\ref{qab92}) and (\ref{qab93}), the additional vacuum
pressures on the spheres are equal to \cite{Sahdis,Sah2shert}
\begin{eqnarray}
p^{(ab)}(a) &=&-\sum_{l=1}^{\infty }\frac{2l+1}{8\pi ^{2}a^{4}}%
\int_{0}^{\infty }dz\,z\sum_{\lambda =0,1}(-1)^{\lambda }\frac{\left[
l(l+1)/z^{2}+1\right] ^{\lambda }e_{l}^{(\lambda )}(bz/a)/e_{l}^{(\lambda
)}(z)}{e_{l}^{(\lambda )}(z)s_{l}^{(\lambda )}(bz/a)-e_{l}^{(\lambda
)}(bz/a)s_{l}^{(\lambda )}(z)},  \label{intforcin9} \\
p^{(ba)}(b) &=&-\sum_{l=1}^{\infty }\frac{2l+1}{8\pi ^{2}b^{4}}%
\int_{0}^{\infty }dz\,z\sum_{\lambda =0,1}(-1)^{\lambda }\frac{\left[
l(l+1)/z^{2}+1\right] ^{\lambda }s_{l}^{(\lambda )}(az/b)/s_{l}^{(\lambda
)}(z)}{e_{l}^{(\lambda )}(az/b)s_{l}^{(\lambda )}(z)-e_{l}^{(\lambda
)}(z)s_{l}^{(\lambda )}(az/b)}.  \label{intforcout9}
\end{eqnarray}%
In \cite{Hoye01}, the vacuum forces acting on boundaries in the geometry of
concentric conducting spherical shells are investigated by making use of
local Green function method. The results of the numerical evaluations of
quantities $\Delta q(a,b,r)$, $q=\varepsilon ,p,p_{\perp }$, as well as
those for $p^{(ab)}(a)$, $p^{(ba)}(b)$ are presented in \cite%
{Sahdis,Sah2shert}. Note that, as follows from the results of these
calculations, the quantities (\ref{intforcin9}) and (\ref{intforcout9}) are
always negative, and therefore the interaction forces between two spheres
are always attractive (as in the parallel plate configuration). Note that
the interaction forces can also be obtained from the corresponding part of
the total Casimir energy in the region between the spheres by standard
differentiation with respect to the sphere radii \cite{Hoye01}.

\section{Vacuum polarization induced by a cylindrical boundary in the cosmic
string spacetime}

\label{sec:CosStCyl}

Cosmic strings generically arise within the framework of grand unified
theories and could be produced in the early universe as a result of symmetry
braking phase transitions \cite{Kibb80,Vile94}. Although the recent
observational data on the cosmic microwave background radiation have ruled
out cosmic strings as the primary source for primordial density
perturbations, they are still candidates for the generation of a number
interesting physical effects such as the generation of gravitational waves,
high energy cosmic strings, and gamma ray bursts. In the simplest
theoretical model describing the infinite straight cosmic string the
spacetime is locally flat except on the string where it has a delta shaped
curvature tensor. In quantum field theory the corresponding non-trivial
topology leads to non-zero VEVs for physical observables. In this section we
will consider the vacuum polarization induced by a cylindrical boundary
coaxial with a cosmic string assuming that on the bounding surface the field
obeys Robin boundary condition \cite{Beze06cyl}. The cylindrical problem for
the electromagnetic field in the Minkowski bulk with perfectly conducting
conditions was first considered in \cite{Miltoncyl} (see also \cite%
{Milt02,Romeocyl,Miltoncyl1}). While the earliest studies have focused on
global quantities, such as the total energy and stress on a shell, the local
characteristics of the corresponding electromagnetic vacuum are considered
in \cite{Sah2} for the interior and exterior regions of a conducting
cylindrical shell, and in \cite{Sah3} for the region between two coaxial
shells (see also \cite{Sahdis}). The vacuum forces acting on the boundaries
in the geometry of two cylinders are also considered in \cite{Mazz02}. The
vacuum densities for a Robin cylindrical boundary in the Minkowski
background are investigated in \cite{Rome01}. In \cite{Khus99} a cylindrical
boundary with Dirichlet boundary condition is introduced in the bulk of the
cosmic string as an intermediate stage for the calculation of the ground
state energy of a massive scalar field in (2+1)-dimensions.

\subsection{Wightman function}

\label{sec:WightFunc}

We consider a scalar field $\varphi $ with curvature coupling parameter $%
\zeta $ propagating on the $(D+1)$-dimensional background spacetime with a
conical-type singularity described by the line element%
\begin{equation}
ds^{2}=g_{ik}dx^{i}dx^{k}=dt^{2}-dr^{2}-r^{2}d\phi
^{2}-\sum_{i=1}^{N}dz_{i}{}^{2},  \label{ds21}
\end{equation}%
with the cylindrical coordinates $(x^{1},x^{2},\ldots ,x^{D})=(r,\phi
,z_{1},\ldots ,z_{N})$, where $N=D-2$, $0\leqslant \phi \leqslant \phi _{0}$%
, and the spatial points $(r,\phi ,z_{1},\ldots ,z_{N})$ and $(r,\phi +\phi
_{0},z_{1},\ldots ,z_{N})$ are to be identified. In the standard $D=3$
cosmic string case the planar angle deficit is related to the mass per unit
length of the string $\mu $ by $2\pi -\phi _{0}=8\pi G\mu $, where $G$ is
the Newton gravitational constant. We assume that the field obeys Robin
boundary condition (\ref{mrobcond}) on the cylindrical surface of radius $a$%
, coaxial with the string.

In the region inside the cylindrical surface the eigenfunctions satisfying
the periodicity condition are specified by the set of quantum numbers $%
\sigma =(n,\gamma ,\mathbf{k})$, $n=0,\pm 1,\pm 2,\ldots $, $\mathbf{k}%
=(k_{1},\ldots ,k_{N})$, $-\infty <k_{j}<\infty $, and have the form
\begin{eqnarray}
\varphi _{\sigma }(x) &=&\beta _{\sigma }J_{q\left\vert n\right\vert
}(\gamma r)\exp \left( iqn\phi +i\mathbf{kr}_{\parallel }-i\omega t\right) ,
\label{eigfunccirc} \\
\omega &=&\sqrt{\gamma ^{2}+k_{m}^{2}},\quad k_{m}^{2}=k^{2}+m^{2},\;q=2\pi
/\phi _{0},  \label{qu}
\end{eqnarray}%
where $\mathbf{r}_{\parallel }=(z_{1},\ldots ,z_{N})$. The eigenvalues for
the quantum number $\gamma $ are quantized by boundary condition (\ref%
{mrobcond}) on the cylindrical surface $r=a$. From this condition it follows
that for a given $n$ the possible values of $\gamma $ are determined by the
relation
\begin{equation}
\gamma =\lambda _{n,l}/a,\quad l=1,2,\ldots ,  \label{ganval}
\end{equation}%
where $\lambda _{n,l}$ are the positive zeros of the function $\bar{J}%
_{q|n|}(z)$, $\bar{J}_{q|n|}(\lambda _{n,l})=0$, arranged in ascending
order, $\lambda _{n,l}<\lambda _{n,l+1}$, $n=0,1,2,\ldots $. Here, for a
given function $f(z)$, we use the barred notation (\ref{efnot1}) with the
coefficients%
\begin{equation}
A=\tilde{A},\;B=-\tilde{B}/a.  \label{fbar}
\end{equation}%
In the following we will assume the values of $A/B$ for which all zeros are
real.

The coefficient $\beta _{\sigma }$ in (\ref{eigfunccirc}) is determined from
the normalization condition with the integration over the region inside the
cylindrical surface and is equal to
\begin{equation}
\beta _{\sigma }^{2}=\frac{\lambda _{n,l}T_{qn}(\lambda _{n,l})}{(2\pi
)^{N}\omega \phi _{0}a^{2}},  \label{betalf}
\end{equation}%
with the notation $T_{\nu }(z)$ from (\ref{teka}). Substituting the
eigenfunctions (\ref{eigfunccirc}) into the mode-sum formula (\ref%
{mmodesumWF}) with a set of quantum numbers $\sigma =(nl\mathbf{k})$, for
the positive frequency Wightman function one finds
\begin{eqnarray}
W(x,x^{\prime }) &=&2\int d^{N}\mathbf{k}\,e^{i\mathbf{k}\Delta \mathbf{r}%
_{\parallel }}\sideset{}{'}{\sum}_{n=0}^{\infty }\cos (qn\Delta \phi )
\notag \\
&&\times \sum_{l=1}^{\infty }\beta _{\sigma }^{2}J_{qn}(\gamma
r)J_{qn}(\gamma r^{\prime })e^{-i\omega \Delta t}|_{\gamma =\lambda
_{n,l}/a},  \label{Wf1}
\end{eqnarray}%
where $\Delta \mathbf{r}_{\parallel }=\mathbf{r}_{\parallel }-\mathbf{r}%
_{\parallel }^{\prime }$, $\Delta \phi =\phi -\phi ^{\prime }$, $\Delta
t=t-t^{\prime }$, and the prime in the sum means that the summand with $n=0$
should be taken with the weight 1/2. As we do not know the explicit
expressions for the eigenvalues $\lambda _{n,l}$ as functions on $n$ and $l$%
, and the summands in the series over $l$ are strongly oscillating functions
for large values of $l$, this formula is not convenient for the further
evaluation of the VEVs of the field square and the EMT. In addition, the
expression on the right of (\ref{Wf1}) is divergent in the coincidence limit
and some renormalization procedure is needed to extract finite result for
the VEVs of the field square and the EMT. To obtain an alternative form for
the Wightman function we will apply to the sum over $l$ summation formula (%
\ref{sumJ1}). In this formula as a function $f(z)$ we choose
\begin{equation}
f(z)=\frac{zJ_{qn}(zr/a)J_{qn}(zr^{\prime }/a)}{\sqrt{k_{m}^{2}+z^{2}/a^{2}}}%
\exp (-i\Delta t\sqrt{k_{m}^{2}+z^{2}/a^{2}}),  \label{ftosum}
\end{equation}%
where $k_{m}^{2}=k^{2}+m^{2}$. The condition to formula (\ref{sumJ1}) to be
satisfied is $r+r^{\prime }+|\Delta t|<2a$. In particular, this is the case
in the coincidence limit $t=t^{\prime }$ for the region under consideration,
$r,r^{\prime }<a$. Formula (\ref{sumJ1}) allows to present the Wightman
function in the form \cite{Beze06cyl}%
\begin{equation}
W(x,x^{\prime })=W_{s}(x,x^{\prime })+\langle \varphi (x)\varphi (x^{\prime
})\rangle _{a},  \label{Wf2}
\end{equation}%
where%
\begin{eqnarray}
W_{s}(x,x^{\prime }) &=&\frac{1}{\phi _{0}}\int \frac{d^{N}\mathbf{k}}{(2\pi
)^{N}}e^{i\mathbf{k}\Delta \mathbf{r}_{\parallel }}\int_{0}^{\infty }dz\frac{%
ze^{-i\Delta t\sqrt{z^{2}+k_{m}^{2}}}}{\sqrt{z^{2}+k_{m}^{2}}}  \notag \\
&&\times \sideset{}{'}{\sum}_{n=0}^{\infty }\cos (qn\Delta \phi
)J_{qn}(zr)J_{qn}(zr^{\prime }),  \label{Wf00}
\end{eqnarray}%
and%
\begin{eqnarray}
\langle \varphi (x)\varphi (x^{\prime })\rangle _{a} &=&-\frac{2}{\pi \phi
_{0}}\int \frac{d^{N}\mathbf{k}}{(2\pi )^{N}}e^{i\mathbf{k}\Delta \mathbf{r}%
_{\parallel }}\int_{k_{m}}^{\infty }dz\frac{z\cosh (\Delta t\sqrt{%
z^{2}-k_{m}^{2}})}{\sqrt{z^{2}-k_{m}^{2}}}  \notag \\
&&\times \sideset{}{'}{\sum}_{n=0}^{\infty }\cos (qn\Delta \phi
)I_{qn}(zr)I_{qn}(zr^{\prime })\frac{\bar{K}_{qn}(za)}{\bar{I}_{qn}(za)}.
\label{Wfa0}
\end{eqnarray}%
In the limit $a\rightarrow \infty $ for fixed $r,r^{\prime }$, the term $%
\langle \varphi (x)\varphi (x^{\prime })\rangle _{a}$ vanishes and, hence,
the term $W_{s}(x,x^{\prime })$ is the Wightman function for the geometry of
a cosmic string without the cylindrical boundary (the integral
representation of the corresponding Green function for a massive scalar
field is considered in Refs. \cite{Line87}). Consequently, the term $\langle
\varphi (x)\varphi (x^{\prime })\rangle _{a}$ is induced by the presence of
the cylindrical boundary. For points away from the cylindrical surface this
term is finite in the coincidence limit and the renormalization is needed
only for the part coming from (\ref{Wf00}). As it has been shown in Ref.
\cite{Beze06cyl}, the boundary induced part of the Wightman function in the
exterior region, $r>a$, is obtained from the corresponding part in the
interior region by the replacements $I\rightleftarrows K$.

\subsection{Vacuum expectation values of the field square and the
energy-momentum tensor inside a cylindrical shell}

\label{sec:inside}

Taking the coincidence limit $x^{\prime }\rightarrow x$ in formula (\ref{Wf2}%
) for the Wightman function and integrating over $\mathbf{k}$, the VEV of
the field square is presented as a sum of two terms:%
\begin{equation}
\langle 0|\varphi ^{2}|0\rangle =\langle 0_{s}|\varphi ^{2}|0_{s}\rangle
+\langle \varphi ^{2}\rangle _{a},  \label{phi2a}
\end{equation}%
where $|0_{s}\rangle $ is the amplitude for the vacuum state in the geometry
when the cylindrical shell is absent. The second term on the right of this
formula is induced by the cylindrical boundary and is given by the formula
\cite{Beze06cyl}%
\begin{equation}
\langle \varphi ^{2}\rangle _{a}=-\frac{A_{D}}{\phi _{0}}\sideset{}{'}{\sum}%
_{n=0}^{\infty }\int_{m}^{\infty }dz\,z\left( z^{2}-m^{2}\right) ^{\frac{D-3%
}{2}}\frac{\bar{K}_{qn}(za)}{\bar{I}_{qn}(za)}I_{qn}^{2}(zr),  \label{phi2a1}
\end{equation}%
where we have introduced the notation%
\begin{equation}
A_{D}=\frac{2^{3-D}\pi ^{\frac{1-D}{2}}}{\Gamma \left( \frac{D-1}{2}\right) }%
.  \label{AD}
\end{equation}%
The formula for the VEV of the field square in the region outside the
cylindrical shell is obtained from (\ref{phi2a1}) by the replacements $%
I\rightleftarrows K$. For points away from the cylindrical surface, $r<a$,
the integral in (\ref{phi2a1}) is exponentially convergent in the upper
limit and the boundary-induced part in the VEV of the field square is
finite. In particular, this part is negative for Dirichlet scalar and is
positive for Neumann scalar. Near the string, $r\ll a$, the main
contribution to $\langle \varphi ^{2}\rangle _{a}$ comes from the summand
with $n=0$ and one has%
\begin{equation}
\langle \varphi ^{2}\rangle _{a}\approx -\frac{A_{D}}{2a^{D-1}\phi _{0}}%
\int_{ma}^{\infty }dz\,\,z\left( z^{2}-m^{2}a^{2}\right) ^{\frac{D-3}{2}}%
\frac{\bar{K}_{0}(z)}{\bar{I}_{0}(z)}.  \label{phi2ar0}
\end{equation}%
As the boundary-free renormalized VEV diverges on the string, we conclude
from here that near the string the main contribution to the VEV of the field
square comes from this part.

The part $\langle \varphi ^{2}\rangle _{a}$ diverges on the cylindrical
surface $r=a$. Near this surface the main contribution into (\ref{phi2a1})
comes from large values of $n$. Introducing a new integration variable $%
z\rightarrow nqz$, replacing the modified Bessel functions by their uniform
asymptotic expansions for large values of the order (see, for instance, \cite%
{abramowiz}), and expanding over $a-r$, to the leading order one finds%
\begin{equation}
\langle \varphi ^{2}\rangle _{a}\approx \frac{(1-2\delta _{B0})\Gamma \left(
\frac{D-1}{2}\right) }{(4\pi )^{\frac{D+1}{2}}(a-r)^{D-1}}.
\label{Phi2neara}
\end{equation}%
This leading behavior is the same as that for a cylindrical surface of
radius $a$ in the Minkowski spacetime. As the boundary-free part is finite
at $r=a$, near the boundary the total renormalized VEV of the field square
is dominated by the boundary-induced part and is negative for Dirichlet
scalar. Combining this with the estimation for the region near the string,
we come to the conclusion that in this case the VEV\ of the field square
vanishes for some intermediate value of $r$.

Now we turn to the investigation of the boundary-induced VEV given by (\ref%
{phi2a1}), in the limiting cases of the parameter $q$. Firstly consider the
limit when the parameter $q$ is large which corresponds to small values of $%
\phi _{0}$ and, hence, to a large planar angle deficit. In this limit the
order of the modified Bessel functions for the terms with $n\neq 0$ in (\ref%
{phi2a1}) is large and we can replace these functions by their uniform
asymptotic expansions. On the base of these expansions it can be seen that
to the leading order the contribution of the terms with $n\neq 0$ is
suppressed by the factor $q^{(D-1)/2}(r/a)^{2q}$ and the main contribution
to the VEV of the field square comes from the $n=0$ term:%
\begin{equation}
\langle \varphi ^{2}\rangle _{a}\approx -\frac{A_{D}}{2\phi _{0}}%
\int_{m}^{\infty }dz\,z\left( z^{2}-m^{2}\right) ^{\frac{D-3}{2}}\frac{\bar{K%
}_{0}(za)}{\bar{I}_{0}(za)}I_{0}^{2}(zr),\;q\gg 1,  \label{phi2largeq}
\end{equation}%
with the linear dependence on $q$. In the same limit the boundary-free part
in the VEV of the field square behaves as $q^{D-1}$ and, hence, its
contribution dominates in comparison with the boundary-induced part. In the
opposite limit when $q\rightarrow 0$, the series over $n$ in Eq. (\ref%
{phi2a1}) diverges and, hence, for small values of $q$ the main contribution
comes from large values $n$. In this case, to the leading order, we can
replace the summation by the integration: $\sum_{n}f(qn)\rightarrow
(1/q)\int_{0}^{\infty }dxf(x)$. As a consequence, we obtain that in the
limit $q\rightarrow 0$ the boundary-induced VEV in the field square tends to
a finite limiting value:%
\begin{equation}
\langle \varphi ^{2}\rangle _{a}\approx -\frac{A_{D}}{2\pi }\int_{0}^{\infty
}dx\int_{m}^{\infty }dz\,z\left( z^{2}-m^{2}\right) ^{\frac{D-3}{2}}\frac{%
\bar{K}_{x}(za)}{\bar{I}_{x}(za)}I_{x}^{2}(zr).  \label{phi2asmallq}
\end{equation}

Now we consider the VEV of the EMT for the situation when the cylindrical
boundary is present. Similar to the case of the field square, this VEV is
written in the form%
\begin{equation}
\langle 0|T_{ik}|0\rangle =\langle 0_{s}|T_{ik}|0_{s}\rangle +\langle
T_{ik}\rangle _{a},  \label{Tika}
\end{equation}%
where the part $\langle 0_{s}|T_{ik}|0_{s}\rangle $ corresponds to the
geometry of a cosmic string without boundaries and $\langle T_{ik}\rangle
_{a}$ is induced by the cylindrical boundary. The first term for a
conformally coupled $D=3$ massless scalar field was evaluated in Ref. \cite%
{Hell86}. The case of an arbitrary curvature coupling is considered in Refs.
\cite{Frol87, Beze06cyl}. The term induced by the cylindrical shell is
obtained from the corresponding part in the Wightman function, acting by the
appropriate differential operator and taking the coincidence limit (see
formula (\ref{mTikVEV})). For points away from the cylindrical surface this
limit gives a finite result. For the corresponding components of the EMT one
obtains (no summation over $i$) \cite{Beze06cyl}%
\begin{equation}
\langle T_{i}^{i}\rangle _{a}=\frac{A_{D}}{\phi _{0}}\sideset{}{'}{\sum}%
_{n=0}^{\infty }\int_{m}^{\infty }dzz^{3}\left( z^{2}-m^{2}\right) ^{\frac{%
D-3}{2}}\frac{\bar{K}_{qn}(za)}{\bar{I}_{qn}(za)}F_{qn}^{(i)}\left[
I_{qn}(zr)\right] ,  \label{Tiia21}
\end{equation}%
with the notations%
\begin{eqnarray}
F_{qn}^{(0)}\left[ f(y)\right] &=&\left( 2\zeta -\frac{1}{2}\right) \left[
f^{\prime 2}(y)+\left( 1+\frac{q^{2}n^{2}}{y^{2}}\right) f^{2}(y)\right] +%
\frac{y^{2}-m^{2}r^{2}}{(D-1)y^{2}}f^{2}(y),  \label{ajpm} \\
F_{qn}^{(1)}\left[ f(y)\right] &=&\frac{1}{2}f^{\prime 2}(y)+\frac{2\zeta }{y%
}f(y)f^{\prime }(y)-\frac{1}{2}\left( 1+\frac{q^{2}n^{2}}{y^{2}}\right)
f^{2}(y),  \label{ajpm1} \\
F_{qn}^{(2)}\left[ f(y)\right] &=&\left( 2\zeta -\frac{1}{2}\right) \left[
f^{\prime 2}(y)+\left( 1+\frac{q^{2}n^{2}}{y^{2}}\right) f^{2}(y)\right]
\notag \\
&&+\frac{q^{2}n^{2}}{y^{2}}f^{2}(y)-\frac{2\xi }{y}f(y)f^{\prime }(y),
\label{ajpm2}
\end{eqnarray}%
and $F_{qn}^{(i)}\left[ f(y)\right] =F_{qn}^{(0)}\left[ f(y)\right] $ for$%
\;i=3,\ldots ,D$. The formula for the VEV of the EMT in the region outside
the cylindrical shell is obtained from (\ref{Tiia21}) by the replacements $%
I\rightleftarrows K$. It can be checked that the expectation values (\ref%
{Tiia21}) satisfy the continuity equation for the EMT. The boundary-induced
part in the VEV of the EMT given by Eq. (\ref{Tiia21}) is finite everywhere
except at points on the boundary and at points on the string in the case $%
q<1 $. Unlike to the surface divergences, the divergences on the string are
integrable.

In the case $q>1$, near the string, $r\rightarrow 0$, the main contribution
to the boundary part (\ref{Tiia21}) comes from the summand with $n=0$ and
one has
\begin{equation}
\langle T_{i}^{i}\rangle _{a}\approx \frac{A_{D}}{2\phi _{0}a^{D+1}}%
\int_{ma}^{\infty }dzz^{3}\left( z^{2}-m^{2}a^{2}\right) ^{\frac{D-3}{2}}%
\frac{\bar{K}_{0}(z)}{\bar{I}_{0}(z)}F^{(i)}(z),  \label{Tii21r0}
\end{equation}%
with the notations%
\begin{equation}
F^{(0)}(z)=2\zeta -\frac{1}{2}+\frac{1-m^{2}a^{2}/z^{2}}{D-1}%
,\;F^{(i)}(z)=\zeta -\frac{1}{2},\;i=1,2.  \label{F0z}
\end{equation}%
For $q<1$ the main contribution to the boundary-induced part for points near
the string comes from $n=1$ term and in the leading order one has%
\begin{equation}
\langle T_{i}^{i}\rangle _{a}\approx \frac{q^{2}A_{D}r^{2q-2}F_{1}^{(i)}}{%
2^{q}\pi \Gamma ^{2}(q+1)}\int_{m}^{\infty }dzz^{2q+1}\left(
z^{2}-m^{2}\right) ^{\frac{D-3}{2}}\frac{\bar{K}_{q}(za)}{\bar{I}_{q}(za)},
\label{Tii21r0q}
\end{equation}%
where%
\begin{equation}
F_{1}^{(0)}=q(2\zeta -1/2),\;F_{1}^{(1)}=\zeta ,\;F_{1}^{(2)}=(2q-1)\zeta .
\label{Fi1}
\end{equation}%
As we see, in this case the VEVs for the EMT diverge on the string. This
divergence is integrable. In particular, the corresponding contribution to
the energy in the region near the string is finite.

As in the case of the field square, in the limit $q\gg 1$ the contribution
of the terms with $n\neq 0$ to the VEV of the EMT is suppressed by the
factor $q^{(D-1)/2}(r/a)^{2q}$ and the main contribution comes from the $n=0$
term with the linear dependence on $q$. In the same limit, the boundary-free
part in the VEV\ of the EMT behaves as $q^{D+1}$ and, hence, the total
energy-momentum tensor is dominated by this part. In the opposite limit,
when $q\ll 1$, by the way similar to that used before for the VEV of the
field square, it can be seen that the boundary-induced part in the vacuum
EMT tends to a finite limiting value which is obtained from (\ref{Tiia21})
replacing the summation over $n$ by the integration.

The boundary part $\left\langle T_{i}^{k}\right\rangle _{a}$ diverges on the
cylindrical surface $r=a$. Introducing a new integration variable $%
z\rightarrow nqz$ and taking into account that near the surface $r=a$ the
main contribution comes from large values of $n$, we can replace the
modified Bessel functions by their uniform asymptotic expansions for large
values of the order. To the leading order this gives%
\begin{equation}
\langle T_{i}^{i}\rangle _{a}\approx \frac{D(\zeta -\zeta _{D})(2\delta
_{B0}-1)}{2^{D}\pi ^{(D+1)/2}(a-r)^{D+1}}\Gamma \left( \frac{D+1}{2}\right)
,\quad i=0,2,\ldots ,D.  \label{T00asra2}
\end{equation}%
This leading divergence does not depend on the parameter $q$ and coincides
with the corresponding one for a cylindrical surface of radius $a$ in the
Minkowski bulk. For the radial component to the leading order one has $%
\langle T_{1}^{1}\rangle _{a}\sim (a-r)^{-D}$. In particular, for a
minimally coupled scalar field the corresponding energy density is negative
for Dirichlet boundary condition and is positive for non-Dirichlet boundary
conditions. For a conformally coupled scalar the leading term vanishes and
it is necessary to keep the next term in the corresponding asymptotic
expansion. As the boundary-free part in the VEV of the EMT is finite on the
cylindrical surface, for points near the boundary the vacuum EMT is
dominated by the boundary-induced part. Taking $q=1$, from the formulae
given in this section we obtain the corresponding results for the geometry
of a cylindrical boundary in the Minkowski bulk \cite{Rome01}.

The problem considered in this section is closely related to the problem of
the investigation of the vacuum densities in the geometry of a wedge with
the opening angle $\phi _{0}$ and with the cylindrical boundary of radius $a$
\cite{Reza02,Saha05wedSc}. For a scalar field with Dirichlet boundary
condition the corresponding eigenfunctions in the region inside the
cylindrical shell are determined by formula (\ref{eigfunccirc}) with the
replacement $\exp (iqn\phi )\rightarrow \sin (qn\phi )$, $n=1,2,\ldots $,
and $q=\pi /\phi _{0}$. The eigenvalues for $\gamma $ are given by relation (%
\ref{ganval}), where now $\lambda _{n,l}$ are the zeros of the function $%
J_{qn}(x)$. The corresponding Wightman function and the VEVs of the field
square and the EMT are evaluated by making use of summation formula (\ref%
{sumJ1}) in the way similar to that used in this section for the geometry of
a cosmic string. The corresponding formulae can be found in \cite%
{Saha05wedSc}. Note that in the case of a wedge the VEVs of the field square
and the EMT depend on the angle $\phi $.

\section{Electromagnetic Casimir densities induced by a conducting
cylindrical shell in the cosmic string spacetime}

\label{sec:CylElCos}

In this section we consider the application of the GAPF for the
investigation of the polarization of the electromagnetic vacuum by a
perfectly conducting cylindrical shell coaxial with the cosmic string \cite%
{Beze06cylEl}. This geometry can be viewed as a simplified model for the
superconducting string, in which the string core in what concerns its
superconducting effects is taken to be an ideal conductor. The background
spacetime is described by the $N=1$ version of the line element (\ref{ds21}).

\subsection{Vacuum expectation values of the field square inside a
cylindrical shell}

In the region inside the cylindrical shell we have two different types of
the eigenfunctions corresponding to the waves of the electric and magnetic
types. In the Coulomb gauge, the vector potentials for these waves are given
by the formulae%
\begin{equation}
\mathbf{A}_{\sigma }=\beta _{\sigma }\left\{
\begin{array}{cc}
(1/i\omega )\left( \gamma ^{2}\mathbf{e}_{3}+ik\nabla _{t}\right)
J_{q|n|}(\gamma r)\exp \left[ i\left( qn\phi +kz-\omega t\right) \right] , &
\lambda =0 \\
-\mathbf{e}_{3}\times \nabla _{t}\left\{ J_{q|n|}(\gamma r)\exp \left[
i\left( qn\phi +kz-\omega t\right) \right] \right\} , & \lambda =1%
\end{array}%
\right. ,  \label{Aalpha}
\end{equation}%
where $\mathbf{e}_{3}$ is the unit vector along the cosmic string, $\nabla
_{t}$ is the part of the nabla operator transverse to the string, and%
\begin{equation}
\omega ^{2}=\gamma ^{2}+k^{2},\;q=2\pi /\phi _{0},\;n=0,\pm 1,\pm 2,\ldots .
\label{omegaEl}
\end{equation}%
Here and in what follows $\lambda =0$ and $\lambda =1$ correspond to the
cylindrical waves of the electric (transverse magnetic (TM)) and magnetic
(transverse electric (TE)) types, respectively. The normalization
coefficient in (\ref{Aalpha}) is found from the orthonormalization condition
(\ref{normElintSph}), where the integration goes over the region inside the
shell. From this condition, by using the standard integral involving the
square of the Bessel function, one finds%
\begin{equation}
\beta _{\sigma }^{2}=\frac{qT_{q|n|}(\gamma a)}{\pi \omega a\gamma },
\label{betalfCyl}
\end{equation}%
with $T_{\nu }(x)$ defined by (\ref{teka}).

The eigenvalues for the quantum number $\gamma $ are determined by standard
boundary conditions (\ref{ElBCcond}) for the electric and magnetic fields on
the cylindrical shell. From boundary conditions we see that these
eigenvalues are solutions of the equation
\begin{equation}
J_{q|n|}^{(\lambda )}(\gamma a)=0,\quad \lambda =0,1,  \label{modes1}
\end{equation}%
where $a$ is the radius of the cylindrical shell, $J_{\nu }^{(0)}(x)=J_{\nu
}(x)$ and $J_{\nu }^{(1)}(x)=J_{\nu }^{\prime }(x)$. We will denote the
corresponding eigenmodes by $\gamma a=j_{n,l}^{(\lambda )}$, $l=1,2,\ldots $%
. As a result the eigenfunctions are specified by the set of quantum numbers
$\sigma =(k,n,\lambda ,l)$.

Substituting the eigenfunctions into the corresponding mode-sum formula, for
the VEVs of the squares of the electric and magnetic fields inside the shell
we find%
\begin{eqnarray}
\langle 0|F^{2}|0\rangle &=&\sum_{\sigma }\mathbf{F}_{\sigma }\cdot \mathbf{F%
}_{\sigma }^{\ast }=\frac{2q}{\pi a^{3}}\sideset{}{'}{\sum}_{n=0}^{\infty
}\int_{-\infty }^{+\infty }dk\sum_{\lambda =0,1}\sum_{l=1}^{\infty
}j_{n,l}^{(\lambda )3}  \notag \\
&&\times \frac{T_{qn}(j_{n,l}^{(\lambda )})}{\sqrt{j_{n,l}^{(\lambda
)2}+k^{2}a^{2}}}g_{qn}^{(\eta _{F\lambda })}[k,J_{qn}(j_{n,l}^{(\lambda
)}r/a)],  \label{F2}
\end{eqnarray}%
where $F=E,B$ with $\eta _{E\lambda }=\lambda $, $\eta _{B\lambda
}=1-\lambda $, and the prime in the summation means that the term $n=0$
should be halved. In (\ref{F2}), for a given function $f(x)$, we have used
the notations
\begin{equation}
g_{\nu }^{(j)}[k,f(x)]=\left\{
\begin{array}{cc}
(k^{2}r^{2}/x^{2})\left[ f^{\prime 2}(x)+\nu ^{2}f^{2}(x)/x^{2}\right]
+f^{2}(x), & j=0 \\
(1+k^{2}r^{2}/x^{2})\left[ f^{\prime 2}(x)+\nu ^{2}f^{2}(x)/x^{2}\right] , &
j=1%
\end{array}%
\right. .  \label{gnulam}
\end{equation}%
The expressions (\ref{F2}) corresponding to the electric and magnetic fields
are divergent. They may be regularized introducing a cutoff function $\psi
_{\mu }(\omega )$ with the cutting parameter $\mu $ which makes the
divergent expressions finite and satisfies the condition $\psi _{\mu
}(\omega )\rightarrow 1$ for $\mu \rightarrow 0$. After the renormalization
the cutoff function is removed by taking the limit $\mu \rightarrow 0$. An
alternative way is to consider the product of the fields at different
spacetime points and to take the coincidence limit after the subtraction of
the corresponding Minkowskian part. Here we will follow the first approach.

In order to evaluate the mode-sum in (\ref{F2}), we apply to the series over
$j$ summation formula (\ref{sumJ1}). As it can be seen, for points away from
the shell the contribution to the VEVs coming from the second integral term
on the right-hand side of (\ref{sumJ1}) is finite in the limit $\mu
\rightarrow 0$ and, hence, the cutoff function in this term can be safely
removed. As a result the VEVs can be written as%
\begin{equation}
\langle 0|F^{2}|0\rangle =\langle 0_{s}|F^{2}|0_{s}\rangle +\langle
F^{2}\rangle _{a},  \label{F21}
\end{equation}%
where%
\begin{eqnarray}
\langle 0_{s}|F^{2}|0_{s}\rangle &=&\frac{q}{\pi }\sideset{}{'}{\sum}%
_{n=0}^{\infty }\int_{-\infty }^{+\infty }dk\int_{0}^{\infty }d\gamma \,%
\frac{\gamma ^{3}\psi _{\mu }(\omega )}{\sqrt{\gamma ^{2}+k^{2}}}  \notag \\
&&\times \left[ \left( 1+2\frac{k^{2}}{\gamma ^{2}}\right) \left(
J_{qn}^{\prime 2}(\gamma r)+\frac{q^{2}n^{2}}{\gamma ^{2}r^{2}}%
J_{qn}^{2}(\gamma r)\right) +J_{qn}^{2}(\gamma r)\right] ,  \label{F2s}
\end{eqnarray}%
and%
\begin{equation}
\langle F^{2}\rangle _{a}=\frac{4q}{\pi ^{2}}\sideset{}{'}{\sum}%
_{n=0}^{\infty }\int_{0}^{\infty }dk\sum_{\lambda =0,1}\int_{k}^{\infty
}dx\,x^{3}\,\frac{K_{qn}^{(\lambda )}(xa)}{I_{qn}^{(\lambda )}(xa)}\frac{%
G_{qn}^{(\eta _{F\lambda })}\left[ k,I_{qn}(xr)\right] }{\sqrt{x^{2}-k^{2}}}.
\label{F2b0}
\end{equation}%
Note that in Eq. (\ref{F2b0}) we used the notations%
\begin{equation}
G_{\nu }^{(j)}\left[ k,f(x)\right] =\left\{
\begin{array}{cc}
(k^{2}r^{2}/x^{2})\left[ f^{\prime 2}(x)+\nu ^{2}f^{2}(x)/x^{2}\right]
+f^{2}(x), & j=0 \\
(k^{2}r^{2}/x^{2}-1)\left[ f^{\prime 2}(x)+\nu ^{2}f^{2}(x)/x^{2}\right] , &
j=1%
\end{array}%
\right. .  \label{Gnuj}
\end{equation}%
The second term on the right-hand side of Eq. (\ref{F21}) vanishes in the
limit $a\rightarrow \infty $. Thus, we can conclude that the term $\langle
0_{s}|F^{2}|0_{s}\rangle $ corresponds to the part in VEVs\ when the
cylindrical shell is absent with the corresponding vacuum state $%
|0_{s}\rangle $. Note that for the geometry without boundaries one has $%
\langle 0_{s}|E^{2}|0_{s}\rangle =\langle 0_{s}|B^{2}|0_{s}\rangle $. Hence,
the application of the GAPF enables us to extract from the VEVs the
boundary-free parts and to write the boundary-induced parts in terms of the
exponentially convergent integrals. The boundary-free parts can be further
simplified with the final result%
\begin{equation}
\langle F^{2}\rangle _{s,\mathrm{ren}}=-\frac{(q^{2}-1)(q^{2}+11)}{180\pi
r^{4}}.  \label{F2sren3}
\end{equation}

Changing the integration variable to $y=\sqrt{x^{2}-k^{2}}$ and introducing
polar coordinates in the $(k,y)$ plane, after the explicit integration over
the angular part, the part in the VEV induced by the cylindrical shell can
be written in the form \cite{Beze06cylEl}
\begin{equation}
\langle F^{2}\rangle _{a}=\frac{q}{\pi }\sideset{}{'}{\sum}_{n=0}^{\infty
}\sum_{\lambda =0,1}\int_{0}^{\infty }dx\,x^{3}\frac{K_{qn}^{(\lambda )}(xa)%
}{I_{qn}^{(\lambda )}(xa)}G_{qn}^{(\eta _{F\lambda })}\left[ I_{qn}(xr)%
\right] ,  \label{F2b}
\end{equation}%
where we have used the notation%
\begin{equation}
G_{\nu }^{(j)}\left[ f(x)\right] =\left\{
\begin{array}{cc}
f^{\prime 2}(x)+\nu ^{2}f^{2}(x)/x^{2}+2f^{2}(x), & j=0 \\
-f^{\prime 2}(x)-\nu ^{2}f^{2}(x)/x^{2}, & j=1%
\end{array}%
\right. .  \label{Gnujtilde}
\end{equation}%
The boundary-induced parts for the electric and magnetic fields are
different and, hence, the presence of the shell breaks the electric-magnetic
symmetry in the VEVs. Of course, this is a consequence of different boundary
conditions for the electric and magnetic fields. The formulae for the VEVs
of the field squares in the region outside the cylindrical shell are
obtained from (\ref{F2b}) by the replacements $I\rightleftarrows K$.

The expression in the right-hand side of (\ref{F2b}) is finite for $0<r<a$
and diverges on the shell with the leading term $\langle E^{2}\rangle
_{a}\approx -\langle B^{2}\rangle _{a}\approx (3/4\pi )(a-r)^{-4}$. Near the
string, $r/a\ll 1$, the asymptotic behavior of the boundary induced part in
the VEVs of the field squares depends on the parameter $q$. For $q\geqslant
1 $, the dominant contribution comes from the lowest mode $n=0$ and to the
leading order one has $\langle E^{2}\rangle _{a}\approx 0.32q/a^{4}$, $%
\langle B^{2}\rangle _{a}\approx -0.742q/a^{4}$. For $q<1$ the main
contribution comes form the mode with $n=1$ and the boundary-induced parts
diverge on the string. The leading terms are given by%
\begin{equation}
\langle E^{2}\rangle _{a}\approx -\langle B^{2}\rangle _{a}\approx \frac{%
(r/a)^{2(q-1)}}{2^{2(q-1)}\pi \Gamma ^{2}(q)a^{4}}\int_{0}^{\infty
}dx\,x^{2q+1}\left[ \frac{K_{q}(x)}{I_{q}(x)}-\frac{K_{q}^{\prime }(x)}{%
I_{q}^{\prime }(x)}\right] .  \label{E2bnearcent2}
\end{equation}%
As for points near the shell, here the leading divergence is cancelled in
the evaluation of the vacuum energy density. In accordance with (\ref%
{F2sren3}), near the string the total VEV is dominated by the boundary-free
part. Here we have considered the VEV for the field square. The VEVs for the
bilinear products of the fields at different spacetime points may be
evaluated in a similar way.

Now, we turn to the investigation of the behavior of the boundary-induced
VEVs in the asymptotic regions of the parameter $q$. For small values of
this parameter, $q\ll 1$, the main contribution into (\ref{F2b}) comes from
large values of $n$. In this case, we can replace the summation over $n$ by
an integration in accordance with the correspondence%
\begin{equation}
\sideset{}{'}{\sum}_{n=0}^{\infty }h(qn)\rightarrow \frac{1}{q}%
\int_{0}^{\infty }dx\,h(x).  \label{replSumInt}
\end{equation}%
By making this replacement, we can see from (\ref{F2b}) that, in this
situation, the boundary induced VEVs tend to a finite value. Note that the
same is the case for the boundary-free part (\ref{F2sren3}). In the limit $%
q\gg 1$, the order of the modified Bessel functions is large for $n\neq 0$.
By using the corresponding asymptotic formulae it can be seen that the
contribution of these term is suppressed by the factor $\exp [-2qn\ln (a/r)]$%
. As a result, the main contribution comes from the lowest mode $n=0$ and
the boundary induced VEVs behave like $q$.

\subsection{Vacuum expectation value for the energy-momentum tensor}

\label{subsec:EMTint}

In this subsection we consider the vacuum EMT in the region inside the
cylindrical shell. Substituting the eigenfunctions (\ref{Aalpha}) into the
corresponding mode-sum formula, we obtain (no summation over $i$)%
\begin{equation}
\langle 0|T_{i}^{k}|0\rangle =\frac{q\delta _{i}^{k}}{4\pi ^{2}a^{3}}%
\sideset{}{'}{\sum}_{n=0}^{\infty }\int_{-\infty }^{+\infty }dk\sum_{\lambda
=0,1}\sum_{l=1}^{\infty }\frac{j_{n,l}^{(\lambda )3}T_{qn}(j_{n,l}^{(\lambda
)})}{\sqrt{j_{n,l}^{(\lambda )2}+k^{2}a^{2}}}%
f_{qn}^{(i)}[k,J_{qn}(j_{n,l}^{(\lambda )}r/a)],  \label{Tik}
\end{equation}%
where we have introduced the notations%
\begin{eqnarray}
f_{\nu }^{(i)}[k,f(x)] &=&(-1)^{i}\left( 2k^{2}/\gamma ^{2}+1\right) \left[
f^{\prime 2}(x)+\nu ^{2}f^{2}(x)/x^{2}\right] +f^{2}(x),  \label{f0} \\
f_{\nu }^{(j)}[k,f(x)] &=&(-1)^{i}f^{\prime 2}(x)-\left[ 1+(-1)^{i}\nu
^{2}/x^{2}\right] f^{2}(x),  \label{fi}
\end{eqnarray}%
with $\;i=0,3$ and $\;j=1,2$. As in the case of the field square, we apply
summation formula (\ref{sumJ1}) to rewrite the sum over $l$. This enables us
to present the VEV as the sum of boundary-free and boundary-induced parts as
follows
\begin{equation}
\langle 0|T_{i}^{k}|0\rangle =\langle 0_{s}|T_{i}^{k}|0_{s}\rangle +\langle
T_{i}^{k}\rangle _{a}.  \label{TikDecCyl}
\end{equation}%
The part induced by the cylindrical shell may be written in the form (no
summation over $i$) \cite{Beze06cylEl}
\begin{equation}
\langle T_{i}^{k}\rangle _{a}=\frac{q\delta _{i}^{k}}{4\pi ^{2}}%
\sideset{}{'}{\sum}_{n=0}^{\infty }\sum_{\lambda =0,1}\int_{0}^{\infty
}dxx^{3}\frac{K_{qn}^{(\lambda )}(xa)}{I_{qn}^{(\lambda )}(xa)}%
F_{qn}^{(i)}[I_{qn}(xr)],  \label{Tikb}
\end{equation}%
with the notations
\begin{eqnarray}
F_{\nu }^{(0)}[f(y)] &=&F_{\nu }^{(3)}[f(y)]=f^{2}(y),  \label{Fnu0} \\
F_{\nu }^{(i)}[f(y)] &=&-(-1)^{i}f^{\prime 2}(y)-\left[ 1-(-1)^{i}\nu
^{2}/y^{2}\right] f^{2}(y),\;i=1,2.  \label{Fnui}
\end{eqnarray}%
As it can be easily checked, this tensor is traceless and satisfies the
covariant continuity equation $\nabla _{k}\langle T_{i}^{k}\rangle _{a}=0$.
By using the inequalities $[I_{\nu }(y)K_{\nu }(y)]^{\prime }<0$ and $I_{\nu
}^{\prime }(y)<\sqrt{1+\nu ^{2}/y^{2}}I_{\nu }(y)$, and the recurrence
relations for the modified Bessel functions, it can be seen that the
boundary-induced parts in the vacuum energy density and axial stress are
negative, whereas the corresponding radial and azimuthal stresses are
positive. The VEV of the EMT in the region outside the cylindrical shell is
obtained from (\ref{Tikb}) by the replacements $I\rightleftarrows K$.

The renormalized VEV of the EMT for the geometry without the cylindrical
shell is obtained by using the corresponding formulae for the field square (%
\ref{F2sren3}). For the corresponding energy density one finds \cite{Frol87}
\begin{equation}
\langle T_{0}^{0}\rangle _{s,\mathrm{ren}}=-\frac{(q^{2}-1)(q^{2}+11)}{%
720\pi ^{2}r^{4}}.  \label{T00s}
\end{equation}%
Other components are found from the tracelessness condition and the
continuity equation.

Now, let us discuss the behavior of the boundary-induced part in the VEV of
the EMT in the asymptotic region of the parameters. Near the cylindrical
shell the main contribution comes from large values of $n$. Thus, using the
uniform asymptotic expansions for the modified Bessel functions for large
values of the order, up to the leading order, we find%
\begin{equation}
\langle T_{0}^{0}\rangle _{\mathrm{b}}\approx -\frac{1}{2}\langle
T_{2}^{2}\rangle _{\mathrm{b}}\approx -\frac{(a-r)^{-3}}{60\pi ^{2}a}%
,\;\langle T_{1}^{1}\rangle _{\mathrm{b}}\approx \frac{(a-r)^{-2}}{60\pi
^{2}a^{2}}.  \label{TiknearCyl}
\end{equation}%
These leading terms do not depend on the planar angle deficit in the cosmic
string geometry. Near the cosmic string the main contribution comes from the
mode $n=0$ and we have%
\begin{equation}
\langle T_{0}^{0}\rangle _{\mathrm{b}}\approx -\langle T_{1}^{1}\rangle _{%
\mathrm{b}}\approx -\langle T_{2}^{2}\rangle _{\mathrm{b}}\approx \frac{q}{%
8\pi ^{2}a^{4}}\int_{0}^{\infty }dx\,x^{3}\left[ \frac{K_{0}(x)}{I_{0}(x)}-%
\frac{K_{1}(x)}{I_{1}(x)}\right] =-0.0168\frac{q}{a^{4}}.  \label{T00nearstr}
\end{equation}%
Therefore, differently from the VEV for the field square, the
boundary-induced part in the vacuum EMT is finite on the string for all
values of $q$.

The behavior of the boundary-induced part in the VEV of the EMT in the
asymptotic regions of the parameter $q$ is investigated in a way analogous
to that for the field square. For $q\ll 1$, we replace the summation over $n$
by the integration in accordance with (\ref{replSumInt}) and the VEV tends
to a finite limiting value which does not depend on $q$. Note that as the
spatial volume element is proportional to $1/q$, in this limit the global
quantities such as the integrated vacuum energy behave as $1/q$. In the
limit $q\gg 1$, the contribution of the modes with $n\geqslant 1$ is
suppressed by the factor $\exp [-2qn\ln (a/r)]$ and the main contribution
comes from the $n=0$ mode with the behavior $\propto q$. Though in this
limit the vacuum densities are large, due to the factor $1/q$ in the spatial
volume the corresponding global quantities tend to finite value. The VEVs
for the EMT\ of the electromagnetic field induced by perfectly conducting
cylindrical shell in the Minkowski spacetime considered in \cite{Sah2}, are
obtained from the formulae given in this section taking $q=1$.

The VEVs of the field square and the EMT for electromagnetic field inside a
perfectly conducting wedge with the opening angle $\phi _{0}$ and with a
conducting cylindrical boundary of radius $a$ can be evaluated by the way
similar to that used in this section for the geometry of a cosmic string.
The corresponding results are given in \cite{Saha07wedEl}. In the geometry
of a wedge the eigenfunctions for the vector potential in the region inside
the cylindrical shell are given by formula (\ref{Aalpha}) with the
replacements $\exp (iqn\phi )\rightarrow \sin (qn\phi )$, $n=1,2,\ldots $,
for TM modes ($\lambda =0$) and $\exp (iqn\phi )\rightarrow \cos (qn\phi )$,
$n=1,2,\ldots $, for TE modes ($\lambda =1$), where now $q=\pi /\phi _{0}$.
In the case of a wedge the VEVs of the field square and the EMT depend on
the angle $\phi $. The vacuum energy density induced by the cylindrical
shell is negative for the interior region and the corresponding vacuum
forces acting on the wedge sides are always attractive.

\section{Vacuum densities in the region between two coaxial cylindrical
surfaces}

\label{sec:TwoCyl}

\subsection{Scalar field}

\label{subsec:TwoCylSc}

In this subsection, we consider the positive frequency Wightman function,
the VEVs of the field square and the EMT for a massive scalar field with
general curvature coupling parameter in the region between two coaxial
cylindrical surfaces with radii $a$ and $b$, $a<b$, on background of the $%
(D+1)$-dimensional Minkowski spacetime \cite{Saha06cyl}. In an appropriately
chosen cylindrical system of coordinates the corresponding line element has
the form (\ref{ds21}) with $0\leqslant \phi \leqslant 2\pi $. We will assume
that on the bounding surfaces the field obeys the boundary conditions%
\begin{equation}
\left( \tilde{A}_{j}+\tilde{B}_{j}n_{(j)}^{i}\nabla _{i}\right) \varphi
\left( x\right) \big|_{r=j}=0,\;j=a,b,  \label{BCtwoCyl}
\end{equation}%
with $\tilde{A}_{j}$ \ and $\tilde{B}_{j}$ being constants, $n_{(j)}^{i}$ \
is the inward-pointing normal to the bounding surface $r=j$. For the region
between the surfaces, $a\leqslant r\leqslant b$, one has $%
n_{(j)}^{i}=n_{j}\delta _{1}^{i}$ with the notations $n_{a}=1$ and $n_{b}=-1$%
. The eigenfunctions are specified by the set of quantum numbers $\sigma
=(\gamma ,n,\mathbf{k})$, $n=0,\pm 1,\pm 2,\ldots $, and have the form%
\begin{equation}
\varphi _{\sigma }(x)=\beta _{\sigma }g_{\left\vert n\right\vert }(\gamma
a,\gamma r)\exp (in\phi +i\mathbf{kr}_{\parallel }-i\omega t),
\label{TCeigfunc1}
\end{equation}%
with%
\begin{equation}
g_{n}(\gamma a,\gamma r)=\bar{Y}_{n}^{(a)}(\gamma a)J_{n}(\gamma r)-\bar{J}%
_{n}^{(a)}(\gamma a)Y_{n}(\gamma r),  \label{TCgn}
\end{equation}%
and the other notations are the same as in Section \ref{sec:CosStCyl}. The
barred notation is defined by formula (\ref{barjnot}) with the coefficients%
\begin{equation}
A_{j}=\tilde{A}_{j},\;B_{j}=n_{j}\tilde{B}_{j}/j,\;\ j=a,b.  \label{TCCoef}
\end{equation}%
From the boundary condition on the surface $r=b$ it follows that the
possible values of $\gamma $ are solutions to the equation ($\eta =b/a$)%
\begin{equation}
C_{n}^{ab}(\eta ,\gamma a)\equiv \bar{J}_{n}^{(a)}(\gamma a)\bar{Y}%
_{n}^{(b)}(\gamma b)-\bar{Y}_{n}^{(a)}(\gamma a)\bar{J}_{n}^{(b)}(\gamma
b)=0.  \label{TCmodeeq}
\end{equation}%
The corresponding positive roots we will denote by $\gamma a=\gamma _{n,l}$,
$l=1,2,\ldots $, assuming that they are arranged in the ascending order, $%
\gamma _{n,l}<\gamma _{n,l+1}$.

From the orthonormality condition for the eigenfunctions, for the
coefficient $\beta _{\sigma }$ one finds%
\begin{equation}
\beta _{\sigma }^{2}=\frac{\pi ^{2}\gamma T_{n}^{ab}(\gamma a)}{4\omega
a(2\pi )^{D-1}},  \label{TCbetalf}
\end{equation}%
with the notation from (\ref{tekaAB}). Substituting eigenfunctions into the
mode-sum formula (\ref{mmodesumWF}), for the positive frequency Wightman
function one finds%
\begin{eqnarray}
W(x,x^{\prime }) &=&\frac{\pi ^{2}}{2a}\int d^{N}\mathbf{k}%
\sideset{}{'}{\sum}_{n=0}^{\infty }\sum_{l=1}^{\infty }\frac{%
zg_{n}(z,zr/a)g_{n}(z,zr^{\prime }/a)}{(2\pi )^{D-1}\sqrt{z+k_{m}^{2}a^{2}}}
\notag \\
&&\times T_{n}^{ab}(z)\cos (n\Delta \phi )\exp (i\mathbf{k}\Delta \mathbf{r}%
_{\parallel }-i\omega \Delta t)\big|_{z=\gamma _{n,l}},  \label{TCW2}
\end{eqnarray}%
where, as before, $k_{m}^{2}=k^{2}+m^{2}$ and the prime on the summation
sign means that the summand with $n=0$ should be halved. For the further
evaluation of this VEV we apply to the sum over $l$ summation formula (\ref%
{cor3form}) with%
\begin{equation}
h(x)=\frac{xg_{n}(x,xr/a)g_{n}(x,xr^{\prime }/a)}{\sqrt{x^{2}+k_{m}^{2}a^{2}}%
}\exp (-i\Delta t\sqrt{x^{2}/a^{2}+k_{m}^{2}}).  \label{TChx}
\end{equation}%
The corresponding conditions are satisfied if $r+r^{\prime }+|\Delta t|<2b$.
In particular, this is the case in the coincidence limit $t=t^{\prime }$ for
the region under consideration. Now we can see that the application of
formula (\ref{cor3form}) allows to present the Wightman function in the form%
\begin{eqnarray}
W(x,x^{\prime }) &=&\frac{1}{(2\pi )^{D-1}}\sideset{}{'}{\sum}_{n=0}^{\infty
}\cos (n\Delta \phi )\int d^{N}\mathbf{k}\,e^{i\mathbf{k}\Delta \mathbf{r}%
_{\parallel }}  \notag \\
&&\times \Bigg[\frac{1}{a}\int_{0}^{\infty }dz\frac{h(z)}{\bar{J}%
_{n}^{(a)2}(z)+\bar{Y}_{n}^{(a)2}(z)}-\frac{2}{\pi }\int_{k_{m}}^{\infty }dz%
\frac{x\Omega _{a\nu }(az,bz)}{\sqrt{z^{2}-k_{m}^{2}}}  \notag \\
&&\times G_{n}^{(a)}(az,zr)G_{n}^{(a)}(az,zr^{\prime })\cosh (\Delta t\sqrt{%
z^{2}-k_{m}^{2}})\Bigg],  \label{TCW3}
\end{eqnarray}%
with the notations from (\ref{Geab}).

In the limit $b\rightarrow \infty $ the second term in figure braces on the
right of (\ref{TCW3}) vanishes, whereas the first term does not depend on $b$%
. It follows from here that the part with the first term presents the
Wightman function in the region outside of a single cylindrical shell with
radius $a$. As a result the Wightman function is presented in the form \cite%
{Saha06cyl}%
\begin{eqnarray}
W(x,x^{\prime }) &=&W_{\mathrm{M}}(x,x^{\prime })+\left\langle \varphi
(x)\varphi (x^{\prime })\right\rangle _{a}  \notag \\
&&-\frac{2^{2-D}}{\pi ^{D}}\sideset{}{'}{\sum}_{n=0}^{\infty }\cos (n\Delta
\phi )\int d^{N}\mathbf{k}\,e^{i\mathbf{k}\Delta \mathbf{r}_{\parallel
}}\int_{k_{m}}^{\infty }dz\,z\frac{\Omega _{an}(az,bz)}{\sqrt{z^{2}-k_{m}^{2}%
}}  \notag \\
&&\times G_{n}^{(a)}(az,zr)G_{n}^{(a)}(az,zr^{\prime })\cosh (\Delta t\sqrt{%
z^{2}-k_{m}^{2}}),  \label{TCW4}
\end{eqnarray}%
where $W_{\mathrm{M}}(x,x^{\prime })$ is the Wightman function for a scalar
field in\ the unbounded Minkowskian spacetime, and%
\begin{eqnarray}
\left\langle \varphi (x)\varphi (x^{\prime })\right\rangle _{a} &=&-\frac{%
2^{2-D}}{\pi ^{D}}\sideset{}{'}{\sum}_{n=0}^{\infty }\cos (n\Delta \phi
)\int d^{N}\mathbf{k}\,e^{i\mathbf{k}\Delta \mathbf{r}_{\parallel
}}\int_{k_{m}}^{\infty }dz\,z  \notag \\
&&\times \frac{\bar{I}_{n}^{(a)}(az)}{\bar{K}_{n}^{(a)}(az)}\frac{%
K_{n}(zr)K_{n}(zr^{\prime })}{\sqrt{z^{2}-k_{m}^{2}}}\cosh (\Delta t\sqrt{%
z^{2}-k_{m}^{2}}),  \label{TCWa}
\end{eqnarray}%
is the part of the Wightman function induced by a single cylindrical shell
with radius $a$ in the region $r>a$. Hence, the last term on the right of (%
\ref{TCW4}) is induced by the presence of the second shell with radius $b$.
It can be seen that the Wightman function can also be presented in an
equivalent form which is obtained from (\ref{TCW4}) by the replacement $%
a\rightarrow b$ except in the argument of the function $\Omega _{an}(az,bz)$%
. In this representation%
\begin{eqnarray}
\left\langle \varphi (x)\varphi (x^{\prime })\right\rangle _{b} &=&-\frac{%
2^{2-D}}{\pi ^{D}}\sideset{}{'}{\sum}_{n=0}^{\infty }\cos (n\Delta \phi
)\int d^{N}\mathbf{k}\,e^{i\mathbf{k}\Delta \mathbf{r}_{\parallel
}}\int_{k_{m}}^{\infty }dz\,z  \notag \\
&&\times \frac{\bar{K}_{n}^{(b)}(bz)}{\bar{I}_{n}^{(b)}(bz)}\frac{%
I_{n}(zr)I_{n}(zr^{\prime })}{\sqrt{z^{2}-k_{m}^{2}}}\cosh (\Delta t\sqrt{%
z^{2}-k_{m}^{2}}),  \label{TCWb}
\end{eqnarray}%
is the part induced by a single cylindrical shell with radius $b$ in the
region $r<b$. This formula is also directly obtained from (\ref{Wfa0})
taking $\phi _{0}=2\pi $.

By making use of the formulae for the Wightman function and taking the
coincidence limit of the arguments, for the VEV of the field square one finds%
\begin{eqnarray}
\langle 0|\varphi ^{2}|0\rangle &=&\langle \varphi ^{2}\rangle _{\mathrm{M}%
}+\langle \varphi ^{2}\rangle _{j}-B_{D}\sideset{}{'}{\sum}_{n=0}^{\infty
}\int_{m}^{\infty }du\,u  \notag \\
&&\times \left( u^{2}-m^{2}\right) ^{\frac{D-3}{2}}\Omega
_{jn}(au,bu)G_{n}^{(j)2}(ju,ru),  \label{TCphi21}
\end{eqnarray}%
where $j=a$ and $j=b$ provide two equivalent representations and%
\begin{equation}
B_{D}=\frac{2^{2-D}}{\pi ^{\frac{D+1}{2}}\Gamma \left( \frac{D-1}{2}\right) }%
.  \label{TCAD}
\end{equation}%
For points away from the boundaries the last two terms on the right of
formula (\ref{TCphi21}) are finite and, hence, the subtraction of the
Minkowskian part without boundaries is sufficient to obtain the renormalized
value for the VEV: $\langle \varphi ^{2}\rangle _{\mathrm{ren}}=\langle
0|\varphi ^{2}|0\rangle -\langle \varphi ^{2}\rangle _{\mathrm{M}}$. In
formula (\ref{TCphi21}) the part $\langle \varphi ^{2}\rangle _{j}$ is
induced by a single cylindrical surface with radius $j$ when the second
surface is absent. The formulae for these terms are obtained from the
Wightman function in the coincidence limit. For $j=a$ one has%
\begin{equation}
\langle \varphi ^{2}\rangle _{a}=-B_{D}\sideset{}{'}{\sum}_{n=0}^{\infty
}\int_{m}^{\infty }du\,u\left( u^{2}-m^{2}\right) ^{\frac{D-3}{2}}\frac{\bar{%
I}_{n}^{(a)}(au)}{\bar{K}_{n}^{(a)}(au)}K_{n}^{2}(ru),  \label{TCphi2a}
\end{equation}%
and the formula for $\langle \varphi ^{2}\rangle _{b}$ is obtained from here
by the replacements $a\rightarrow b$, $I\rightleftarrows K$. The last term
on the right of formula (\ref{TCphi21}) is induced by the presence of the
second cylindrical surface.

The VEV for the EMT is obtained by using the formulae for the Wightman
function and the VEV of the field square:%
\begin{eqnarray}
\langle 0|T_{i}^{k}|0\rangle &=&\langle T_{i}^{k}\rangle ^{(0)}+\langle
T_{i}^{k}\rangle _{j}+B_{D}\delta _{i}^{k}\sideset{}{'}{\sum}_{n=0}^{\infty
}\int_{m}^{\infty }du\,u^{3}  \notag \\
&&\times \left( u^{2}-m^{2}\right) ^{\frac{D-3}{2}}\Omega
_{jn}(au,bu)F_{n}^{(i)}[G_{n}^{(j)}(ju,ru)],  \label{TCvevemt2}
\end{eqnarray}%
where the notations $F_{\nu }^{(i)}[f(z)]$ are defined by formulae (\ref%
{ajpm})-(\ref{ajpm2}) with $f(z)=G_{n}^{(j)}(ju,z)$ and $%
F_{n}^{(i)}[f(z)]=F_{n}^{(0)}[f(z)]$ for $i=3,\ldots ,D-1$. In formula (\ref%
{TCvevemt2}), the term $\langle T_{i}^{k}\rangle ^{(j)}$ is induced by a
single cylindrical surface with radius $j$. These parts for both interior
and exterior regions are investigated in \cite{Rome01} and are obtained from
formulae of subsection \ref{sec:inside} taking $\phi _{0}=2\pi $. The
formula for the case $j=a$ is obtained from (\ref{TCphi2a}) by the
replacement $K_{n}^{2}(ru)\rightarrow u^{2}F_{n}^{(i)}[K_{n}(ru)]$.

The vacuum force per unit surface of the cylinder at $r=j$ is determined by
the ${}_{1}^{1}$ -- component of the vacuum EMT at this point. Similar to
the case of spherical geometry, for the region between two surfaces the
corresponding effective pressures can be presented as the sum of self-action
and interaction parts, formula (\ref{pjsphere}). The interaction forces are
determined from the last term on the right of formula (\ref{TCvevemt2}) with
$i=k=1$ taking $r=j$:
\begin{eqnarray}
p_{\mathrm{(int)}}^{(j)} &=&\frac{A_{D}}{2j^{2}}\sideset{}{'}{\sum}%
_{n=0}^{\infty }\int_{m}^{\infty }du\,u\left( u^{2}-m^{2}\right) ^{\frac{D-3%
}{2}}\Omega _{jn}(au,bu)  \notag \\
&&\times \left[ \left( n^{2}/j^{2}+u^{2}\right) B_{j}^{2}+4\xi
n_{j}A_{j}B_{j}/j-A_{j}^{2}\right] .  \label{TCpjint0}
\end{eqnarray}%
The expression on the right of this formula is finite for all non-zero
distances between the shells. It can be seen that the vacuum effective
pressures are negative for both Dirichlet and Neumann scalars and, hence,
the corresponding interaction forces are attractive. For the general Robin
case the interaction force can be either attractive or repulsive in
dependence on the coefficients in the boundary conditions. By using the
properties of the modified Bessel functions, the interaction forces per unit
surface can also be presented in another equivalent form \cite{Saha06cyl}:%
\begin{eqnarray}
p_{\mathrm{(int)}}^{(j)} &=&\frac{A_{D}n_{j}}{2j}\sideset{}{'}{\sum}%
_{n=0}^{\infty }\int_{m}^{\infty }du\,u\left( u^{2}-m^{2}\right) ^{\frac{D-3%
}{2}}  \notag \\
&&\times \left[ 1+(4\xi -1)\frac{n_{j}A_{j}B_{j}}{jB_{jn}(u)}\right] \frac{%
\partial }{\partial j}\ln \left\vert 1-\frac{\bar{I}_{n}^{(a)}(au)\bar{K}%
_{n}^{(b)}(bu)}{\bar{I}_{n}^{(b)}(bu)\bar{K}_{n}^{(a)}(au)}\right\vert .
\label{pjint1cyl}
\end{eqnarray}%
The relation between the interaction forces and the corresponding bulk and
surface Casimir energies in the geometry under consideration is discussed in
\cite{Saha06cyl}.

\subsection{Electromagnetic field}

\label{subsec:TCEl}

In this subsection we consider the VEV for the EMT of the electromagnetic
field in the region between two coaxial cylindrical surfaces with radii $a$
and $b$, $a<b$ \cite{Sah3}. The corresponding eigenfunctions satisfying the
boundary condition on the surface $r=a$ have the form%
\begin{equation}
\mathbf{A}_{\sigma }=\beta _{\sigma }\left\{
\begin{array}{cc}
(1/i\omega )\left( \gamma ^{2}\mathbf{e}_{3}+ik\nabla _{t}\right) P_{\lambda
n}(\gamma a,\gamma r)\exp \left[ i\left( n\phi +kz-\omega t\right) \right] ,
& \lambda =0 \\
-\mathbf{e}_{3}\times \nabla _{t}\left\{ P_{\lambda n}(\gamma a,\gamma
r)\exp \left[ i\left( n\phi +kz-\omega t\right) \right] \right\} , & \lambda
=1%
\end{array}%
\right. ,  \label{TCAalpha}
\end{equation}%
where
\begin{equation}
P_{\lambda n}(x,y)=J_{n}(y)Y_{n}^{(\lambda )}(x)-Y_{n}(y)J_{n}^{(\lambda
)}(x),  \label{radfout}
\end{equation}%
and the other notations are the same as in (\ref{Aalpha}). From the boundary
conditions on $r=b$ one obtains that the eigenvalues for $\gamma $ have to
be solutions to the following equations
\begin{equation}
\partial _{r}^{\lambda }P_{\lambda n}(\gamma a,\gamma r)|_{r=b}=0.
\label{cyloutcond1}
\end{equation}%
These equations have an infinite number of simple real solutions. The
eigenvalue equation (\ref{cyloutcond1}) can be written in terms of function (%
\ref{bescomb1}) as
\begin{equation}
C_{n}^{ab}(\eta ,\gamma b)=0,\quad A_{a}=A_{b}=1-\lambda
,\,B_{a}=B_{b}=\lambda ,\quad \lambda =0,1  \label{cyloutcondCAB}
\end{equation}%
(see notation (\ref{efnot1})). Now the normalization coefficient $\beta
_{\sigma }$ is given by the formulae
\begin{equation}
\beta _{\sigma }^{2}=\frac{\pi }{4\omega }\left\{
\begin{array}{ll}
\left[ J_{n}^{2}(z)/J_{n}^{2}(z\eta )-1\right] ^{-1}, & \lambda =0 \\
\left[ \left( 1-n^{2}/z^{2}\eta ^{2}\right) J_{n}^{^{\prime
}2}(z)/J_{n}^{^{\prime }2}(z\eta )-1+n^{2}/z^{2}\right] ^{-1}, & \lambda =1%
\end{array}%
\right. ,  \label{cylnormout}
\end{equation}%
where $z=\gamma a$, $\eta =b/a$. Note that this coefficients can be
expressed in terms of function (\ref{tekaAB}) as
\begin{equation}
\beta _{\alpha }^{2}=\frac{\pi z^{2\lambda -1}}{4\omega }T_{n}^{ab}(\eta ,z).
\label{cylnormoutTAB}
\end{equation}

{} Using these relations and introducing the cutoff function $\psi _{\mu }$,
the vacuum EMT can be written in the form of the following finite
integrosums (no summation over $i$)
\begin{eqnarray}
\langle 0|T_{i}^{k}|0\rangle &=&\frac{\delta _{i}^{k}}{16a^{3}}%
\sideset{}{'}{\sum}_{n=0}^{\infty }\int_{-\infty }^{+\infty }dk\sum_{\lambda
=0,1}\sum_{l=1}^{\infty }\frac{\psi _{\mu }(\gamma _{n,l}^{(\lambda )}/a)}{%
\sqrt{\gamma _{n,l}^{(\lambda )2}+k^{2}a^{2}}}  \notag \\
&&\times T_{m}^{ab}(\eta ,\gamma _{n,l}^{(\lambda )})\gamma _{n,l}^{(\lambda
)}{}^{3+2\lambda }f_{n}^{(i)}[k,P_{\lambda n}(\gamma _{n,l}^{(\lambda
)},\gamma _{n,l}^{(\lambda )}r/a)],  \label{cylqout1}
\end{eqnarray}%
where $\gamma a=\gamma _{n,l}^{(\lambda )}$, $l=1,2,\ldots $, are the
solutions to eigenvalue equations (\ref{cyloutcond1}) and the expressions
for the functions $f_{n}^{(i)}[k,P_{\lambda n}(\gamma a,x)]$ are obtained
from (\ref{f0}) and (\ref{fi}) taking $f(x)=P_{\lambda n}(\gamma a,x)$. By
choosing in formula (\ref{cor3form})
\begin{equation}
h(z)=\frac{z^{3+2\lambda }\psi _{\mu }(z/a)}{\sqrt{z^{2}+k^{2}a^{2}}}%
f_{n}^{(i)}[k,P_{\lambda n}(z,zx)],  \label{cyloutfg}
\end{equation}%
one obtains that the VEV of the electromagnetic EMT in the region between
two coaxial conducting cylindrical surfaces is presented in the form \cite%
{Sah3}
\begin{equation}
\langle 0|T_{i}^{k}|0\rangle =\langle 0|T_{i}^{k}|0\rangle ^{(a)}+\langle
T_{i}^{k}\rangle _{ab},\quad a<r<b.  \label{cylab12}
\end{equation}%
In this formula the first term on the right is given by (no summation over $%
i $)
\begin{equation}
\langle 0|T_{i}^{k}|0\rangle ^{(a)}=\frac{\delta _{i}^{k}}{8\pi ^{2}}%
\sideset{}{'}{\sum}_{n=0}^{\infty }\int_{-\infty }^{+\infty
}dk\int_{0}^{\infty }dz\sum_{\lambda =0,1}\frac{z^{3}\psi _{\mu }(z)}{\sqrt{%
k^{2}+z^{2}}}\frac{f_{n}^{(i)}[k,P_{\lambda n}(az,rz)]}{J_{\nu }^{(\lambda
)2}(az)+Y_{\nu }^{(\lambda )2}(az)}\,,  \label{cylqout2}
\end{equation}%
and for the second term we have
\begin{equation}
\langle T_{i}^{k}\rangle _{ab}=\frac{\delta _{i}^{k}}{4\pi ^{2}}%
\sideset{}{'}{\sum}_{n=0}^{\infty }\int_{0}^{\infty }dz\sum_{\lambda
=0,1}z^{3}\frac{F_{n}^{(i)}[Q_{\lambda n}(az,rz)]K_{n}^{(\lambda
)}(bz)/K_{n}^{(\lambda )}(az)}{K_{n}^{(\lambda )}(az)I_{n}^{(\lambda
)}(bz)-K_{n}^{(\lambda )}(bz)I_{n}^{(\lambda )}(az)},  \label{cylqab122}
\end{equation}%
with the notation $F_{\nu }^{(i)}[f(y)]$ from (\ref{Fnu0}), (\ref{Fnui}) and
\begin{equation}
Q_{\lambda n}(z,y)=K_{n}^{(\lambda )}(z)I_{n}(y)-I_{n}^{(\lambda
)}(z)K_{n}(y).  \label{Qlamm}
\end{equation}%
Note that the part (\ref{cylqab122}) is finite for $a\leqslant r<b$ and in
this part we have removed the cutoff function.

In the limit $b\rightarrow \infty $ the second term on the right of (\ref%
{cylab12}) tends to zero, whereas the first one does not depend on $b$. From
here we conclude that the term $\langle 0|T_{i}^{k}|0\rangle ^{(a)}$ is the
VEV in the region outside a single cylindrical shell with radius $a$. For
the renormalization of this term we subtract the contribution of unbounded
Minkowski spacetime which can be presented in the form:
\begin{equation}
\langle 0|T_{i}^{k}|0\rangle _{\mathrm{M}}=\frac{\delta _{i}^{k}}{4\pi ^{2}}%
\sideset{}{'}{\sum}_{n=0}^{\infty }\int_{-\infty }^{+\infty
}dk\int_{0}^{\infty }dz\frac{z^{3}\psi _{\mu }(z)}{\sqrt{k^{2}+z^{2}}}%
f_{n}^{(i)}[k,J_{n}(rz)].  \label{qMinkcyl}
\end{equation}%
By using the identity
\begin{equation}
\frac{f_{n}^{(i)}[k,P_{\lambda n}(az,rz)]}{J_{\nu }^{(\lambda )2}(az)+Y_{\nu
}^{(\lambda )2}(az)}-f_{n}^{(i)}[k,J_{n}(rz)]=-\frac{1}{2}\sum_{s=1,2}\frac{%
J_{n}^{(\lambda )}(az)}{H_{n}^{(s)(\lambda )}(az)}%
f_{n}^{(i)}[k,H_{n}^{(s)}(rz)],  \label{reloucyl2}
\end{equation}%
and rotating the integration contour for $z$ by angle $\pi /2$ for $s=1$ and
by angle $-\pi /2$ for $s=2$, for the renomalized components we obtain
\begin{equation}
\langle T_{i}^{k}\rangle _{a}=\frac{\delta _{i}^{k}}{4\pi ^{2}}%
\sideset{}{'}{\sum}_{n=0}^{\infty }\int_{0}^{\infty }dz\,z^{3}\left[ \frac{%
I_{n}(az)}{K_{n}(az)}+\frac{I_{n}^{\prime }(az)}{K_{n}^{\prime }(az)}\right]
F_{n}^{(i)}[K_{n}(rz)].  \label{cylqout3}
\end{equation}%
Note that the corresponding expressions in the region inside a single
cylindrical shell are obtained from (\ref{Tikb}) taking $q=1$ and differ
from (\ref{cylqout3}) by the replacements $I\rightleftarrows K$. As in the
case of the interior components, the renormalized VEV (\ref{cylqout3}) is
divergent when $r\rightarrow a$ with the leading terms given by formulae (%
\ref{TiknearCyl}). The corresponding asymptotic behavior at large distances
from the cylinder, $r\gg a$, can be found from (\ref{cylqout3}) introducing
a new integration variable $y=rz$ and expanding the integrands over $a/r$.
In this limit the main contribution comes from the lowest order mode with $%
n=0$ and one obtains (no summation over $i$)
\begin{equation}
\langle T_{i}^{k}\rangle _{a}\approx \frac{\delta _{i}^{k}c^{(i)}}{8\pi
^{2}r^{4}\ln (r/a)},\quad c^{(0)}=c^{(1)}=\frac{1}{3},\,\,c^{(2)}=-1,\quad
r\gg a.  \label{asympfaraxis}
\end{equation}%
Here, compared to the spherical case, the corresponding quantities tend to
zero more slowly.

{}From the continuity equation for the vacuum EMT one has the following
integral relation
\begin{equation}
\langle T_{1}^{1}\rangle _{a}=\frac{2}{r^{2}}\int_{r}^{\infty }dr\,r\langle
T_{0}^{0}\rangle _{a}=\frac{E_{\mathrm{cyl}}^{\mathrm{out}}(r)}{\pi r^{2}},
\label{cylcontintout}
\end{equation}%
where $E_{\mathrm{cyl}}^{\mathrm{out}}(r)$ is the total energy (per unit
length) outside the cylindrical surface of radius $r$. Combining this
relation with the similar relation in the interior region, for the total
vacuum energy of the cylindrical shell per unit length we obtain
\begin{equation}
E_{\mathrm{cyl}}=E_{\mathrm{cyl}}^{\mathrm{in}}(a)+E_{\mathrm{cyl}}^{\mathrm{%
out}}(a)=\pi a^{2}\left[ \langle T_{1}^{1}\rangle _{a}(a+)-\langle
T_{1}^{1}\rangle _{a}(a-)\right] .  \label{cyltotenergy}
\end{equation}%
By taking into account the corresponding expressions for the radial
component this yields
\begin{eqnarray}
E_{\mathrm{cyl}} &=&\frac{-1}{4\pi a^{2}}\sideset{}{'}{\sum}_{n=0}^{\infty
}\int_{0}^{\infty }dz\,\chi _{\mu }(z/a)\big(\ln [I_{n}(z)K_{n}(z)]\big)%
^{\prime }\left[ z^{2}+\left( z^{2}+n^{2}\right) \frac{I_{n}(z)K_{n}(z)}{%
I_{n}^{\prime }(z)K_{n}^{\prime }(z)}\right]  \notag \\
&=&\frac{-1}{4\pi a^{2}}\sideset{}{'}{\sum}_{n=0}^{\infty }\int_{0}^{\infty
}dz\,\chi _{\mu }(z/a)z^{2}\frac{d}{dz}\ln \left[ 1-z^{2}\big(%
I_{n}(z)K_{n}(z)\big)^{\prime 2}\right] .  \label{cyltotenergy1}
\end{eqnarray}%
In the last expression, integrating by part and omitting the boundary term
we obtain the Casimir energy in the form used in numerical calculations. The
corresponding results are presented in \cite{Miltoncyl,Romeocyl,Miltoncyl1}.
Note that in the evaluation of the Casimir energy for a perfectly conducting
cylindrical shell by the Green function method to perform the complex
frequency rotation procedure an additional cutoff function has to be
introduced (see \cite{Miltoncyl}). This is related to the above mentioned
divergence of the integrals over $z$ for $r=a$. The results of the numerical
evaluations for the energy density and pressures distributions (formula (\ref%
{cylqout3})) are presented in \cite{Sahdis,Sah2}. The energy density, $%
\langle T_{0}^{0}\rangle _{a}$, and azimuthal pressure, $-\langle
T_{2}^{2}\rangle _{a}$, in the exterior region are positive, and the radial
pressure, $-\langle T_{1}^{1}\rangle _{a}$, is negative. The ratio of the
energy density to the azimuthal pressure is a decreasing function on $r$,
and $1/3\leqslant -\langle T_{0}^{0}\rangle _{a}/\langle T_{2}^{2}\rangle
_{a}\leqslant 0.5$. Note that this ratio is a continuous function for all $r$
and monotonically decreases from 1 at the cylinder axis to 1/3 at infinity.

The quantities (\ref{cylab12}) with (\ref{cylqab122}) and (\ref{cylqout3})
represent the renormalized VEV of the EMT in the region between two coaxial
conducting cylindrical surfaces. Let us consider the limiting cases of the
term (\ref{cylqab122}). First let $a/r,\,a/b\ll 1$. After replacing $%
z\rightarrow bz$ and expanding the integrand over $a/r$ and $a/b$ it can be
seen that
\begin{equation}
\langle T_{i}^{k}\rangle _{ab}\approx \langle T_{i}^{k}\rangle _{b},\quad
a/r,\,a/b\ll 1,\,r<b,  \label{cylablim1}
\end{equation}%
where $\langle T_{i}^{k}\rangle _{b}$ is the vacuum EMT inside a single
cylindrical shell with radius $b$. When $a\rightarrow b$ the sum over $n$ in
(\ref{cylqab122}) diverges. Consequently for $b-a\ll b$ the main
contribution comes from large $n$. By using the uniform asymptotic
expansions for the modified Bessel functions, in this limit one obtains the
corresponding quantity for the Casimir parallel plate configuration.

{}From (\ref{cylab12}) and (\ref{cylqab122}) it can be seen that the VEV of
the EMT can also be written in the form
\begin{equation}
\langle 0|T_{i}^{k}|0\rangle =\langle 0|T_{i}^{k}|0\rangle ^{(b)}+\langle
T_{i}^{k}\rangle _{ba},\quad a<r<b,  \label{cylinb12}
\end{equation}%
where
\begin{equation}
\langle T_{i}^{k}\rangle _{ba}=\frac{\delta _{i}^{k}}{4\pi ^{2}}%
\sideset{}{'}{\sum}_{n=0}^{\infty }\int_{0}^{\infty }dz\sum_{\lambda
=0,1}z^{3}\frac{F_{n}^{(i)}[Q_{\lambda n}(bz,rz)]I_{n}^{(\lambda
)}(az)/I_{n}^{(\lambda )}(bz)}{K_{n}^{(\lambda )}(az)I_{n}^{(\lambda
)}(bz)-K_{n}^{(\lambda )}(bz)I_{n}^{(\lambda )}(az)}.  \label{cylabqin12}
\end{equation}%
The quantities (\ref{cylabqin12}) are finite for all $a<r\leqslant b$ and
diverge on the surface $r=a$.

{}Now we turn to the interaction forces between the cylindrical surfaces due
to the vacuum fluctuations. The vacuum force acting per unit surface at $r=j$
is determined by the ${}_{1}^{1}$--component of the EMT evaluated at this
point. The corresponding effective pressures are presented as a sum of the
self-action and interaction parts, similar to (\ref{pjint1cyl}).
Substituting into formulae (\ref{cylqab122}) and (\ref{cylabqin12}) with $%
i=k=1$ the values $r=a$ and $r=b$, respectively, and using the Wronskian for
the modified Bessel functions, for the interaction parts of the vacuum
pressures on the cylindrical surfaces one has \cite{Sah3}:
\begin{eqnarray}
p_{\mathrm{(int)}}^{(a)} &=&\frac{-1}{4\pi ^{2}a^{4}}\sideset{}{'}{\sum}%
_{n=0}^{\infty }\sum_{\lambda =0,1}(-1)^{\lambda }\int_{0}^{\infty }dz\,%
\frac{z\left( n^{2}/z^{2}+1\right) ^{\lambda }K_{n}^{(\lambda
)}(bz/a)/K_{n}^{(\lambda )}(z)}{K_{n}^{(\lambda )}(z)I_{n}^{(\lambda
)}(bz/a)-K_{n}^{(\lambda )}(bz/a)I_{n}^{(\lambda )}(z)},  \label{forceincyl1}
\\
p_{\mathrm{(int)}}^{(b)} &=&\frac{-1}{4\pi ^{2}b^{4}}\sideset{}{'}{\sum}%
_{n=0}^{\infty }\sum_{\lambda =0,1}(-1)^{\lambda }\int_{0}^{\infty }dz\,%
\frac{z\left( n^{2}/z^{2}+1\right) ^{\lambda }I_{n}^{(\lambda
)}(az/b)/I_{n}^{(\lambda )}(z)}{K_{n}^{(\lambda )}(az/b)I_{n}^{(\lambda
)}(z)-K_{n}^{(\lambda )}(z)I_{n}^{(\lambda )}(az/b)}.  \label{forceoutcyl1}
\end{eqnarray}%
In (\ref{forceincyl1}) and (\ref{forceoutcyl1}) the summands with $\lambda
=0 $ ($\lambda =1$) come from the electric (magnetic) waves contribution. By
using the properties of the modified Bessel functions it can be seen that
the quantities $p_{\mathrm{(int)}}^{(j)}$ are negative and, hence the
corresponding interaction forces are attractive. Note that interaction
forces (\ref{forceincyl1}), (\ref{forceoutcyl1}) can also be obtained by
differentiating the corresponding part in the vacuum energy with respect to
the radii \cite{Mazz02}.

\section{Polarization of the Fulling-Rindler vacuum by a uniformly
accelerated mirror}

\label{sec:FuRi}

It is well known that the uniqueness of the vacuum state is lost
when we work within the framework of quantum field theory in a
general curved spacetime or in non--inertial frames. In
particular, the use of general coordinate transformations in
quantum field theory in flat spacetime leads to an infinite number
of unitary inequivalent representations of the commutation
relations. Different inequivalent representations will in general
give rise to different vacuum states. For instance, the vacuum
state for a uniformly accelerated observer, the Fulling-Rindler
vacuum, turns out to be inequivalent to that for an inertial
observer, the familiar Minkowski vacuum. An interesting topic in
the investigations of the Casimir effect is the dependence of the
vacuum characteristics on the type of the vacuum. In this section,
by using the GAPF, we will study the scalar vacuum polarization
brought about by the presence of infinite plane boundary moving by
uniform acceleration through the Fulling-Rindler vacuum. The
corresponding VEVs of the EMT were studied by Candelas and Deutsch \cite%
{CandD} for conformally coupled 4D Dirichlet and Neumann massless scalar and
electromagnetic fields. In this paper only the region of the right Rindler
wedge to the right of the barrier is considered. In \cite{SahaRind1} we have
investigated the Wightman function, the VEVs of the field square and the EMT
for a massive scalar field with general curvature coupling parameter,
satisfying Robin boundary condition on infinite plate in an arbitrary number
of spacetime dimensions and for the electromagnetic field. The both regions,
including the one between the barrier and Rindler horizon are considered.

\subsection{Wightman function}

\label{subsec:FuRiWF}

In the accelerated reference frame it is convenient to introduce Rindler
coordinates $(\tau ,\xi ,\mathbf{x})$ related to the Minkowski ones, $%
(t,x^{1},\mathbf{x})$ by the formulae $t=\xi \sinh \tau ,\quad x^{1}=\xi
\cosh \tau $, and $\mathbf{x}=(x^{2},\ldots ,x^{D})$ denotes the set of
coordinates parallel to the plate. In these coordinates the Minkowski line
element takes the form%
\begin{equation}
ds^{2}=\xi ^{2}d\tau ^{2}-d\xi ^{2}-d\mathbf{x}^{2},  \label{metricRin}
\end{equation}%
and a world-line defined by $\xi ,\mathbf{x}=\mathrm{const}$ describes an
observer with constant proper acceleration $\xi ^{-1}$. The Rindler time
coordinate $\tau $ is proportional to the proper time along a family of
uniformly accelerated trajectories which fill the Rindler wedge, with the
proportionality constant equal to the acceleration.

We will assume that the plate is located at $\xi =a$, with $a^{-1}$ being
the proper acceleration, and the field satisfies boundary condition (\ref%
{mrobcond}). We will consider the region on the right from the boundary, $%
\xi \geqslant a$. In Rindler coordinates the boundary condition takes the
form
\begin{equation}
(\tilde{A}+\tilde{B}\partial _{\xi })\varphi =0,\quad \xi =a.
\label{boundRind}
\end{equation}%
In the region $\xi >a$, a complete set of solutions that are of positive
frequency with respect to $\partial /\partial \tau $ and bounded as $\xi
\rightarrow \infty $ is
\begin{equation}
\varphi _{\sigma }(x)=\beta _{\sigma }K_{i\omega }(\lambda \xi )e^{i\mathbf{%
kx}-i\omega \tau },\;\lambda =\sqrt{k^{2}+m^{2}},\;\sigma =(\omega ,\mathbf{k%
}).  \label{sol2}
\end{equation}%
From boundary condition (\ref{boundRind}) we find that the possible values
for $\omega $ have to be zeros of the function $\bar{K}_{i\omega }(\lambda
a) $, where the barred notation is defined by (\ref{Kizbar}) with the
coefficients%
\begin{equation}
A=\tilde{A},\;B=\tilde{B}/a.  \label{R1AB}
\end{equation}%
We will denote these zeros by $\omega =\omega _{n}=\omega _{n}(k)$, $%
n=1,2,...$, arranged in ascending order: $\omega _{n}<\omega _{n+1}$. The
coefficient $\beta _{\sigma }$ in (\ref{sol2}) is determined by the
normalization condition with respect to the standard Klein-Gordon inner
product, with the $\xi $-integration over the region $(a,\infty )$. From
this condition one finds
\begin{equation}
\beta _{\sigma }^{2}=\frac{1}{(2\pi )^{D-1}}\frac{\bar{I}_{i\omega
_{n}}(\lambda a)}{\partial _{\omega }\bar{K}_{i\omega }(\lambda a)\mid
_{\omega =\omega _{n}}}.  \label{normc}
\end{equation}%
Substituting the eigenfunctions (\ref{sol2}) into (\ref{mmodesumWF}), for
the Wightman function we obtain
\begin{equation}
W(x,x^{\prime })=\int d{}\mathbf{k}\,\frac{e^{i{}\mathbf{k}\Delta \mathbf{x}}%
}{(2\pi )^{D-1}}\sum_{n=1}^{\infty }\frac{\bar{I}_{i\omega }(\lambda a)}{%
\partial _{\omega }\bar{K}_{i\omega }(\lambda a)}K_{i\omega }(\lambda \xi
)K_{i\omega }(\lambda \xi ^{\prime })e^{-i\omega \Delta \tau }|_{\omega
=\omega _{n}},  \label{emtdiag}
\end{equation}%
with $\Delta \mathbf{x=x}-\mathbf{x}^{\prime }$, $\Delta \tau =$\ $\tau
-\tau ^{\prime }$. For the further evaluation of VEV (\ref{emtdiag}) we
apply to the sum over $n$ summation formula (\ref{sumformKi2}). As a
function $F(z)$ in this formula we choose
\begin{equation}
F(z)=K_{iz}(\lambda \xi )K_{iz}(\lambda \xi ^{\prime })e^{-iz\Delta \tau }.
\label{Fztoform}
\end{equation}%
Using the asymptotic formulae for the modified Bessel functions it can be
seen that condition (\ref{condFKi}) is satisfied if $a^{2}e^{|\Delta \tau
|}<\xi \xi ^{\prime }$. In particular, this is the case in the coincidence
limit $\tau =\tau ^{\prime }$ for the points in the region under
consideration, $\xi ,\xi ^{\prime }>a$. With $F(z)$ from (\ref{Fztoform}),
the contribution corresponding to the integral term on the left of formula (%
\ref{sumformKi2}) is the Wightman function for the Fulling-Rindler vacuum
without boundaries:
\begin{equation}
W_{\mathrm{R}}(x,x^{\prime })=\frac{1}{\pi ^{2}}\int d{}\mathbf{k}\,\frac{%
e^{i{}\mathbf{k}\Delta \mathbf{x}}}{(2\pi )^{D-1}}\int_{0}^{\infty }d\omega
\sinh (\pi \omega )e^{-i\omega \Delta \tau }K_{i\omega }(\lambda \xi
)K_{i\omega }(\lambda \xi ^{\prime }).  \label{emtRindler}
\end{equation}%
Taking into account this and applying summation formula (\ref{sumformKi2})
to Eq. (\ref{emtdiag}), we receive \cite{SahaRind1}
\begin{equation}
W(x,x^{\prime })=W_{\mathrm{R}}(x,x^{\prime })+\langle \varphi (x)\varphi
(x^{\prime })\rangle _{a},  \label{WRb}
\end{equation}%
where the second term on the right is induced by the barrier:
\begin{equation}
\langle \varphi (x)\varphi (x^{\prime })\rangle _{a}=-\frac{1}{\pi }\int d{}%
\mathbf{k}\,\frac{e^{i{}\mathbf{k}\Delta \mathbf{x}}}{(2\pi )^{D-1}}%
\int_{0}^{\infty }d\omega \frac{\bar{I}_{\omega }(\lambda a)}{\bar{K}%
_{\omega }(\lambda a)}K_{\omega }(\lambda \xi )K_{\omega }(\lambda \xi
^{\prime })\cosh (\omega \Delta \tau ),  \label{Wb}
\end{equation}%
and is finite for $\xi >a$. The divergences in the coincidence limit are
contained in the first term corresponding to the Fulling-Rindler vacuum
without boundaries. The boundary-induced part in the Wightman function in
the region $0<\xi <a$ is obtained from (\ref{Wb}) by the replacements $%
I\rightleftarrows K$ \cite{SahaRind1} (see below).

\subsection{Vacuum expectation values of the field square and the
energy-momentum tensor}

\label{subsec:FuRiVev}

In the coincidence limit, from formula (\ref{Wb}) for the boundary part of
the field square we have
\begin{equation}
\langle \varphi ^{2}\rangle _{a}=\frac{-2^{2-D}}{\pi ^{(D+1)/2}\Gamma \left(
\frac{D-1}{2}\right) }\int_{0}^{\infty }dk\,k^{D-2}\int_{0}^{\infty }d\omega
\,\frac{\bar{I}_{\omega }(\lambda a)}{\bar{K}_{\omega }(\lambda a)}K_{\omega
}^{2}(\lambda \xi ),\quad \xi >a.  \label{phi2br}
\end{equation}%
This quantity is monotone increasing negative function on $\xi $ for
Dirichlet scalar and monotone decreasing positive function for Neumann
scalar. Substituting the function (\ref{WRb}) into Eq. (\ref{mTikVEV}) and
taking into account Eqs. (\ref{Wb}) and (\ref{phi2br}), for the VEV of the
EMT in the region $\xi >a$ one finds
\begin{equation}
\langle 0|T_{i}^{k}|0\rangle =\langle 0|T_{i}^{k}|0\rangle _{\mathrm{R}%
}+\langle T_{i}^{k}\rangle _{a},  \label{emtrig}
\end{equation}%
where the first term on the right is the VEV for the Fulling-Rindler vacuum
without boundaries,%
\begin{equation}
\langle 0|T_{i}^{k}|0\rangle _{\mathrm{R}}=\frac{-2^{2-D}\delta _{i}^{k}}{%
\pi ^{(D+3)/2}\Gamma \left( \frac{D-1}{2}\right) }\int_{0}^{\infty
}dkk^{D-2}\lambda ^{2}\int_{0}^{\infty }d\omega e^{-\pi \omega }f^{(i)}\left[
K_{i\omega }(\lambda \xi )\right] ,  \label{R1TikR}
\end{equation}%
and the second term is the contribution brought by the presence of the
barrier:
\begin{equation}
\langle T_{i}^{k}\rangle _{a}=\frac{-2^{2-D}\delta _{i}^{k}}{\pi
^{(D+1)/2}\Gamma \left( \frac{D-1}{2}\right) }\int_{0}^{\infty
}dkk^{D-2}\lambda ^{2}\int_{0}^{\infty }d\omega \frac{\bar{I}_{\omega
}(\lambda a)}{\bar{K}_{\omega }(\lambda a)}F^{(i)}\left[ K_{\omega }(\lambda
\xi )\right] .  \label{emtbound}
\end{equation}%
In formula (\ref{R1TikR}) we have introduced the notations
\begin{eqnarray}
f^{(0)}[g(z)] &=&\left( \frac{1}{2}-2\zeta \right) \left\vert \frac{dg(z)}{dz%
}\right\vert ^{2}+\frac{\zeta }{z}\frac{d}{dz}\left\vert g(z)\right\vert
^{2}+\left[ \frac{1}{2}-2\zeta +\left( \frac{1}{2}+2\zeta \right) \frac{%
\omega ^{2}}{z^{2}}\right] \left\vert g(z)\right\vert ^{2},  \notag \\
f^{(1)}[g(z)] &=&-\frac{1}{2}\left\vert \frac{dg(z)}{dz}\right\vert ^{2}-%
\frac{\zeta }{z}\frac{d}{dz}\left\vert g(z)\right\vert ^{2}+\frac{1}{2}%
\left( 1-\frac{\omega ^{2}}{z^{2}}\right) \left\vert g(z)\right\vert ^{2},
\label{fq} \\
f^{(i)}[g(z)] &=&\left( \frac{1}{2}-2\zeta \right) \left[ \left\vert \frac{%
dg(z)}{dz}\right\vert ^{2}+\left( 1-\frac{\omega ^{2}}{z^{2}}\right)
\left\vert g(z)\right\vert ^{2}\right] -\frac{\lambda ^{2}-m^{2}}{%
(D-1)\lambda ^{2}}\left\vert g(z)\right\vert ^{2},  \notag
\end{eqnarray}%
with $i=2,3,...,D$, and the expressions for the functions $F^{(i)}[g(z)]$
are obtained by the replacement $\omega \rightarrow i\omega $:
\begin{equation}
F^{(i)}[g(z)]=f^{(i)}[g(z),\omega \rightarrow i\omega ].  \label{Ffunc}
\end{equation}%
The boundary-induced parts in the VEVs of the field square and the EMT for
the region $\xi <a$ are obtained from (\ref{phi2br}) and (\ref{emtbound})
with the replacements $I\rightleftarrows K$. It can be easily checked that
both summands on the right of Eq. (\ref{emtrig}) satisfy the continuity
equation $\nabla _{k}T_{i}^{k}=0$, which for the geometry under
consideration takes the form $\partial _{\xi }(\xi T_{1}^{1})-T_{0}^{0}=0$.

The purely Fulling-Rindler part (\ref{R1TikR}) of the EMT is investigated in
a large number of papers (see, for instance, references given in \cite%
{Avag02}). The most general case of a massive scalar field in an arbitrary
number of spacetime dimensions has been considered in Ref. \cite{Hill} for
conformally and minimally coupled cases and in Ref. \cite{SahaRind1} for
general values of the curvature coupling parameter. For a massless scalar
the VEV for the Rindler part without boundaries can be presented in the form
\begin{eqnarray}
\langle T_{i}^{k}\rangle _{\mathrm{R,sub}} &=&\langle 0|T_{i}^{k}|0\rangle _{%
\mathrm{R}}-\langle 0|T_{i}^{k}|0\rangle _{\mathrm{M}}  \notag \\
&=&-\frac{2\delta _{i}^{k}\xi ^{-D-1}}{(4\pi )^{D/2}\Gamma \left( D/2\right)
}\int_{0}^{\infty }d\omega \frac{\omega ^{D}g^{(i)}(\omega )}{e^{2\pi \omega
}+(-1)^{D}}\,,  \label{subRindm0}
\end{eqnarray}%
where the expressions for the functions $g^{(i)}(\omega )$ are presented in
Ref. \cite{SahaRind1}, and $\langle 0|T_{i}^{k}|0\rangle _{\mathrm{M}}$ is
the EMT for Minkowski vacuum without boundaries. Expression (\ref{subRindm0}%
) corresponds to the absence from the vacuum of thermal distribution with
standard temperature $T=(2\pi \xi )^{-1}$. In general, the corresponding
spectrum has non-Planckian form: the density of states factor is not
proportional to $\omega ^{D-1}d\omega $. The spectrum takes the Planckian
form for conformally coupled scalars in $D=1,2,3$ with $g^{(0)}(\omega
)=-Dg^{(i)}(\omega )=1$, $i=1,2,\ldots D$. It is of interest to note that
for even values of spatial dimension the distribution is Fermi-Dirac type
(see also \cite{Taga85}). For a massive scalar the energy spectrum is not
strictly thermal and the corresponding quantities do not coincide with ones
for the Minkowski thermal bath.

Boundary part (\ref{emtbound}) is finite for all values $\xi >a$ and
diverges at the plate surface $\xi =a$. To extract the leading part of the
boundary divergence, note that near the boundary the main contribution into
the $\omega $-integral comes from large values of $\omega $ and we can use
the uniform asymptotic expansions for the modified Bessel functions \cite%
{abramowiz}. Introducing a new integration variable $k\rightarrow \omega k$
and replacing the modified Bessel functions by their uniform asymptotic
expansions, it can be seen that the leading terms do not depend on the mass
and the Robin coefficients and are the same as for the plate in Minkowski
spacetime.

Now let us consider the asymptotic behavior of the boundary part (\ref%
{emtbound}) for large $\xi $, $\xi \gg a$. Introducing in Eq. (\ref{emtbound}%
) a new integration variable $y=\lambda \xi $ and using the asymptotic
formulae for the modified Bessel functions for small values of the argument,
we see that the integrand is proportional to $(ya/2\xi )^{2\omega }$. It
follows from here that the main contribution into the $\omega $-integral
comes from small values of $\omega $. Expanding with respect to $\omega $,
in the leading order we obtain
\begin{equation}
\langle T_{i}^{k}\rangle _{a}\sim -\frac{\delta _{i}^{k}(-1)^{\delta
_{A0}}\xi ^{-D-1}A_{0}^{(i)}(m\xi )}{2^{D}3\pi ^{(D-1)/2}(1+\delta
_{A0})\Gamma \left( \frac{D-1}{2}\right) \ln ^{2}(2\xi /a)},
\label{asimpfar}
\end{equation}%
where
\begin{equation}
A_{0}^{(i)}(x)=\int_{x}^{\infty
}dy\,y^{3}(y^{2}-x^{2})^{(D-3)/2}F^{(i)}[K_{\omega }(y)]|_{\omega =0}.
\label{Ai0}
\end{equation}%
If, in addition, one has $m\xi \gg 1$, the integral in this formula can be
evaluated replacing the McDonald function by its asymptotic for large values
of the argument. In the leading order this yields
\begin{equation}
A_{0}^{(0)}\approx A_{0}^{(i)}\approx -2m\xi A_{0}^{(1)}\approx \pi
(1/4-\zeta )\Gamma \left( \frac{D-1}{2}\right) (m\xi )^{(D+1)/2}e^{-2m\xi },
\label{Aiklargem}
\end{equation}%
with $i=2,3,\cdots ,D$. For massless case the integral in Eq. (\ref{Ai0})
may be evaluated and one obtains
\begin{equation}
A_{0}^{(0)}\approx -DA_{1}^{(1)}\approx \frac{D}{D-1}A_{2}^{(2)}\approx
\frac{2^{D}D(\zeta _{c}-\zeta )}{(D-1)^{2}\Gamma (D)}\Gamma ^{4}\left( \frac{%
D+1}{2}\right) ,\quad m=0.  \label{A00}
\end{equation}%
For a conformally coupled scalar the leading terms vanish and the VEVs are
proportional to $\xi ^{-D-1}\ln ^{-3}(2\xi /a)$ for $\xi /a\gg 1$. From Eq. (%
\ref{asimpfar}) we see that for a given $\xi $ the boundary part tends to
zero as $a\rightarrow 0$ (the barrier coincides with the right Rindler
horizon) and the corresponding VEVs of the EMT are the same as for the
Fulling-Rindler vacuum without boundaries. Hence, the barrier located at the
Rindler horizon does not alter the vacuum EMT.

And finally, we turn to the asymptotic $a,\xi \rightarrow \infty ,\,\,\xi -a=%
\mathrm{const}$. In this limit $\xi /a\rightarrow 1$ and $\omega $-integrals
in Eq. (\ref{emtbound}) are dominated by large $\omega $. Replacing the
modified Bessel functions by their uniform asymptotic expansions, keeping
the leading terms only and introducing a new integration variable $\nu
=\omega /a$, we can see that the VEVs coincide with the corresponding VEVs
induced by a single plate in the Minkowski spacetime with Robin boundary
condition on it.

\subsection{Electromagnetic field}

\label{subsec:FuRiEl1pl}

We now turn to the case of the electromagnetic field in the region $\xi >a$
for the case $D=3$. We will assume that the mirror is a perfect conductor
with the standard boundary conditions of vanishing of the normal component
of the magnetic field and the tangential components of the electric field,
evaluated at the local inertial frame in which the conductor is
instantaneously at rest. As it has been shown in \cite{CandD}, the
corresponding eigenfunctions for the vector potential $A_{l}(x)$ may be
resolved into the transverse electric (TE) and transverse magnetic (TM)
(with respect to $\xi $-direction) modes $A_{\alpha \sigma l}$, $\sigma
=(\omega ,\mathbf{k})$:
\begin{eqnarray}
A_{0\sigma l}(x) &=&(0,0,-ik_{3},ik_{2})\varphi _{0\sigma }(x),\;\alpha =0,
\label{apot0} \\
A_{1\sigma l}(x) &=&\left( -\xi \partial _{\xi },i\omega /\xi ,0,0\right)
\varphi _{1\sigma }(x),\;\alpha =1,  \label{apot1}
\end{eqnarray}%
where the eigenfunctions $\varphi _{\sigma \alpha }(x)$ are given by formula
(\ref{sol2}) with $m=0$, $\mathbf{k}=(k_{2},k_{3})$, and $\alpha =0,1$
correspond to the TE and TM waves, respectively. From the perfect conductor
boundary conditions on the vector potential we obtain the corresponding
boundary conditions for the scalar modes $\varphi _{\sigma \alpha }(x)$:%
\begin{equation}
\varphi _{0\sigma }=0,\;\partial _{\xi }\varphi _{1\sigma }=0,\;\xi =a.
\label{FuRiElbc1pl}
\end{equation}%
As a result the TE/TM modes correspond to Dirichlet/Neumann scalars. In the
corresponding expressions for the eigenfunctions $A_{\alpha \sigma l}(x)$
the normalization coefficient is determined from the orthonormality
condition
\begin{equation}
\int d\mathbf{x}\int_{a}^{\infty }\frac{d\xi }{\xi }A_{\alpha \sigma
}^{l}A_{\alpha ^{\prime }\sigma ^{\prime }l}^{\ast }=-\frac{2\pi }{\omega }%
\delta _{\alpha \alpha ^{\prime }}\delta _{\sigma \sigma ^{\prime }}.
\label{FuRielecnorm}
\end{equation}%
The eigenvalues for $\omega $ are the zeros of the function $K_{i\omega
}(ka) $, $k=|\mathbf{k}|$, for the TE\ modes and the zeros of the function $%
K_{i\omega }^{\prime }(ka)$ for the TM modes.

Substituting the eigenfunctions for $A_{\alpha \sigma l}(x)$ into mode-sum
formula (\ref{emtgform}) with the standard bilinear form for the
electromagnetic field EMT and applying to the sums over $\omega _{n}$
formula (\ref{sumformKi}), we find
\begin{equation}
\langle 0|T_{i}^{k}|0\rangle =\langle 0|T_{i}^{k}|0\rangle _{\mathrm{R}}-%
\frac{\delta _{i}^{k}}{4\pi ^{2}}\int_{0}^{\infty }dk\,k^{3}\int_{0}^{\infty
}d\omega \left[ \frac{I_{\omega }(ka)}{K_{\omega }(ka)}+\frac{I_{\omega
}^{\prime }(ka)}{K_{\omega }^{\prime }(ka)}\right] F_{\mathrm{em}}^{(i)}%
\left[ K_{\omega }(k\xi )\right] ,  \label{FuRiEl1pl}
\end{equation}%
where%
\begin{equation}
\langle 0|T_{i}^{k}|0\rangle _{\mathrm{R}}=\langle 0|T_{i}^{k}|0\rangle _{%
\mathrm{M}}-\frac{1}{\pi ^{2}\xi ^{4}}\int_{0}^{\infty }d\omega \frac{\omega
^{3}+\omega }{e^{2\pi \omega }-1}\,\mathrm{diag}(1,-\frac{1}{3},-\frac{1}{3}%
,-\frac{1}{3})  \label{FuRiEl0pl}
\end{equation}%
is the VEV for the Fulling-Rindler vacuum without boundaries \cite{CandD}.
In formula (\ref{FuRiEl1pl}) the notations%
\begin{eqnarray}
F_{\mathrm{em}}^{(i)}\left[ g(z)\right] &=&(-1)^{i}g^{\prime
2}(z)+(1-(-1)^{i}\omega ^{2}/z^{2})g^{2}(z),\;i=0,1,  \label{FuRiElF0} \\
F_{\mathrm{em}}^{(i)}\left[ g(z)\right] &=&-g^{2}(z),\;i=2,3,
\label{FuRiElF2}
\end{eqnarray}%
are introduced. Formula (\ref{FuRiEl1pl}) is derived in \cite{CandD} by
using the Green function method. Note that the $\omega $-integral in (\ref%
{FuRiEl0pl}) is equal to 11/240. The VEV for the EMT of the electromagnetic
field in the region $\xi <a$ is obtained from (\ref{FuRiEl1pl}) by the
replacements $I_{\omega }\rightleftarrows K_{\omega }$ \cite{SahaRind1}. By
using the properties of the modified Bessel functions it can be seen that%
\begin{equation}
\frac{I_{\omega }(ka)}{K_{\omega }(ka)}+\frac{I_{\omega }^{\prime }(ka)}{%
K_{\omega }^{\prime }(ka)}>0,\;\frac{K_{\omega }(ka)}{I_{\omega }(ka)}+\frac{%
K_{\omega }^{\prime }(ka)}{I_{\omega }^{\prime }(ka)}<0,  \label{ineqBessMod}
\end{equation}%
and $F_{\mathrm{em}}^{(0)}\left[ K_{\omega }(z)\right] >0$, $F_{\mathrm{em}%
}^{(0)}\left[ I_{\omega }(z)\right] >0$, $F_{\mathrm{em}}^{(1)}\left[
K_{\omega }(z)\right] <0$, $F_{\mathrm{em}}^{(1)}\left[ I_{\omega }(z)\right]
>0$. In particular, from these inequalities it follows that the energy
density induced by the plate is positive in the region $\xi <a$ and negative
in the region $\xi >a$. The corresponding graphs are given in \cite%
{SahaRind1,CandD}. The behavior of the plate-induced VEVs in various
asymptotic regions of the parameters can be found in \cite{SahaRind1,CandD}.
In Ref. \cite{SahSet04} the formulae derived in this section were used to
generate the vacuum densities for a conformally coupled massless scalar
field in de Sitter spacetime in presence of a curved brane on which the
field obeys the Robin boundary conditions with coordinate dependent
coefficients.

\section{Vacuum densities in the region between two plates uniformly
accelerated through the Fulling-Rindler vacuum}

\label{sec:FulRin2pl}

\subsection{Wightman function}

Consider two parallel plates moving with uniform proper accelerations
assuming that the quantum scalar field is prepared in the Fulling-Rindler
vacuum \cite{Avag02,SahaRind2}. We will let the surfaces $\xi =a$ and $\xi
=b $, $b>a>0$, represent the trajectories of the plates, which therefore
have proper accelerations $a^{-1}$ and $b^{-1}$. We consider the case of a
scalar field satisfying Robin boundary conditions on the surfaces of the
plates:
\begin{equation}
\left. \left( \tilde{A}_{j}+\tilde{B}_{j}\partial _{\xi }\right) \varphi
\right\vert _{\xi =j}=0,\quad j=a,b,  \label{Dboundcond}
\end{equation}%
with constant coefficients $\tilde{A}_{j}$ and $\tilde{B}_{j}$. The plates
divide the right Rindler wedge into three regions: $0<\xi <a$, $\xi >b$, and
$a<\xi <b$. The VEVs in two first regions are the same as those induced by
single plates located at $\xi =a$ and $\xi =b$, respectively. In the region
between the plates the eigenfunctions satisfying boundary condition on the
plate $\xi =b$ have the form%
\begin{equation}
\varphi _{\sigma }(x)=\beta _{\sigma }G_{i\omega }^{(b)}(\lambda b,\lambda
\xi )e^{i\mathbf{kx}-i\omega \tau },  \label{R2eigfunc}
\end{equation}%
with the function $G_{\nu }^{(j)}(x,y)$ defined by formula (\ref{Geab}).
Note that the function $G_{i\omega }^{(b)}(\lambda b,\lambda \xi )$ is real,
$G_{-i\omega }^{(b)}(\lambda b,\lambda \xi )=G_{i\omega }^{(b)}(\lambda
b,\lambda \xi )$. From the boundary condition on the plate $\xi =a$ we find
that the possible values for $\omega $ are roots to the equation
\begin{equation}
Z_{i\omega }(\lambda a,\lambda b)=0,  \label{Deigfreq}
\end{equation}%
where the function $Z_{i\omega }(u,v)$ is defined by formula (\ref{Zi}), and
the barred notations are defined by formula (\ref{barjnot}) with $A_{j}=%
\tilde{A}_{j}$, $B_{j}=\tilde{B}_{j}/j$. For a fixed $\lambda $, the
equation (\ref{Deigfreq}) has an infinite set of real solutions with respect
to $\omega $. We will denote them by $\Omega _{n}=\Omega _{n}(\lambda
a,\lambda b)$, $\Omega _{n}>0$, $n=1,2,\ldots $, and will assume that they
are arranged in ascending order $\Omega _{n}<\Omega _{n+1}$. In the
consideration below we will assume the values of the coefficients in Eq. (%
\ref{Dboundcond}) for which the imaginary solutions are absent and the
vacuum is stable.

The coefficient $\beta _{\sigma }$ in formula (\ref{R2eigfunc}) is
determined from the orthonormality condition. Taking into account the
boundary conditions, for this coefficient one finds
\begin{equation}
\beta _{\sigma }^{2}=\left. \frac{\left( 2\pi \right) ^{1-D}\bar{I}_{i\omega
}^{(a)}(\lambda a)}{\bar{I}_{i\omega }^{(b)}(\lambda b)\partial _{\omega
}Z_{i\omega }(\lambda a,\lambda b)}\right\vert _{\omega =\Omega _{n}}.
\label{Dnormc}
\end{equation}%
Now, substituting the eigenfunctions into mode-sum formula (\ref{mmodesumWF}%
), for the positive frequency Wightman function one finds
\begin{eqnarray}
W(x,x^{\prime }) &=&\int \frac{d\mathbf{k\,}e^{i\mathbf{k}\Delta \mathbf{x}}%
}{(2\pi )^{D-1}}\sum_{n=1}^{\infty }\frac{\bar{I}_{i\omega }^{(a)}(\lambda
a)e^{-i\omega \Delta \tau }}{\bar{I}_{i\omega }^{(b)}(\lambda b)\partial
_{\omega }Z_{i\omega }(\lambda a,\lambda b)}  \notag \\
&&\times \left. G_{i\omega }^{(b)}(\lambda b,\lambda \xi )G_{i\omega
}^{(b)}(\lambda b,\lambda \xi ^{\prime })\right\vert _{\omega =\Omega _{n}}.
\label{Wigh1}
\end{eqnarray}%
For the further evaluation of this VEV we apply to the sum over
$n$ summation formula (\ref{Dsumformula}). As a function $F(z)$ in
this formula we choose
\begin{equation}
F(z)=\frac{G_{i\omega }^{(b)}(\lambda b,\lambda \xi )G_{i\omega
}^{(b)}(\lambda b,\lambda \xi ^{\prime })}{\bar{I}_{iz}^{(b)}(\lambda b)\bar{%
I}_{-iz}^{(b)}(\lambda b)}e^{-iz\Delta \tau }.  \label{FtoAPF}
\end{equation}%
Condition (\ref{condforAPF2pl}) for this function is satisfied if $%
a^{2}e^{|\Delta \tau |}<\xi \xi ^{\prime }$. By using formula (\ref%
{Dsumformula}), for the Wightman function one obtains the expression \cite%
{SahaRind2}%
\begin{eqnarray}
W(x,x^{\prime }) &=&W^{(b)}(x,x^{\prime })-\int \frac{d\mathbf{k\,}e^{i%
\mathbf{k}\Delta \mathbf{x}}}{\pi (2\pi )^{D-1}}\int_{0}^{\infty }d\omega
\,\Omega _{b\omega }(\lambda a,\lambda b)  \notag \\
&&\times G_{\omega }^{(b)}(\lambda b,\lambda \xi )G_{\omega }^{(b)}(\lambda
b,\lambda \xi ^{\prime })\cosh (\omega \Delta \tau ),  \label{Wigh3}
\end{eqnarray}%
where we have used the notation (\ref{Omegatilde}). In Eq. (\ref{Wigh3}),
\begin{eqnarray}
W^{(b)}(x,x^{\prime }) &=&\int \frac{d\mathbf{k\,}e^{i\mathbf{k}\Delta
\mathbf{x}}}{\pi ^{2}(2\pi )^{D-1}}\int_{0}^{\infty }d\omega \sinh (\pi
\omega )  \notag \\
&&\times e^{-i\omega \Delta \tau }\frac{G_{i\omega }^{(b)}(\lambda b,\lambda
\xi )G_{i\omega }^{(b)}(\lambda b,\lambda \xi ^{\prime })}{\bar{I}_{i\omega
}^{(b)}(\lambda b)\bar{I}_{-i\omega }^{(b)}(\lambda b)},  \label{Wigh1pl}
\end{eqnarray}%
is the Wightman function in the region $\xi <b$ for a single plate at $\xi
=b $. This function is investigated in Ref. \cite{SahaRind1} and can be
presented in the form
\begin{equation}
W^{(b)}(x,x^{\prime })=W_{\mathrm{R}}(x,x^{\prime })+\left\langle \varphi
(x)\varphi (x^{\prime })\right\rangle _{b},  \label{G+2}
\end{equation}%
where $W_{\mathrm{R}}(x,x^{\prime })$ is the Wightman function for the right
Rindler wedge without boundaries and the part%
\begin{eqnarray}
\left\langle \varphi (x)\varphi (x^{\prime })\right\rangle _{b} &=&-\int
\frac{d\mathbf{k\,}e^{i\mathbf{k}\Delta \mathbf{x}}}{\pi (2\pi )^{D-1}}%
\int_{0}^{\infty }d\omega \frac{\bar{K}_{\omega }^{(b)}(\lambda b)}{\bar{I}%
_{\omega }^{(b)}(\lambda b)}  \notag \\
&&\times I_{\omega }(\lambda \xi )I_{\omega }(\lambda \xi ^{\prime })\cosh
(\omega \Delta \tau )  \label{phi212}
\end{eqnarray}%
is induced in the region $\xi <b$ by the presence of the plate at $\xi =b$.
Note that the representation (\ref{G+2}) with (\ref{phi212}) is valid under
the assumption $\xi \xi ^{\prime }<b^{2}e^{|\Delta \tau |}$.

By using the identity%
\begin{eqnarray}
&&\frac{\bar{K}_{\omega }^{(b)}(\lambda b)}{\bar{I}_{\omega }^{(b)}(\lambda
b)}I_{\omega }(\lambda \xi )I_{\omega }(\lambda \xi ^{\prime })=\frac{\bar{I}%
_{\omega }^{(a)}(\lambda a)}{\bar{K}_{\omega }^{(a)}(\lambda a)}K_{\omega
}(\lambda \xi )K_{\omega }(\lambda \xi ^{\prime })  \notag \\
&&+\sum_{j=a,b}n^{(j)}\Omega _{j\omega }(\lambda a,\lambda b)G_{\omega
}^{(j)}(\lambda j,\lambda \xi )Z_{\omega }^{(j)}(\lambda j,\lambda \xi
^{\prime }),  \label{ident1Rind}
\end{eqnarray}%
with $n^{(a)}=1$, $n^{(b)}=-1$, the Wightman function can also be presented
in the form%
\begin{eqnarray}
W(x,x^{\prime }) &=&W^{(a)}(x,x^{\prime })-\int \frac{d\mathbf{k\,}e^{i%
\mathbf{k}\Delta \mathbf{x}}}{\pi (2\pi )^{D-1}}\int_{0}^{\infty }d\omega
\,\Omega _{a\omega }(\lambda a,\lambda b)  \notag \\
&&\times G_{\omega }^{(a)}(\lambda a,\lambda \xi )G_{\omega }^{(a)}(\lambda
a,\lambda \xi ^{\prime })\cosh (\omega \Delta \tau ).  \label{Wigh31}
\end{eqnarray}%
In this formula $W^{(a)}(x,x^{\prime })$ is the Wightman function in the
region $\xi >a$ for a single plate at $\xi =a$, and is investigated in the
previous section. In the coincidence limit the second term on the right of
formula (\ref{Wigh3}) is finite on the plate $\xi =b$ and diverges on the
plate at $\xi =a$, whereas the second term on the right of Eq. (\ref{Wigh31}%
) is finite on the plate $\xi =a$ and is divergent for $\xi =b$.
Consequently, the form (\ref{Wigh3}) ((\ref{Wigh31})) is convenient for the
investigations of the VEVs near the plate $\xi =b$ ($\xi =a$).

\subsection{Scalar Casimir densities}

\label{sec:VEVEMT}

Here we will consider the VEVs of the field square and the EMT in the region
between the plates \cite{SahaRind2}. In the coincidence limit from the
formulae for the Wightman function one obtains two equivalent forms for the
VEV\ of the field square,%
\begin{eqnarray}
\left\langle 0\left\vert \varphi ^{2}\right\vert 0\right\rangle
&=&\left\langle 0\left\vert \varphi ^{2}\right\vert 0\right\rangle _{\mathrm{%
R}}+\left\langle \varphi ^{2}\right\rangle _{j}  \notag \\
&&-B_{D}\int_{0}^{\infty }dk\,k^{D-2}\int_{0}^{\infty }d\omega \,\Omega
_{j\omega }(\lambda a,\lambda b)G_{\omega }^{(j)2}(\lambda j,\lambda \xi ),
\label{phi2sq1}
\end{eqnarray}%
corresponding to $j=a$ and $j=b$, and $B_{D}$ is defined by (\ref{TCAD}). In
Eq. (\ref{phi2sq1}) the part $\left\langle \varphi ^{2}\right\rangle _{j}$
is induced by a single plate at $\xi =j$ when the second plate is absent.
The last term on the right of formula (\ref{phi2sq1}) is finite on the plate
at $\xi =j$ and diverges for points on the other plate. Extracting the
contribution from this plate, we can write the expression (\ref{phi2sq1})
for the VEV in the symmetric form
\begin{equation}
\left\langle 0\left\vert \varphi ^{2}\right\vert 0\right\rangle
=\left\langle 0\left\vert \varphi ^{2}\right\vert 0\right\rangle _{\mathrm{R}%
}+\sum_{j=a,b}\left\langle \varphi ^{2}\right\rangle _{j}+\Delta
\left\langle \varphi ^{2}\right\rangle ,  \label{phi2sq2n}
\end{equation}%
with the interference part%
\begin{eqnarray}
\Delta \left\langle \varphi ^{2}\right\rangle &=&-B_{D}\int_{0}^{\infty
}dk\,k^{D-2}\int_{0}^{\infty }d\omega \bar{I}_{\omega }^{(a)}(\lambda a)
\notag \\
&&\times \left[ \frac{G_{\omega }^{(b)2}(\lambda b,\lambda \xi )}{\bar{I}%
_{\omega }^{(b)}(\lambda b)G_{\omega }^{ab}(\lambda a,\lambda b)}-\frac{%
K_{\omega }^{2}(\lambda \xi )}{\bar{K}_{\omega }^{(a)}(\lambda a)}\right] .
\label{phi2int}
\end{eqnarray}%
An equivalent form for this part is obtained with the replacements $%
a\rightleftarrows b$ and $I\rightleftarrows K$ in the integrand.
Interference term (\ref{phi2int}) is finite for all values of $\xi $ in the
range $a\leqslant \xi \leqslant b$, including the points on the boundaries.
The surface divergences are contained in the single plate parts only. In the
limit $a\rightarrow b$ with fixed values of the coefficients in the boundary
conditions, the interference part (\ref{phi2int}) is divergent and for small
values of $b/a-1$ the main contribution comes from large values of $\omega $%
. Introducing an integration variable $x=k/\omega $ and replacing the
modified Bessel functions by their uniform asymptotic expansions, we can see
that to the leading order $\Delta \langle \varphi ^{2}\rangle $ coincides
with the VEV\ of the field square in the region between two parallel plates
in the Minkowski bulk.

Large values of the proper accelerations for the plates correspond to the
limit $a,b\rightarrow 0$. In this limit the plates are close to the Rindler
horizon. From formulae (\ref{phi2br}), (\ref{phi2int}) we see that for fixed
values of the ratios $a/b$, $\xi /b$, both single plate and interference
parts behave as $b^{1-D}$ in the limit $b\rightarrow 0$. In the limit $%
a\rightarrow 0$ for fixed values $\xi $ and $b$, the left plate tends to the
Rindler horizon for a fixed world-line of the right plate. The main
contribution into the $\omega $-integral in Eq. (\ref{phi2int}) comes from
small values $\omega $, $\omega \lesssim 1/\ln (2/\lambda a)$. Using the
formulae for the modified Bessel functions for small arguments, it can be
seen that interference part (\ref{phi2int}) vanishes as $\ln ^{-2}(2b/a)$.

In the limit of small accelerations of the plates, $a,b\rightarrow \infty $,
with fixed values $b-a$, $B_{j}/A_{j}$, and $m$, the main contribution comes
from large values $\omega $. Using the uniform asymptotic formulae for the
modified Bessel functions, it can be seen that $\left\langle \varphi
^{2}\right\rangle _{j}$ and $\Delta \langle \varphi ^{2}\rangle $ coincide
with the corresponding expressions for the geometry of two parallel plates
on the Minkowski bulk. In this limit, $\xi $ corresponds to the Cartesian
coordinate perpendicular to the plates which are located at $\xi =a$ and $%
\xi =b$.

For large values of the mass, $ma\gg 1$, we introduce a new integration
variable $y=\lambda /m$. The main contribution into the $\omega $-integral
comes from the values $\omega \sim \sqrt{ma}$. By using the uniform
asymptotic expansions for the modified Bessel functions for large values of
the order and further expanding over $\omega /ma$, for the single plate
parts to the leading order one finds%
\begin{equation}
\left\langle \varphi ^{2}\right\rangle _{j}\approx -\frac{m^{\frac{D}{2}%
-1}e^{-2m|\xi -j|}\sqrt{j/\xi }}{2(4\pi )^{\frac{D}{2}}c_{j}(m)|\xi -j|^{%
\frac{D}{2}}},\;j=a,b,  \label{phi2largemass}
\end{equation}%
where we have introduced notations%
\begin{equation}
c_{j}(y)=\frac{A_{j}-n^{(j)}\tilde{B}_{j}y}{A_{j}+n^{(j)}\tilde{B}_{j}y}.
\label{cjy}
\end{equation}%
In the similar way, for the interference part we obtain the formula%
\begin{equation}
\Delta \langle \varphi ^{2}\rangle \approx \frac{m^{\frac{D}{2}-1}e^{2m(a-b)}%
\sqrt{ab}}{(4\pi )^{\frac{D}{2}}\xi c_{a}(m)c_{b}(m)(b-a)^{\frac{D}{2}}}.
\label{phi2largemassab}
\end{equation}%
As we could expect, the both single plate and interference parts are
exponentially suppressed for large values of the mass.

Making use of the formulae for the Wightman function and the field square,
one obtains two equivalent forms for the VEV of the EMT, corresponding to $%
j=a$ and $j=b$ (no summation over $i$) \cite{SahaRind2} (see also \cite%
{Avag02} for the case of Dirichlet and Neumann scalars):
\begin{eqnarray}
\langle 0|T_{i}^{k}|0\rangle &=&\langle 0|T_{i}^{k}|0\rangle _{\mathrm{R}%
}+\langle T_{i}^{k}\rangle _{j}-B_{D}\delta _{i}^{k}\int dk\,k^{D-2}  \notag
\\
&&\times \lambda ^{2}\int_{0}^{\infty }d\omega \,\Omega _{j\omega }(\lambda
a,\lambda b)F^{(i)}\left[ G_{\omega }^{(j)}(\lambda j,\lambda \xi )\right] .
\label{Tik1}
\end{eqnarray}%
In this formula, $\langle 0|T_{i}^{k}|0\rangle _{\mathrm{R}}$ is the
corresponding VEV for the Fulling-Rindler vacuum without boundaries, and the
term $\langle T_{i}^{k}\rangle _{j}$ is induced by the presence of a single
plane boundary located at $\xi =j$ in the region $\xi >a$ for the case $j=a$
and in the region $\xi <b$ for $j=b$. In formulae (\ref{Tik1}), for a given
function $g(z)$ we use the notations $F^{(i)}[g(z)]$ from (\ref{Ffunc}). For
the last term on the right of Eq. (\ref{Tik1}) we have to substitute $%
g(z)=G_{\omega }^{(j)}(\lambda j,z)$. It can be easily seen that for a
conformally coupled massless scalar the EMT is traceless.

Now let us present the VEV (\ref{Tik1}) in the form%
\begin{equation}
\langle 0|T_{i}^{k}|0\rangle =\langle 0|T_{i}^{k}|0\rangle _{\mathrm{R}%
}+\sum_{j=a,b}\langle T_{i}^{k}\rangle _{j}+\Delta \langle T_{i}^{k}\rangle ,
\label{Tikdecomp}
\end{equation}%
where (no summation over $i$)
\begin{eqnarray}
\Delta \langle T_{i}^{k}\rangle &=&-A_{D}\delta _{i}^{k}\int_{0}^{\infty
}dk\,k^{D-2}\lambda ^{2}\int_{0}^{\infty }d\omega \bar{I}_{\omega
}^{(a)}(\lambda a)  \notag \\
&&\times \left[ \frac{F^{(i)}[G_{\omega }^{(b)}(\lambda b,\lambda \xi )]}{%
\bar{I}_{\omega }^{(b)}(\lambda b)G_{\omega }^{ab}(\lambda a,\lambda b)}-%
\frac{F^{(i)}[K_{\omega }(\lambda \xi )]}{\bar{K}_{\omega }^{(a)}(\lambda a)}%
\right]  \label{intterm1}
\end{eqnarray}%
is the interference term. The surface divergences are contained in the
single boundary parts and this term is finite for all values $a\leqslant \xi
\leqslant b$. An equivalent formula for $\Delta \langle T_{i}^{k}\rangle $
is obtained from Eq. (\ref{intterm1}) by replacements $a\rightleftarrows b$,
$I\rightleftarrows K$. Both single plate and interference parts separately
satisfy the continuity equation. For a conformally coupled massless scalar
field they are traceless and we have an additional relation $\left\langle
T_{i}^{i}\right\rangle =0$.

In the limit $a\rightarrow b$ expression (\ref{intterm1}) is divergent and
for small values of $b/a-1$ the main contribution comes from large values of
$\omega $. Introducing a new integration variable $x=k/\omega $ and
replacing the modified Bessel functions by their uniform asymptotic
expansions for large values of the order, at the leading order one receives%
\begin{eqnarray}
\Delta \langle T_{i}^{i}\rangle &\approx &-\frac{(4\pi )^{-\frac{D}{2}}}{%
\Gamma \left( \frac{D}{2}+1\right) }\int_{0}^{\infty }dy\frac{y^{D}}{%
k_{a}k_{b}e^{2y(b-a)}-1}  \notag \\
&&\times \left[ 1+2D\left( 1-\delta _{1}^{i}\right) (\zeta -\zeta
_{D})\sum_{j=a,b}k_{j}e^{-2y|\xi -j|}\right] ,  \label{Tiiclose}
\end{eqnarray}%
where $k_{j}=1-2\delta _{0B_{j}}$. In the limit of large proper
accelerations for the plates, $a,b\rightarrow 0$, for fixed values $a/b$ and
$\xi /b$, the world-lines of both plates are close to the Rindler horizon.
In this case the single plate and interference parts grow as $b^{-D-1}$. The
situation is essentially different when the world-line of the left plate
tends to the Rindler horizon, $a\rightarrow 0$, whereas $b$ and $\xi $ are
fixed. In the way similar to that for the case of the field square, it can
be seen that in this limit the interference part (\ref{intterm1}) vanishes
as $\ln ^{-2}(2b/a)$.

In the limit of small proper accelerations, $a,b\rightarrow \infty $ with
fixed values $b-a$, $B_{j}/A_{j}$, and $m$, the main contribution comes from
large values of $\omega $. Using the asymptotic formulae for the modified
Bessel functions, to the leading order one obtains the corresponding
expressions for the geometry of two parallel plates on the Minkowski
background investigated in \cite{Rome02}. In particular, in this limit the
single boundary terms vanish for a conformally coupled massless scalar.

For large values of the mass, $ma\gg 1$, by the method similar to that used
in the previous subsection for the field square, it can be seen that the
both single plate and interference parts are exponentially suppressed (no
summation over $i$): $\langle T_{i}^{i}\rangle _{j}\sim m^{D/2+1}\exp
[-2m|\xi -j|]$, $j=a,b$, for single plate parts and $\Delta \langle
T_{i}^{i}\rangle \sim m^{D/2+1}\exp [2m(a-b)]$ for the interference part.

Now we turn to the interaction forces between the plates due to the vacuum
fluctuations. The vacuum force acting per unit surface of the plate at $\xi
=j$ is determined by the ${}_{1}^{1}$--component of the vacuum EMT evaluated
at this point. The corresponding effective pressures can be presented as a
sum of self-action and interactions terms, $p^{(j)}=p_{1}^{(j)}+p_{\mathrm{%
(int)}}^{(j)}$, $j=a,b$. The first term on the right is the pressure for a
single plate at $\xi =j$ when the second plate is absent. The first term on
the right is the pressure for a single plate at $\xi =j$ when the second
plate is absent. This term is divergent due to the surface divergences in
the subtracted VEVs and needs additional renormalization. This can be done,
for example, by applying the generalized zeta function technique to the
corresponding mode-sum. This procedure is similar to that used in \cite%
{Saha04FuRi} for the evaluation of the total Casimir energy in the cases of
Dirichlet and Neumann boundary conditions and in \cite{SahSet04b} for the
evaluation of the surface energy for a single Robin plate. The term $p_{%
\mathrm{(int)}}^{(j)}$ is the pressure induced by the presence of the second
plate. This term is finite for all nonzero distances between the plates and
is not affected by the renormalization procedure. For the plate at $\xi =j$
the interaction term is due to the third summand on the right of Eq. (\ref%
{Tik1}). Substituting into this term $\xi =j$ and using the Wronskian for
the modified Bessel functions one finds \cite{SahaRind2}
\begin{equation}
p_{\mathrm{(int)}}^{(j)}=\frac{A_{D}A_{j}^{2}}{2j^{2}}\int_{0}^{\infty
}dk\,k^{D-2}\int_{0}^{\infty }d\omega \left[ \left( \lambda ^{2}j^{2}+\omega
^{2}\right) \beta _{j}^{2}+4\zeta \beta _{j}-1\right] \,\Omega _{j\omega
}(\lambda a,\lambda b),  \label{pint2}
\end{equation}%
with $\beta _{j}=B_{j}/(jA_{j})$. In dependence of the values for the
coefficients in the boundary conditions, the effective pressures (\ref{pint2}%
) can be either positive or negative, leading to repulsive or attractive
forces. It can be seen that for Dirichlet boundary condition on one plate
and Neumann boundary condition on the other one has $p_{\mathrm{(int)}%
}^{(j)}>0$ and the interaction forces are repulsive for all distances
between the plates. Note that for Dirichlet or Neumann boundary conditions
on both plates the interaction forces are always attractive \cite{Avag02}.
The interaction forces can also be written in another equivalent form
\begin{eqnarray}
p_{\mathrm{(int)}}^{(j)} &=&n^{(j)}\frac{A_{D}}{2j}\int_{0}^{\infty
}dk\,k^{D-2}\int_{0}^{\infty }d\omega \left[ 1+\frac{\left( 4\zeta -1\right)
\beta _{j}}{\left( \lambda ^{2}j^{2}+\omega ^{2}\right) \beta _{j}^{2}+\beta
_{j}-1}\right]  \notag \\
&&\times \frac{\partial }{\partial j}\ln \left\vert 1-\frac{\bar{I}_{\omega
}^{(a)}(\lambda a)\bar{K}_{\omega }^{(b)}(\lambda b)}{\bar{I}_{\omega
}^{(b)}(\lambda b)\bar{K}_{\omega }^{(a)}(\lambda a)}\right\vert .
\label{pint3}
\end{eqnarray}%
For Dirichlet and Neumann scalars the second term in the square brackets is
zero.

The results obtained above can be applied to the geometry of two parallel
plates near the $D=3$ 'Rindler wall.'\ This wall is described by the static
plane-symmetric distribution of the matter with the diagonal EMT $T_{i}^{k}=%
\mathrm{diag}(\varepsilon _{m},-p_{m},-p_{m},-p_{m})$ (see Ref. \cite{Avak01}%
). Below we will denote by $x$ the coordinate perpendicular to the wall and
will assume that the plane $x=0$ is at the center of the wall. If the plane $%
x=x_{s}$ is the boundary of the wall, when the external ($x>x_{s}$) line
element with the time coordinate $t$ can be transformed into form (\ref%
{metricRin}) with
\begin{equation}
\xi (x)=x-x_{s}+\frac{1}{2\pi \sigma _{s}},\quad \tau =2\pi \sigma _{s}\sqrt{%
g_{00}(x_{s})}t.  \label{ksiRw}
\end{equation}%
In this formula the parameter $\sigma _{s}$ is the mass per unit surface of
the wall and is determined by the distribution of the matter:%
\begin{equation}
\sigma _{s}=2\int_{0}^{x_{s}}\left( \varepsilon _{m}+3p_{m}\right) \left[
g(x)/g(x_{s})\right] ^{1/2}dx.  \label{sigmas1}
\end{equation}%
For the 'Rindler wall'\ one has $g_{22}^{\prime }(x)|_{x=0}<0$ \cite{Avak01}
(the external solution for the case $g_{22}^{\prime }(x)|_{x=0}>0$ is
described by the standard Taub metric). Hence, the Wightman function, the
VEVs for the field square and the EMT in the region between two plates
located at $x=x_{a}$ and $x=x_{b}$, $x_{j}>x_{s}$ near the 'Rindler wall'\
are obtained from the results given above substituting $j=\xi (x_{j})$, $%
j=a,b$ and $\xi =\xi (x)$. For $\sigma _{s}>0$, $x\geqslant x_{s}$ one has $%
\xi (x)\geqslant \xi (x_{s})>0$ and the Rindler metric is regular everywhere
in the external region.

\subsection{Electromagnetic field}

As in the case of a single plate considered in subsection \ref%
{subsec:FuRiEl1pl}, in the region between two uniformly accelerated
perfectly conducting plates the eigenfunctions for the electromagnetic
vector potential are resolved int TE and TM modes in accordance with (\ref%
{apot0}), (\ref{apot1}), with Dirichlet and Neumann boundary conditions:
\begin{equation}
\varphi _{0\sigma }|_{\xi =a}=\varphi _{0\sigma }|_{\xi =b}=0,\quad \partial
_{\xi }\varphi _{1\sigma }|_{\xi =a}=\partial _{\xi }\varphi _{1\sigma
}|_{\xi =b}=0.  \label{elecbound}
\end{equation}%
The corresponding eigenvalues for $\omega $ are the zeros of the function%
\begin{equation}
Z_{\alpha ,i\omega }(ka,kb)=I_{i\omega }^{(\alpha )}(kb)K_{i\omega
}^{(\alpha )}(ka)-K_{i\omega }^{(\alpha )}(kb)I_{i\omega }^{(\alpha )}(ka),
\label{Zalfomeg}
\end{equation}%
where $\alpha =0,1$ for the TE and TM modes respectively, and $f_{i\omega
}^{(0)}(x)=f_{i\omega }(x)$, $f_{i\omega }^{(1)}(x)=f_{i\omega }^{\prime
}(x) $ for $f=I,K$. The eigenfunctions for the separate scalar modes $%
\varphi _{\alpha \sigma }(x)$ are given by formula (\ref{R2eigfunc}) with $%
m=0$, $D=3 $ (for the case of an arbitrary $D$ see \cite{Saha04FuRi}) and
with the replacement $G_{i\omega }^{(b)}(\lambda b,\lambda \xi )\rightarrow
G_{\alpha ,i\omega }(k\xi ,kb)$, $\alpha =0,1$, where%
\begin{equation}
G_{\alpha ,\omega }(x,y)=I_{\omega }^{(\alpha )}(y)K_{\omega }(x)-K_{\omega
}^{(\alpha )}(y)I_{\omega }(x).  \label{Galfom}
\end{equation}%
The corresponding normalization coefficients are determined from (\ref%
{FuRielecnorm}) where now the $\xi $-integration goes over the interval $%
(a,b)$.

Substituting the normalized eigenfunctions into the mode-sum formula for the
VEV of the EMT and applying formula (\ref{Dsumformula}) for the summation
over the eigenvalues of $\omega $, similar to the case of a scalar field one
finds \cite{Avag02}
\begin{equation}
\langle 0|T_{i}^{k}|0\rangle =\langle T_{i}^{k}\rangle ^{(b)}-\frac{\delta
_{i}^{k}}{4\pi ^{2}}\int_{0}^{\infty }dk\,k^{3}\int_{0}^{\infty }d\omega
\sum_{\alpha =0,1}\frac{I_{\omega }^{(\alpha )}(ka)F_{{\mathrm{em}}%
}^{(i)}[G_{\alpha ,\omega }(k\xi ,kb)]}{I_{\omega }^{(\alpha )}(kb)Z_{\alpha
,\omega }(ka,kb)},  \label{FuRiEMTelec2pl}
\end{equation}%
where $\langle T_{i}^{k}\rangle ^{(b)}$ is the VEV in the region $\xi <b$
corresponding to the geometry of a single plate with $\xi =b$ and the
functions $F_{{\mathrm{em}}}^{(i)}[g(z)]$ are defined by formulae (\ref%
{FuRiElF0}), (\ref{FuRiElF2}) with $g(z)=G_{\alpha ,\omega }(z,kb)$. An
alternative form for the vacuum EMT in the region between two plates is
\begin{equation}
\langle 0|T_{i}^{k}|0\rangle =\langle T_{i}^{k}\rangle ^{(a)}-\frac{\delta
_{i}^{k}}{4\pi ^{2}}\int_{0}^{\infty }dk\,k^{3}\int_{0}^{\infty }d\omega
\sum_{\alpha =0,1}\frac{K_{\omega }^{(\alpha )}(kb)F_{{\mathrm{em}}%
}^{(i)}[G_{\alpha ,\omega }(k\xi ,ka)]}{K_{\omega }^{(\alpha )}(ka)Z_{\alpha
,\omega }(ka,kb)},  \label{FuRiEMTelec2pl1}
\end{equation}%
where $\langle T_{i}^{k}\rangle ^{(a)}$ is the vacuum EMT in the region $\xi
>a$ corresponding to the geometry of a single boundary at $\xi =a$.

For the interaction force $p_{{\mathrm{em(int)}}}^{(j)}$, $j=a,b$, per unit
area of the plate at $\xi =j$, from Eqs. (\ref{FuRiEMTelec2pl}) and (\ref%
{FuRiEMTelec2pl1}) one obtains \cite{Avag02}
\begin{eqnarray}
p_{{\mathrm{em(int)}}}^{(a)} &=&-\frac{1}{4\pi ^{2}a^{2}}\int_{0}^{\infty
}dk\,k\int_{0}^{\infty }d\omega \sum_{\alpha =0,1}(-1)^{\alpha }\frac{%
K_{\omega }^{(\alpha )}(kb)}{K_{\omega }^{(\alpha )}(ka)}\frac{(1+\omega
^{2}/k^{2}a^{2})^{\alpha }}{Z_{\alpha ,\omega }(ka,kb)},  \label{pem1} \\
p_{{\mathrm{em(int)}}}^{(b)} &=&-\frac{1}{4\pi ^{2}b^{2}}\int_{0}^{\infty
}dk\,k\int_{0}^{\infty }d\omega \sum_{\alpha =0,1}(-1)^{\alpha }\frac{%
I_{\omega }^{(\alpha )}(ka)}{I_{\omega }^{(\alpha )}(kb)}\frac{(1+\omega
^{2}/k^{2}b^{2})^{\alpha }}{Z_{\alpha ,\omega }(ka,kb)}.  \label{pem2}
\end{eqnarray}%
Recalling that $(-1)^{\alpha }Z_{\alpha ,\omega }(ka,kb)>0$ we see that the
electromagnetic interaction forces are attractive. Note that $p_{{\mathrm{%
em(int)}}}^{(j)}=p_{D{\mathrm{(int)}}}^{(j)}+p_{N{\mathrm{(int)}}}^{(j)}$,
where $p_{D{\mathrm{(int)}}}^{(j)}$ and $p_{N{\mathrm{(int)}}}^{(j)}$ are
the interaction forces for Dirichlet and Neumann scalars. In the limit $%
a\rightarrow b$, to the leading order of $1/(b-a)$ from these expressions
the electromagnetic Casimir interaction force between two parallel plates in
the Minkowski spacetime is obtained. Note that the interaction forces (\ref%
{pem1}), (\ref{pem2}) can also be obtained by the differentiation the
corresponding Casimir energy \cite{Saha04FuRi}.

\section{Wightman function and Casimir densities for branes on AdS bulk}

\label{sec:Branes}

Anti-de Sitter (AdS) spacetime is one of the simplest and most interesting
spacetimes allowed by general relativity. Quantum field theory in this
background has been discussed by several authors. Much of early interest to
AdS spacetime was motivated by the questions of principle nature related to
the quantization of fields propagating on curved backgrounds. The importance
of this theoretical work increased when it was realized that AdS spacetime
emerges as a stable ground state solution in extended supergravity and
Kaluza-Klein models and in string theories. The appearance of the AdS/CFT
correspondence and braneworld models of Randall-Sundrum type \cite{Rand99a}
has revived interest in this subject considerably. Motivated by the problems
of the radion stabilization and the generation of cosmological constant, the
role of quantum effects in braneworlds has attracted great deal of attention
(see, for instance, \cite{Saha06a} for relevant references). In this section
we apply the GAPF for the investigation of the positive frequency Wightman
function and VEV of the EMT for a massive scalar field with general
curvature coupling parameter subject to Robin boundary conditions on two
parallel branes located on $(D+1)$-dimensional AdS background \cite{Saha05b}%
. The general case of different Robin coefficients on separate branes is
considered.

\subsection{Wightman function}

\label{sec:WF}

Consider a scalar field $\varphi (x)$ on background of a $(D+1)$-dimensional
plane-symmetric spacetime with the line element
\begin{equation}
ds^{2}=g_{ik}dx^{i}dx^{k}=e^{-2\sigma (y)}\eta _{\mu \nu }dx^{\mu }dx^{\nu
}-dy^{2},  \label{metric}
\end{equation}%
and with $\eta _{\mu \nu }=\mathrm{diag}(1,-1,\ldots ,-1)$ being the metric
for the $D$-dimensional Minkowski spacetime. Here and below in this section $%
i,k=0,1,\ldots ,D$, and $\mu ,\nu =0,1,\ldots ,D-1$. By making a coordinate
transformation
\begin{equation}
z=\int e^{\sigma (y)}dy,  \label{zcoord}
\end{equation}%
metric (\ref{metric}) is written in a conformally-flat form $%
ds^{2}=e^{-2\sigma (y)}\eta _{ik}dx^{i}dx^{k}$.

We will study quantum vacuum effects brought about by the presence of
parallel infinite plane boundaries (branes), located at $y=a$ and $y=b$, $%
a<b $, with mixed boundary conditions
\begin{equation}
\left( \tilde{A}_{y}+\tilde{B}_{y}\partial _{y}\right) \varphi (x)=0,\quad
y=a,b,  \label{boundcond}
\end{equation}%
and constant coefficients $\tilde{A}_{y}$, $\tilde{B}_{y}$. As a first stage
we will consider the positive frequency Wightman function. To apply the
mode-sum formula we need the eigenfunctions for the problem under
consideration. On the base of the plane symmetry of the problem these
functions can be presented in the form
\begin{equation}
\varphi _{\alpha }(x^{i})=\frac{e^{-i\eta _{\mu \nu }k^{\mu }x^{\nu }}}{%
\sqrt{2\omega (2\pi )^{D-1}}})f_{n}(y),\;k^{\mu }=(\omega ,\mathbf{k}),
\label{eigfunc1B}
\end{equation}%
where $\omega =\sqrt{k^{2}+m_{n}^{2}}$, $k=|\mathbf{k}|$, and the separation
constants $m_{n}$ are determined by the boundary conditions. Substituting
eigenfunctions (\ref{eigfunc1B}) into the field equation, for the function $%
f_{n}(y)$ one obtains the following equation
\begin{equation}
-e^{D\sigma }\partial _{y}\left( e^{-D\sigma }\partial _{y}f_{n}\right)
+\left( m^{2}+\zeta R\right) f_{n}=m_{n}^{2}e^{2\sigma }f_{n}.
\label{eqforfn}
\end{equation}%
For the AdS geometry one has $\sigma (y)=k_{D}y$, $z=e^{\sigma (y)}/k_{D}$,
and $R=-D(D+1)k_{D}^{2}$, where the AdS curvature radius is given by $%
1/k_{D} $. In this case the solution to equation (\ref{eqforfn}) for the
region $a<y<b$ satisfying the boundary condition at $y=b$ is given by
\begin{equation}
f_{n}(y)=c_{n}e^{D\sigma /2}g_{\nu }(m_{n}z_{a},m_{n}z),  \label{fny}
\end{equation}%
where the function $g_{\nu }(u,v)$ is defined by formula (\ref{TCgn}),
\begin{equation}
\nu =\sqrt{(D/2)^{2}-D(D+1)\zeta +m^{2}/k_{D}^{2}},\quad z_{j}=e^{\sigma
(j)}/k_{D},  \label{nu}
\end{equation}%
and we use the barred notation (\ref{barjnot}) with
\begin{equation}
A_{j}=\tilde{A}_{j}+\tilde{B}_{j}k_{D}D/2,\quad B_{j}=\tilde{B}%
_{j}k_{D},\quad j=a,b.  \label{notbar}
\end{equation}%
We will assume values of the curvature coupling parameter for which $\nu $
is real. For imaginary $\nu $ the ground state becomes unstable \cite{Brei82}%
. Note that for a conformally coupled massless scalar one has $\nu =1/2$ and
the cylinder functions in Eq. (\ref{fny}) are expressed via the elementary
functions. From the boundary condition on the brane $y=b$ we receive that
the eigenvalues $m_{n}$ have to be solutions to the equation
\begin{equation}
C_{\nu }^{ab}(z_{b}/z_{a},m_{n}z_{a})\equiv \bar{J}_{\nu }^{(a)}(m_{n}z_{a})%
\bar{Y}_{\nu }^{(b)}(m_{n}z_{b})-\bar{Y}_{\nu }^{(a)}(m_{n}z_{a})\bar{J}%
_{\nu }^{(b)}(m_{n}z_{b})=0.  \label{cnu}
\end{equation}%
The eigenvalues for $m_{n}$ are related to the zeros of the function $C_{\nu
}^{ab}(\eta ,z)$ as
\begin{equation}
m_{n}=k_{D}\gamma _{\nu ,n}e^{-\sigma (a)}=\gamma _{\nu ,n}/z_{a}.
\label{mntogam}
\end{equation}%
The coefficient $c_{n}$ in Eq. (\ref{fny}) is determined from the
orthonormality condition
\begin{equation}
\int_{a}^{b}dye^{(2-D)\sigma }f_{n}(y)f_{n^{\prime }}(y)=\delta _{nn^{\prime
}},  \label{ortcond}
\end{equation}%
and is equal to
\begin{equation}
c_{n}^{2}=\frac{\pi ^{2}u}{2k_{D}z_{a}^{2}}T_{\nu }^{ab}\left( \eta
,u\right) ,\quad u=\gamma _{\nu ,n},\quad \eta =z_{b}/z_{a}.  \label{cn}
\end{equation}%
Note that, as we consider the quantization in the region between the branes,
$z_{a}\leqslant z\leqslant z_{b}$, the modes defined by (\ref{fny}) are
normalizable for all real values of $\nu $ from Eq. (\ref{nu}).

Substituting the eigenfunctions (\ref{eigfunc1B}) into mode-sum (\ref%
{mmodesumWF}), for the expectation value of the field product one finds
\begin{equation}
W(x,x^{\prime })=\frac{k_{D}^{D-1}(zz^{\prime })^{D/2}}{2^{D+1}\pi
^{D-3}z_{a}^{2}}\int d\mathbf{k\,}e^{i\mathbf{k}\Delta \mathbf{x}%
}\sum_{n=1}^{\infty }h_{\nu }(\gamma _{\nu ,n})T_{\nu }^{ab}\left( \eta
,\gamma _{\nu ,n}\right) ,  \label{W11}
\end{equation}%
where $\mathbf{x}=(x^{1},x^{2},\ldots ,x^{D-1})$ represents the coordinates
in $(D-1)$-hyperplane parallel to the branes, $\Delta \mathbf{x=x}-\mathbf{x}%
^{\prime }$, and%
\begin{equation}
h_{\nu }(u)=\frac{ue^{-i\Delta t\sqrt{u^{2}/z_{a}^{2}+k^{2}}}}{\sqrt{%
u^{2}/z_{a}^{2}+k^{2}}}g_{\nu }(u,uz/z_{a})g_{\nu }(u,uz^{\prime }/z_{a}),
\label{habbrane}
\end{equation}%
with $\Delta t=t-t^{\prime }$.

To sum over $n$ we will use summation formula (\ref{cor3form}). Using the
asymptotic formulae for the Bessel functions for large arguments when $\nu $
is fixed (see, e.g., \cite{abramowiz}), we can see that for the function $%
h_{\nu }(u)$ from Eq. (\ref{habbrane}) the condition (\ref{cond31}) is
satisfied if $z+z^{\prime }+|\Delta t|<2z_{b}$. As for $|u|<k$ one has $%
h_{\nu }(ue^{\pi i})=-h_{\nu }(u)$, the condition (\ref{cor3cond1}) is also
satisfied for the function $h_{\nu }(u)$. Note that $h_{\nu }(u)\sim
u^{1-\delta _{k0}}$ for $u\rightarrow 0$ and the residue term on the right
of formula (\ref{cor3form}) vanishes. Applying to the sum over $n$ in Eq. (%
\ref{W11}) formula (\ref{cor3form}), one obtains
\begin{eqnarray}
W(x,x^{\prime }) &=&\frac{k_{D}^{D-1}(zz^{\prime })^{D/2}}{2^{D}\pi ^{D-1}}%
\int d\mathbf{k\,}e^{i\mathbf{k}\Delta \mathbf{x}}  \notag \\
&&\Bigg\{\frac{1}{z_{a}^{2}}\int_{0}^{\infty }\frac{h_{\nu }(u)du}{\bar{J}%
_{\nu }^{(a)2}(u)+\bar{Y}_{\nu }^{(a)2}(u)}-\frac{2}{\pi }\int_{k}^{\infty
}du\,u\frac{\Omega _{a\nu }(uz_{a},uz_{b})}{\sqrt{u^{2}-k^{2}}}  \notag \\
&&\times G_{\nu }^{(a)}(uz_{a},uz)G_{\nu }^{(a)}(uz_{a},uz^{\prime })\cosh
(\Delta t\sqrt{u^{2}-k^{2}})\Bigg\},  \label{W13}
\end{eqnarray}%
where we use notations (\ref{Omegaanu}),(\ref{Geab}). Note that we have
assumed values of the coefficients $A_{j}$ and $B_{j}$ for which all zeros
for Eq. (\ref{cnu}) are real and have omitted the residue terms. In the
following we will consider this case only.

In the way similar to that used before, it can be seen that the first
integral in the figure braces in Eq. (\ref{W13}) is presented in the form
\begin{eqnarray}
\int_{0}^{\infty }\frac{h_{\nu }(u)du}{\bar{J}_{\nu }^{(a)2}(u)+\bar{Y}_{\nu
}^{(a)2}(u)} &=&z_{a}^{2}\int_{0}^{\infty }duu\frac{e^{-i\Delta t\sqrt{%
u^{2}+k^{2}}}}{\sqrt{x^{2}+k^{2}}}J_{\nu }(uz)J_{\nu }(uz^{\prime })  \notag
\\
&&-\frac{2z_{a}^{2}}{\pi }\int_{k}^{\infty }du\,\ u\frac{\bar{I}_{\nu
}^{(a)}(uz_{a})}{\bar{K}_{\nu }^{(a)}(uz_{a})}\frac{K_{\nu }(uz)K_{\nu
}(uz^{\prime })}{\sqrt{u^{2}-k^{2}}}  \notag \\
&&\times \cosh (\Delta t\sqrt{u^{2}-k^{2}}).  \label{rel1term}
\end{eqnarray}%
Substituting this into formula (\ref{W13}), the Wightman function can be
written in the form \cite{Saha05b}
\begin{eqnarray}
W(x,x^{\prime }) &=&W^{(a)}(x,x^{\prime })-\frac{k_{D}^{D-1}(zz^{\prime
})^{D/2}}{2^{D-1}\pi ^{D}}\int d\mathbf{k\,}e^{\mathbf{k}\Delta \mathbf{x}%
}\int_{k}^{\infty }duu\frac{\Omega _{a\nu }(uz_{a},uz_{b})}{\sqrt{u^{2}-k^{2}%
}}  \notag \\
&&\times G_{\nu }^{(a)}(uz_{a},uz)G_{\nu }^{(a)}(uz_{a},uz^{\prime })\cosh
(\Delta t\sqrt{u^{2}-k^{2}}).  \label{W15}
\end{eqnarray}%
Here the term
\begin{equation}
W^{(a)}(x,x^{\prime })=W_{\mathrm{S}}(x,x^{\prime })+\langle \varphi
(x)\varphi (x^{\prime })\rangle _{a}  \label{WaAds}
\end{equation}%
does not depend on the parameters of the boundary at $y=b$ and is the
Wightman function for a single brane located at $y=a$. In formula (\ref%
{WaAds}), the term
\begin{equation}
W_{\mathrm{S}}(x,x^{\prime })=\frac{k_{D}^{D-1}(zz^{\prime })^{D/2}}{%
2^{D}\pi ^{D-1}}\int d\mathbf{k\,}e^{\mathbf{k}\Delta \mathbf{x}%
}\int_{0}^{\infty }du\,u\frac{e^{-i\Delta t\sqrt{u^{2}+k^{2}}}}{\sqrt{%
u^{2}+k^{2}}}J_{\nu }(uz)J_{\nu }(uz^{\prime }),  \label{WAdS}
\end{equation}%
does not depend on the boundary conditions and is the Wightman function for
the AdS space without boundaries, and the second term on the right,
\begin{eqnarray}
\langle \varphi (x)\varphi (x^{\prime })\rangle _{a} &=&-\frac{%
k_{D}^{D-1}(zz^{\prime })^{D/2}}{2^{D-1}\pi ^{D}}\int d\mathbf{k\,}e^{%
\mathbf{k}\Delta \mathbf{x}}\int_{k}^{\infty }du\,u\frac{\bar{I}_{\nu
}^{(a)}(uz_{a})}{\bar{K}_{\nu }^{(a)}(uz_{a})}  \notag \\
&&\times \frac{K_{\nu }(uz)K_{\nu }(uz^{\prime })}{\sqrt{u^{2}-k^{2}}}{\cosh
}(\Delta t\sqrt{u^{2}-k^{2}}),  \label{1bounda}
\end{eqnarray}%
is induced in the region $z>z_{a}$ by a single brane at $z=z_{a}$ when the
brane $z=z_{b}$ is absent.

The Wightman function in the region $z_{a}\leqslant z\leqslant z_{b}$ can
also be presented in the equivalent form
\begin{eqnarray}
W(x,x^{\prime }) &=&W^{(b)}(x,x^{\prime })-\frac{k_{D}^{D-1}(zz^{\prime
})^{D/2}}{2^{D-1}\pi ^{D}}\int d\mathbf{k\,}e^{\mathbf{k}\Delta \mathbf{x}%
}\int_{k}^{\infty }du\,u\frac{\Omega _{b\nu }(uz_{a},uz_{b})}{\sqrt{%
u^{2}-k^{2}}}  \notag \\
&&\times G_{\nu }^{(b)}(uz_{b},uz)G_{\nu }^{(b)}(uz_{b},uz^{\prime })\cosh
(\Delta t\sqrt{u^{2}-k^{2}}).  \label{W17}
\end{eqnarray}%
In this formula%
\begin{equation}
W^{(b)}(x,x^{\prime })=W_{\mathrm{S}}(x,x^{\prime })+\langle \varphi
(x)\varphi (x^{\prime })\rangle _{b},  \label{WbAds}
\end{equation}%
where
\begin{eqnarray}
\langle \varphi (x)\varphi (x^{\prime })\rangle _{b} &=&-\frac{%
k_{D}^{D-1}(zz^{\prime })^{D/2}}{2^{D-1}\pi ^{D}}\int d\mathbf{k\,}e^{%
\mathbf{k}\Delta \mathbf{x}}\int_{k}^{\infty }du\,u\frac{\bar{K}_{\nu
}^{(b)}(uz_{b})}{\bar{I}_{\nu }^{(b)}(uz_{b})}  \notag \\
&&\times \frac{I_{\nu }(uz)I_{\nu }(uz^{\prime })}{\sqrt{u^{2}-k^{2}}}{\cosh
}(\Delta t\sqrt{u^{2}-k^{2}}),  \label{1boundb}
\end{eqnarray}%
is the Wightman function for a single brane at $y=b$. The term (\ref{1boundb}%
) is the boundary part induced in the region $z<z_{b}$ by the brane at $y=b$%
. Combining two forms, formulae (\ref{W15}) and (\ref{W17}), we see that the
expressions for the Wightman function in the region $z_{a}\leqslant
z\leqslant z_{b}$ is symmetric under the interchange $a\rightleftarrows b$
and $I_{\nu }\rightleftarrows K_{\nu }$. Note that the expression for the
Wightman function is not symmetric with respect to the interchange of the
brane indices. The reason for this is that, though the background AdS
spacetime is homogeneous, the boundaries have nonzero extrinsic curvature
tensors and two sides of the boundaries are not equivalent. In particular,
for the geometry of a single brane the VEVs differ for the regions on the
left and on the right of the brane. Here the situation is similar to that
for the case of a spherical shell on background of the Minkowski spacetime.

\subsection{Casimir densities for a single brane}

\label{sec:CD1b}

In this subsection we will consider the VEV of the EMT for a scalar field in
the case of a single brane located at $z=z_{a}$. As it has been shown in the
previous subsection the Wightman function for this geometry is presented in
the form (\ref{WaAds}). The brane induced part $\langle \varphi (x)\varphi
(x^{\prime })\rangle _{a}$ is given by formula (\ref{1bounda}) in the region
$z>z_{a}$ and by formula (\ref{1boundb}) with replacement $z_{b}\rightarrow
z_{a}$ in the region $z<z_{a}$. For points away from the brane this part is
finite in the coincidence limit and in the corresponding formulae for the
Wightman function we can directly put $x=x^{\prime }$. Introducing a new
integration variable $v=\sqrt{u^{2}-k^{2}}$, transforming to polar
coordinates in the plane $(v,k)$ and integrating over angular part, the
following formula can be derived
\begin{equation}
\int_{0}^{\infty }dkk^{D-2}\int_{k}^{\infty }du\frac{uf(u)}{\sqrt{u^{2}-k^{2}%
}}=\frac{\sqrt{\pi }\Gamma \left( \frac{D-1}{2}\right) }{2\Gamma (D/2)}%
\int_{0}^{\infty }duu^{D-1}f(u).  \label{rel3brane}
\end{equation}%
By using this formula and Eq. (\ref{1bounda}), the boundary induced VEV for
the field square in the region $z>z_{a}$ is presented in the form \cite%
{Saha05b}
\begin{equation}
\langle \varphi ^{2}\rangle _{\mathrm{b}}^{(a)}=-\frac{k_{D}^{D-1}z^{D}}{%
2^{D-1}\pi ^{D/2}\Gamma (D/2)}\int_{0}^{\infty }du\,u^{D-1}\frac{\bar{I}%
_{\nu }^{(a)}(uz_{a})}{\bar{K}_{\nu }^{(a)}(uz_{a})}K_{\nu }^{2}(uz),\quad
z>z_{a}.  \label{phi2spl}
\end{equation}%
The corresponding formula in the region $z<z_{a}$ is obtained from Eq. (\ref%
{1boundb}) by a similar way and differs from Eq. (\ref{phi2spl}) by
replacements $I_{\nu }\rightleftarrows K_{\nu }$.

The VEV of the EMT can be evaluated by substituting expressions for the
positive frequency Wightman function and VEV of the field square into Eq. (%
\ref{mTikVEV}). First of all we will consider the region $z>z_{a}$. The
vacuum EMT is diagonal and can be presented in the form
\begin{equation}
\langle 0|T_{i}^{k}|0\rangle =\langle 0|T_{i}^{k}|0\rangle _{\mathrm{S}%
}+\langle T_{i}^{k}\rangle _{a},  \label{EMT41pl}
\end{equation}%
where
\begin{equation}
\langle 0|T_{i}^{k}|0\rangle _{\mathrm{S}}=\frac{k_{D}^{D+1}\delta _{i}^{k}}{%
2^{D}\pi ^{D/2}}\Gamma \left( 1-\frac{D}{2}\right) \int_{0}^{\infty
}duu^{D-1}f^{(i)}[J_{\nu }(u)],  \label{EMTAdS}
\end{equation}%
is the VEV for the EMT in the AdS background without boundaries, and the
term \cite{Saha05b}
\begin{equation}
\langle T_{i}^{k}\rangle _{a}=-\frac{k_{D}^{D+1}z^{D}\delta _{i}^{k}}{%
2^{D-1}\pi ^{D/2}\Gamma (D/2)}\int_{0}^{\infty }duu^{D-1}\frac{\bar{I}_{\nu
}^{(a)}(uz_{a})}{\bar{K}_{\nu }^{(a)}(uz_{a})}F^{(i)}[K_{\nu }(uz)],
\label{EMT1bounda}
\end{equation}%
is induced in the region $z>z_{a}$ by a single boundary at $z=z_{a}$. For a
given function $g(v)$ the functions $F^{(i)}[g(v)]$ in formula (\ref%
{EMT1bounda}) are defined as
\begin{eqnarray}
F^{(i)}[g(v)] &=&\left( \frac{1}{2}-2\zeta \right) \left[ v^{2}g^{\prime
2}(v)+\left( D+\frac{4\zeta }{4\zeta -1}\right) vg(v)g^{\prime }(v)+\right.
\notag \\
&&+\left. \left( \nu ^{2}+v^{2}+\frac{2v^{2}}{D(4\zeta -1)}\right) g^{2}(v)%
\right] ,\quad i=0,1,\ldots ,D-1,  \label{Fi} \\
F^{(D)}[g(v)] &=&-\frac{v^{2}}{2}g^{\prime }{}^{2}(v)+\frac{D}{2}\left(
4\zeta -1\right) vg(v)g^{\prime }(v)+  \notag \\
&&+\frac{1}{2}\left[ v^{2}+\nu ^{2}+2\zeta D(D+1)-D^{2}/2\right] g^{2}(v),
\label{FD}
\end{eqnarray}%
and the expressions for the functions $f^{(i)}\left[ g(v)\right] $ are
obtained from those for $F^{(i)}\left[ g(v)\right] $ by the replacement $%
v\rightarrow iv$. Note that the boundary-induced part (\ref{EMT1bounda}) is
finite for points away the brane and, hence, the renormalization procedure
is needed for the boundary-free part only. The latter is well investigated
in literature.

In a similar way, for the VEVs induced by a single brane in the region $%
z<z_{a}$, by making use of expression (\ref{1boundb}) (with replacement $%
z_{b}\rightarrow z_{a}$), one obtains
\begin{equation}
\langle T_{i}^{k}\rangle _{a}=-\frac{k_{D}^{D+1}z^{D}\delta _{i}^{k}}{%
2^{D-1}\pi ^{D/2}\Gamma (D/2)}\int_{0}^{\infty }duu^{D-1}\frac{\bar{K}_{\nu
}^{(a)}(uz_{a})}{\bar{I}_{\nu }^{(a)}(uz_{a})}F^{(i)}[I_{\nu }(uz)].
\label{Tik1plnewleft}
\end{equation}%
Note that VEVs (\ref{EMT1bounda}), (\ref{Tik1plnewleft}) depend only on the
ratio $z/z_{a}$ which is related to the proper distance from the brane by
the equation
\begin{equation}
z/z_{a}=e^{k_{D}(y-a)}.  \label{propdisz}
\end{equation}%
As we see, for the part of the EMT corresponding to the coordinates in the
hyperplane parallel to the branes one has $\langle T_{\mu \nu }\rangle
^{(a)}\sim \eta _{\mu \nu }$. Of course, we could expect this result from
the problem symmetry. It can be seen that the VEVs obtained above obey the
continuity equation $\nabla _{k}T_{i}^{k}=0$, which for the AdS metric takes
the form
\begin{equation}
z^{D+1}\partial _{z}\left( z^{-D}T_{D}^{D}\right) +DT_{0}^{0}=0.
\label{conteq}
\end{equation}

For a conformally coupled massless scalar $\nu =1/2$, and by making use of
the expressions for the modified Bessel functions, it can be seen that $%
\langle T_{i}^{k}\rangle ^{(a)}=0$ in the region $z>z_{a}$ and
\begin{equation}
\langle T_{D}^{D}\rangle ^{(a)}=-D\langle T_{0}^{0}\rangle ^{(a)}=-\frac{%
(k_{D}z/z_{a})^{D+1}}{(4\pi )^{D/2}\Gamma (D/2)}\int_{0}^{\infty }\frac{%
t^{D}\,dt}{\frac{B_{a}(t-1)+2A_{a}}{B_{a}(t+1)-2A_{a}}e^{t}+1}
\label{conf1pl}
\end{equation}%
in the region $z<z_{a}$. Note that the corresponding energy-momentum tensor
for a single Robin plate in the Minkowski bulk vanishes \cite{Rome02} and
the result for the region $z>z_{a}$ is obtained by a simple conformal
transformation from that for the Minkowski case. In the region $z<z_{a}$
this procedure does not work as in the AdS problem one has $0<z<z_{a}$
instead of $-\infty <z<z_{a}$ in the Minkowski problem and, hence, the part
of AdS under consideration is not a conformal image of the corresponding
manifold in the Minkowski spacetime.

The brane-induced VEVs given by equations (\ref{EMT1bounda}) and (\ref%
{Tik1plnewleft}), in general, can not be further simplified and need
numerical calculations. Relatively simple analytic formulae can be obtained
in limiting cases. First of all, as a partial check, in the limit $%
k_{D}\rightarrow 0$ the corresponding formulae for a single plate on the
Minkowski bulk are obtained (see Ref. \cite{Rome02}). This can be seen
noting that in this limit $\nu \sim m/k_{D}$ is large and by introducing the
new integration variable in accordance with $u=\nu y$, we can replace the
modified Bessel functions by their uniform asymptotic expansions for large
values of the order. The Minkowski result is obtained in the leading order.

In the limit $z\rightarrow z_{a}$ for a fixed $k_{D}$ expressions (\ref%
{EMT1bounda}) and (\ref{Tik1plnewleft}) diverge. In accordance with (\ref%
{propdisz}), this corresponds to small proper distances from the brane. Near
the brane the main contribution into the integral over $u$ in Eqs. (\ref%
{EMT1bounda}), (\ref{Tik1plnewleft}) comes from large values of $u$ and we
can replace the modified Bessel functions by their asymptotic expressions
for large values of the argument when the order is fixed (see, for instance,
\cite{abramowiz}). To the leading order this yields
\begin{equation}
\langle T_{0}^{0}\rangle ^{(a)}\approx \frac{\langle T_{D}^{D}\rangle ^{(a)}%
}{1-z_{a}/z}\approx -\Gamma \left( \frac{D+1}{2}\right) \frac{%
Dk_{D}^{D+1}(\zeta -\zeta _{D})k_{a}}{2^{D}\pi ^{(D+1)/2}|1-z_{a}/z|^{D+1}},
\label{Tik1near}
\end{equation}%
where $k_{a}$ is defined after formula (\ref{Tiiclose}). Note that the
leading terms for the components with $i=0,1,\ldots ,D-1$ are symmetric with
respect to the brane, and the $_{D}^{D}$ -- component has opposite signs for
different sides of the brane. Near the brane the vacuum energy densities
have opposite signs for the cases of Dirichlet ($B_{a}=0$) and non-Dirichlet
($B_{a}\neq 0$) boundary conditions. Recall that for a conformally coupled
massless scalar the vacuum EMT vanishes in the region $z>z_{a}$ and is given
by expression (\ref{conf1pl}) in the region $z<z_{a}$. The latter is finite
everywhere including the points on the brane. For large proper distances
from the brane compared with the AdS curvature radius, $k_{D}|y-a|\gg 1$,
the boundary induced EMT vanishes as $\exp [2\nu k_{D}(a-y)]$ in the region $%
y>a$ and as $\exp [k_{D}(2\nu +D)(y-a)]$ in the region $y<a$. The same
behavior takes place for a fixed $y-a$ and large values of the parameter $%
k_{D}$. In the large mass limit, $m\gg k_{D}$, the boundary parts are
exponentially suppressed.

\subsection{Two-brane geometry}

\label{sec:2branes}

In this subsection we will investigate the VEVs for the field square and the
EMT in the region between two branes. Taking the coincidence limit in the
formulae for the Wightman function and using formula (\ref{rel3brane}), for
the VEV of the field square one finds \cite{Saha05b}%
\begin{eqnarray}
\langle 0|\varphi ^{2}|0\rangle &=&\langle 0|\varphi ^{2}|0\rangle _{\mathrm{%
S}}+\langle \varphi ^{2}\rangle _{j}-\frac{k_{D}^{D-1}z^{D}}{2^{D-1}\pi
^{D/2}\Gamma \left( D/2\right) }  \notag \\
&&\times \int_{0}^{\infty }du\,u^{D-1}\Omega _{j\nu }(uz_{a},uz_{b})G_{\nu
}^{(j)2}(uz_{j},uz),  \label{phi22brAdS}
\end{eqnarray}%
where $j=a,b$ provide two equivalent representations. The last term on the
right of this formula is finite for points on the brane at $y=j$ and
diverges on the other brane. In the similar way, substituting the
corresponding Wightman function from Eq. (\ref{W15}) into the mode-sum
formula (\ref{mTikVEV}), we obtain
\begin{eqnarray}
\langle 0|T_{i}^{k}|0\rangle &=&\langle 0|T_{i}^{k}|0\rangle _{\mathrm{S}%
}+\langle T_{i}^{k}\rangle _{j}-\frac{k_{D}^{D+1}z^{D}\delta _{i}^{k}}{%
2^{D-1}\pi ^{D/2}\Gamma \left( D/2\right) }  \notag \\
&&\times \int_{0}^{\infty }duu^{D-1}\Omega _{j\nu
}(uz_{a},uz_{b})F^{(i)}[G_{\nu }^{(j)}(uz_{j},uz)],  \label{Tik1int}
\end{eqnarray}%
with the functions $F^{(i)}[g(v)]$ from Eqs. (\ref{Fi}), (\ref{FD}). The
formulae for the cases $j=a,b$ are obtained from each other by the
replacements $a\rightleftarrows b$, $I_{\nu }\rightleftarrows K_{\nu }$. The
VEVs in the region between the branes are not symmetric under the
interchange of indices of the branes. As it has been mentioned above, the
reason for this is that, though the background spacetime is homogeneous, due
to the non-zero extrinsic curvature tensors for the branes, the regions on
the left and on the right of the brane are not equivalent. By the same way
as for the case of a single brane, it can be seen that in the limit $%
k_{D}\rightarrow 0$ from formulae (\ref{phi22brAdS}) and (\ref{Tik1int}) the
corresponding results for two plates geometry in the Minkowski bulk. For a
conformally coupled massless scalar field $\nu =1/2$ and the formulae for
the VEVs in the region between the branes can also be obtained from the
corresponding formulae for parallel Robin plates in the Minkowski bulk by
conformal transformation \cite{Saha03ConfBr}.

Now we turn to the interaction forces between the branes. The vacuum force
acting per unit surface of the brane at $z=z_{j}$ is determined by the $%
{}_{D}^{D}$ -- component of the vacuum EMT at this point. The corresponding
effective pressures can be presented as the sum $p^{(j)}=p_{1}^{(j)}+p_{{%
\mathrm{(int)}}}^{(j)}$, $j=a,b$, where the first term on the right is the
pressure for a single brane at $z=z_{j}$ when the second brane is absent.
This term is divergent due to the surface divergences in the VEVs. The
second term is determined by the last term on the rhs in (\ref{Tik1int})
evaluated at $z=z_{j}$. This term is the pressure induced by the presence of
the second brane. Using the relations $G_{\nu }^{(j)}(u,u)=-B_{j}$, $G_{\nu
}^{(j)\prime }(u,u)=A_{j}/u$, one finds
\begin{eqnarray}
p_{{\mathrm{(int)}}}^{(j)} &=&\frac{k_{D}^{D+1}}{2^{D}\pi ^{D/2}\Gamma
\left( \frac{D}{2}\right) }\int_{0}^{\infty }dx\,x^{D-1}\Omega _{j\nu
}\left( xz_{a}/z_{j},xz_{b}/z_{j}\right)  \notag \\
&&\times \left[ \left( x^{2}-\nu ^{2}+2m^{2}/k_{D}^{2}\right)
B_{j}^{2}-D(4\zeta -1)A_{j}B_{j}-A_{j}^{2}\right] .  \label{pintj}
\end{eqnarray}%
In dependence of the values for the coefficients in the boundary conditions,
these effective pressures can be either positive or negative, leading to
repulsive or attractive forces. Note that due to the asymmetry in the VEV of
the EMT, the interaction forces acting on the branes are not symmetric under
the interchange of the brane indices. It can be seen that the vacuum
effective pressures are negative for Dirichlet scalar and for a scalar with $%
A_{a}=A_{b}=0$ and, hence, the corresponding interaction forces are
attractive for all values of the interbrane distance.

Let us consider the limiting cases for the interaction forces described by
Eq. (\ref{pintj}). For small distances compared with the AdS curvature
radius, $k_{D}(b-a)\ll 1$, the leading terms are the same as for the plates
in Minkowski bulk. In particular, in this limit the interaction forces are
repulsive for Dirichlet boundary condition on one brane and non-Dirichlet
boundary condition on the other, and are attractive for all other cases. For
large distances between the branes, $k_{D}(b-a)\gg 1$ (this limit is
realized in the Randall-Sundrum model), by using the expressions for the
modified Bessel functions for small values of the argument, one can see that
$p_{{\mathrm{(int)}}}^{(a)}\sim \left( z_{a}/2z_{b}\right) ^{D+2\nu }$ and $%
p_{{\mathrm{(int)}}}^{(b)}\sim \left( z_{a}/2z_{b}\right) ^{2\nu }$.

Now we consider the application of the results described in this section to
the Randall-Sundrum braneworld models \cite{Rand99a} based on the AdS
geometry with one extra dimension. The fifth dimension $y$ is compactified
on an orbifold, $S^{1}/Z_{2}$ of length $L$, with $-L\leqslant y\leqslant L$%
. The orbifold fixed points at $y=a=0$ and $y=b$ are the locations of two
3-branes. We will allow these submanifolds to have an arbitrary dimension $D$%
. The metric in the Randall-Sundrum model has the form (\ref{metric}) with $%
\sigma (y)=k_{D}|y|$. The corresponding boundary conditions for an untwisted
scalar field are in form (\ref{boundcond}) with \cite{Saha05b,Gher00}
\begin{equation}
\tilde{A}_{j}/\tilde{B}_{j}=-(n^{(j)}c_{j}+4D\zeta k_{D})/2,
\label{AtildeRS}
\end{equation}%
and respectively
\begin{equation}
A_{j}/B_{j}=\left[ D(1-4\zeta )-n^{(j)}c_{j}/k_{D}\right] /2,  \label{ARS}
\end{equation}%
where $c_{j}$ are the brane mass terms. For a twisted scalar $\tilde{B}%
_{j}=0 $, which corresponds to Dirichlet boundary conditions on
both branes. Recently the EMT in the Randall--Sundrum braneworld
for a bulk scalar with zero brane mass terms $c_{1}$ and $c_{2}$
is considered in \cite{Knap04}, where a general formula is given
for the unrenormalized VEV in terms of the differential operator
acting on the Green function. In our approach the application of
the GAPF allowed to extract manifestly the part due to the AdS
bulk without boundaries and for the points away from the
boundaries the renormalization procedure is the same as for the
boundary-free parts. In addition, the boundary-induced parts are
presented in terms of exponentially convergent integrals
convenient for numerical calculations.

From the point of view of embedding the Randall-Sundrum type braneworld
models into a more fundamental theory, such as string/M theory, one may
expect that a more complete version of this scenario must admit the presence
of additional extra dimensions compactified on an internal manifold. The
results discussed in this section can be generalized for the geometry of two
parallel branes of codimension one on background of $(D+1)$ -dimensional
spacetime with topology $AdS_{D_{1}+1}\times \Sigma $ and the line element
\begin{equation}
ds^{2}=e^{-2k_{D}y}\eta _{\mu \sigma }dx^{\mu }dx^{\sigma
}-e^{-2k_{D}y}\gamma _{ik}dX^{i}dX^{k}-dy^{2},  \label{HighDimRSmetric}
\end{equation}%
where $\eta _{\mu \sigma }=\mathrm{diag}(1,-1,\ldots ,-1)$ is the metric
tensor for $D_{1}$-dimensional Minkowski spacetime $R^{(D_{1}-1,1)}$, and
the coordinates $X^{i}$, $i=1,\ldots ,D_{2}$, cover the manifold $\Sigma $.
The quantum effective potential and the problem of moduli stabilization in
the orbifolded version of this model with zero mass parameters on the branes
were discussed recently in \cite{Flac03}. In particular, it has been shown
that one-loop effects induced by bulk scalar fields generate a suitable
effective potential which can stabilize the hierarchy between the
gravitational and electroweak scales. In \cite{Saha06a} the Wightman
function, the VEVs of the field square and the EMT for a scalar field
brought about by the presence of two parallel infinite branes, located at $%
y=a$ and $y=b$ with boundary conditions (\ref{boundcond}), are investigated
in a way similar to that used in this section.

\section{Casimir densities for spherical branes in Rindler-like spacetimes}

\label{sec:Rindbrane}

In previous section we have considered braneworld models on background of
AdS spacetime. It seems interesting to generalize the study of quantum
effects to other types of bulk spacetimes. In particular, bulk geometries
generated by higher-dimensional black holes are of special interest. In
these models, the tension and the position of the brane are tuned in terms
of black hole mass and cosmological constant and brane gravity trapping
occurs in just the same way as in the Randall-Sundrum model. Though, in the
generic black hole background the investigation of brane-induced quantum
effects is technically complicated, the exact analytical results can be
obtained in the near horizon and large mass limit when the brane is close to
the black hole horizon. In this limit the black hole geometry may be
approximated by the Rindler-like manifold (for some investigations of
quantum effects on background of Rindler-like spacetimes see \cite%
{Byts96,Saha05RindBr} and references therein). In this section, by using the
GAPF, we investigate the Wightman function, the VEVs of the field square and
the EMT for a scalar field with an arbitrary curvature coupling parameter
for two spherical branes in the bulk $Ri\times S^{D-1}$, where $Ri$ is a
two-dimensional Rindler spacetime \cite{Saha07RindBr}.

\subsection{Wightman function}

Consider a scalar field $\varphi (x)$ propagating on background of $(D+1)$%
-dimensional Rindler-like spacetime $Ri\times S^{D-1}$. The corresponding
metric is described by the line element%
\begin{equation}
ds^{2}=\xi ^{2}d\tau ^{2}-d\xi ^{2}-r_{H}^{2}d\Sigma _{D-1}^{2},
\label{ds22RB}
\end{equation}%
with the Rindler-like $(\tau ,\xi )$ part and $d\Sigma _{D-1}^{2}$ is the
line element for the space with positive constant curvature with the Ricci
scalar $R=n(n+1)/r_{H}^{2}$, $n=D-2$. Line element (\ref{ds22RB}) describes
the near horizon geometry of $(D+1)$-dimensional topological black hole with
the line element
\begin{equation}
ds^{2}=A_{H}(r)dt^{2}-\frac{dr^{2}}{A_{H}(r)}-r^{2}d\Sigma _{D-1}^{2},
\label{ds21RB}
\end{equation}%
where $A_{H}(r)=k+r^{2}/l^{2}-r_{0}^{D}/l^{2}r^{n}$ and the parameter $k$
classifies the horizon topology, with $k=0,-1,1$ corresponding to flat,
hyperbolic, and elliptic horizons, respectively. The parameter $l$ is
related to the bulk cosmological constant and the parameter $r_{0}$ depends
on the mass of the black hole. In the non-extremal case the function $%
A_{H}(r)$ has a simple zero at $r=r_{H}$, and in the near horizon limit,
introducing new coordinates $\tau $ and $\rho $ in accordance with%
\begin{equation}
\tau =A_{H}^{\prime }(r_{H})t/2,\quad r-r_{H}=A_{H}^{\prime }(r_{H})\xi
^{2}/4,  \label{tau}
\end{equation}%
the line element is written in the form (\ref{ds22RB}). Note that for a $%
(D+1)$-dimensional Schwarzschild black hole one has $%
A_{H}(r)=1-(r_{H}/r)^{n} $ and, hence, $A_{H}^{\prime }(r_{H})=n/r_{H}$.

We will assume that the field satisfies Robin boundary conditions (\ref%
{Dboundcond}) on the hypersurfaces $\xi =a$ and $\xi =b$, $a<b$. In
accordance with \ (\ref{tau}), the hypersurface $\xi =j$ corresponds to the
spherical shell near the black hole horizon with the radius $%
r_{j}=r_{H}+A_{H}^{\prime }(r_{H})j^{2}/4$. In the corresponding braneworld
scenario based on the orbifolded version of the model the region between the
branes is employed only and the ratio $\tilde{A}_{j}/\tilde{B}_{j}$ for
untwisted bulk scalars is related to the brane mass parameters $c_{j}$\ of
the field by the formula \cite{Saha05RindBr}
\begin{equation}
\tilde{A}_{j}/\tilde{B}_{j}=\frac{1}{2}\left( c_{j}-\zeta /j\right) ,\;j=a,b.
\label{ABbraneworld}
\end{equation}%
For a twisted scalar Dirichlet boundary conditions are obtained on both
branes.

In hyperspherical coordinates (see Section \ref{sec:GlobMonSc} for the
notations) the eigenfunctions in the region between the branes can be
written in the form%
\begin{equation}
\varphi _{\sigma }(x)=\beta _{\sigma }G_{i\omega }^{(b)}(\lambda
_{l}b,\lambda _{l}\xi )Y(m_{k};\vartheta ,\phi )e^{-i\omega \tau },
\label{eigfunc1}
\end{equation}%
where the function $G_{\nu }^{(j)}(x,y)$ is defined by formula (\ref{Geab})
with the barred notations (\ref{barjnot}), where $A_{j}=\tilde{A}_{j}$, $%
B_{j}=\tilde{B}_{j}/j$, and
\begin{equation}
\quad \lambda _{l}=\frac{1}{r_{H}}\sqrt{l(l+n)+\zeta n(n+1)+m^{2}r_{H}^{2}}%
\,.  \label{lambdal}
\end{equation}%
From the boundary conditions we find that the eigenvalues for $\omega $ are
roots to the equation $Z_{i\omega }(\lambda _{l}a,\lambda _{l}b)=0$, where
the function $Z_{i\omega }(u,v)$ is defined by formula (\ref{Zi}), and,
hence, $\omega =\Omega _{s}=\Omega _{s}(\lambda _{l}a,\lambda _{l}b)$, $%
s=1,2,\ldots $. For the normalization coefficient one has%
\begin{equation}
\beta _{\sigma }^{2}=\left. \frac{r_{H}^{1-D}\bar{I}_{i\omega
}^{(a)}(\lambda _{l}a)}{N(m_{k})\bar{I}_{i\omega }^{(b)}(\lambda
_{l}b)\partial _{\omega }Z_{i\omega }(\lambda _{l}a,\lambda _{l}b)}%
\right\vert _{\omega =\Omega _{s}}.  \label{Calfa}
\end{equation}

Substituting eigenfunctions (\ref{eigfunc1}) into the mode-sum formula, for
the Wightman function in the region between the branes one finds
\begin{eqnarray}
W(x,x^{\prime }) &=&\frac{r_{H}^{1-D}}{nS_{D}}\sum_{l=0}^{\infty
}(2l+n)C_{l}^{n/2}(\cos \theta )\sum_{s=1}^{\infty }\frac{\bar{I}_{i\omega
}^{(a)}(\lambda _{l}a)e^{-i\omega \Delta \tau }}{\bar{I}_{i\omega
}^{(b)}(\lambda _{l}b)\partial _{\omega }Z_{i\omega }(\lambda _{l}a,\lambda
_{l}b)}  \notag \\
&&\times \left. G_{i\omega }^{(b)}(\lambda _{l}b,\lambda _{l}\xi )G_{i\omega
}^{(b)}(\lambda _{l}b,\lambda _{l}\xi ^{\prime })\right\vert _{\omega
=\Omega _{s}},  \label{Wigh1RB}
\end{eqnarray}%
with $\Delta \tau =$ $\tau -\tau ^{\prime }$. The application of summation
formula (\ref{Dsumformula}) to the sum over $s$ leads to the result%
\begin{eqnarray}
W(x,x^{\prime }) &=&W_{0}(x,x^{\prime })+\langle \varphi (x)\varphi
(x^{\prime })\rangle ^{(j)}-\frac{r_{H}^{1-D}}{\pi nS_{D}}\sum_{l=0}^{\infty
}(2l+n)C_{l}^{n/2}(\cos \theta )  \notag \\
&&\times \int_{0}^{\infty }d\omega \,\Omega _{j\omega }(\lambda
_{l}a,\lambda _{l}b)G_{\omega }^{(j)}(\lambda _{l}j,\lambda _{l}\xi
)G_{\omega }^{(j)}(\lambda _{l}j,\lambda _{l}\xi ^{\prime })\cosh (\omega
\Delta \tau ),  \label{Wigh3RB}
\end{eqnarray}%
with $j=b$, where $W_{0}(x,x^{\prime })$ is the Wightman function for the
geometry without boundaries and the part%
\begin{eqnarray}
\langle \varphi (x)\varphi (x^{\prime })\rangle ^{(b)} &=&-\frac{r_{H}^{1-D}%
}{\pi nS_{D}}\sum_{l=0}^{\infty }(2l+n)C_{l}^{n/2}(\cos \theta
)\int_{0}^{\infty }d\omega \frac{\bar{K}_{\omega }^{(b)}(\lambda _{l}b)}{%
\bar{I}_{\omega }^{(b)}(\lambda _{l}b)}  \notag \\
&&\times I_{\omega }(\lambda _{l}\xi )I_{\omega }(\lambda _{l}\xi ^{\prime
})\cosh (\omega \Delta \tau )  \label{phi212RB}
\end{eqnarray}%
is induced in the region $\xi <b$ by the presence of the brane at $\xi =b$.
Note that the representation (\ref{Wigh3RB}) is valid under the condition $%
a^{2}e^{|\Delta \tau |}<\xi \xi ^{\prime }<b^{2}e^{|\Delta \tau |}$. As it
has been shown in \cite{Saha05RindBr}, the Wightman function for the
boundary-free geometry may be written in the form%
\begin{eqnarray}
W_{0}(x,x^{\prime }) &=&\tilde{W}_{0}(x,x^{\prime })-\frac{r_{H}^{1-D}}{\pi
^{2}nS_{D}}\sum_{l=0}^{\infty }(2l+n)C_{l}^{n/2}(\cos \theta )  \notag \\
&&\times \int_{0}^{\infty }d\omega e^{-\omega \pi }\cos (\omega \Delta \tau
)K_{i\omega }(\lambda _{l}\xi )K_{i\omega }(\lambda _{l}\xi ^{\prime }),
\label{GM1}
\end{eqnarray}%
where $\tilde{W}_{0}(x,x^{\prime })$ is the Wightman function for the bulk
geometry $R^{2}\times S^{D-1}$. Outside the horizon the divergences in the
coincidence limit of the expression on the right of (\ref{GM1}) are
contained in the first term. It can be seen that the Wightman function in
the region between the branes can also be presented in the form (\ref%
{Wigh3RB}) with $j=a$. In this representation $\langle \varphi (x)\varphi
(x^{\prime })\rangle ^{(a)}$ is induced in the region $\xi >a$ by the
presence of the brane at $\xi =a$ and the corresponding expression is
obtained from (\ref{phi212RB}) by the replacements $b\rightarrow a$, $%
I_{\omega }\rightleftarrows K_{\omega }$.

\subsection{Vacuum densities}

In the coincidence limit, from the formulae for the Wightman function one
obtains two equivalent forms for the VEV\ of the field square:%
\begin{eqnarray}
\langle 0|\varphi ^{2}|0\rangle &=&\langle 0_{0}|\varphi ^{2}|0_{0}\rangle
+\langle \varphi ^{2}\rangle ^{(j)}  \notag \\
&&-\frac{r_{H}^{1-D}}{\pi S_{D}}\sum_{l=0}^{\infty }D_{l}\int_{0}^{\infty
}d\omega \,\Omega _{j\omega }(\lambda _{l}a,\lambda _{l}b)G_{\omega
}^{(j)2}(\lambda _{l}j,\lambda _{l}\xi ),  \label{phi2sq1RB}
\end{eqnarray}%
corresponding to $j=a$ and $j=b$, and $|0_{0}\rangle $ is the amplitude for
the vacuum without boundaries. The coefficient $D_{l}$ in this formula is
defined by relation (\ref{Dlang}). The VEV $\langle 0_{0}|\varphi
^{2}|0_{0}\rangle $ is obtained from the corresponding Wightman function
given by (\ref{GM1}). For the points outside the horizon, the
renormalization procedure is needed for the first term on the right only,
which corresponds to the VEV in the geometry $R^{2}\times S^{D-1}$. This
procedure is realized in \cite{Saha05RindBr} on the base of the zeta
function technique.

In (\ref{phi2sq1RB}), the part $\langle \varphi ^{2}\rangle ^{(j)}$ is
induced by a single brane at $\xi =j$ when the second brane is absent. For
the geometry of a single brane at $\xi =b$, from (\ref{phi212RB}) one has
\begin{equation}
\langle \varphi ^{2}\rangle ^{(b)}=-\frac{r_{H}^{1-D}}{\pi S_{D}}%
\sum_{l=0}^{\infty }D_{l}\int_{0}^{\infty }d\omega \frac{\bar{K}_{\omega
}^{(b)}(\lambda _{l}b)}{\bar{I}_{\omega }^{(b)}(\lambda _{l}b)}I_{\omega
}^{2}(\lambda _{l}\xi ).  \label{phi21plb}
\end{equation}%
The expression for $\langle \varphi ^{2}\rangle ^{(a)}$ is obtained from (%
\ref{phi21plb}) by the replacements $b\rightarrow a$, $I_{\omega
}\rightleftarrows K_{\omega }$. The last term on the right of formula (\ref%
{phi2sq1}) is induced by the presence of the second brane. It is finite on
the brane at $\xi =j$ and diverges for points on the other brane. By taking
into account the relation $G_{\omega }^{(j)}(u,u)=B_{j}/j$, we see that for
the Dirichlet boundary condition this term vanishes on the brane $\xi =j$.

By using the formulae for the Wightman function and the VEV of the field
square, one obtains two equivalent forms for the VEV of the EMT,
corresponding to $j=a$ and $j=b$ (no summation over $i$):
\begin{eqnarray}
\langle 0|T_{i}^{k}|0\rangle &=&\langle 0_{0}|T_{i}^{k}|0_{0}\rangle
+\langle T_{i}^{k}\rangle ^{(j)}-\delta _{i}^{k}\frac{r_{H}^{1-D}}{\pi S_{D}}%
\sum_{l=0}^{\infty }D_{l}\lambda _{l}^{2}  \notag \\
&&\times \int_{0}^{\infty }d\omega \,\Omega _{j\omega }(\lambda
_{l}a,\lambda _{l}b)F^{(i)}[G_{\omega }^{(j)}(\lambda _{l}j,\lambda _{l}\xi
)].  \label{Tik1RB}
\end{eqnarray}%
The functions $F^{(i)}[g(z)]$ in this formula are defined by relations (\ref%
{fq}) and (\ref{Ffunc}) with $\lambda \rightarrow \lambda _{l}$ and $%
g(z)=G_{\omega }^{(j)}(\lambda _{l}j,z)$. In formula (\ref{Tik1RB}),
\begin{equation}
\langle 0_{0}|T_{i}^{k}|0_{0}\rangle =\delta _{i}^{k}\frac{r_{H}^{1-D}}{\pi
^{2}S_{D}}\sum_{l=0}^{\infty }D_{l}\lambda _{l}^{2}\int_{0}^{\infty }d\omega
\sinh \pi \omega \,f^{(i)}[K_{i\omega }(\lambda _{l}\xi )],  \label{DFR}
\end{equation}%
with $f^{(i)}[g(z)]$ defined by Eq. (\ref{fq}), is the corresponding VEV for
the vacuum without boundaries. The term $\langle T_{i}^{k}\rangle ^{(j)}$ is
induced by the presence of a single spherical brane located at $\xi =j$. For
the brane at $\xi =b$ and in the region $\xi <b$ one has (no summation over $%
i$)
\begin{equation}
\langle T_{i}^{k}\rangle ^{(b)}=-\delta _{i}^{k}\frac{r_{H}^{1-D}}{\pi S_{D}}%
\sum_{l=0}^{\infty }D_{l}\lambda _{l}^{2}\int_{0}^{\infty }d\omega \frac{%
\bar{K}_{\omega }^{(b)}(\lambda _{l}b)}{\bar{I}_{\omega }^{(b)}(\lambda
_{l}b)}F^{(i)}[I_{\omega }(\lambda _{l}\xi )].  \label{D1plateboundb}
\end{equation}%
For the geometry of a single brane at $\xi =a$, the corresponding expression
in the region $\xi >a$ is obtained from (\ref{D1plateboundb}) by the
replacements $b\rightarrow a$, $I_{\omega }\rightleftarrows K_{\omega }$. It
can be easily seen that for a conformally coupled massless scalar the
boundary induced part in the EMT is traceless.

Vacuum forces acting on the branes are presented in the form of the sum of
self-action and interaction terms. The vacuum effective pressures
corresponding to the interaction forces are obtained from the third summand
on the right of Eq. (\ref{Tik1RB}), taking $i=k=1$, $\xi =j$ \cite%
{Saha07RindBr}:
\begin{equation}
p_{\mathrm{(int)}}^{(j)}=\frac{\tilde{A}_{j}^{2}}{2j^{2}}\frac{r_{H}^{1-D}}{%
\pi S_{D}}\sum_{l=0}^{\infty }D_{l}\int_{0}^{\infty }d\omega \left[ \left(
\lambda _{l}^{2}j^{2}+\omega ^{2}\right) \beta _{j}^{2}+4\zeta \beta _{j}-1%
\right] \,\Omega _{j\omega }(\lambda _{l}a,\lambda _{l}b),  \label{pint2RB}
\end{equation}%
with $\beta _{j}=\tilde{B}_{j}/j\tilde{A}_{j}$. The interaction force acts
on the surface $\xi =a+0$ for the brane at $\xi =a$ and on the surface $\xi
=b-0$ for the brane at $\xi =b$. In dependence of the values for the
coefficients in the boundary conditions, effective pressures (\ref{pint2})
can be either positive or negative, leading to repulsive or attractive
forces, respectively. For Dirichlet or Neumann boundary conditions on both
branes the interaction forces are always attractive. For Dirichlet boundary
condition on one brane and Neumann boundary condition on the other one has $%
p_{\mathrm{(int)}}^{(j)}>0$ and the interaction forces are
repulsive for all distances between the branes. The investigation
of the vacuum densities in
various asymptotic regions of the parameters can be forund in \cite%
{Saha07RindBr,Saha05RindBr}.

\section{Radiation from a charge moving along a helical orbit inside a
dielectric cylinder}

\label{sec:RadDiel}

The radiation from a charged particle moving along a helical orbit in vacuum
has been widely discussed in literature. This type of electron motion is
used in helical undulators for generating electromagnetic radiation in a
narrow spectral interval at frequencies ranging from radio or millimeter
waves to X-rays (see, for instance, \cite{Hofm04}). The unique
characteristics, such as high intensity and high collimation, have resulted
in extensive applications of this radiation in a wide variety of experiments
and in many disciplines. In this section we apply the GAPF for the
investigation of the radiation on the lowest azimuthal mode by a charged
particle moving along a helical orbit inside a dielectric cylinder \cite%
{Saha07Helic}.

Consider a dielectric cylinder of radius $\rho _{1}$ and dielectric
permittivity $\varepsilon _{0}$ and a point charge $q$ moving along the
helical trajectory of radius $\rho _{0}<\rho _{1}$. We assume that the
system is immersed in a homogeneous medium with permittivity $\varepsilon
_{1}$. The velocities of the charge along the axis of the cylinder and in
the perpendicular plane we will denote by $v_{\parallel }$ and $v_{\perp }$,
respectively. In a properly chosen cylindrical coordinate system ($\rho
,\phi ,z$) with the $z$-axis along the cylinder axis, the vector potential
of the electromagnetic field is presented in the form of the Fourier
expansion
\begin{equation}
A_{l}(\mathbf{r},t)=\sum_{m=-\infty }^{\infty }e^{im(\phi -\omega
_{0}t)}\int_{-\infty }^{\infty }dk_{z}e^{ik_{z}(z-v_{\parallel
}t)}A_{ml}(k_{z},\rho ),  \label{vecpot2}
\end{equation}%
where $\omega _{0}=v_{\perp }/\rho _{0}$ is the angular velocity of the
charge. The radiation field inside the dielectric cylinder is presented as a
sum over the eigenmodes of the cylinder. Unlike to the case of the waveguide
with perfectly conducting walls, here the eigenmodes with $m\neq 0$\ are not
decomposed into independent transverse electric (TE or M-type) and
transverse magnetic (TM or E-type) parts. This decomposition takes place
only for the $m=0$ mode. The radiation on this mode is present under the
condition
\begin{equation}
\beta _{1\parallel }<1<\beta _{0\parallel },\;\beta _{i\parallel
}=v_{\parallel }\sqrt{\varepsilon _{i}}/c,\;i=1,2.  \label{betzug}
\end{equation}%
With these conditions the equation for the corresponding TM eigenmodes for $%
k_{z}$ has the form
\begin{equation}
\varepsilon _{0}|\lambda _{1}|\frac{J_{0}^{\prime }(\lambda _{0}\rho _{1})}{%
J_{0}(\lambda _{0}\rho _{1})}+\varepsilon _{1}\lambda _{0}\frac{%
K_{0}^{\prime }(|\lambda _{1}|\rho _{1})}{K_{0}(|\lambda _{1}|\rho _{1})}%
=0,\;\lambda _{i}=k_{z}\sqrt{\beta _{i\parallel }^{2}-1}.  \label{eigmodesm0}
\end{equation}%
The corresponding eigenvalues for $k_{z}$ we will denote by $\pm k_{s}^{%
\mathrm{(E)}}$, $s=1,2,\ldots $. The eigenmodes of the M-type are solutions
of the equation
\begin{equation}
|\lambda _{1}|\frac{J_{0}^{\prime }(\lambda _{0}\rho _{1})}{J_{0}(\lambda
_{0}\rho _{1})}+\lambda _{0}\frac{K_{0}^{\prime }(|\lambda _{1}|\rho _{1})}{%
K_{0}(|\lambda _{1}|\rho _{1})}=0,  \label{eigmodesm0M}
\end{equation}%
and we will denote them by $k_{z}=\pm k_{s}^{\mathrm{(M)}}$, $s=1,2,\ldots $.

For the radiation intensities on the mode $m=0$ inside the dielectric
waveguide we have
\begin{equation}
I_{m=0}=I_{m=0}^{\mathrm{(E)}}+I_{m=0}^{\mathrm{(M)}},  \label{Im=0}
\end{equation}%
where the contributions of the TM and TE modes are given by the formulae
\cite{Saha07Helic}
\begin{eqnarray}
I_{m=0}^{\mathrm{(E)}} &=&\frac{2q^{2}v_{\parallel }}{\rho _{1}^{2}}\sum_{s}%
\frac{\beta _{0\parallel }^{2}-1}{\varepsilon _{0}-\varepsilon _{1}}\frac{%
J_{0}^{2}(\rho _{0}\lambda _{s}^{\mathrm{(E)}})}{(\varepsilon
_{0}/\varepsilon _{1})J_{0}^{\prime 2}(\rho _{1}\lambda _{s}^{\mathrm{(E)}%
})+(\beta _{0\parallel }^{2}-1)J_{0}^{2}(\rho _{1}\lambda _{s}^{\mathrm{(E)}%
})},  \label{IE} \\
I_{m=0}^{\mathrm{(M)}} &=&\frac{2q^{2}v_{\perp }^{2}}{\rho
_{1}^{2}v_{\parallel }}\sum_{s}\frac{J_{1}^{2}(\rho _{0}\lambda _{s}^{%
\mathrm{(M)}})}{(\varepsilon _{0}-\varepsilon _{1})J_{0}^{\prime 2}(\rho
_{1}\lambda _{s}^{\mathrm{(M)}})}.  \label{IM}
\end{eqnarray}%
Here we have introduced the notation
\begin{equation}
\lambda _{s}^{\mathrm{(F)}}=k_{s}^{\mathrm{(F)}}\sqrt{\beta _{0\parallel
}^{2}-1},\;\mathrm{F=E,M}.  \label{lamb0F}
\end{equation}%
In formulae (\ref{IE}) and (\ref{IM}) the upper limit for the summation over
$s$, which we will denote by $s_{m}$, is determined by the dispersion law of
the dielectric permittivity $\varepsilon _{0}=\varepsilon _{0}(\omega )$
through the condition $\varepsilon _{0}(\omega _{s}^{\mathrm{(F)}%
})>c^{2}/v_{\parallel }^{2}$. The summands with a given $s$ describe the
radiation field with the frequency $\omega _{s}^{\mathrm{(F)}}=v_{\parallel
}k_{s}^{\mathrm{(F)}}$. As the modes $k_{s}^{\mathrm{(F)}}$ are not
explicitly known as functions on $s$, formulae (\ref{IE}), (\ref{IM}) are
not convenient for the evaluation of the corresponding radiation
intensities. More convenient form may be obtained by making use of the GAPF.

Let us derive a summation formula for the series over zeros of the function
\begin{equation}
C_{\alpha }(\eta ,z)=V_{\alpha }\{J_{0}(z),K_{0}(\eta z)\},  \label{Cdiel}
\end{equation}%
where and in what follows for given functions $F(z)$ and $G(z)$ we use the
notation%
\begin{equation}
V\{F(z),G(\eta z)\}=F(z)G^{\prime }(\eta z)+\alpha \eta F^{\prime }(z)G(\eta
z),  \label{Vfgnotdiel}
\end{equation}%
with $\alpha \geqslant 1$ and $\eta $ being real constants. In the GAPF we
substitute%
\begin{equation}
g(z)=if(z)\frac{V_{\alpha }\left\{ Y_{0}(z),K_{0}(\eta z)\right\} }{%
C_{\alpha }(\eta ,z)}.  \label{gdiel}
\end{equation}%
For the combinations of the functions entering in the GAPF one has%
\begin{equation}
f(z)-(-1)^{k}g(z)=f(z)\frac{V_{\alpha }\{H_{0}^{(k)}(z),K_{0}(\eta z)\}}{%
C_{\alpha }(\eta ,z)}.  \label{fmingdiel}
\end{equation}%
Let us denote positive zeros of the function $C_{\alpha }(\eta ,z)$ by $%
k_{s} $, $s=1,2,\ldots $, assuming that these zeros are arranged in the
ascending order. Note that, for $z\gg 1$ we have $C_{\alpha }(\eta
,z)\approx K_{0}^{\prime }(\eta z)-\alpha \eta zK_{0}(\eta z)/2<0$ and,
hence, $k_{s}\gtrsim 1$. By using the asymptotic formulae for the cylinder
functions for large values of the argument, it can be seen that for large
values $s$ one has $k_{s}\approx -\arctan (1/\alpha \eta )+\pi /4+\pi s$.
For the derivative of the function $C_{\alpha }(\eta ,z)$ at the zeros $%
k_{s} $\ one obtains%
\begin{equation}
C_{\alpha }^{\prime }(\eta ,z)=-\eta \frac{K_{0}(\eta z)}{J_{0}(z)}\left[
\alpha (1+\alpha \eta ^{2})J_{0}^{\prime 2}(z)+(\alpha -1)J_{0}^{2}(z)\right]
,\;z=k_{s}.  \label{Cderdiel}
\end{equation}%
In particular, it follows from here that the zeros are simple. Assuming that
the function $f(z)$ is analytic in the right half-plane, for the residue
term in the GAPF one finds%
\begin{equation}
\underset{z=k_{s}}{\mathrm{Res}}g(z)=-\frac{i}{\pi }P_{\alpha
}(k_{s})f(k_{s}),  \label{Resgdiel}
\end{equation}%
where we have introduced the notation%
\begin{equation}
P_{\alpha }(z)=\frac{2\alpha /z}{\alpha (1+\alpha \eta ^{2})J_{0}^{\prime
2}(z)+(\alpha -1)J_{0}^{2}(z)}.  \label{Talf}
\end{equation}%
Substituting the expressions for the separate terms into the GAPF, we obtain
the following result \cite{Saha07Helic}%
\begin{eqnarray}
\lim_{x_{0}\rightarrow \infty }\left[ \sum_{s=1}^{s_{0}}P_{\alpha
}(k_{s})f(k_{s})-\int_{0}^{x_{0}}dx\,f(x)\right] &=&-\frac{1}{\pi }%
\int_{0}^{\infty }dx\,\left[ f(ix)\frac{V_{\alpha
}\{K_{0}(x),H_{0}^{(2)}(\eta x)\}}{V_{\alpha }\{I_{0}(x),H_{0}^{(2)}(\eta
x)\}}\right.  \notag \\
&&\left. +f(-ix)\frac{V_{\alpha }\{K_{0}(x),H_{0}^{(1)}(\eta x)\}}{V_{\alpha
}\{I_{0}(x),H_{0}^{(1)}(\eta x)\}}\right] ,  \label{sumformdiel}
\end{eqnarray}%
where $s_{0}$ is defined by the relation $k_{s_{0}}<x_{0}<k_{s_{0}+1}$. This
formula is valid for functions $f(z)$ obeying the condition
\begin{equation}
|f(z)|<\epsilon (x)e^{c|y|},\;z=x+iy,\;|z|\rightarrow \infty ,
\label{condfz}
\end{equation}%
where $c<2$ and $\epsilon (x)\rightarrow 0$ for $x\rightarrow \infty $.
Formula (\ref{sumformdiel}) is further simplified for functions satisfying
the additional condition $f(-ix)=-f(ix)$:%
\begin{equation}
\lim_{x_{0}\rightarrow \infty }\left[ \sum_{s=1}^{s_{0}}P_{\alpha
}(k_{s})f(k_{s})-\int_{0}^{x_{0}}dx\,f(x)\right] =-\frac{4i\alpha }{\pi ^{2}}%
\int_{0}^{\infty }dx\,\frac{f(ix)}{x^{2}g_{\alpha }(\eta ,x)},
\label{sumformdiel1}
\end{equation}%
where%
\begin{eqnarray}
g_{\alpha }(\eta ,x) &=&I_{0}^{2}(x)\left[ J_{1}^{2}(\eta x)+Y_{1}^{2}(\eta
x)\right] +\alpha ^{2}\eta ^{2}I_{1}^{2}(x)\left[ J_{0}^{2}(\eta
x)+Y_{0}^{2}(\eta x)\right]  \notag \\
&&-2\alpha \eta I_{0}(x)I_{1}(x)\left[ J_{0}(\eta x)J_{1}(\eta x)+Y_{0}(\eta
x)Y_{1}(\eta x)\right] .  \label{denomdiel}
\end{eqnarray}%
Note that we have denoted $g_{\alpha }(\eta
,x)=|V\{I_{0}(z),H_{0}^{(1)}(\eta z)\}|^{2}$ and this function is always
non-negative.

Now, summation formulae for the series over $m=0$ TM and TE\ modes are
obtained taking
\begin{equation}
\eta =\sqrt{\frac{1-\beta _{1\parallel }^{2}}{\beta _{0\parallel }^{2}-1}}.
\label{eta}
\end{equation}%
In formula (\ref{sumformdiel}) we choose $\alpha =\varepsilon
_{0}/\varepsilon _{1}$, $f(z)=zJ_{0}^{2}(z\rho _{0}/\rho _{1})$ for the
waves of the E-type, and $\alpha =1$, $f(z)=zJ_{1}^{2}(z\rho _{0}/\rho _{1})$
for the waves of the M-type. For both types of modes one has $f(-ix)=-f(ix)$
and we can use the version of the summation formula given by (\ref%
{sumformdiel1}). In the intermediate step of the calculations, it is
technically simpler instead of considering the dispersion of the dielectric
permittivity to assume that in formulae (\ref{IE}), (\ref{IM}) a cutoff
function $\psi _{\mu }(\lambda _{s}^{\mathrm{(F)}})$ is introduced with $\mu
$ being the cutoff parameter and $\psi _{0}=1$ (for example, $\psi _{\mu
}(x)=\exp (-\mu x)$), which will be removed after the summation. In this
way, after the application of the summation formula we find the following
results
\begin{eqnarray}
I_{m=0}^{\mathrm{(E)}} &=&q^{2}v_{\parallel }\left[ c^{2}\int_{0}^{\infty }dx%
\frac{x}{\varepsilon _{0}}\psi _{\mu }(x)J_{0}^{2}(\rho _{0}x)+\frac{4}{\pi
^{2}\rho _{1}^{2}}\int_{0}^{\infty }dx\,\frac{I_{0}^{2}(x\rho _{0}/\rho _{1})%
}{\varepsilon _{1}xg_{\varepsilon _{0}/\varepsilon _{1}}(\eta ,x)}\right] ,
\label{IE2} \\
I_{m=0}^{\mathrm{(M)}} &=&\frac{q^{2}v_{\perp }^{2}v_{\parallel }}{c^{2}}%
\left[ \int_{0}^{\infty }dx\frac{x\psi _{\mu }(x)}{\beta _{0\parallel }^{2}-1%
}J_{1}^{2}(\rho _{0}x)-\frac{4}{\pi ^{2}\rho _{1}^{2}}\int_{0}^{\infty }dx\,%
\frac{I_{1}^{2}(x\rho _{0}/\rho _{1})}{(\beta _{0\parallel
}^{2}-1)xg_{1}(\eta ,x)}\right] ,  \label{IM2}
\end{eqnarray}%
where the function $g_{\alpha }(\eta ,x)$ is defined by formula (\ref%
{denomdiel}). In the first terms in the square brackets replacing the
integration variable by the frequency $\omega =v_{\parallel }x/\sqrt{\beta
_{0\parallel }^{2}-1}$ and introducing the physical cutoff \ through the
condition $\beta _{0\parallel }>1$ instead of the cutoff function, we see
that these terms coincide with the radiation intensities on the harmonic $%
m=0 $ for the waves of the E- and M-type in the homogeneous medium with
dielectric permittivity $\varepsilon _{0}$. The second terms in the square
brackets are induced by the inhomogeneity of the medium in the problem under
consideration. Note that, unlike to the terms corresponding to a homogeneous
medium, for $\rho _{0}<\rho _{1}$ the terms induced by the inhomogeneity are
finite also in the case when the dispersion is absent: for large values of $%
x $ the integrands decay as $\exp [-2x(1-\rho _{0}/\rho _{1})]$ (for this
reason we have removed the cutoff function from these terms). In particular,
from the last observation it follows that under the condition $(1-\rho
_{0}/\rho _{1})\gg v_{\parallel }/\omega _{d}$, where $\omega _{d}$ is the
characteristic frequency for the dispersion of the dielectric permittivity,
the influence of the dispersion on the inhomogeneity induced terms can be
neglected. Note that in a homogeneous medium the corresponding radiation
propagates under the Cherenkov angle $\theta _{C}=\arccos (1/\beta
_{0\parallel })$ and has a continuous spectrum, whereas the radiation
described by (\ref{IE2}), (\ref{IM2}) propagates inside the dielectric
cylinder and has a discrete spectrum with frequencies $\omega _{s}^{\mathrm{%
(F)}}$. As the function $g_{\alpha }(\eta ,x)$ is always non-negative, from
formulae (\ref{IE2}),(\ref{IM2}) we conclude that the presence of the
cylinder amplifies the $m=0$ part of the radiation for the waves of the
E-type and suppresses the the radiation for the waves of the M-type. In the
helical undulators one has $v_{\perp }\ll v_{\parallel }$ and the
contribution of the TM waves dominates.

\section{Summary}

\label{sec:Conc}

The Abel-Plana summation formula is a powerful tool for the evaluation of
the difference between a sum and the corresponding integral. This formula
has found numerous physical applications including the Casimir effect for
various bulk and boundary geometries. However, the applications of the APF
in its standard form are restricted to the problems where the normal modes
are explicitly known. In the present paper we have considered a
generalization of the APF, proposed in \cite{Sah1}, which essentially
enlarges the application range and allows to include problems where the
eigenmodes are given implicitly as zeros of a given function. Well known
examples of this kind are the boundary-value problems with spherical and
cylindrical boundaries. The generalized version contains two meromorphic
functions $f(z)$ and $g(z)$ and is formulated in the form of Theorem 1. The
special choice $g(z)=-if(z)\cot \pi z$ gives the APF with additional residue
terms coming from the poles of the function $f(z)$. We have shown that
various generalizations of the APF previously discussed in literature are
obtained from the GAPF by specifying the function $g(z)$. Further we
consider applications of the GAPF to cylinder functions. These applications
include summation formulae for series over the zeros of these functions and
their combinations (Sections \ref{sec:ApplBes},\ref{sec:SumFormBess},\ref%
{sec:SumKi}) and formulae for integrals involving cylinder functions
(Sections \ref{sec:BessInt},\ref{sec:BessInt2},\ref{sec:BessInt4}). In the
second part of the paper, including Sections \ref{sec:Gen}--\ref{sec:RadDiel}%
, we outline the applications of the summation formulae over the zeros of
cylinder functions for the evaluation of the VEVs for local physical
observables in the Casimir effect and the radiation intensity from a charge
moving along a helical trajectory inside a dielectric cylinder.

In Section \ref{sec:ApplBes}, in the GAPF choosing the function $g(z)$ in
the form (\ref{gebessel}), we derive two types of summation formulae, (\ref%
{sumJ1}), (\ref{sumJ2}), for the series over the zeros $\lambda _{\nu ,k}$
of the function $AJ_{\nu }(z)+BzJ_{\nu }^{\prime }(z)$. This type of series
arises in a number of problems of mathematical physics with spherically and
cylindrically symmetric boundaries. As a special case they include
Fourier-Bessel and Dini series (see \cite{Watson}). Using formula (\ref%
{sumJ1}), the difference between the sum over zeros $\lambda _{\nu ,k}$ and
the corresponding integral is presented in terms of an integral involving
modified Bessel functions plus residue terms. For a large class of functions
$f(z)$ the last integral converges exponentially fast and, in particular, is
useful for numerical calculations. The APF is a special case of formula (\ref%
{sumJ1}) with $\nu =1/2$, $A=1$, $B=0$ and for an analytic function $f(z)$.
Choosing in (\ref{sumJ1}) $\nu =1/2$, $A=1$, $B=2$ we obtain the APF in the
form (\ref{apsf2half}) useful for fermionic field calculations. Note that
formula (\ref{sumJ1}) may also be used for some functions having poles and
branch points on the imaginary axis. The second type of summation formula, (%
\ref{sumJ2}), is valid for functions satisfying condition (\ref{caseb}) and
presents the difference between the sum over zeros $\lambda _{\nu ,k}$ and
the corresponding integral in terms of residues over poles of the function $%
f(z)$ in the right half-plane (including purely imaginary ones). This
formula may be used to summarize a large class of series of this type in
finite terms. In particular, the examples we found in literature, when the
corresponding sum may be presented in a closed form, are special cases of
this formula. A number of new series summable by this formula and some
classes of functions to which it can be applied is presented.

As a next application of the GAPF, in Section \ref{sec:SumFormBess} we
consider series over zeros $z=\gamma _{\nu ,k}$ of the function $\bar{J}%
_{\nu }^{(a)}(z)\bar{Y}_{\nu }^{(b)}(\eta z)-\bar{Y}_{\nu }^{(a)}(z)\bar{J}%
_{\nu }^{(b)}(\eta z)$, where the barred notations are defined by relation (%
\ref{barjnot}). The corresponding results are formulated in the form of
Corollary 2 and Corollary 3. Using formula (\ref{cor3form}), the difference
between the sum over $\gamma _{\nu ,k}$ and the corresponding integral can
be expressed as an integral containing modified Bessel functions plus
residue terms. For a large class of functions $h(z)$ this integral converges
exponentially fast. The formula of the second type, (\ref{cor2form}), allows
to summarize a class of series over the zeros $\gamma _{\nu ,k}$ in closed
form. To evaluate the corresponding integral, the formula can be used
derived in section \ref{sec:BessInt4}. This yields to another summation
formula, (\ref{th4form}), containing residue terms only. The examples we
have found in literature when the corresponding sum was evaluated in closed
form are special cases of the formulae considered here. We present new
examples and some classes of functions satisfying the corresponding
conditions. In Section \ref{sec:SumKi}, taking in the GAPF the functions in
the form (\ref{fgK}), we derive summation formula (\ref{sumformKi2}) for
series over the zeros $z=\omega _{k}$ of the function $AK_{iz}(\eta )+B\eta
\partial _{\eta }K_{iz}(\eta )$. These zeros are eigenfrequencies for scalar
and electromagnetic fields in the geometry of a uniformly accelerated plane
boundary and the corresponding VEVs contain summation over them. For the
geometry of two plane boundaries the eigenfrequencies in the region between
the planes are zeros $z=\Omega _{k}$ of the function $\bar{K}_{iz}^{(a)}(u)%
\bar{I}_{iz}^{(b)}(v)-\bar{I}_{iz}^{(a)}(u)\bar{K}_{iz}^{(b)}(v)$. Summation
formula (\ref{Dsumformula}) for series over these zeros is obtained from (%
\ref{th12}) with the functions $f(z)$ and $g(z)$ given by (\ref{DfgtoAP}).

In Section \ref{sec:BessInt} we consider relations for the integrals of type%
\begin{equation*}
\int_{0}^{\infty }F(x)\bar{J}_{\nu }(x)dx,\;\int_{0}^{\infty }F(x)[J_{\nu
}(x)\cos \delta +Y_{\nu }(x)\sin \delta ]dx,
\end{equation*}%
which are obtained from the GAPF. The corresponding formulae have the form (%
\ref{intJform41}), (\ref{intJform42}) and (\ref{intJYform43}). In
particular, formula (\ref{intJform41}) is useful to express the integrals
containing Bessel functions with oscillating integrand through the integrals
of modified Bessel functions with exponentially fast convergence. Formulae (%
\ref{intJform42}) and (\ref{intJYform43}) allow to evaluate a large class of
integrals involving the Bessel function in closed form. In particular, the
results obtained in \cite{Schwartz} are special cases of these formulae.
Illustrating examples of applications of the formulae for integrals are
given in Section \ref{sec:BessInt2} (see (\ref{examp51})-(\ref{examp54ad})
and (\ref{examp56})-(\ref{examp59ad})). Looking at the standard books (see,
e.g., \cite{Erde53}, \cite{Watson}-\cite{Oberhettinger}) one will find many
particular cases which follow from these formulae. A number of new integrals
can be evaluated as well. We also consider two examples of functions having
purely imaginary poles, (\ref{examp510}) and (\ref{exampforpole}), with
corresponding formulae (\ref{intJ5sum}) and (\ref{exampforpoleform}) (for
two special cases of these formulae see \cite{Watson}). By choosing the
functions $f(z)$ and $g(z)$ in accordance with (\ref{fg61}), formulae (\ref%
{intJYth65}) and (\ref{th6form}) for the integrals of the type
\begin{equation*}
\int_{0}^{\infty }\frac{J_{\nu }(x)Y_{\mu }(\lambda x)-J_{\mu }(\lambda
x)Y_{\nu }(x)}{J_{\nu }^{2}(x)+Y_{\nu }^{2}(x)}F(x)dx
\end{equation*}%
can be derived from the GAPF. The corresponding results are formulated in
the form of Theorem 5 and Theorem 6 in Section \ref{sec:BessInt4}. Several
examples for integrals of this type we have been able to find in literature
are particular cases of formula (\ref{intJYth65}). New examples when the
integral is evaluated in finite terms are presented. Some classes of
functions are distinguished to which the corresponding formulae may be
applied.

In the second part of the paper, started from Section \ref{sec:Gen},
physical applications of the summation formulae for series over zeros of
cylinder functions obtained from the GAPF are reviewed. The general
procedure for the evaluation of the VEVs of the field square and the EMT on
manifolds with boundaries based on the point-splitting technique is
described. This VEVs are obtained from the corresponding positive frequency
Wightman function in the coincidence limit by using formulae (\ref{mphi2VEV}%
), (\ref{mTikVEV}). In these formulae instead of the Wightman function any
other two-point function can be chosen. Our choice of the Wightman function
is related to that it also determines the response of the Unruh-DeWitt type
particle detectors. Mode-sum (\ref{mmodesumWF}) contains a summation over
the eigenmodes of the problem under consideration. For curved boundaries
these eigenmodes are given implicitly as zeros of the eigenfunctions for the
corresponding boundary value problem or their combinations. The second
difficulty for the direct evaluation of the mode-sum is related to that it
diverges in the coincidence limit and exhibits very slow convergence
characteristics when the arguments of the Wightman function are close to
each other. In addition, the separate terms in the mode-sum are strongly
oscillatory for higher modes. The application of summation formulae obtained
from the GAPF enables to present the mode-sum in terms of integrals and,
hence, in the corresponding procedure the explicit form of the eigenmodes is
not necessary. These formulae explicitly extract from the VEVs of local
physical observables the parts corresponding to the bulk without boundaries
and the boundary-induced parts are presented in terms of integrals which are
exponentially convergent for the points away from the boundaries. As a
result, in the coincidence limit the renormalization is necessary for the
boundary-free parts only and this procedure is the same as that in quantum
field theory without boundaries. Note that, by using summation formulae
obtained from the GAPF, the mode-sums for more general class of objects such
as the heat and cylinder kernels can be evaluated as well.

First, in Section \ref{sec:Topol} we consider examples where the eigenmodes
are explicitly known and the application of the APF in standard form enables
to obtain the renormalized VEVs. These examples include the problems for the
evaluation of the VEVs in topologically non-trivial space $R^{D}\times S^{1}$
and for the geometry of two parallel plates with Dirichlet and Neumann
boundary conditions on them. On these simple examples we have seen two
important advantages of the application of the APF. First, this formula
enables to extract explicitly from the VEVs the Minkowskian part and second,
to present the parts induced by the non-trivial topology/boundaries in terms
of rapidly convergent integrals. Already in the case of two parallel plates
with Robin boundary conditions the eigenmodes are not explicitly known and
the procedure for the evaluation of the VEVs based on the APF needs a
generalization. The corresponding problem is discussed in Section \ref%
{sec:Robplates}. In this problem the eigenmodes for the projection of the
wave vector perpendicular to the plates are solutions of transcendental
equation (\ref{eigvaluesR}). For the summation of the mode-sums over these
eigenmodes we have derived formula (\ref{sumformula2plR}) by making use of
the GAPF. The application of this summation formula enables to extract from
the Wightman function the part corresponding to the geometry of a single
plate (see formula (\ref{WF2pl1R})). Similar decomposed formulae are
obtained for the VEVs of the field square and the EMT.

In Section \ref{sec:GlobMonSc} we consider the scalar Casimir densities
inside a spherical boundary and in the region between two concentric
spherical boundaries on background of global monopole spacetime described by
line element (\ref{mmetric}). It is assumed that on the bounding surfaces
the scalar field obeys Robin boundary conditions. For the region inside a
single sphere the eigenmodes are zeros of the function $\bar{J}_{\nu
_{l}}(x) $, where the order of the Bessel function is defined by formula (%
\ref{nuel}) and the coefficients in the barred notation are related to the
coefficients in the boundary condition by formula (\ref{eigenmodes}). For
the evaluation of the corresponding mode-sum we have used summation formula (%
\ref{sumJ1}). The term with the integral on the left of this formula
corresponds to the Wightman function for the global monopole bulk without
boundaries and the term with the integral on the right is induced by the
spherical boundary. In the region between two spheres the eigenmodes are
zeroes of the function $C_{\nu }^{ab}(\eta ,z)$ defined by formula (\ref%
{bescomb1}) and for the summation over these eigenmodes formula (\ref%
{cor3form}) is used. We have seen that the term with the integral on the
left of this formula corresponds to the Wightman function in the region
outside a single sphere and the term with the integral on the right is
induced by the presence of the outer sphere.

In section (\ref{sec:GlobMonFerm}) we have investigated the vacuum densities
for a fermionic field induced by spherical boundaries in a global monopole
background. It is assumed that the field obeys bag boundary condition. Both
regions inside a single sphere and between two spheres are considered. For
the first case the eigenmodes are zeros of the function $\tilde{J}_{\nu
_{\sigma }}(x)$ with the tilted notation defined by (\ref{tildenotFerm}). By
special choice of the function $g(z)$ in the GAPF we have derived formula (%
\ref{sumJ1Ferm}) for the summation over these zeros. The application of this
formula to the mode-sum for the VEVs enables to extract the parts
corresponding to the boundary-free bulk without specifying the form of the
cutoff function. In the region between two spheres the eigenmodes are zeros
of the function $C_{\nu _{\sigma }}^{\mathrm{f}}(\eta ,x)$ defined by
formula (\ref{Cnufe2}). By special choice of the functions $f(z)$ and $g(z)$
in the GAPF given by (\ref{gefcombfe2}), in subsection \ref{subsec:vevemtfe2}%
\ we have derived summation formula (\ref{cor3formfe2}) for the series over
these zeros. This formula presents the VEV of the EMT in the form of the sum
of a single sphere and second sphere induced parts. Scalar and fermionic
Casimir densities for spherical boundaries in the Minkowski bulk are
obtained from the corresponding results for the global monopole geometry as
special cases with $\alpha =1$. In Section \ref{sec:ElSpheric}, by making
use of the GAPF we have investigated the local properties of the
electromagnetic vacuum inside a perfectly conducting spherical shell and in
the region between two spherical shells on background of Minkowski spacetime.

Vacuum polarization by cylindrical boundaries is considered in Sections \ref%
{sec:CosStCyl}, \ref{sec:CylElCos}, \ref{sec:TwoCyl}. We start with the case
of a scalar field on background of the cosmic string spacetime with line
element (\ref{ds21}) and with coaxial cylindrical boundary. On the boundary
the field obeys Robin boundary condition. In the region inside the
cylindrical shell the eigenmodes are zeros of the function $\bar{J}%
_{q|n|}(z) $, where the parameter $q$ is determined by the planar angle
deficit and the coefficients in the barred notation are defined by relation (%
\ref{fbar}). The application of the GAPF to the mode-sum of the Wightman
function extracts the part corresponding to the cosmic string geometry
without boundaries and the boundary induced part is presented in the form
which can be directly used for the evaluation of the VEVs for the field
square and the EMT. In Section \ref{sec:CylElCos} similar consideration is
done for the electromagnetic field with perfectly conducting boundary
conditions. The corresponding eigenmodes are the zeros of the function $%
J_{q|n|}(z)$ for the TM waves and the zeros of the function $%
J_{q|n|}^{\prime }(z)$ for the TE waves. The formulae for the VEVs induced
by a cylindrical boundary in the Minkowski bulk are obtained from the
corresponding results for the cosmic string geometry in the special case $%
q=1 $ with zero planar angle deficit. Vacuum densities for both scalar and
electromagnetic fields in the region between two coaxial cylindrical shells
in background of Minkowski spacetime are investigated in Section \ref%
{sec:TwoCyl}.

The use of general coordinate transformations in quantum field theory in
flat spacetime leads to an infinite number of unitary inequivalent
representations of the commutation relations with different vacuum states.
In particular, the vacuum state for a uniformly accelerated observer, the
Fulling-Rindler vacuum, turns out to be inequivalent to that for an inertial
observer, the Minkowski vacuum. In Section \ref{sec:FuRi}, the polarization
of the Fulling-Rindler vacuum for a scalar field by a uniformly accelerated
plate with Robin boundary conditions is investigated by using the GAPF. The
corresponding eigenfrequencies are the zeros of the function $\bar{K}%
_{i\omega }(x)$, where the coefficients in the barred notation (\ref{Kizbar}%
) are related to the Robin coefficients by formula (\ref{R1AB}). On the base
of the summation formula derived in Section \ref{sec:SumKi} for the series
over these zeros, we have extracted from the Wightman function the part
corresponding to the Rindler wedge without boundaries. The part induced by
the plate is given by formula (\ref{Wb}) and is directly used for the
evaluation of the VEVs for the field square and the EMT. In the case of $D=3$
electromagnetic field with perfect conductor boundary condition on the
plate, the corresponding eigenfunctions for the vector potential are
resolved into the TE and TM scalar modes with Dirichlet and Neumann boundary
conditions respectively. The corresponding VEV of the EMT is given by
formula (\ref{FuRiEl1pl}). The Casimir densities in the region between two
infinite parallel plates moving by uniform proper accelerations are
discussed in Section \ref{sec:FulRin2pl} for both scalar and electromagnetic
fields. In the case of scalar field the eigenfrequencies are the zeros $%
\omega =\Omega _{n}$ of the function $Z_{i\omega }(u,v)$ defined by formula (%
\ref{Zi}) (see (\ref{Deigfreq})). After the application of summation formula
(\ref{Dsumformula}), the Wightman function is presented in the form (\ref%
{Wigh3}) or equivalently (\ref{Wigh31}). The first term on the right of
these formulae is the Wightman function for the case of a single plate and
the second term is induced by the presence of the second plate. Similar
decomposition is obtained for the VEVs of the field square and the EMT,
formulae (\ref{phi2sq1}) and (\ref{Tik1}). We have investigated the parts in
the VEVs induced by the presence of the second brane in various asymptotic
limits of the parameters. The vacuum interaction forces between the plates
are considered as well. The VEV of the EMT\ for the electromagnetic field in
the region between the plates is given by formula (\ref{FuRiEMTelec2pl}) or
equivalently by (\ref{FuRiEMTelec2pl1}), and the corresponding interaction
forces are given by formulae (\ref{pem1}), (\ref{pem2}).

Recent proposals of large extra dimensions use the concept of brane as a
sub-manifold embedded in a higher dimensional spacetime, on which the
Standard Model particles are confined. Braneworlds naturally appear in
string/M-theory context and provide a novel setting for discussing
phenomenological and cosmological issues related to extra dimensions. In
Section \ref{sec:Branes} we consider the geometry of two parallel flat
branes in the AdS bulk. In the region between the branes the radial
Kaluza-Klein masses $m=m_{n}$ are zeros of the function $\bar{J}_{\nu
}^{(a)}(mz_{a})\bar{Y}_{\nu }^{(b)}(mz_{b})-\bar{Y}_{\nu }^{(a)}(mz_{a})\bar{%
J}_{\nu }^{(b)}(mz_{b})$, where the coefficients in barred notation (\ref%
{barjnot}) are related to the Robin coefficients by formulae (\ref{notbar}).
The application of summation formula (\ref{cor3form}) allows to extract from
the VEVs the part due to a single brane. Both single brane and second brane
induced parts and vacuum interaction forces between the branes are
investigated. An application to the Randall-Sundrum braneworld with
arbitrary mass terms on the branes is discussed. Similar issues for two
spherical branes in Rindler-like spacetimes are considered in Section \ref%
{sec:Rindbrane}. The corresponding eigenfrequencies $\omega =\Omega _{n}$
are zeros of the function $\bar{K}_{i\omega }^{(a)}(\lambda _{l}a)\bar{I}%
_{i\omega }^{(b)}(\lambda _{l}b)-\bar{I}_{i\omega }^{(a)}(\lambda _{l}a)\bar{%
K}_{i\omega }^{(b)}(\lambda _{l}b)$, where $\ a$ and $b$ are the branes
radii and $\lambda _{l}$ is defined by (\ref{lambdal}). In Section \ref%
{sec:RadDiel} we apply the GAPF for the investigation of the radiation
intensity from a charge moving along a helical trajectory inside a
dielectric cylinder. Summation formula (\ref{sumformdiel}) is derived for
the series over the zeros of function (\ref{Cdiel}). This formula is used
for the summation of the series over the TE and TM eigenmodes of the
dielectric cylinder appearing in the expressions for the corresponding
radiation intensities on the lowest azimuthal harmonic.

Of course, the applications of the summation formulae obtained from GAPF are
not restricted by the Casimir effect only. Similar types of series will
arise in considerations of various physical phenomenon near the boundaries
with spherical and cylindrical symmetries, for example in calculations of
the electron self-energy and the electron anomalous magnetic moment (for
similar problems in the plane boundary case see, e.g., \cite{Fischbach} and
references therein). The dependence of these quantities on boundaries
originates from the modification of the photon propagator due to the
boundary conditions imposed by the walls of the cavity.

\section*{Acknowledgements}

I am indebted to Prof. E. Chubaryan and Prof. A. Mkrtchyan for general
encouragement and to my collaborators for stimulating discussions. I would
also like to acknowledge the hospitality of the Abdus Salam International
Centre for Theoretical Physics, Trieste, Italy. The work was supported in
part by the Armenian Ministry of Education and Science Grant No. 0124.

\end{document}